\documentclass{aastex701}
\usepackage[utf8]{inputenc}
\usepackage{natbib}

\shortauthors{Matthews}
\shorttitle{Radio Stars 3 Conference Summary}

\begin{document}
\newcommand{\ang}{\rm \AA}
\newcommand{\msun}{M$_\odot$}
\newcommand{\lsun}{L$_\odot$}
\newcommand{\days}{$d$}
\newcommand{\degree}{$^\circ$}
\newcommand{\ud}{{\rm d}}
\newcommand{\as}[2]{$#1''\,\hspace{-1.7mm}.\hspace{.0mm}#2$}
\newcommand{\am}[2]{$#1'\,\hspace{-1.7mm}.\hspace{.0mm}#2$}
\newcommand{\ad}[2]{$#1^{\circ}\,\hspace{-1.7mm}.\hspace{.0mm}#2$}
\newcommand{\lsim}{~\rlap{$<$}{\lower 1.0ex\hbox{$\sim$}}}
\newcommand{\gsim}{~\rlap{$>$}{\lower 1.0ex\hbox{$\sim$}}}
\newcommand{\HA}{H$\alpha$}
\newcommand{\HII}{\mbox{H\,{\sc ii}}}
\newcommand{\kms}{\mbox{km s$^{-1}$}}
\newcommand{\HI}{\mbox{H\,{\sc i}}}
\newcommand{\KI}{\mbox{K\,{\sc i}}}
\newcommand{\nan}{Nan\c{c}ay}
\newcommand{\jks}{Jy~km~s$^{-1}$}

\title{Radio Stars in the Era of New Observatories}

\author[orcid=0000-0002-3728-8082]{Lynn D. Matthews}
\affiliation{Massachusetts Institute of Technology Haystack Observatory, 99 Millstone Road, Westford, MA
  01886 USA}
\email[show]{lmatthew@mit.edu}

\begin{abstract}
An international
conference  {\it Radio Stars in the Era of New Observatories} was held at the Massachusetts Institute of Technology
Haystack Observatory on 2024 April 17--19. The conference brought together more than 60 researchers from around
the world, united by an interest in using radio wavelength observations to explore
the physical processes that operate in stars (including the Sun), how stars evolve and interact with
their environments, and the role of radio stars as probes of our Galaxy. 
Topics discussed at the meeting included radio emission from cool and ultracool dwarfs, extrasolar space weather,
stellar masers, thermal radio emission from evolved stars, circumstellar chemistry,
low frequency observations of the Sun, radio emission from hot stars, applications of
very long baseline interferometry techniques to stellar astrophysics, stellar explosive events, the detection of
radio stars in the latest generation of widefield sky surveys, the importance of radio stars for understanding
the structure and evolution of the Milky Way, and the anticipated applications for stellar astrophysics of
future radio observatories on the ground and in space.
This article summarizes research topics and results featured at the conference, along with some background
and contextual information.
It also
highlights key outstanding questions in stellar astrophysics 
where new insights
are anticipated from the next generation of observational facilities operating at meter through
submillimeter wavelengths.

\end{abstract}

\keywords{meeting summary, Stars --- stars: AGB and
post-AGB -- stars: winds, outflows -- circumstellar matter --
radio lines: stars}  

\section{Background and Motivation for the Workshop}
The detection and characterization of electromagnetic emission at
meter through submillimeter wavelengths (hereafter collectively referred to as ``radio'' emission)
have the ability to provide a broad range of
unique insights into the physical processes that govern the workings of
stars spanning virtually every type and evolutionary phase
(e.g., \citealt{1985ARA&A..23..169D}; \citealt{Hjellming1988}; \citealt{Linsky1996}; \citealt{2000riss.conf...86W};
\citealt{2002ARA&A..40..217G}; \citealt{2005EAS....15..187P}). Stellar radio emission can be either thermal or nonthermal in nature
and may arise from a range of different  mechanisms, including bremsstrahlung (free-free), 
gyromagnetic radiation, plasma emission, electron cyclotron maser (ECM)
instabilities, and spectral lines [including hyperfine atomic transitions, rotational
  transitions of molecules, and radio recombination lines
(RRLs)]. 
While these various types of 
radio emission typically comprise only a small fraction of a star's total luminosity, radio wavelength studies provide
unique diagnostics of a wide range of
phenomena in stars and stellar systems that cannot be probed by any other means.

As illustration of this, during the past decade the current generation of radio telescopes has been
pivotal in enabling a rapid pace of exciting discoveries in numerous branches of
stellar astrophysics, ranging from
the identification of new classes of hot, magnetic stars to unveiling
evidence of the ubiquitous role of low-mass companions to shaping evolved star winds,
to  the discover of radiation belts beyond the solar system and the unveiling of
new clues to the longstanding puzzle
of how the solar corona is heated. 
These discoveries are at the same time feeding a growing anticipation for the scientific potential
of the next-generation radio facilities currently being built or planned.
Motivated by this, a 3-day workshop ``Radio
Stars in the Era of New Observatories'' was convened at the Massachusetts
Institute of Technology (MIT) Haystack Observatory in Westford, Massachusetts from 2024 April 17--19
(hereafter ``Radio Stars 3'' or ``RS3''). As described in \cite{Matthews2013} and \cite{Matthews2019},
previous  Radio Stars conferences were hosted by Haystack Observatory in 2012  and 2017, respectively,
each showcasing the latest advances in solar and stellar radio
astrophysics.

RS3 sought to
bring together stellar astrophysicists (including observers, theorists, modelers,
software developers, and instrument builders) from a variety of
sub-disciplines to share the latest developments in the study of stars across
the Hertzsprung-Russell (H-R) diagram, with a focus on discoveries that exploit the unique
potential of the latest generation of radio instruments. Additional underlying goals were taking
a forward look at what observations with the next generation of radio instruments are likely to
enable for solar and stellar astrophysics and the identification of
areas that will require research and development in order
to maximally exploit new facilities for solar and stellar science.

\section{Meeting Overview}
The RS3 conference was attended by 61 registered participants 
representing 12 countries.  
The science program comprised 12 invited talks, 25 contributed oral presentations,
and 22 posters.  The complete program and
presentation abstracts can be viewed on the meeting 
web site\footnote{{\url{https://www.haystack.mit.edu/radio-stars-2024}}}.
Copies of many of the presentation slides and posters are also accessible through the meeting web page.

G. Umana (Instituto Nazionale Di Astrofisica, Italy)  opened the conference with an invited review that provided
a  framework for the conference and introduced  many of the topics and
themes that recurred throughout the meeting. Subsequent
oral and invited presentations were organized into
twelve additional sessions: (I) Circumstellar Chemistry; (II) Stellar Interactions and
Environments; (III) Evolved Star Atmospheres; (IV) Stellar Masers as  Tool for Galactic Science and
Stellar Astrophysics; (V) Radio Stars with the Highest Angular Resolution; (VI) The Sun as a Radio
Star; (VII) Space and Space Weather; (VIII) Ultracool Dwarfs; (IX) Radio Emission from Hot Stars; (X)
Radio Star Surveys; (XI) Stellar Explosions; (XII) Looking Back and Looking Ahead. In addition,
during several short oral sessions, poster authors delivered brief summaries of their
presentations. 

The current review is  intended to capture a snapshot of the state of stellar radio astronomy by
summarizing many of the research topics and highlights
presented at the meeting. I also attempt to 
identify some of the common themes and topical synergies that emerged.
In addition, I draw attention to some of the outstanding
questions in stellar astrophysics that were identified at the conference and
where key insights are anticipated from the next generation
of radio wavelength instruments. To maintain cohesiveness, the content is organized
by subject area rather than strictly by scientific session.
For the benefit of the more general reader, some brief contextual and
historical information is included throughout. 

\section{The Sun as a Radio Star\protect\label{sun-as-star}}
For decades, observations of radio emission of the Sun have been providing unique diagnostics of
the dynamic solar atmosphere,  the Sun's magnetic field, coronal mass ejections (CMEs), space weather,
and other phenomena (e.g., \citealt{1985ARA&A..23..169D}; \citealt{2023ARA&A..61..427G}). Multiple mechanisms are known to produce solar
radio emission emission, including thermal
bremsstrahlung, thermal gyroresonance, nonthermal gyrosynchrotron, (nonthermal) plasma emission, and
ECM emission. As illustrated in a series of RS3 presentations, 
the collective insights gleaned from
observations of each of these various types of emission
have increased markedly in the past few years
thanks to recent advances in instrumentation, data processing
techniques, and modeling.

Radio observations of the Sun at different frequencies probe  different layers of the
solar atmosphere and are
thus complementary in advancing our understanding of various solar phenomena.
As described below, discussion of the radio Sun at RS3  primarily focused on
GHz and MHz frequencies  (i.e., centimeter and meter
wavelengths).  Solar radio
emission in the GHz range samples the chromosphere, transition region, and lower corona, and arises from
  a combination of optically thin thermal bremsstrahlung from the corona, optically thick chromospheric emission \citep[e.g.,][]{1991ApJ...370..779Z}, together with
  possible gyrosynchrotron and gyroresonant emission from active
regions. Solar emission at MHz frequencies is primarily a
diagnostic of the solar corona and arises from thermal bremsstrahlung (from the ``quiet'' Sun), as well as
plasma emission \footnote{Plasma emission is produced when high-speed
electrons propagate into dense plasma, creating beam-driven instabilities that lead to the production of Langmuir
waves and electromagnetic transverse waves, whose frequency scales as the
square root of the plasma density.} associated with solar activity.

\subsection{Advances in Low-Frequency Solar Radio Imaging\protect\label{sunpix}}
Historically the Sun has been a notoriously difficult source to image using radio interferometers.
Among the myriad challenges are the need to sample a
wide variety of spatial scales (from a few arseconds, to a degree or more), coupled with the time-varying nature
of the solar atmosphere, which generally precludes the use of earth rotation synthesis to improve
$u$-$v$ sampling.
In addition, images of high dynamic range are needed to enable the simultaneous study of both weak and strong
emission components. The limited angular resolution of most radio images to date
has also hindered detailed comparisons with
data at other wavelengths [ultraviolet (UV), optical, X-ray], while accurate flux calibration has remained
challenging owing to the high signal strengths, which may introduce nonlinearities in the signal chain
(e.g., \citealt{Kansabanik2022a}a).
As a consequence of these factors, many solar radio studies historically focused on the
analysis of dynamic spectra rather than images,
with an emphasis on the study of bright bursts from the active Sun.   However, this situation has evolved dramatically over the
past decade \cite[e.g.,][]{2023ARA&A..61..427G}, and as showcased in the solar astronomy
session at RS3, the past few years have sustained significant improvements in solar radio imaging capabilities.

The session on the Sun opened with an
invited talk by D. Oberoi (National Centre for Radio Astrophysics/Tata Institute of Fundamental Research),
who provided an overview of recent progress in the domain of low-frequency ($<$300~MHz) solar physics, with a focus
on results from the Murchison Widefield Array (MWA). The MWA is an interferometric array
of 128 elements (``tiles''), each comprising 16
dipole  antennas, located in the Outback of
Western Australia. It operates in the frequency range 80--300~MHz \citep{2013JPhCS.440a2033T}.
As described by Oberoi, the MWA (and its 32-tile
predecessor) has been instrumental in demonstrating the power of low frequency radio observations for
revealing dynamic behaviors on the Sun
(in both time and frequency), even during periods where it appears ``quiet'' at visible, UV, or X-ray
wavelengths  (e.g., \citealt{2011ApJ...728L..27O}).

Given the wide bandwidths and high time and frequency resolution of MWA solar data, they are
extraordinarily rich in information content, but this also presents challenges.
For example, Oberoi noted that imaging each time and frequency slice
from just 5 minutes of an MWA solar observations obtained with 0.5~s time resolution
could in principle spawn more than half
a million images. Additionally, the range of brightness temperatures manifested by phenomena observable
in the MWA wavebands---ranging from thermal bremsstrahlung emission from the quiet Sun to plasma emission from active
phenomena---span
roughly ten orders of magnitude ($T_{\rm B}\sim10^{3}$~K to $10^{13}$~K). At the same time, the range of circular polarization
can range from  $\sim 0-100$\% (see \citealt{Kansabanik2022a}). Consequently, one needs high-fidelity, high
dynamic range, spectro-polarimetric snapshot images to fully exploit the scientific potential of these data sets.

An important advantage of the MWA is its excellent instantaneous $u$-$v$ coverage, which in turn enables
snapshot imaging of rapidly time-varying phenomena. A key breakthrough in leveraging this
has been the development of automated
imaging pipelines which are able to routinely achieve dynamic ranges of a few hundred, to as high as $\sim10^{5}$
\citep{2019ApJ...875...97M}. This marks an improvement of 1--2 orders of magnitude compared with
previous solar images at similar frequencies.
Recently, polarimetric imaging has also been incorporated into these pipelines
(\citealt{Kansabanik2022a}; \citealt{Kansabanik2022b}b, \citeyear{Kansabanik2023a}).

\begin{figure*}[ht!] 
\plotone{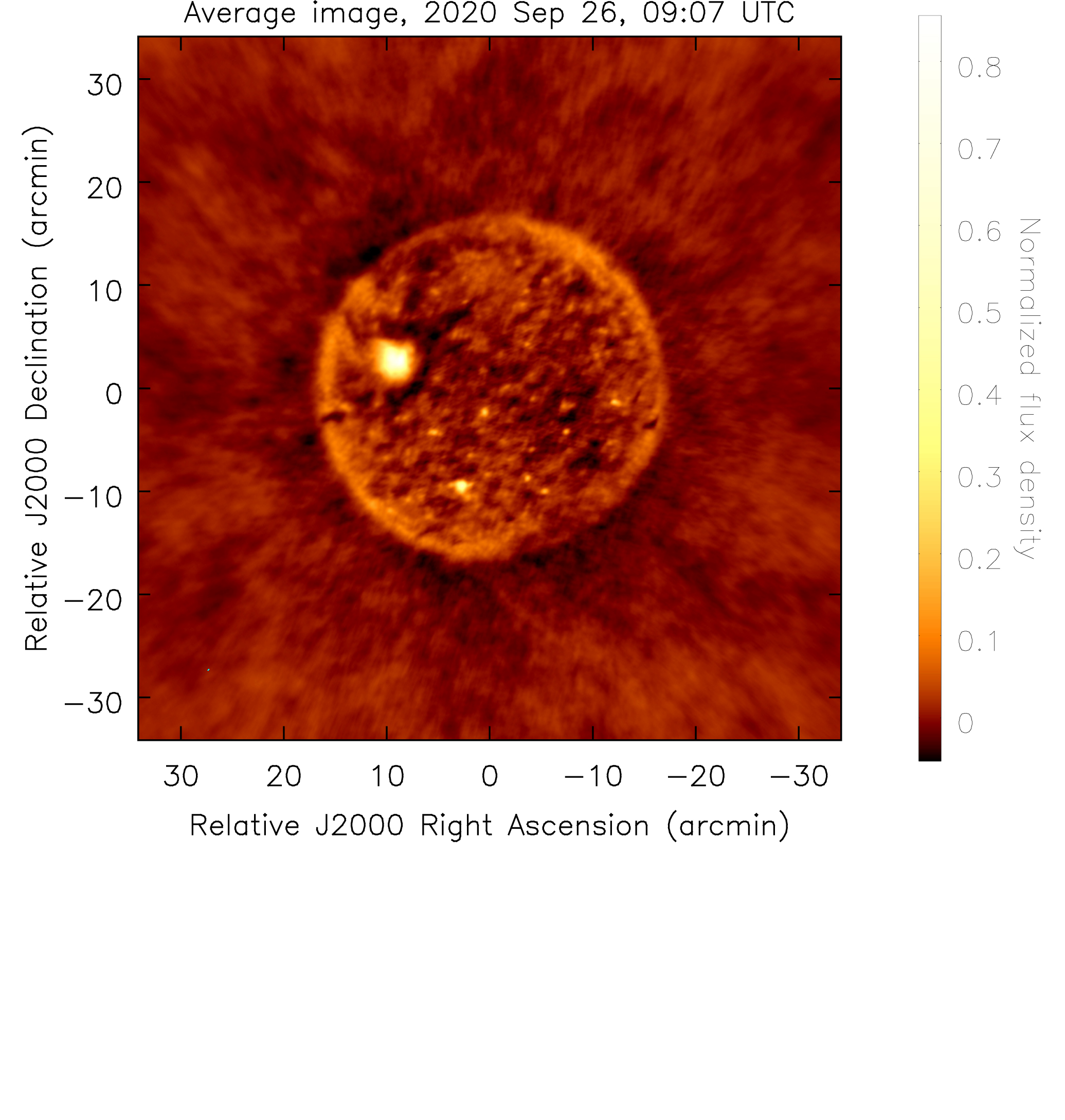}
\caption{Snapshot image of the Sun obtain with MeerKAT on 2020 September 26 at 09:07~UTC. The image represents a
  normalized average over a 880--1670~MHz  band. The angular resolution is $\sim8''$, represented by a
  tiny  dot in  the bottom left corner of the panel. From \cite{Kansabanik2024b}. 
\label{fig:kansabanik-meerkat}}
\end{figure*}

As showcased in the poster presentation of
D. Kansabanik (Johns Hopkins University/National Centre for Radio Astrophysics/Tata Institute
of Fundamental Research),
another general purpose radio array well-suited to solar imaging is MeerKAT, situated in the Karoo region of South Africa.
MeerKAT comprises 64 antennas, each 13.5~m in diameter, with baselines up to 
$\sim$8~km. Kansabanik's team has recently been exploring MeerKAT's potential to obtain high-fidelity
spectroscopic snapshot imaging of the solar corona across the frequency range 880--1670~MHz 
(\citealt{Kansabanik2024b}; Figure~\ref{fig:kansabanik-meerkat}). They have already performed MeerKAT commissioning
observation with
the Sun in the sidelobes of the primary beam (as a means to attenuate the strong solar signal) and are
investigating strategies to introduce
suitable attenuation in the signal chain to enable  accurately flux-calibrated observations for the case where the Sun
is within the primary beam.

\subsection{New Insights into the Coronal Heating Problem: WINQSEs\protect\label{coronalheating}}
Another advance highlighted by Oberoi was the progress in our ability to study weak nonthermal
emission from the Sun. He reported that recent studies have successfully exploited a continuous
wavelet transform-based method to identify and image features with
brightnesses of order a {\em milli-SFU} \footnote{A solar flux unit (SFU) is equal to $10^{4}$~Jy.}
Such sensitivity levels have been achieved only rarely in the past \citep[e.g.,][]{2000SoPh..191..341K}, with more
typical detection limits being $\sim$10~SFU in late 1990s \citep[e.g.,][]{1997ApJ...474L..65M} and $\sim$1~SFU  a decade ago
\citep[e.g.,][]{2013ApJ...762...89R}. 
These newly detected emissions have been given the moniker
``weak impulsive narrowband quiet Sun emissions'' (WINQSEs) and
are thought to be the radio counterparts of the nanoflares which have long been postulated
to explain the heating of the solar corona by providing a vehicle to extract the required energy from the
solar magnetic field  (\citealt{2021AGUFMSH15E2059M}, \citeyear{2022tess.conf40808M},
\citeyear{Mondal2023a}a; \citealt{2022ApJ...937...99S}; \citealt{2023ApJ...954...39B}). 

One advantageous features of WINQSEs is that the associated radio emission is coherent,
providing a much more readily observable  means to explore weaker energies (down to ``picoflare''
levels) than is
accessible with traditional nanoflare studies based on
X-ray or UV radiation. Typical
brightness temperatures of the WINQSEs
lie in the range $\sim10^{3}-10^{4}$~K and counterparts have been identified
in the extreme UV \citep{2021SoPh..296..131M}. However, Oberoi noted that WINQSEs seem to be ubiquitous even when the
Sun is deemed ``quiet'' based on other such traditional diagnostics. Because the WINQSEs  that occur at a given time
may number in the thousands, Oberoi and his collaborators have developed machine learning methods for
the detection and characterization of these events \citep{2023ApJ...954...39B}. Surprisingly, WINQSEs
appear spatially resolved in images, despite the a priori expectation that they should be
compact; this is likely a consequence of scattering (see \citealt{2023ApJ...954...39B}). 

\subsection{ECM Emission from the Sun\protect\label{solarECM}}
Among the different emission mechanisms known to contribute to solar radio emission,
ECM emission has remained the least well-studied.
A poster by S. White (Air Force Research Laboratory) presented results from a recent
study to assess the prevalence
of ECM emission from solar radio bursts at GHz frequencies (see also \citealt{2024ApJ...969....3W}).
This work was motivated
by the desire to better understand the impact of solar bursts and space weather on bands
where Global Navigation Satellite Systems (GNSS) operate. White's team used multi-frequency
data from over 3000 bursts obtained using the Nobeyama
Radio Polarimeters between 1998 and 2023. Since optically thick (and incoherent) synchrotron emission
has a spectrum that rises with increasing frequency below 5~GHz,
cases where the 1.0~GHz flux exceeded  the flux at 2.0 or 3.75~GHz were flagged as
incompatible with a synchrotron mechanism
and identified as likely arising from (coherent) ECM emission.
\citeauthor{2024ApJ...969....3W}  estimated that $\sim$75\% of solar bursts at 1.0~GHz and
$\sim$50\% of bursts at 2.0~GHz are dominated by coherent emission---a significantly
larger fraction than previously believed. For 23 flares,
brightness temperatures exceeded $10^{11}$~K, definitively ruling out an incoherent emission mechanism. 

The putative ECM emission is seen to be strongly variable,
sometimes lasting for hours after the initial impulsive burst
phase. Generally, it is highly circularly polarized (though rarely as high as 100\%).
White's team
also reported evidence that solar ECM emission may be dominated by ``spikes'' (i.e., forests of short-duration,
narrow frequency burst events; see e.g., \citealt{1985ARA&A..23..169D}), making them distinctly different from the ECM
phenomena  associated with solar system planets. To put this in context, the brightest of the bursts studied by
\cite{2024ApJ...969....3W} 
would produce mJy-level fluxes at a distance of 10~pc. This work raises the
question of whether the ECM emission seen associated
with brown dwarfs and magnetically active OB stars (Sections~\ref{UCDs}, \ref{mag-OB}) is in fact more analogous
to the ECM emission seen in solar spike bursts as opposed to that in solar system planets.

\subsection{Solar Coronal Mass Ejections (CMEs) and Space Weather\protect\label{solarCMEs}}
Solar CMEs are explosive events that result in the ejection of
large quantities of magnetized plasma. They are also the major drivers
of space weather in the solar system. In some cases the ejected plasma 
reaches the Earth's atmosphere and can have a significant effect on the near-Earth environment.
The study of CMEs at radio wavelengths is uniquely powerful
since radio observations provide the only means available to constrain the magnetic fields of CME plasma via
remote-sensing techniques. Such measurements of the magnetic field are critical for
understanding the propagation and evolution of these events and for forecasting
their geoeffectiveness.

Traditionally, the only means of characterizing the
magnetic field of the CME plasma has been through in situ measurements using satellites. However, this is
limited to the near-Earth environment where satellites are located, leaving
very little advance warning for forecasting possible impacts of these events on Earth. As discussed at the meeting, this
has motivated the development
of methods to measure CMEs much closer to the Sun using radio techniques, with the ultimate goal of tracking the
magnetic field energy all the way from the low corona into interplanetary space. 

\subsubsection{Plasma Emission from CME-Driven Shocks\protect\label{CMEplasma}}
Some solar CMEs travel at speeds approaching (or even exceeding) the Alfv\'en speed, driving
shocks that accelerate electrons to nonthermal energies,
and giving rise to outbursts of
plasma emission at both the fundamental frequency and the first harmonic (so-called type~II radio
bursts; e.g., \citealt{1950AuSRA...3..399W}; \citealt{2020ApJ...897L..15M}). 
Each harmonic is  sometimes seen to be
split into two bands (e.g., \citealt{2020ApJ...897L..15M}),
although the explanation for this has remained controversial. Two
competing models attribute this to either
independent emission sites \citep[e.g.,][]{1967PASA....1...47M} or to emission arising from
slightly different plasma densities in the upstream and downstream regions of the shock front, respectively
\citep[e.g.,][]{2018ApJ...868...79C}.  As described by D. Oberoi,
recent MWA imaging results \citep{2023A&A...670A.169B} have provided
rare but compelling 
evidence that at least in some cases
the band splitting is caused by emission from independent emission sites
in the shock. 

\subsubsection{Gyromagnetic Emission from CMEs\protect\label{CMEgyro}}
In addition to plasma emission associated with accompanying type~II bursts (Section~\ref{CMEplasma}),
CMEs give rise to gyrosynchrotron emission, produced by the interaction of mildly
relativistic electrons with the ambient magnetic field.  However,
owing to its relative weakness, gyrosynchrotron emission can easily be overpowered by
the presence of much brighter plasma emission from the active Sun in solar images with limited dynamic
range, making it difficult to detect.
Indeed, following the first successful detection of the gyrosynchrotron emission from a CME
by \cite{2001ApJ...558L..65B}  using the \nan\ Radioheliograph, there have been only a handful of subsequent
detections reported, and all were rare, highly energetic events with speeds in excess of 1000~\kms.
However, as described at the RS3 meeting, this has begun to change dramatically, thanks to the high
dynamic range images now routinely achievable with the MWA (see \citealt{Kansabanik2023b},
\citeyear{Kansabanik2024a} and references therein).

S. Mondal (New Jersey
Institute of Technology) reported that the first detection of
gyrosynchrotron emission associated with a CME using the MWA  was a ``regular and unremarkable''
event, with a propagation speed of only
$\sim$400~\kms---significantly slower than the handful of highly energetic events
detectable with previous generations of instruments \citep{2020ApJ...893...28M}.
As discussed by Mondal, and in a poster presentation by D. Kansabanik, the high dynamic range
images now possible with the
MWA raise the hope of being able to make such measurements regularly (\citealt{Kansabanik2023a}).

Despite its advantages for CME studies, a drawback of the MWA is that it does not observe the Sun consistently,
making it difficult
to amass statistics on CME events. However, as discussed by Mondal, 
the new solar-dedicated backend on the Owens Valley Radio  Observatory's Long Wavelength Array (OVRO-LWA)
is helping to remedy this.
An automated imaging pipeline is now producing daily images of the Sun from OVRO-LWA data, enabling the study of
gyrosynchrotron
emission from CMEs on a routine basis (\citealt{Mondal2023b}). Mondal also cited one recently
discovered case where the OVRO-LWA spectrum of the CME and the quiet solar disk are distinctly
different, and moreover, where the spectral energy distribution (SED) of the CME appears to be consistent with {\it thermal
gyroresonance emission}. This represents the first time thermal gyroresonance has been detected from a CME. 

Another breakthrough in the study of gyrosynchrotron emission described by D. Oberoi 
comes from the ability to now use both Stokes~I and Stokes~V information
simultaneously to constrain models (\citealt{Kansabanik2023b}, \citeyear{Kansabanik2024a}).
Recently,  there has been a positive Stokes~V
detection of CME for the first time
(at $\nu\sim$100~MHz; \citealt{Kansabanik2024a}), a result that challenges current models and
their underlying assumptions, namely that the
plasma has a homogeneous  distribution along the line-of-sight and that
the electron distribution can be characterized by an isotropic pitch angle.

\subsubsection{Faraday Rotation Measurements\protect\label{Faraday}}
At heights of $\gsim10~R_{\sun}$ from the  solar surface,
gyrosynchrotron emission from CMEs  starts to become too faint to be detectable. However,
in this domain, measurements of the Faraday rotation imposed by the CME plasma on the emission
from  background sources 
offers an alternative method for measuring coronal magnetic fields. A poster by D.
Kansabanik highlighted work that he and his colleagues have done to explore this topic,
including the possibility of exploiting the wide MWA field-of-view for Faraday rotation measurements
(see also \citealt{2023JApA...44...40O}). Unfortunately, one current challenge is the low source density of suitable
 background objects at MWA frequencies (currently $\sim$ 0.05 sources
 deg$^{-2}$).

 As described in the presentations by S. Mondal and D. Oberoi, a key future goal is to
 not only measure the
magnetic field at specific locations in the corona, but over a wide range of coronal heights by
using a combination
 multi-frequency measurements from different instruments to create a
 ``movie'' of the outward propagation of a CME.
For example, the current Karl G. Jansky Very Large Array
(VLA) operates at GHz frequencies and
can be used to measure the magnetic field at $\sim5-20R_{\odot}$ via
Faraday rotation measurements (e.g., \citealt{2017SoPh..292...56K}, \citeyear{2021SoPh..296...11K};
\citealt{Mondal2023a}), while the
 OVRO-LWA is capable of providing complementary detections of
 gyrosynchrotron from coronal plasma at $\lsim5R_{\odot}$ (see
 Section~\ref{CMEgyro}). Crucially, such measurements would supply constraints on the
 vector magnetic field and 
 significantly advance our ability to forecast space weather.

\subsubsection{Advances in CME Modeling}
While the quality of radio wavelength CME data has improved dramatically in the past decade,
the derivation of magnetic field properties from these measurements
is strongly dependent on the accurate modeling of the observed emission, which remains challenging.
A poster presentation by D. Kansabanik
showcased a recent breakthrough in using spectral modeling of MWA measurements of
gyrosynchrotron emission to characterize the magnetic field of CME plasma within $\lsim10~R_{\sun}$
\citep{Kansabanik2023b}.
His team's work has also demonstrated that polarimetric observations can be used as a tool for resolving the
degeneracies that traditionally result from the large numbers of degrees of freedom in gyrosynchrotron
models (\citealt{Kansabanik2024a}).


\subsection{Other Solar Phenomena\protect\label{othersun}}
\subsubsection{Coronal Holes}
As described by D. Oberoi, recent
MWA observations have revealed new information about so-called coronal holes. Owing to their low
densities, these regions are expected to appear as dark structures in radio images. However, at MWA
frequencies, multiple examples have been found that transition from dark at higher MWA frequencies to
bright at frequencies $\lsim$145~MHz \citep{2019SoPh..294....7R}.
According to Oberoi, this is likely the result of refraction effects.

\subsubsection{Propagation Effects in the Solar Plasma}
Another powerful application  of low-frequency radio observations discussed by Oberoi is
to constrain propagation effects (refraction and scattering) in the solar plasma. He cited a
multi-frequency MWA study
of an active region by \cite{2020ApJ...903..126S}
 that found the measured position  of the region to shift as a function of
frequency, and where the nonlinearities of these shifts suggest significant inhomogeneities
in the plasma. Propagation
effects can also be constrained by the study of changes in the surface area of an active  source as a function of time,
and Oberoi was part of a team that discovered what appear to be
quasi-periodic pulsations in intensity in
type~III solar bursts\footnote{Type III events are short-lived bursts of $\lsim$10~s duration, often linked
with open magnetic field lines.}
that are anti-correlated with variations in the size of the region \citep{2019ApJ...875...98M}.
These quasi-periodic pulsations are thought to arise from
oscillations in the magnetic field.

\subsubsection{Linear Polarization of Solar Bursts\protect\label{linearsun}}
Perhaps one of the most surprising solar results presented at the RS3 meeting was the announcement by
D. Kansabanik of the first robust detection of linearly polarized emission from the Sun at meter wavelengths \citep{2022AGUFMSH24A..01D}.
As noted above, the meter-wavelength radio emission detected from solar bursts primarily
emanates from the corona via the plasma emission mechanism.
It has long been assumed that any polarization signatures
associated with such bursts would be circularly polarized, since linear polarization is expected be erased by
differential Faraday rotation.

Low-frequency polarization studies of solar bursts have been carried out since the 1950s, but
nearly all have been based on dynamic spectra from non-imaging instruments. And
while there have been a handful of
published claims for the detection of linear polarization based on those studies (e.g.,
\citealt{1961ApJ...133..258A}; \citealt{1964ApJ...139.1312B}), these were greatly outnumbered by non-detections (e.g., \citealt{1973SoPh...29..149G}; \citealt{1975A&A....40...55B}),
leading to the emergence of a consensus that  linear polarization signatures
detected from solar bursts were the result
of instrumental polarization leakage. Indeed, this interpretation had become so well established
that it formed part of the basis for the calibration of solar instruments
(cf. \citealt{2019SoPh..294..106M}; \citealt{2022SoPh..297...47M}). However,
for the MWA, Kansabanik and collaborators recently developed a new, robust polarization calibration
pipeline that does not depend on this assumption (\citealt{Kansabanik2022b}, \citeyear{Kansabanik2023a}). This work was motivated by
the detection of variable Stokes~Q emission from a solar burst observed with the MWA, a result that cannot be
readily explained by calibration errors.

As described by Kansabanik,  multiple lines of evidence point to
the MWA-detected linearly polarized signal being real. 
First, the putatively polarized emission is spatially confined to compact active regions, with no
linearly polarized emission seen over the quiet portions of the solar disk. Second, the linear polarization fraction is both
large ($\sim$20--25\%) and variable in time and frequency. Third, the linear polarization angles
are distinct in two observed active regions.

To obtain additional verification, Kansabanik and his team obtained observations of the Sun using
two different instruments at the same time and frequency: the MWA and the Giant Metrewave Telescope (GMRT). They detected a
type~III burst during this campaign, with linear polarization independently detected by
both telescopes. Kansabanik also
reported that to date, a few additional examples of type~I, type~II, and type~III
solar bursts have been examined, and
all show evidence of linear polarization.
He stressed that a next step is to understand what mechanism is responsible for producing this linearly polarized
emission. Such work by his team is ongoing.

\subsection{The Next Decade of Solar Radio Science}
Oberoi noted that owing to both practical and financial reasons, it
is unclear that a next-generation dedicated solar radio imaging array
will be built in the next decade. Nonetheless, as was amply illustrated by the recent results presented
at RS3, the outlook for solar radio science appears to be bright, thanks to the current generation of
solar instruments (see, e.g., Table~1 of \citealt{2023ARA&A..61..427G}), along with continuing efforts to exploit
 current and planned general-purpose radio instruments for solar science.

To increase the scientific impact of these
facilities for solar work, Oberoi advocated for the expanded use of triggered observations
to make more efficient use of telescope time. He also suggested building and distributing
solar imaging pipelines to the wider science community as a means to make solar radio
data more accessible and more widely used. 
Additional items on Oberoi's ``wish-list'' for future solar work included:
(i) high-fidelity polarimetric imaging of the Sun (both
quiescent and active emission); 
(ii) more sensitive studies of gyrosynchrotron emission and Faraday rotation associated with CMEs; (iii)
further
study and modeling of slowly varying (hours to days) solar emission, including coronal holes and streamers; (iv)
more sophisticated modeling to better understand propagation effects in the solar plasma.


\section{Stellar Activity\protect\label{flares}}
\subsection{Active Binaries\protect\label{activebinaries}}
So-called ``active binaries''---a category that includes RS~CVn and Algol-type systems---are well-known
emitters of non-thermal radio continuum that undergo periods of solar-like magnetic activity, including
strong flares \citep[e.g.,][]{1990ASIC..319..363F}. 
However, as stressed in the presentation by G. Umana, we still lack a full understanding of the physical processes
underlying their time-varying radio emission, including the relation between their ``quiescent'' and
``active'' phases.

Historically, modeling of the radio wavelength
SEDs, coupled with high-resolution imaging using very long baseline interferometry
(VLBI) techniques,
have served as important means of exploring the physics of active binaries (e.g.,
\citealt{1985ApJ...289..262M}, \citeyear{1987AJ.....93.1220M}). 
An example of the latest in such efforts was described by
W. Golay [Center for Astrophysics $|$ Harvard \& Smithsonian (CfA)/University of Iowa],
who spoke about recent radio observations of the active binary HR~1099,
a chromospherically active, RS~CVn type system in which
Zeeman Doppler imaging had previously revealed large
spots on the subgiant
primary star (spectral type K1~IV; \citealt{1999ApJS..121..547V}).
Using the VLA, \cite{2023MNRAS.522.1394G} measured the  radio SED of HR~1099 between 15--45~GHz and
found evidence for the presence of a $B$-field of strength 240$\pm$50~G. 
However, the binary was unresolved by the VLA, leaving open the question of exactly where in the system
the radio emission arises. To answer this, Golay and collaborators obtained 6 epochs of
data with the Very Long Baseline Array (VLBA) at 22.2~GHz over the course of 3 months, scheduled so as to
sample different phases of the binary's 2.8~day orbit. The data rule out the emission arising from the
inter-binary region and instead show it is most likely linked to the magnetically active primary
\citep{2024ApJ...965...86G}.  Using their individual VLBA epochs, they also divided the data into time slices and fit a linear
velocity model in the corotating frame of the binary to search for plasma motions consistent with
a stellar CME. Five of six epochs show no evidence for motion, while
the sixth (the only
epoch containing a flare) shows tentative evidence for motion at a level of 3$\sigma$, but the result
is inconclusive, leaving the definitive detection of extrasolar CMEs still elusive
(see also Section~\ref{exoCMEs}).

\subsection{The G\"udel-Benz Relation\protect\label{gudelbenz}}
An empirical relation between the 5~GHz radio luminosity and the soft X-ray luminosity
known as the G\"udel-Benz relation (\citealt{1993ApJ...405L..63G};
\citealt{1994A&A...285..621B})
has long been known to hold for active stars and solar flares.
However, after several decades, the physics underpinning this relation is still not fully understood.
Further, as described at the RS3 meeting, some additional
puzzles have recently emerged. For example, as described by G. Umana,
observations at 144~MHz with the LOw Frequency Array (LOFAR) have
led to the identification of a sample of active binaries with {\em coherent} radio emission that
unexpectedly follows
the G\"udel-Benz relation \citep{2022ApJ...926L..30V}. On the other hand, certain recently detected millimeter flare stars
appear to deviate from G\"udel-Benz (see Section~\ref{mmflares}).
Umana predicted that the expanded frequency coverage provided by a
combination of the Square Kilometer Array (SKA), the Next Generation VLA
(ngVLA), and an upgraded ALMA (see Sections~\ref{SKA}, \ref{ngVLA}, \ref{ALMA2030}, respectively)
will help to more comprehensively model the SEDs of these stars to better
constrain the plasma properties and the energy densities of the emitting particles.
These facilities should also provide
the sensitivity to identify radio coronae across a wider range of stellar types than is accessible currently and enable
variability studies of large new samples of stars, including searches for variability cycles similar to those
of the Sun. 

\subsection{VLBI Studies of M Dwarfs}
The typical high levels of magnetic activity of M dwarfs
leads to their frequent detection as nonthermal radio emitters.
Fortuitously, this nonthermal emission often exhibits sufficiently high brightness temperatures
to enable studies of these stars using VLBI techniques, including
measurements of their space motions. As an illustration of this, P. Boven (Joint Institute for VLBI
European Research
Infrastructure Consortium/University of Leiden) described recent astrometric VLBI work on M dwarfs using
a technique known as MultiView. MultiView compensates for the
phase corruption caused by the Earth's atmosphere  through the use multiple phase reference
calibrators distributed around the target of interest, coupled with
two-dimensional interpolation of phase corrections to the position of the target
\citep{2017AJ....153..105R}. 
At centimeter wavelengths, MultiView  enables decreasing by at least an order of magnitude the
milliarcsecond-level astrometric errors caused by the ionosphere that would occur if only a single
calibrator were used, thereby allowing  astrometric accuracy
comparable with {\it Gaia} (see \citealt{2023elvb.confE..50B}). 

Boven discussed two examples of VLBI studies of M star systems that exploited MultiView. One was
GJ3789A/B, a system originally discovered in the 8.4~GHz
astrometric survey of \cite{2009ApJ...701.1922B},  but
for which the data had remained
unpublished because of the unsolved puzzle of their unusual
astrometric residuals. Using new follow-up VLBA observations, Boven
and colleagues obtained multi-epoch astrometry for GJ3789A/B  from which they were able to determine
a precise astrometric solution and a set of
orbital parameters. The new measurements also show that the system lies at a larger distance than originally
thought and that the radio emission originates from the secondary star. However, an outstanding puzzle is
that the deprojected radial velocities 
inferred from the radio astrometric solution are much higher than independently measured radial velocities.

Boven concluded by suggesting that
an ``ideal'' next-generation telescope array for future astrometric VLBI studies of M
dwarfs would be comprised of phased arrays of small dishes equipped with wideband receivers. This could
provide improved sensitivity (and thus access to more calibrators with close proximity to the
target), coupled with
the ability to observe both
calibrators and science target simultaneously within the field-of-view.

\subsection{Quiescent Coronal Emission from Zero Age Main Sequence Stars}
J. Climent (University of Valencia) presented
VLBI measurements of the coronally active zero age main sequence star
AB~Dor~A, a young, rapidly rotating K dwarf that is part of
a quadruple system. Motivated by the previous VLBI observations of \cite{2020A&A...641A..90C},
\cite{2024MNRAS.530.2442B} were able to product a 3D model of the star's coronal magnetic field,
and from this,
create synthetic radio images over the course of  the stellar rotation period. The models are able to
reproduce the  morphology and extent of the coronal emission seen in 8.4~GHz VLBI maps, which reaches
$\sim 8-10R_{\star}$, making it significantly more extended than the solar corona. 

\subsection{Coherent Emission}
In addition to the ``quiescent'' radio emission from active stars, which is
generally attributed to gyrosynchrotron radiation (see above),
a growing numbers of such stars are seen to exhibit activity in the form of coherent bursts.
These bursts are generally attributed to ECM instabilities 
and exhibit short duration (pulse-like) emission profiles that are
narrow in frequency and display
high levels of circular polarization (up to 100\%). Importantly, the radio frequency at which these
coherent bursts are observed
is directly proportional to the magnetic field strength, providing a means of quantifying the magnetic
field. Back in 2021,
\citeauthor{2021MNRAS.507.1979L} published an empirical relation for early-type magnetic stars (see Section~\ref{mag-OB}),
linking the nonthermal radio luminosity with the ratio of
the magnetic flux to the rotation period. As discussed at the RS3 meeting, this
correlation has now been shown to also hold for a wide
range of objects with stable, dipole-dominated magnetospheres, including ultracool dwarfs (UCDs; see Section~\ref{UCDs})
and the planet Jupiter. Indeed, 
the topic of coherent emission from active stars
was raised in multiple presentations and sessions at RS3
(see Sections~\ref{solarECM}, \ref{mag-OB}, \ref{UCDs}),
highlighting commonalities in the underlying physics
across different classes of stars exhibiting coherent bursts, including
the presence of strong magnetic fields to aid the production of energetic particles.

\subsection{Flaring at Millimeter Wavelengths\protect\label{mmflares}}
To date, most of the radio wavelength work on stellar flaring has focused on the centimeter
bands. In contrast, as pointed out by
C. Tandoi (University of Illinois), stellar flaring in the millimeter regime is relatively unexplored, and until
recently, only $\sim$30 stellar millimeter flares were documented in the literature (see
\citealt{2021ApJ...916...98G} and references therein).
However, this has begun to change, thanks in part to datasets available from Cosmic Microwave Background (CMB)
experiments, which simultaneously provide the frequency coverage, aerial coverage,
temporal sampling, and angular resolution
required for blind transient searches, including the identification of stellar flares.
For example, Tandoi reported on the detection of 111 millimeter flares from a sample of
66 stars based on
data from the South Pole Telescope (SPT; \citealt{2024ApJ...972....6T}), including main sequence stars, evolved stars, and
interacting binaries.
He argued that the emission is most likely to be synchrotron generated by the initial impulsive phase of the
flare during which particles are accelerated, although the measured spectral indices of individual sources
show a range of values.
In cross-comparing with other wavelengths, Tandoi's team
found that nearly all of their detected stars are X-ray active,
although they do not appear to follow a G\"udel-Benz type of correlation
between their radio and X-ray fluxes (see Section~\ref{gudelbenz}).
The poster presentation by E. Biermann (University of Pittsburgh) also demonstrated how the high sensitivity
and large aerial
coverage provided by data from
the Atacama Cosmology Telescope (ACT) have been leveraged to perform blind transient searches at
millimeter wavelengths and provide a powerful means of identifying flares and transient emission from stellar
sources including M dwarfs, RS~CVn variables, and even a classical novae
\citep{2025ApJ...986....7B}. 

As described by G. Umana, the recent detection of millimeter wavelength flares from M dwarfs using ALMA has also opened
a new window for understanding stellar activity (\citealt{2018ApJ...855L...2M},
\citeyear{2020ApJ...891...80M}, \citeyear{2021ApJ...911L..25M}),
Crucially, the ALMA wavelength bands provide access to high-energy
(MeV) particles immediately after they are accelerated in the atmosphere. Such particles are
not accessible with centimeter-wavelength observations. Building a more comprehensive understanding
of the properties of flares detected in the millimeter regime is thus expected to be a rich area
for future study.

\subsection{Future Prospects in Active Star Research}
Umana predicted that future research on active stars will benefit enormously  from the broad frequency
coverage ($\sim$300~MHz to $\gsim$100~GHz)
that will be provided by a combination of current and planned radio
facilities such LOFAR, the upgraded GMRT (uGMRT), the VLA, ALMA, the SKA (both SKA1-Low and SKA1-Mid), and the
ngVLA (see Section~\ref{futureisbright}).
She stressed the value of this kind of multi-frequency approach, as
the lower-frequency bands (up to a few GHz) probe the coherent emission
from stellar bursts (enabling characterization of the plasma density and magnetic field strength);
the intermediate frequencies ($\sim$10-40~GHz) sample the gyrosynchrotron emission
emitted by accelerated particles; finally, the ALMA frequency bands (35--950~GHz), 
along with the higher-frequency ngVLA bands (up to 116~GHz), supply access to
the most strongly accelerated
particles (including possible synchrotron emission) while  simultaneously providing
the sensitivity to study thermal emission from the stellar chromospheres.

\section{Radio Emission from Ultracool Dwarfs (UCDs)\protect\label{UCDs}}
UCDs 
encompass low-mass objects ranging from late-type M dwarfs to brown dwarfs (i.e.,
objects with spectral types later than M7). UCDs thus bridge the transition between stars and gas-giant planets, and
along this sequence of objects the onset of planet-like auroral/magnetospheric emission is seen.
As discussed in the invited review by M. Kao (Lowell Observatory), as well as by several other speakers,
radio wavelength studies have emerged as one of the most powerful means of improving our
understanding of these objects.

Brown dwarfs were first discovered to be radio emitters by  \cite{2001Natur.410..338B}, who detected 
bursting emission in the 4--8~GHz range that violated the G\"udel-Benz relation
(see Section~\ref{gudelbenz})
by more than 4 orders of magnitude. Now, more than 20 years later, such behavior has been established as
relatively common among brown dwarfs and other UCDs \citep[e.g.,][]{2014ApJ...785....9W}.
However, it remains a longstanding
puzzle as to  how UCDs, which are fully convective, can support
the strong magnetic fields inferred  from their radio emission.

As noted by J. Climent, as of 2023 there were approximately 30 UCDs that had been
detected at GHz frequencies (\citealt{2001Natur.410..338B}; \citealt{2022RAA....22f5013T}
 and references therein; \citealt{2022ApJ...932...21K}; \citealt{2023ApJ...951L..43R}),
plus two more at MHz frequencies (see Section~\ref{BDaurora}).
He stressed, however, that presently these
represent only a relatively small fraction of
the total number of targets searched (see, e.g., Figure~3 of \citealt{2022AJ....163...15C}). 
Furthermore, as of the RS3 meeting there had not yet been any confirmed,
direct detection of radio emission from bona fide exoplanets (Section~\ref{exodirect})
or star-exoplanet
interactions (see Sections~\ref{exodirect}, \ref{star-planet}), although several of the aforementioned
radio-detected UCDs are in the $\sim12-70~M_{\rm Jupipter}$ mass range.

\subsection{Brown Dwarf Auroral Emission\protect\label{BDaurora}}
Radio-detected brown dwarfs tend to be rapid rotators, and the dominant component of their
radio emission is typically rotationally modulated and
strongly circularly polarized. These traits are  now interpreted as the hallmarks of ECM emission arising from
powerful aurorae, linked to kG strength magnetic fields
(e.g., \citealt{2015Natur.523..568H}; \citealt{2016ApJ...818...24K};
\citealt{2017ApJ...846...75P}).  As noted by Kao, the dynamo regions
in these objects are thought to be similar to scaled-up analogs of those in exoplanets
(see also Section~\ref{exoplanets}).

 One recent development in the radio studies of brown dwarfs is
the detection of two extremely cold ($<$1000~K) brown dwarfs at MHz frequencies using LOFAR, one of which represents
the first direct  discovery of a brown dwarf using radio observations
(\citealt{2020A&A...639L...7V}, \citeyear{2023A&A...675L...6V}). 
 A persistent puzzle, however, is that the inferred strength
magnetic fields of these cold brown dwarfs are significantly weaker than predicted by the prevailing
dynamo model paradigm. 

T. W. H. Yiu (ASTRON/Netherlands Institute
for Radio Astronomy) described  a recent follow-up study of 
one of the LOFAR-detected brown dwarfs (the T dwarf binary WISEP~J101905.63+652984.2;
\citealt{2023A&A...675L...6V})
in an attempt to identify
a spectral cut-off in its periodic emission.
For this purpose, Yiu and collaborators obtained observations with the VLA, GMRT, and LOFAR over multiple epochs.
Surprisingly,  during 10 epochs of LOFAR observations
no pulse was detected  at a strength comparable to the originally detected
level of $\sim$10~mJy  (see \citealt{2023A&A...675L...6V}). Moreover,
9 of 10 epochs showed no evidence for any type of pulse. There was also no detected radio emission
in either the VLA or the GMRT data. Despite this, after combining the available data from all
epochs, Yiu et al. identified a new 0.8~hr periodicity in the data, unrelated to the originally detected pulses.
The nature of this periodic emission is unclear, although
Yiu noted that if it is interpreted as being linked to rotational period, it is interestingly
close to the predicted breakup limit for the brown dwarf.

\subsection{Radiation Belts\protect\label{belts}}
In addition to auroral emission (Section~\ref{BDaurora}), radio-detected brown dwarfs are seen to display
a weaker, quiescent component of radio emission with little or no polarization, and
whose origin and nature had remained
enigmatic \citep[e.g.,][]{2013ApJ...767L..30W}.  Presentations at the meeting showcased recent observational
evidence from two independent groups that this emission arises from
{\it radiation belts}, analogous
to those observed in all of the strongly magnetized planets in our Solar System. This is the
first empirical confirmation of an idea described at the previous Radio Stars conference
by P. Williams (see \citealt{2018haex.bookE.171W}; \citealt{Matthews2019}).

\begin{figure*}[ht!] 
\plotone{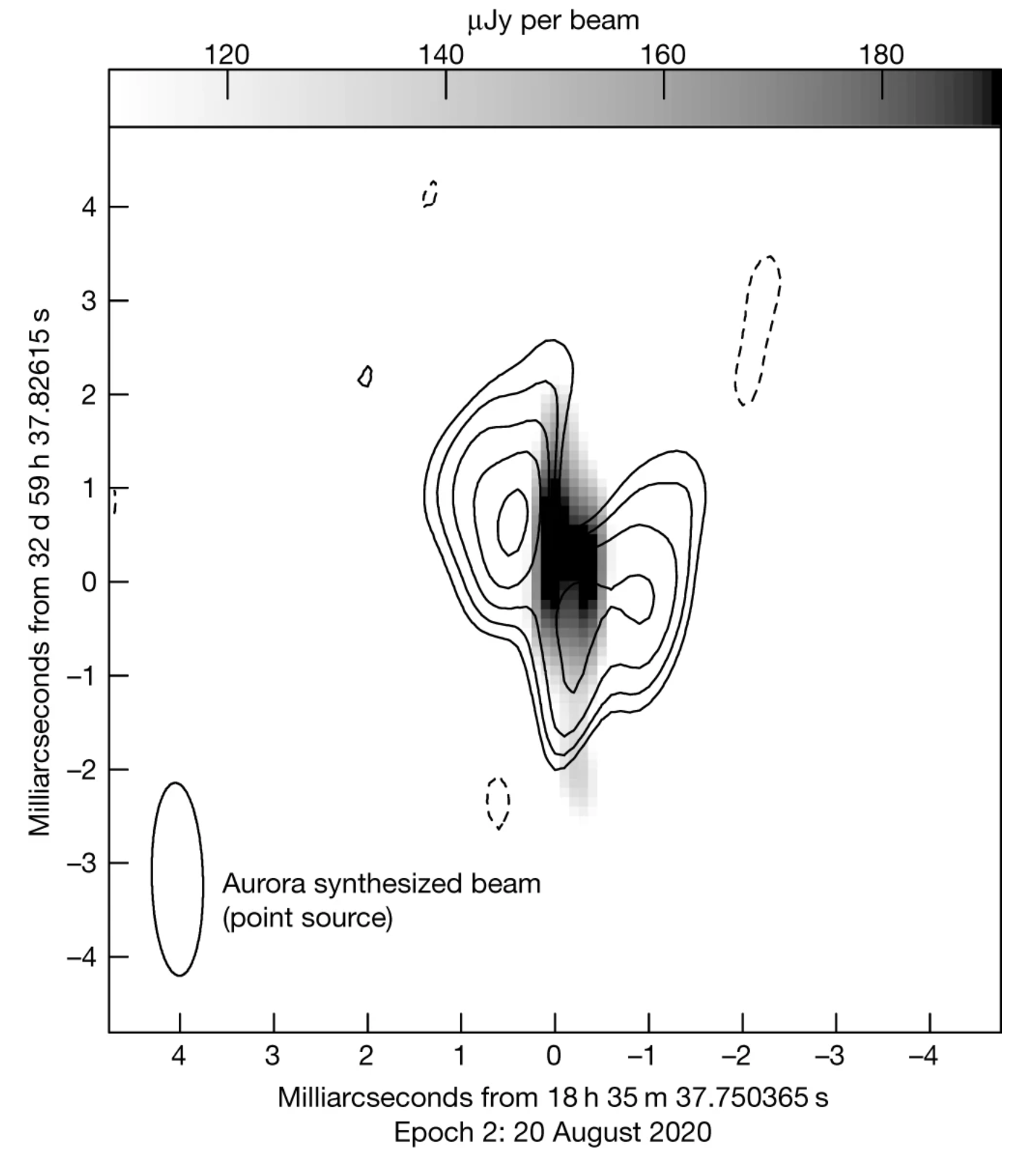}
\caption{Right-circularly polarized auroral emission (greyscale) from the UCD LSR~J1835+3259 as observed
  with the HSA at 8.4~GHz, overlaid on the
  separately imaged quiescent emission at the same frequency (contours). The contour levels are  
  $(-1, 1, \sqrt{2}, 2, 2\sqrt{2}, 4)\times\sigma_{\rm RMS}$, where $\sigma_{\rm RMS}\approx12-13\mu$Jy beam$^{-1}$.
  The synthesized beam
  for the auroral observation is indicated by an ellipse in the lower left corner.  The aurora appears centered within
  the double-lobed structure traced by the quiescent emission.
  Credit: \cite{2023Natur.619..272K}.
\label{fig:kao2023}}
\end{figure*}

Kao and her collaborators recently used the High Sensitivity Array (HSA), a global VLBI
array of radio telescopes spread across Europe and North America, to spatially resolve
the radio emission of the UCD LSR~J1835+3259 at 8.4~GHz \citep{2023Natur.619..272K}. The
team obtained three images over the course
of a year with angular resolution $\sim$0.5~mas, revealing a double-lobed, axisymmetric
structure with an extent of
$\sim 24R_{\rm Jupiter}$ and a morphology similar to the Jovian radiation belts (Figure~\ref{fig:kao2023}).
This provided compelling evidence that the steady, quiescent emission associated with this system is synchrotron,
arising from 
the presence of radiation belts of magnetically
charged particles and a dipole-ordered magnetic field.
Additionally, during 2 out of 3 observing
epochs,
circularly polarized ``burst'' emission (attributed to an aurora) was detected by Kao's team
from a location between
the two radio lobes (Figure~\ref{fig:kao2023}). LSR~J1835+3259 is rapidly rotating,
with a period
of $\sim$2.8~hr, and this burst emission was seen to be rotationally modulated. The authors
derived a magnetic field strength of 3~kG at
$\sim 12R_{\rm Jupiter}$, decaying
in strength as a function of radial distance, $r$, as $\sim r^{-3}$. Kao noted that the origin of such
strong magnetic fields remains a puzzle, as it cannot readily be explained by convected thermal energy.

As reported by J. Climent, LSR~J1835+3259 
was also independently observed at 5~GHz by \cite{2023Sci...381.1120C} using the European VLBI Network (EVN).
The \citeauthor{2023Sci...381.1120C} observations confirmed the double-lobed structure seen by \cite{2023Natur.619..272K},
 but the lower
frequency of their observations allowed sampling a distinct population of electrons in the
radiation belt. Climent's team also detected a second
auroral-like emission component whose properties appear consistent with the earlier model predictions
of \cite{2021MNRAS.507.1979L}. Climent noted that overall the observed radio emission components in LSR~J1835+3259 appear
well described by the type of model presented in the poster by R. Kavanagh,
namely an oblique rotator coupled with
an auroral ring  (see Section~\ref{cmsurveys}). 

Kao estimated that $\sim$30\% of T dwarfs have radiation belts, but noted that a persistent puzzle is
where their energetic electrons are originating from. Younger, hotter objects are not found to be
more likely to exhibit evidence of radiation belts compared with older and colder UCDs, in contrast to earlier predictions.
However, binarity {\it does} enhance the likelihood of detecting radiation belt emission
\citep{2025MNRAS.539.2292K}.
Kao predicted that future observations, especially observations with facilities like the ngVLA (Section~\ref{ngVLA})
are likely to reveal that radiation belts are far more ubiquitous among different classes of stellar objects
than previously thought, including in magnetic massive stars (see Section~\ref{mag-OB}).
She also pointed out that spatially resolved observations of additional brown dwarfs to directly search for radiation
belts  (see Section~\ref{belts}) would be both challenging and time-consuming and suggested instead that
future radio surveys of spatially {\it unresolved} emission could serve as a useful means of
compiling better statistics and improving more generally our understanding of the magnetic field properties of UCDs \citep{2024MNRAS.527.6835K}.

\subsection{Measuring Physical Parameters of UCDs from VLBI Astrometry}
Another topic discussed by
J. Climent  was how astrometric VLBI techniques can benefit the study
of UCDs. He noted that prior to 2023, among radio-detected UCDs only 3 had been detected using VLBI 
(\citealt{2009ApJ...706L.205F}; \citealt{Forbrich2016a}; \citealt{2020ApJ...897...11Z}).  Nonetheless,
the results were particularly powerful, enabling dynamical mass measurements,
parallax (distance) and proper motion measurements, and in one case, the discovery of an
extrasolar planet (\citealt{2020AJ....160...97C};  see also Section~\ref{VLBI}). 
Climent presented new preliminary results from additional, ongoing VLBI studies of UCDs,
including an astrometric monitoring campaign of the substellar T6 dwarf WISEP~J112254.73+255021.5.
By measuring the radio light curve, Climent and collaborators aim to pin down the rotational period
of this object, which has, until now, been a subject of controversy in the literature (cf.
\citealt{2017ApJ...834..117W}).

\section{Radio Emission from and Exoplanets and their Host Stars\protect\label{exoplanets}}
During the 2017 Haystack Radio Stars meeting  \citep{Matthews2019},
radio wavelength searches for exoplanets were called out as an exciting new
frontier in astrophysics. 
This area of research has seen substantial growth over the past several years.
Although as of the RS3 meeting
there still had not been a definitive detection of exoplanet radio emission (see below), several interesting,
related developments were reported.

Attempts to detect radio emission associated with exoplanets can largely be divided into two categories:
{\it direct detection} of an exoplanet, or detection of the {\it signatures of star-planet interaction}.
Consensus was that
the latter is probably easier and likely to happen sooner---and may even be within the reach of
existing facilities, but efforts targeting both approaches were described at the meeting. 

\subsection{Efforts to Directly Detect Radio Emission from Exoplanets\protect\label{exodirect}}
C.-M. Cordun (ASTRON) described her team's effort to detect low-frequency ($<$40~MHz) emission associated
with ECM-generated aurorae in hot Jupiters using LOFAR. Such a detection
would represent a major breakthrough by allowing the possibility to directly measure the magnetic field strength of an exoplanet.
One of the most promising candidates for detection is thought to be $\tau$~Bo\"otis~b, a gas giant of
$\sim$6 Jupiter masses orbiting an F-type main sequence star at a separation of 0.049~au. Unfortunately, ionospheric
scintillation, human-made radio frequency interference (RFI), contamination from sidelobes of other bright
sources, and cosmic noise from galactic background sources all pose challenges for imaging observations at
these low frequencies. The deepest images of $\tau$~Bo\"otis~b obtained to date by Cordun's team reach RMS noise levels of
225~mJy, 80~mJy, and 40~mJy at 15.2~MHz, 27.0~MHz, and 37.7~MHz, respectively. The exoplanet remains undetected,
but an important outcome is that the results demonstrate the feasibility of imaging observations
at frequencies $<$40~MHz \citep{2025A&A...693A.162C}. An ongoing
LOFAR upgrade  (see, e.g., \citealt{2024SPIE13098E..0RO}) is
expected to make the array a factor of two more sensitive and further bolster the chance of detection.

The poster presented by K. Ortiz Ceballos (CfA)
described another direct search for exoplanet radio emission, in this case at frequencies between
4--8~GHz, using new and archival observations from
the VLA \citep{2024AJ....168..127O}.   Ortiz Ceballos and his colleagues examined
a total of 77 nearby stellar systems ($d<$17.5~pc)
known to host exoplanets (140 known in total).
Radio emission was
detected from one target (an M4 dwarf), but based on the radio luminosity and
corresponding X-ray luminosity,
the emission appears to be  stellar in origin. Radio luminosity upper limits
for the remaining targets are $L\lsim10^{12.5}$ erg s$^{-1}$ Hz$^{-1}$ (3$\sigma$),
comparable to the lowest radio luminosities of UCDs (see Section~\ref{UCDs}).

\subsection{Searches for Radio Signatures of Star-Planet Interactions\protect\label{star-planet}}
While direct radio detections of exoplanets have remained elusive,
the past several years have also seen a rapid growth in interest in detecting the
auroral-like signatures of star-planet interactions at radio wavelengths.  As
described by M. Kao,
such detections would in principle have a unique
advantage over the observations of star-planet interactions at optical or other wavelengths by
enabling direct measurements of magnetic field strengths, viz.
$\nu_{[{\rm MHz}]}\approx 2.8B_{\rm planet} [{\rm Gauss}]$ \citep[e.g.,][]{1998JGR...10320159Z}, and
thereby providing crucial tests of magnetic dynamo models
for exoplanets (see e.g., \citealt{2021MNRAS.504.1511K}; \citealt{2021A&A...645A..77P}; \citealt{2023NatAs...7..569P}; \citealt{2023arXiv230500809T}; \citealt{2024NatAs...8.1359C}).

The leading dynamo model \citep{2009Natur.457..167C}
unifies planetary and stellar dynamos by assuming that the
thermal energy from the deep interior is what determines the energy density in the dynamo region.
In this model the strength of the radio emission is expected to be proportional to the apparent
size of the emitting
object, but additionally depends on  the
properties of the stellar wind that impacts it \citep{2013A&A...552A.119S}.  Kao emphasized that a consequence is that the
magnetic fields of certain hot Jupiters are predicted to be
10--100 times higher than the one seen in Jupiter itself
(\citealt{2017ApJ...849L..12Y}; \citealt{2019NatAs...3.1128C}).
However, despite numerous attempts, these predictions have yet to be observationally
confirmed; as of
the RS3 meeting, no definitive cases of star-planet interactions had yet been unambiguously detected
\citep[e.g.,][]{2022AJ....163...15C}. Previously, \cite{2021A&A...645A..59T}
 reported a tentative detection of radio
emission from the $\tau$~Bo\"otis system in the
14--21~MHz range, but follow-up observations have so far failed to confirm this detection
\citep[e.g.,][]{2024A&A...688A..66T},  suggesting that
detectability may depend on the phase of the stellar activity cycle or other, as-yet-to-be determined factors.
YZ~Ceti is another system where radio emission that may be consistent with star-planet
interaction  has been reported, although
so far the evidence remains inconclusive (\citealt{2023NatAs...7..569P};
\citealt{2024ESS.....561713P}; \citealt{2025AAS...24541807V}). 

Kao predicted that
when the ngVLA comes online (Section~\ref{ngVLA}), it will become possible to routinely
directly observe satellite-induced aurorae on brown dwarfs and other UCDs
through characteristic features in their dynamic spectra (cf. \citealt{2008GeoRL..3513107H}). Once
such cases are confirmed, these observations will help to anchor modeling of UCD magnetic
fields and dynamo theories for both the high- and low-mass ends of the UCD population. 

\section{Extrasolar Space Weather\protect\label{weather}}
\subsection{Extrasolar CMEs\protect\label{exoCMEs}}
\subsubsection{CMEs from Main Sequence Stars}
The quest to directly detect CME-like events and the ejections of energetic particles from stars other
than the Sun was discussed at both previous Haystack
Radio Stars meetings (\citealt{Matthews2013}, \citeyear{Matthews2019}). However, as of the time of
the RS3
conference there still had not yet been an unambiguous detection of these phenomena beyond our solar system
(see also Section~\ref{activebinaries}).
One challenge has been simultaneously achieving the necessary sensitivity (particularly in the presence
of the strong terrestrial RFI environment), in combination with the time
and frequency resolution required to identify signatures of such events. Additional factors may include
the lack of coordinated, multi-wavelength data, and the fact that many searches
to date have targeted M dwarfs, where the magnetic field may suppress associated radio emission at frequencies
detectable from Earth-based radio observatories (\citealt{2016IAUS..320..196D}; \citealt{2018ApJ...862...93A}).
Yet another possibility discussed in the literature is that large-scale overlying magnetic fields in active
  stars may suppress CME emergence (\citealt{2018ApJ...862...93A}; \citealt{2022MNRAS.509.5075S}).
  
I. Davis (California Institute of Technology) described her team's
ongoing Space Weather Around Young Suns (SWAYS)
project, aimed at overcoming these challenges. SWAYS is a dedicated, multiwavelength effort to
monitor space weather from nearby solar-type stars
and M dwarfs using the recently commissioned OVRO-LWA \citep{Davis2024b}.
This instrument comprises 352 dipoles spread over an area spanning 2.5~km and operates in the
 13--87~MHz range with 24~kHz spectral resolution
(see also Section~\ref{CMEgyro}). 

Davis emphasized that at present, the
particle environments of active stars are very poorly constrained, which in turn significantly impacts our
ability to understand processes such as abiotic chemistry in exoplanet environments and the retention
of exoplanetary atmospheres. New efforts are therefore underway to use low-frequency radio emission as a tool
for characterizing particle fluxes from stars known to host planets. In particular, the goal is to detect the
extrasolar analog of type~II and III radio bursts, whose radio emissions are produced by the plasma emission
mechanism (see also Sections~\ref{solarCMEs}, \ref{othersun}). An advantageous characteristics of plasma emission
is that the frequency at which
it arises is related to the plasma density. Additionally, the structure of the emission seen in a dynamic spectrum will
depend on the speed of the particles. 

For their searches, Davis's team is employing the beam-forming mode of OVRO-LWA, which provides
1~ms time resolution, coupled with simultaneously obtained (and more sensitive)
``slow-visibility'' mode data with 10~s time resolution to aid
in identification of candidate events. Events of interest are de-dispersed in the higher time resolution
data in a manner analogous to pulsar searches. In parallel with the OVRO-LWA observations,
candidate stars are observed at optical wavelengths using Flarescope, a fully-automated
0.5~m telescope designed to achieve photometric monitoring of bright stars
with sub-millimagnitude precision on 5~minute timescales \citep{2023AAS...24134607D}.
At the time of the conference,
reduction pipelines for the multi-wavelength data were being finalized.

Another effort to detect extrasolar CMEs using low-frequency observations was described by D. Konijn
(ASTRON/Kapteyn Institute). He presented a progress report on an effort to identify signatures of such
events in data from the LOFAR Two-Metre Sky Survey (LoTSS; \citealt{2019A&A...622A...1S}).
LoTSS is a deep (83~$\mu$Jy beam$^{-1}$), widefield (6035 deg$^{2}$)
survey with high angular resolution (6$''$) over the frequency
range 120--168~MHz and is expected to yield over 5 million sources. At the time of the meeting
this already included data for
$\sim$250,000 stars within 100~pc for which dynamic spectra with 8~s time resolution have been produced.
Currently searches are being performed on Stokes~I data so as not to bias against detection of
unpolarized bursts.

Konijn presented one particularly interesting
case [originally identified by meeting participant J. Callingham (ASTRON/Leiden University)]
in which a bright burst was detected with an apparent drift rate of $-0.35$~MHz s$^{-1}$ and
emission that is 90\% circularly polarized.  The host star is a high proper motion M-type dwarf
lying at $d\sim$40~pc. The implied ejection speed would be $\sim$1000~\kms, making
this a tantalizing candidate for a long-sought extrasolar analog to a type~II solar burst (see Section~\ref{CMEplasma}). However, 
some caveats have prevented this event being unambiguously linked to a type~II-like event. First, the harmonic
is not seen in the data (though it may plausibly lie  outside the observed band).
Second, the degree of circular polarization is 
higher than is typical of solar type~II bursts. Unexpectedly, linear polarization is also observed---though
new results presented at the conference by
D. Kansabanik now suggest that this may be unsurprising; see Section~\ref{linearsun}. 

Konijn also described efforts to search for type~III-like bursts in the LoTSS data. No such events have so
far been identified, but as these are typically short-duration events ($\lsim$10~s),
a re-processing of the data to 1~s time resolution is underway to enhance the search.
Konijn estimates that in total, $\sim$40 extrasolar type~III bursts should be detectable at $\ge10\sigma$
in the LoTSS survey (see also \citealt{2020A&A...639L...7V}). 

\subsubsection{CMEs from Young Stellar Objects (YSOs)\protect\label{YSOCMEs}}
An additional program to search for evidence of extrasolar CMEs---in this case associated with young stellar objects
(YSOs)---was presented by J. Forbrich (University of
Hertfordshire). Forbrich reminded us that highly energetic processes arise even in the earliest stages of star
formation, making YSOs among the brightest stellar sources in both radio emission and X-rays. As was reported
at the second Haystack Radio Stars meeting (see \citealt{Matthews2019}), Forbrich and his collaborators have used high angular
resolution observations with the VLA at 4--8~GHz to observe a single field in the Orion Nebula Cluster (ONC)
for $\sim$30~hr, reaching an RMS noise level of 3$\mu$Jy beam$^{-1}$ and allowing them to identify
nearly 700 sources
(\citealt{Forbrich2016b}). Parallel X-ray observations were also obtained using {\it Chandra}
\citep{2017ApJ...844..109F}. 
At the RS3 meeting, Forbrich described follow-up work to obtain multi-epoch VLBA observations of {\it all}
sources in their sample, providing the means to investigate emission on sub-au scales within the ONC
\citep{2021ApJ...906...23F}. 
Forbrich stressed that this effort has benefited substantially from recent VLBA
upgrades (see Section~\ref{VLBA}), including the ability to perform software correlation of multiple sources
within the primary beam, eliminating the need to limit focus to just one or two objects per pointing.\footnote{Typically
the field-of-view of VLBI observations is limited to $\sim10^{-4}$ times the primary beam size as a result
of time and frequency smearing away from the adopted phase center; see \cite{1989ASPC....6..247B}.}

Forbrich pointed out that while the rate of occurrence of X-ray megaflares from YSOs is reasonably
well constrained (e.g., \citealt{2024ApJ...976..195G}  and references therein), the fraction
of megaflares generating CMEs is still unknown, as are the parameters that may impact their detectability,
such as viewing angle, plasma density, and timescales. 
His team is therefore using their VLBA data to search for CME signatures within a time window of a
few days for targets exhibiting strong X-ray
flares. Specifically, expected
CME signatures associated with these megaflares are expected to include a rise and fall of the radio flux over several
days after the X-ray event, and a spatial offset of the emission from the star
on scales $\gsim$1~au. However, at the time of the meeting, despite having amassed a volume of VLBA data
exceeding 20~TB, there was still no evidence of an unambiguous CME. A caveat noted by Forbrich is that the 5~GHz VLBA
band in which they are observing (selected to maximize achievable sample size)
is expected to be sensitive only to the most energetic CMEs owing to the high
Lorentz factors needed to produce emission at this frequency. 

\subsection{Winds from Cool Main Sequence Stars\protect\label{coolwinds}}
S. Bloot (ASTRON/Kapteyn Astronomical Institute) described recent work leveraging LOFAR to
study the winds and space weather of M dwarfs. The stars included in her team's
sample span the full range of M dwarfs, not just the
very latest objects ($\le$M7) categorized as UCDs (see Section~\ref{UCDs}).
As noted by Bloot, stellar
winds are important in the context of these stars, since they can  determine the degree to which a planet
interacts with its host star, whether orbiting planets can maintain an atmosphere, and whether planets may be able to
produce detectable radio signatures. Stellar winds are also of vital importance from a stellar evolutionary
perspective, carrying away angular momentum, impacting changes in rotation speed, and returning mass to the
interstellar medium. However, 
the winds of low-mass main sequence stars are in general extremely tenuous, making them difficult to detect,
particularly in the case of M dwarfs.

As described by Bloot, historically, the most successful method for characterizing
M dwarf winds 
has been the study of atmospheric absorption lines in the UV toward the interstellar medium
(ISM; e.g., \citealt{2002ApJ...574..412W}, \citeyear{2021ApJ...915...37W}).
Drawbacks of this method are that it requires a detailed understanding of the surrounding ISM and that it
is limited
to stars lying within $\lsim$7~pc. The alternative approach used by Bloot and her collaborators instead utilizes free-free
absorption at radio wavelengths to place upper limits on the mass-loss rate (see also
\citealt{1996ApJ...462L..91L}). This in
turn provides a direct upper limit on the density of the stellar wind. Since free-free absorption
is stronger at lower
frequencies, Bloot's team has used LOFAR detections of coherent (ECM) emission from M dwarfs 
at $\nu\sim$120~MHz, in combination
with a simple stellar wind model that includes an estimate of the stellar magnetic field
\citep{2025A&A...695A.176B}. 
At the time of the meeting they had obtained
mass-loss upper limits for 19 M dwarfs and were able to reach a sensitivity of 4$\times$ the
solar-wind  mass-loss rate, independent of distance.

\section{Radio Emission from Hot Stars}
\subsection{Thermal Emission from Hot Massive Stars\protect\label{thermalOB}}
Hot massive stars have winds powered by radiation pressure on lines from metallic elements
\citep{1975ApJ...195..157C}  and
the groundwork for the use of thermal radio emission to study their resulting mass-loss rates
dates back to the seminal papers by \cite{1975A&A....39....1P} and \cite{1975MNRAS.170...41W}.
With an underlying assumption of spherical symmetry,\footnote{A nonspherical but
axisymmetric wind will produce the same SED slope, but with a change in resulting flux level
\citep{1982A&A...108..169S}.}
these analytic models lead to the classical prediction
that the SEDs of OB stars should follow $S_{\lambda}\propto \lambda^{-0.6}$, where $S_{\lambda}$ is the
flux density at a given radio wavelength, $\lambda$.
A net result is that in the
case of free-free opacity, longer wavelengths
(lower frequencies) ``see'' a larger photosphere. Thus observing a range in wavelengths effectively ``scans''
the geometric layers of the atmosphere/wind.
The presentation by R. Ignace (East Tennessee State University) focused on ongoing efforts to address the
added complexity of the deviations from spherical symmetry that occur in the envelopes of hot massive
stars and how these ultimately shape the winds and outflows.

One example discussed by Ignace is the case
of so-called corotating interaction regions (CIRs; e.g., \citealt{1996ApJ...462..469C}).
CIRs are expected to occur when bright or dark spots occur on the surface
of a massive star, leading, respectively, to the formation of low- and high-speed streams.
Rotation of the star  causes these flows
to be accelerated at different
rates, leading to the formation of corotating structures which then collide.
Ignace and his colleagues
have explored the question of how far out these regions persist and how  radio observations may be able to
provide constraints through the comparison of radio flux variations over time with geometric models
\citep{2020MNRAS.497.1127I}.

In an update from the 2017 Radio Stars meeting (see \citealt{Matthews2019}),
Ignace pointed to the study by \cite{2022ApJ...932...12E}
that models how the combination of synchrotron emission (occurring the presence of a strong magnetic field)
and free-free absorption from colliding wind
massive binaries can effect changes in their radio SEDs.  He also discussed the role of wind clumping in biasing
the derived mass-loss rates for hot massive stars. 
Ignace noted that the presence of clumping will not necessarily change the SED {\it shape}; however, depending on the
porosity, there may be wavelength-dependent variability. Work on this topic remains ongoing.

\subsection{Magnetic Massive Stars\protect\label{mag-OB}}
Approximately 10\% of massive (O and B) main sequence stars are now known to possess stable,
large-scale, kG
level surface magnetic fields \citep[e.g.,][]{2012AIPC.1429...67G}. Often these fields are found to be simple, axisymmetric
dipolar configurations aligned with the
rotation axis of the star. As discussed in the
invited presentation by B. Das (Commonwealth Scientific and Industrial Research Organisation Space \&
Astronomy), recent research has increasingly shown that
the presence of these magnetic fields significantly impacts the evolution of these stars relative
to non-magnetized OB stars, underscoring the importance of studying this class of object from a stellar
evolutionary perspective (e.g., \citealt{2020MNRAS.493..518K}, \citeyear{2021MNRAS.504.2474K},
\citeyear{2024MNRAS.533.3457K}). More generally, since magnetic OB
stars are bright and
their fields very stable, they provide excellent laboratories for studying magnetospheric physics.

Historically, the H$\alpha$ emission line has been used as a diagnostic of
the magnetic properties of OB stars. However,
this line becomes undetectable in magnetic OB stars with effective temperatures $T_{\rm eff}\lsim$15,000~K, 
which has led to some controversy over whether the cooler end of such stars can even support magnetic fields.
Fortunately, as described by Das,
radio emission is detectable from magnetic massive stars over a broader range of temperature.
Interest in this radio emission is further heightened by the discovery of key similarities with the
radio emission
observed from UCDs (Section~\ref{UCDs}).

Similar to UCDs, magnetic OB stars are found to exhibit both quiescent (slowly varying, rotationally modulated)
radio emission and pulsed, period emission believed to arise from to an ECM mechanism. The periodicity
of the latter emission
may be impacted by either the stellar rotational period and/or the orbital period in the case of a
binary\footnote{Das reported that at present no magnetic massive stars are confirmed binaries.}
\citep[e.g.,][]{2023MNRAS.523.5155B}. Additionally, radio flares have been detected
from such stars (see Section~\ref{OBflaring}).
Also analogous to UCDs, magnetic massive stars are generally seen to be overluminous in the radio relative to the
G\"udel-Benz relation (Section~\ref{gudelbenz})
by up to several orders of magnitude
(e.g., \citealt{2017MNRAS.467.2820L}, \citeyear{2018MNRAS.476..562L}). 

\subsubsection{Quiescent (Incoherent) Radio Emission\protect\label{OBquiescent}}
For the quiescent component of the emission from magnetic OB stars (primary gyrosynchrotron in origin), an early
study by \cite{1992ApJ...393..341L} proposed
an empirical relationship between the radio luminosity, the
magnetic field
strength, and the stellar mass-loss rate. However, more recent studies using larger samples ($\sim$30--50 stars)
have now shown that the radio luminosity is a strong function of both magnetic flux and stellar
rotation period---with no dependence on either stellar temperature or mass-loss rate
(\citealt{2021MNRAS.507.1979L}; \citealt{2022MNRAS.513.1429S}). This result led \cite{2021MNRAS.507.1979L} to propose a new model for the production
of radio emission in magnetic massive stars, namely that the (incoherent) radio emission is produced from a
``radiation belt'' (or ``shell''), similar to what is now found for UCDs (see Section~\ref{belts}).
In this scenario, the radio emission arises interior to the
closed magnetosphere, hence the wind in not expected to play an important role. As to what produces the required
energetic electrons in this shell, as related by Das,
the emerging consensus is that the responsible mechanism is centrifugal
breakout (CBO), and \cite{2022MNRAS.513.1449O}
have provided a theoretical framework that predicts that the
radio luminosity should be proportional to
the power provided by CBO in a co-rotating magnetosphere.
In this picture, all magnetic massive stars have a region in
their magnetosphere where
the outward centrifugal force is stronger than the inward gravitational pull.
When a critical density of material
is reached within the magnetosphere, the magnetic field lines will break open,
leading to the escape of plasma from the star. When the magnetic field lines
reconnect, part of the energy will go into electron acceleration and in turn, radio emission (similar to the
process that occurs in solar/stellar flares). Since these breakouts are not resolved in time by current observations, the resulting radio
emission appears quasi-constant. Interestingly, Das noted that this same picture seems to also work for much cooler and
lower-mass objects, extending down to the planet Jupiter.
However, an outstanding puzzle is that for this scenario to work,
the magnetic field must behave as a
monopole; if it behaves instead as a dipole, there would be an implied dependence of the radio luminosity on
mass-loss rate, which is not observed.

\subsubsection{Coherent Radio Emission\protect\label{OBcoherent}}
\begin{figure*} [ht!] 
\plotone{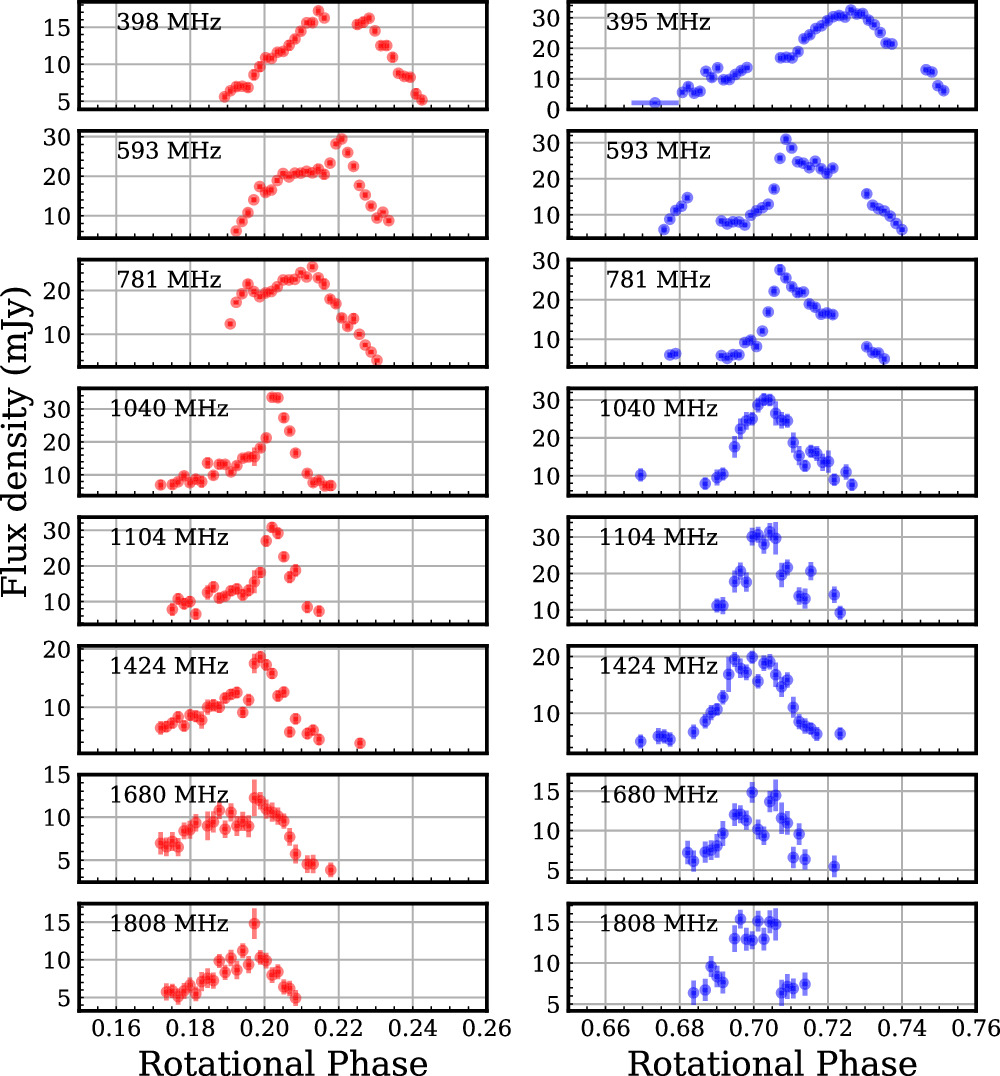}
\caption{ECM pulses observed at several different frequencies ($\sim$400--1800~MHz)
  from the magnetic massive star HD 133880. Data were obtained using the uGMRT and the VLA.
  Right circular polarization emission is shown
  in red (left panels, observed near the magnetic null at rotational phase 0.175) and left
  circular polarization is in blue (right panels).
  The pulses at different frequencies are seen shifted in rotational phase relative to
  one another, as they originate from different regions of the magnetosphere. Credit:
  \cite{2024ApJ...974..267D}. 
\label{fig:das-fig}}
\end{figure*}

The presence of coherent, pulsed radio emission from magnetic OB stars was first discovered
by \cite{2000A&A...362..281T}  in
the late B-type star CU~Vir. This emission was seen to be nearly 100\% circularly polarized and was observed
close to the magnetic nulls (i.e., when the modulating line-of-sight magnetic field was close to zero).
The origin of these pulses was identified as ECM emission (\citealt{2000A&A...362..281T},
\citeyear{2008MNRAS.384.1437T}), the same mechanism
as is believed to be responsible for the auroral radio emission from UCDs (see Section~\ref{BDaurora}).
As in  the
case with UCDs, the radio emission frequency is proportional to the
magnetic field strength, and emission at higher frequencies is
produced closer to the star than emission at lower frequencies (e.g., Figure~\ref{fig:das-fig}).
The emission is also highly beamed,
and similar to
the case of pulsars, can only be seen with the beam sweeps through the observer's line-of-sight. 
\cite{2021ApJ...921....9D}  dubbed these stars ``main-sequence radio pulse emitters'' (MRPs), and this phenomenon is now
believed to be ubiquitous among magnetic massive stars (\citealt{Das2022a}).

As reported by Das, 
more extensive observations of MRPs over the past several years have uncovered a number of new and unexpected
phenomena. For example, observations of the B star HD~133880 by \cite{Das2020a}
over a wide range of frequencies (300--4000~MHz)
using the GMRT and the VLA showed that the RCP and LCP components of the pulses (which are
present near the magnetic nulls that appear twice during each rotation phase
and arise, respectively, from the two different hemispheres of the star)
appear closer together in rotation
phase at higher frequencies, as  predicted by current working models (\citealt{2011ApJ...739L..10T};
\citealt{2016MNRAS.459.1159L}).  However, the {\it shapes} of the observed pulses are different during the two magnetic nulls. In
the case of the star CU~Vir, \cite{2021ApJ...921....9D} observed the star over the entire rotational period (not just
near the nulls) and found secondary highly circularly polarized pulses to sometimes occur away from the
primary pulses arising at the magnetic nulls.

Das noted that we are likely still missing several pieces of important physical information in understanding
magnetic massive stars. One such factor is obliquity---i.e., the misalignment between the rotation axis and the
magnetic axis the star. When these two axes are aligned, most of the magnetized plasma is expected
to be confined to
the equatorial regions. However, a significant misalignment in these axes may result in a much wider 
angular extent of the plasma, forcing ECM pulses to pass through regions of high-density plasma en route to the
observer and explaining the appearance of secondary pulses that do not occur in pairs (see
\citealt{Das2020b}, \citeyear{2024ApJ...974..267D}).
Das  expressed hope that eventually this concept, coupled with observational constraints on the presence
of secondary pulses (or lack thereof), can be used to constrain the plasma density in the magnetosphere.

Another trend discussed by Das is the role of stellar effective temperature on the relative
strength of coherent versus incoherent radio emission. She
noted that for magnetic massive stars with temperatures $T_{\rm eff}\lsim$18,000~K, the luminosity of the two types of
emission is reasonably well correlated, but for hotter stars, the ECM emission seems to be suppressed
\citep{Das2022b}.
The explanation for this is currently unclear; indeed, the incoherent emission itself appears to show no
dependence on temperature.

\subsubsection{Radio Flares\protect\label{OBflaring}}
In addition to quasi-stable quiescent emission (Section~\ref{OBquiescent}) and pulsed, coherent
emission (Section~\ref{OBcoherent}), magnetic massive stars have been found to exhibit a third
category of  radio emission: flares. Recent and ongoing
work on this topic was presented by E. Polisensky (Naval Research Laboratory).

Because magnetic massive stars possess very stable magnetospheres,
previously detected optical and X-ray
flares were generally attributed to low-mass companions. However, it was subsequently recognized that
flaring could also result from CBO events (\citealt{2005MNRAS.357..251T};
see also Section~\ref{OBquiescent}). In this
picture, the Kepler radius lies interior to the Alfv\'en radius, leading to trapping of the stellar wind
plasma. This trapped plasma is then forced to co-rotate with the star and collapses into a disk
\citep{2022MNRAS.513.1429S}.  Once
centrifugal force overcomes gravity, plasma can move outward to the point that a reconnection event occurs,
leading to a flare. 

Although the possibility of CBO-induced flaring had been theoretically predicted
\cite[e.g.,][]{2022MNRAS.513.1429S},
there had been little observational
evidence of such flares prior to the study of \cite{2021ApJ...921....9D}, which reported for the first time
{\it transient} radio emission in the magnetic massive star
CU~Vir based on 500--800~MHz data from the GMRT. The detected flares exhibited durations of $\sim$4--8~min
and hallmarks of coherent emission ($T_{B}>10^{14}$~K, in addition to circular polarization).
As described by Polisensky,
this study inspired a follow-up search for additional transient events in magnetic massive stars using data from
the VLA Low-band Ionosphere and Transient Experiment (VLITE) Commensal Sky Survey (VCSS; \citealt{2015fers.confE..19C}).  
VLITE commensally operates
during most other scheduled VLA experiments conducted at higher frequencies, and at the time of the meeting, it
had been operating for over 9 years on a subset of 18 of the VLA antennas,
accumulating
data at a frequency of $\nu\approx$340~MHz and with a bandwidth of 40~MHz. 

Polisensky and his team recently focused on the VLITE data obtained in parallel with observations performed as
part of the VLA Sky Survey (VLASS; see \citealt{2020PASP..132c5001L}), which was conducted at $\nu\sim$3~GHz.
Using data from the VCSS, which has a 5$\sigma$ sensitivity
limit of $\sim$50~mJy, they searched for counterparts to 
761 catalogued magnetic O, B, and A stars from Shultz et al. (in prep.) and found 3 matches.
Each cross-matched detection was seen  with a flux density $\sim$100~mJy in only a single data
epoch \citep{2023ApJ...958..152P}. The detected
emission was coherent and persisted for of order a few minutes. All three of the detected stars are
believed to have centrifugal magnetospheres, and the luminosities of the flares
($>10^{18}$--19$^{19}$~erg s$^{-1}$ Hz$^{-1}$)
appear to rule out an origin tied to low-mass companions. Polisensky noted that
while these data are not yet sufficient to show
definitively that magnetic massive stars undergo CBO flares, the three cases so far provide interesting
candidates for follow-up monitoring. Searches for new detections were also planned
for additional epochs of VCSS data expected  to  soon be available.

\subsection{Future Work and Unsolved Puzzles}
As reported by Das, current radio facilities are  continuing
to deliver important new results in the area of magnetic massive star research, including the
recent discovery
of 20 new such objects in datasets obtained by the
Variables and Slow Transients Survey (VAST) with the Australian Square Kilometre Array Pathfinder (ASKAP).
However, to gain new insights into many of the unsolved puzzles concerning magnetic massive stars,
Das stressed the need for more extensive, multi-frequency radio observations that
extend to both higher  ($>$50~GHz) and lower ($<$200~MHz) frequencies than most current studies, along with
spatially resolved VLBI
observations. The latter were first attempted in the 1980s \citep{1988Natur.334..329P}, 
but with limited success owing to
insufficient sensitivities of VLBI arrays of this era. This is expected to change with the planned new
VLBI capabilities of the ngVLA
(Section~\ref{ngVLA}). 

\section{Evolved Stars\protect\label{evolved}}
\subsection{Red Giants and Hypergiants\protect\label{redgiants}}
Radio observations have long provided a powerful tool for studying cool, mass-losing giants, 
including red supergiant (RSG) stars (the late-stage descendents of stars  with masses $\sim10-30M_{\odot}$) and
asymptotic giant branch (AGB) stars
(descendents of stars with initial masses in the range
$\sim$0.8--8.0~$M_{\odot}$).  The cool temperatures of these red giants
($\sim$2000-3000~K for AGB stars and $\sim$3500--4500~K for RSGs)  are
conducive to the formation of dust and a wide range of molecular species in their atmospheres.
These stars are also typically characterized by
high rates of mass loss (${\dot M}\sim10^{-8}$ to $10^{-4}~M_{\odot}$ yr$^{-1}$) through dense,
low-velocity winds which produce enormous circumstellar
envelopes (CSEs) of gas and dust. These CSEs can in turn be probed through a multitude of molecular lines
at centimeter through (sub)millimeter wavelengths  (e.g., \citealt{1988gera.book..200T};
 \citealt{2018A&ARv..26....1H};  see also Section~\ref{chemistry}). Furthermore, some of these molecular lines may exhibit
masing behavior, enabling ultra-high angular resolution studies of the outer atmosphere using VLBI
techniques
(see also Section~\ref{masers}), as well as a means to detect these stars out to significant distances
for use as probes of the structure and dynamics of the Galaxy (see Section~\ref{probes}).

The extended atmospheres of AGB and RSG stars, which can have diameters of up to several au,
also give rise to thermal continuum emission, detectable across centimeter and (sub)millimeter bands. 
For AGB stars, this thermal (free-free) emission arises from a region known as the radio
photosphere, lying at $\sim2R_{\star}$, where $R_{\star}$ is the classical photospheric radius
\citep{1997ApJ...476..327R},
while for RSG stars the thermal continuum (particularly at centimeter wavelengths)
arises predominantly from the
chromosphere, and samples a cooler component of the chromospheric
gas than the emission lines traditionally observed in the optical and UV
(\citealt{1998Natur.392..575L}; \citealt{2020A&A...638A..65O}; \citealt{2022ApJ...934..131M}).
As described in the presentation
by A. Richards (Jodrell Bank Centre for Astrophysics/University of Manchester), in both AGB and RGB stars, to a first approximation,
the observed brightness temperature is correlated with frequency owing to the wavelength dependence of the opacity
 \citep{1997ApJ...476..327R};
 higher frequencies (shorter wavelength)
effectively probe deeper layers of the star, hence the stars appear
larger at longer radio wavelengths and smaller at shorter wavelengths.
However, as highlighted in the poster presentation by  B. Bojnordi Arbab
(Chalmers University),
the relationship between the radius and stellar
brightness temperature is complex and not yet fully understood (see also Section~\ref{AGB}). 

G. Umana reminded attendees that
both the VLA and ALMA currently have the sensitivity to detect thermal radio emission from AGB and RSG
stars within a few kpc, and the longest baseline configurations of these two arrays are able to
modestly spatially resolve the radio-emitting atmospheres of such stars within $\lsim$200~pc (e.g.,
\citealt{1997ApJ...476..327R}, \citeyear{2007ApJ...671.2068R}; \citealt{2018AJ....156...15M};
 \citealt{2020A&A...638A..65O}), although at
wavelengths $\lambda\lsim$3~mm the angular
resolution is rather limited (typically $\lsim$2--3 synthesized beams across the stellar disk). In the future, however, Umana
noted that the combined thermal sensitivity and
angular resolution of ngVLA (Section~\ref{ngVLA}) should make it
possible to readily identify the presence of asymmetries, hot spots, and other surface features in a large sample of red giants (see also \citealt{2018ASPC..517..281M})---and make ``movies'' of how
these features evolve during the course
of the stellar pulsation cycle \citep{2019arXiv191000013A}.  Importantly, the frequency ranges
covered by ALMA+ngVLA will be complementary, since owing to the wavelength
dependence of the opacity, different wavelengths probe different depths in the atmosphere, enabling
the characterization of temperature with depth (e.g.,
\citealt{1998Natur.392..575L}; \citealt{2017A&A...602L..10O}, \citeyear{2020A&A...638A..65O}).
Additionally, though SKA-Mid  (which is expected to have a maximum angular
resolution of $\sim$27~mas; Section~\ref{SKA}) will not match the maximum spatial resolution of ngVLA ($\sim$1~mas), it is expected
to  contribute crucial radio light curves and
multi-frequency measurements of SEDs at $\nu\lsim$15~GHz to help
constrain atmospheric models of AGB and RSG stars and their mass-loss mechanisms \citep[e.g.,][]{2004NewAR..48.1349M}. Such measurements are 
extremely difficult  to carry out with current instruments owing to a combination of sensitivity limitations
and logistical challenges (cf. \citealt{1997ApJ...476..327R}).

\subsection{Recent Results from Interferometric Surveys\protect\label{redgiantsurveys}}
C. Gottlieb (CfA) presented an overview the an ALMA Large Project
dubbed ATOMIUM (ALMA Tracing the Origins of Molecules In dUst-forming oxygen-rich M-type stars) in which
he has been a key team member (\citealt{2022Msngr.189....3D}; \citealt{2022A&A...660A..94G}). The overarching goals of ATOMIUM were to improve our quantitative understanding of
the physical and chemical  processes  that govern the inner winds and outflows of oxygen-rich evolved stars, including
the formation of dust.
As described by Gottlieb, the
project targeted a sample of 14 AGB and 3 RSG stars spanning a range in mass-loss rates and other properties
using wideband (214--270~GHz) observations with ALMA. The use of multiple ALMA configurations provided access to spatial
information on scales ranging from $\sim$25~mas to $\sim 8''$, aiding the study of the
complex and spatially extended CSEs of nearby evolved giants. The ATOMIUM team found a
large morphological diversity in the
molecular ejecta of their sample stars, which they interpreted as strong evidence for the role of a companion in shaping
the ejecta of AGB stars \citep{2020Sci...369.1497D}---i.e., well before the onset of the planetary nebula (PN) stage, where
companions have long been assumed to play a key role in shaping the ejecta. The ATOMIUM data also provided
a wealth of information for studying the astrochemistry of the circumstellar environments of each
source (see Section~\ref{chemistry}).

A. Richards  presented some additional
results from ATOMIUM, including a statistical analysis of the derived stellar diameters
at $\nu\sim$250~GHz based on uniform elliptical disk fits to the visibility data. She noted that 
in terms of their diameters, the stars appear to fall into groups corresponding to their luminosity class.
However, two RSGs in the sample (AH~Sco and VX~Sgr) have 250~GHz
brightness temperatures comparable to the majority of the AGB stars, despite the
higher effective photospheric temperatures of the RSGs. Additionally
she reported that at the frequencies targeted by ATOMIUM, the emission from the AGB stars seem to require a
chromospheric contribution in addition to that from the radio photosphere (see Section~\ref{redgiants}).

Richards also previewed some work on
the analysis of the SiO masers and high-excitation H$_{2}$O masers
detectable within the 215--265~GHz band covered by ATOMIUM (see also
\citealt{2024IAUS..380..386P} and Section~\ref{masers}). The ATOMIUM data
have sufficient spatial resolution to allow tracing the motions of some individual SiO maser-emitting features,
revealing evidence for both inflow and outflow. Likewise, the high-excitation H$_{2}$O lines,
which arise at $\sim$2--3$R{\star}$ show evidence of both infall and outflow motions
(\citealt{2022IAUS..366..199E}; \citealt{2023A&A...674A.125B}). 

\subsection{Studies of Individual Stars}
\subsubsection{Red Supergiants (RSGs)\protect\label{RSGs}}
\paragraph{Betelgeuse} Richards presented 
new results that have emerged from the recent spatially resolved observations of the nearby RSG
star Betelgeuse ($\alpha$~Orionis)
at frequencies ranging from $\sim$6--485~GHz based on observations from ALMA and the enhanced Multi-Element
Remotely Linked Interferometer Network (eMERLIN). Many of the new
observations were performed  in 2023, motivated in part
by the desire to obtain an accurate position for the star in advance of its occultation by
the asteroid Leona, which occurred on December 12, 2023 \citep[e.g.,][]{2023ATel16374....1C}.
At centimeter wavelengths ($\nu\sim$6~GHz) Richards
reported that the radio emission exhibits significant variations over time, though the exact
cause of the variations is unknown. One possibility raised by Richards is carbon recombination.
Consistent with this, recent work presented in the
poster from W. Dent (Joint ALMA Observatory) showed evidence of very little ionized carbon
in the atmosphere of Betelgeuse. This is based on analysis and modeling by Dent, Richards, and collaborators of a recent detection of the
H30$\alpha$ Rydberg line at 231.905~GHz in the atmosphere of Betelgeuse, along with a second
line at 232.025~GHz (X30$\alpha$) that is attributed to a blend of Rydberg transitions from
metals with low first ionization potentials \citep{2024ApJ...966L..13D}.

At higher frequencies, new images obtained by Richards's team show that the ``hot
spot'' previously seen in the northeast quadrant of the star
in 338~GHz data from 2015  \citep{2017A&A...602L..10O}
still appears to be at a similar location $\sim$8~yr later, although the contrast between the hot spot
and the surrounding emission is less pronounced in the more recent data. New ALMA images presented
by Richards also allowed for the first time an estimate of the spectral index of the hot
spot; preliminary indications are that at 233--338~GHz it is steeper ($\alpha\sim1.9$) than
the mean spectral index of the star (see below).

Richards reported that in a global sense,
the millimeter and submillimeter emission of Betelgeuse is reasonably fitted by a uniform elliptical disk, but divergences 
from this simple model were seen on various scales; on small scales this is suggestive of
the presence of one or more hot spots, while on larger scales there is evidence of an additional
layer around the star. The exact nature of the latter component is uncertain. One possibility
is emission from dust, although Richards
suggested that based on its uniformity, it is perhaps more likely to be a cool free-free-emitting plasma component.

Another finding reported for Betelgeuse by Richards is that the  brightness temperatures derived from
multi-wavelength radio observations appear to be
systematically diminished by of order a few hundred Kelvin
subsequent to the so-called ``Great Dimming''   that
occurred in late 2019/early 2020 \citep{2019ATel13341....1G} and which led to a historically large drop
in brightness at visible wavelengths (see e.g., \citealt{2020ApJ...899...68D}).  This echoed complementary results
presented in the poster by L. Matthews (MIT Haystack Observatory) based on ALMA and VLA measurements
(see also \citealt{2022ApJ...934..131M}; \citealt{2024AAS...24340901M} and in prep.). The presentations from Richards and Matthews
also highlighted how the disk-averaged brightness temperature of Betelgeuse rises from $\sim6R_{\star}$
(sampled by centimeter-wave observations) to $\sim2R_{\star}$ (sampled by millimeter-wave measurements), but
then falls near $\sim1.2R_{\star}$, as probed by submillimeter observations. A similar trend was previously
reported by \cite{2017A&A...602L..10O},  but additional measurements from ALMA more clearly define
the changes in temperature as a function of radial distance
and should help to better constrain future models.
Richards pointed out that it is also intriguing
that despite changes in brightness temperature over time, the spectral index
of Betelgeuse between $\sim$5--500~GHz has remained remarkably stable ($\alpha\sim1.4$)
over more than 20 years (e.g., \citealt{2015A&A...580A.101O}, \citeyear{2017A&A...602L..10O};
\citealt{2022ApJ...934..131M}; Matthews et al. in prep.).

\begin{figure*} [ht!] 
\plotone{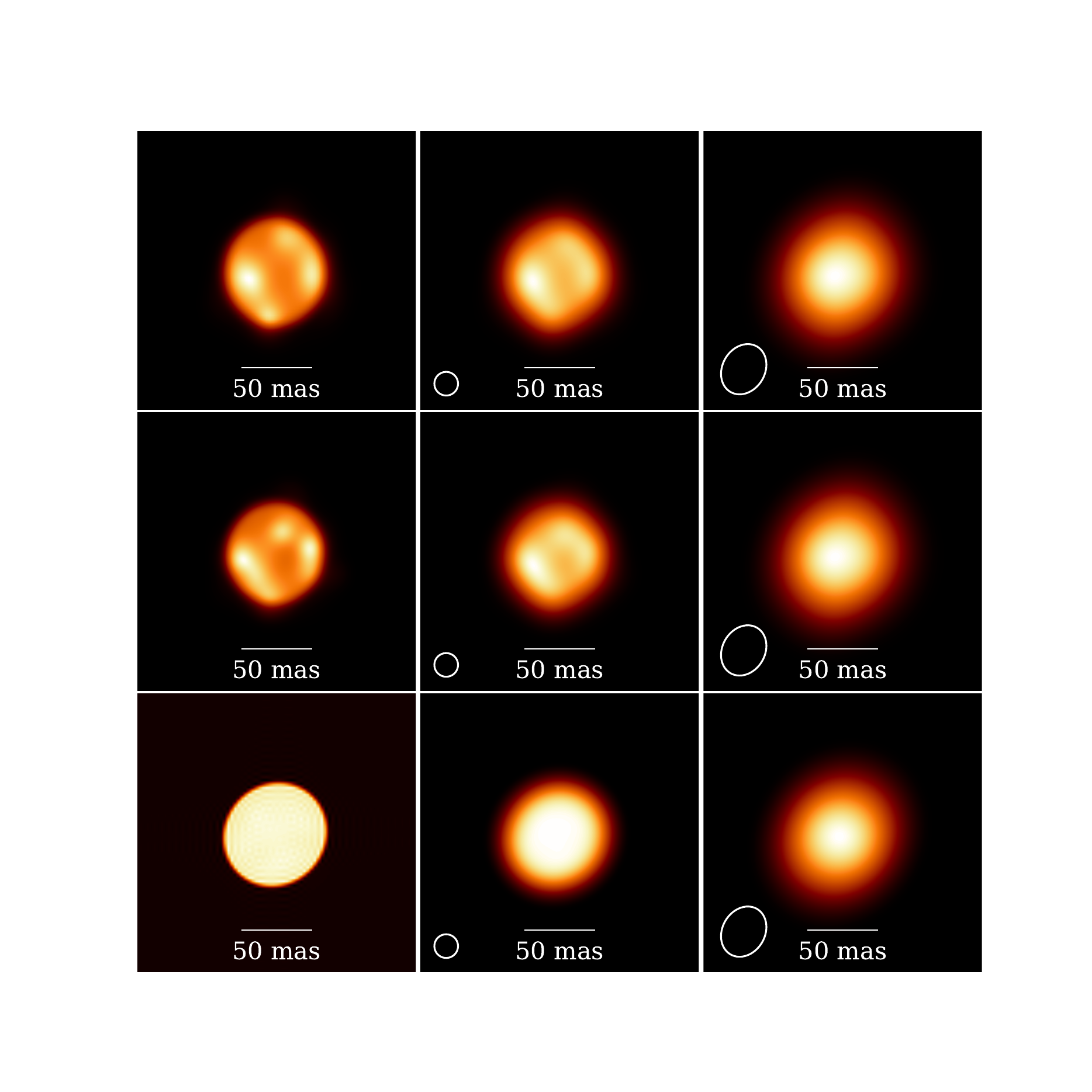}
\caption{Super-resolved ALMA images of Betelgeuse at 107~GHz (top) and 136~GHz (middle row), obtained with the
  Bayesian interferometric imaging package {\sc Comrade.jl} from \cite{2022JOSS....7.4457T}. Mean posterior images are shown, calculated
  using a prior based on a stationary Gaussian Markov random field model. The bottom row shows a reconstruction
  of a uniform disk model (with realistic noise) as a comparison. The center and right panels in
  each row show the reconstructions blurred by Gaussians. From L. D. Matthews, K. Akiyama, et al., in prep.
\label{fig:matthews-alphaOri}}
\end{figure*}

The poster presentation of Matthews also showcased the first spatially resolved
images of Betelgeuse at 107 and 136~GHz ($\lambda$2--3~mm; Figure~\ref{fig:matthews-alphaOri}).
The images were created using a regularized maximum
likelihood imaging technique that allowed moderately super-resolving the star.
Use of this technique has revealed that the radio surface of the star appears markedly
different at wavelengths of $\lambda$2--3~mm compared with $\lambda$0.85--1.3~mm. Given
the qualitative similarities between $\lambda$0.85~mm images obtained several
years apart, the differing appearance in the respective bands 
cannot be readily be attributed entirely to temporal variations. 

Using comparisons with sophisticated 3D modeling, J.-Z. Ma
(Max Planck Institute for Astrophysics) presented results aimed at exploring the question of
whether Betelgeuse is truly rotating with a projected equatorial velocity of
$v_{\rm eq}~{\rm sin}~i\approx5.5$~\kms, as had been claimed previously based on evidence
from  spatially resolved ALMA observations of molecular line (SiO) emission \citep{2018A&A...609A..67K}.
Such a rotational velocity is surprisingly high for a RSG star; even if the main sequence progenitor
was rapidly rotating, conservation of angular momentum during the transition
to the RSG phase (which involves an increase of radius of two orders of magnitude)
is predicted to slow rotation to $\lsim$0.1~\kms---i.e., two orders of magnitude lower than the observations imply \citep{2024ApJ...962L..36M}. 
One previously proposed explanation is that Betelgeuse resulted from a stellar merger (e.g.,
 \citealt{2017MNRAS.465.2654W}; \citealt{2024ApJ...962..168S}), but Ma and collaborators put forward a competing idea. Ma
pointed out that if the bright regions or ``hot spots'' observed in spatially resolved radio
images of RSG stars such as Betelgeuse  (see \citealt{2017A&A...602L..10O}  and above) result from convection,
then it is predicted that these features should change on timescales of months (commensurate with
the convective turnover timescale of the star).

Using 3D hydrodynamic (non-rotating) {\sc CO5BOLD}
simulations of convection in RSG stars (\citealt{2011A&A...528A.120C};
\citealt{2012JCoPh.231..919F}), 
coupled with chemical modeling as prescribed by the {\sc FastChem2} code
of \citeauthor{2018MNRAS.479..865S} (2018, \citeyear{2022MNRAS.517.4070S}),  Ma and his colleagues produced model images of the
continuum and spectral line emission of Betelgeuse, convolved to the angular resolution appropriate
for ALMA. They found that the finite spatial resolution (beam smearing) of the ALMA
observations can mimic the signatures of rotation and produce spurious rotational velocities of $\ge$2~\kms\
in $\sim$90\% of their test cases. Ma noted that a follow-up study of Betelgeuse using ALMA's highest resolution
configuration is needed to confirm this observationally.
He also highlighted the value of future
ALMA observations of Betelgeuse with even higher spatial resolution that would
enable spatially resolved velocity fields of various molecular lines.
Since different molecular
lines (e.g., different transitions of SiO) probe different layers of the atmosphere, this would in
effect allow velocity field tomography of the star and enable searches for the expected signatures
of the different physical processes that impact these various layers (e.g., rotation, shock waves, wind
launching; see \citealt{2024ApJ...962L..36M}). 

\paragraph{VY CMa}
Richards presented a poster  led by collaborator R. Humphreys (University of Minnesota)
that described recent ALMA
spectral line observations  of the red hypergiant VY~CMa covering 230--250~GHz. VY~CMa has been previously
extensively studied at multiple wavelengths
as an archetype for understanding late-stage stellar mass loss
owing to the star's high current mass-loss rate, complex chemistry, and extensive circumstellar ejecta. 
The new ALMA data (with angular resolution of $\sim$200~mas),
combined with archival ALMA measurements and a VLBI-determined distance to the star,
have now enabled the derivation
of line-of-sight velocities and proper motions (over $\sim$6~yr) of four newly discovered
$^{12}$CO-rich clumps near the star. The existence of these clumps, along with other clumps previously detected
with the {\it Hubble Space Telescope} ($HST$), imply that the star underwent at
least 6 outflow events within a 30~yr period, roughly 60--100~yr ago,
resulting in the loss of $\ge0.07~M_{\odot}$ of material (\citealt{2024AJ....167...94H};
see also \citealt{2023ApJ...954L...1S}).

Turning to smaller scales,
Richards also drew attention to ALMA Band~9
(670~GHz) observations of VY~CMa with 9~mas angular resolution \citep{2020ApJS..247...23A}.
Surprisingly, those observations showed primarily dust, with no clear detection of the stellar continuum;
indeed, Richards noted that VY~CMa may be the dustiest known Galactic RSG star.
Richards and colleagues have also used ALMA to study the H$_{2}$O masers around VY~CMa
and found evidence of both inflow and outflow \citep{2020ApJS..247...23A},
as well as radial velocity gradients within
spoke-like features \citep{2024IAUS..380..389R}. Their data also showed 268~GHz H$_{2}$O
masers confined to a bow-shock-like structure, suggesting a recent clump ejection.
Finally, Richards reported the recent detection of a X30$\alpha$ Rydberg (recombination-like)
line arising from
high-excitation metals. However, compared with the Rydberg lines recently detected
in Betelgeuse (see above), those associated with VY~CMa are much more spatially extended and have a much
narrower linewidth. Furthermore, the hydrogen H30$\alpha$ line is so far undetected in this source.

\subsubsection{AGB Stars\protect\label{AGB}}
A poster by B. Bojnordi Arbab summarized recent work aimed at comparing the predictions of
1-D DARWIN atmospheric models of AGB stars from \citeauthor{2016A&A...594A.108H} (2016, \citeyear{2022A&A...657A.109H}) 
with multi-frequency radio measurements from the VLA and ALMA. He showed examples of predictions for key
observable properties of these stars, including
brightness temperature variations as a function
of radius, as well as flux density variations as a function of time and frequency and discussed
how comparisons with multi-frequency radio observations can be used to constrain the role of shocks
and associated opacity changes with increasing height in the atmosphere \citep{2024ApJ...976..138B}.
Additionally the poster  highlighted the prospects for
significant future advances in our ability to
study the time-varying atmospheres of AGB stars out to tens of kpc with the arrival of SKA-Mid,
and especially, the ngVLA---in combination
with ALMA---owing to the unprecedented combination of broad frequency coverage, sensitivity, and angular resolution
that this trio of facilities will bring (see Section~\ref{futureisbright}).

Other recent ALMA results related to AGB stars were highlighted in the presentation
by A. Moullet [National Radio Astronomy Observatory (NRAO)].
Among these was a Band~8--10 (397--908~GHz) study of the nearby carbon-rich AGB star
R~Lep \citep{2023ApJ...958...86A}  that
produced the highest angular resolution  (5~mas) images ever made of a stellar radio
photosphere. Additionally, the observations detected and imaged an HCN maser line
at 890.8~GHz at comparable angular resolution.
Oxygen-rich AGB stars frequently exhibit maser emission from species such as SiO, H$_{2}$O, and OH, providing a 
probe of the dynamic atmospheres of these stars
that can be observed with ultra-high angular resolution  (see also Sections~\ref{masers}), but
these lines are generally not detectable in carbon-rich AGB stars like R~Lep.
The \citeauthor{2023ApJ...958...86A}  study therefore illustrates the potential of submillimeter
HCN masers as a new tool for ultra-high angular resolution studies of carbon AGB star atmospheres
using VLBI techniques.

\subsection{Luminous Blue Variables\protect\label{LBV}}
Luminous blue variables (LBVs) are representatives of a short-lived and poorly understood post-main-sequence
evolutionary phase of very
massive stars (initial masses $\sim60-100M_{\odot}$). LBVs mark the transition from the main
sequence to Wolf-Rayet stars and are
characterized by strong variability and high rates of mass loss ($\sim10^{-6}M_{\odot}$ yr$^{-1}$),
which lead to extended circumstellar nebulae (e.g., \citealt{1994PASP..106.1025H};
\citealt{2011BSRSL..80..335U}). 
As discussed in the presentation by G. Umana, radio observations provide valuable information
on the mass-loss histories of LBVs, including
current mass-loss rates and the mass and structure  of the
circumstellar ejecta.

Among the latest developments in LBV research described by Umana was
a recent 1.3~GHz survey of the Galactic Plane with MeerKAT that
identified several new LBV candidates  and enabled detailed imaging of their extended
ejecta (\citealt{2024MNRAS.531..649G};  Umana et al. in prep.). Radio observations in additional bands 
enabled identification of the central component \citep{2025A&A...695A.144B} and provided
spectral information on the ejecta, which in turn aids in identifying a possible non-thermal emission component
that may arise from the interaction between the stellar wind and the local environment
(see also Section~\ref{interaction}).
Another new finding reported by Umana was the surprising discovery (using ALMA)
of sulfur- and silicon-bearing molecules
in the ejecta of LBVs, despite their harsh environments \citep{2022ApJ...939L..30B}.
This underscores that these stars can serve as laboratories for the
study of molecular chemistry under
extreme physical and chemical conditions. 

\section{Stars Interacting with Their Environments\protect\label{interaction}}
\subsection{Modeling Circumstellar Environments}
An invited talk by S. Mohamed (University of Virginia) showcased the results of recent high-resolution
simulations of the complex environments of evolved, mass-losing stars, with a focus on AGB stars (see also
Section~\ref{AGB}).
Modeling
such systems is enormously challenging owing to the wide range of relevant physical
processes and to the enormous range in scales involved.
Driving home the latter point, Mohamed noted that if the stellar core were
the size of a marble, the tenuous outer envelope would be a kilometer away. 

Mohamed focused in particular on what can be learned from smooth particle hydrodynamics (SPH)
simulations in which each particle represents a bit of the fluid being modeled. The
particles interact with each other according to a set of specified equations describing the relevant
physics [i.e.,
conservation of mass, energy, momentum, and an equation of state, along with equations describing
other physical processes (e.g., dust formation, chemistry, magnetic fields)]. Other
necessary ingredients include a numerical solver,
boundary conditions, a set of initial conditions, and lastly, a set of parameters and approximations to
make the problem tractable.
Interpolation between
particles using a smoothing kernel is then used to approximate a smooth fluid.

Mohamed showed examples of SPH simulations covering three types of processes of relevance to
AGB stars, including
stellar interaction with a nearby binary companion,
wind-wind interactions, and wind-ISM interactions.
She noted that
ALMA has been a game-changer for allowing direct confrontation of such models with observations;
indeed, researchers have been able to use the increasing body of exceptionally detailed ALMA
images of evolved stars and planetary nebulae (PNe) and their circumstellar environments
to constrain models of the wide range of physical processes that underlie such objects,
and also improve our understanding  of stellar evolution and cosmic recycling.

In the case of AGB stars with a binary companion, simulations by Mohamed and her collaborators
have shown that as a companion moves through the AGB star's wind it will
shape the wind into an Archimedes spiral. 
Importantly, spacing between the spiral arms will depend on the orbital period, while 
the mass of the companion impacts the density ratio between the arm and inter-arm regions
of the spiral.
Detection and measurement of such structures therefore provide a means of inferring the
existence of companions of AGB stars that are otherwise difficult or impossible to detect
owing to the enormous luminosity of the AGB star itself---and moreover, to
constrain the mass and orbit of the companion. One spectacular example of this---the case of the AGB
star R~Scl \citep{2012Natur.490..232M}---was presented
at the first Haystack Radio Stars meeting in 2012 (see \citealt{Matthews2013}),
and additional examples have since been
identified with ALMA (e.g., \citealt{2017A&A...605A.126R}; \citealt{2020A&A...633A..13D}).

Recently, \cite{2022MNRAS.513.4405A} modeled the case of 
substellar/planetary mass companions and found that such companions can have an impact on the
mass-loss rate of an AGB star. They also showed that if
there are resonances between the orbital period and the
pulsation period, multiple spiral structures can form. Mohamed prognosticated that with future
higher sensitivity, high angular resolution (sub)millimeter observations (such as will become
possible thanks to ongoing upgrades to ALMA; see Section~\ref{ALMA2030})
it will finally become possible to directly observe
some of these predicted structures---not only in AGB star binaries, but even in binaries
with lower wind densities
where the spiral signatures are expected to be much more subtle  (e.g., Wolf-Rayet stars).

Mohamed also discussed the case of the symbiotic binary Mira~AB, comprising an AGB star
and a white dwarf companion separated by $\sim$60~au. Given the relatively wide
separation of this pair, and the fact that the AGB star does not fill its Roche lobe,
the strong focusing of material from the AGB
wind onto the companion first observed in the UV and
X-rays was initially surprising (\citealt{1997ApJ...482L.175K}, \citeyear{2005ApJ...623L.137K}).
However, \cite{2012BaltA..21...88M} 
showed that this could be explained by a process dubbed ``wind Roche-Lobe
overflow'' (WRLOF). In this case, if the wind has a low velocity and its acceleration zone lies
at a significant fraction of the Roche lobe radius, material
can more readily get channeled into the potential well of the
companion, resulting in a higher accretion rate compared with the classic Bondi-Hoyle accretion
scenario. Evidence of this has since been seen in ALMA observations of other
evolved stars (e.g., \citealt{2018A&A...616A..61R}; \citealt{2020A&A...633A..13D}).
Mohamed noted that WRLOF can also play a role in the formation of certain chemically
peculiar stars \citep[e.g.,][]{2013A&A...552A..26A}.

Mohamed explained that realistic SPH modeling of a system as complex as Mira~AB requires the inclusion of
a broad range of physical processes including the 
wind acceleration, radiation pressure on
the dust, as well as  cooling and accretion. In such a case the resulting spiral structure
is no longer 3D, but more confined to the orbital plane.
Various arcs and cavities are also predicted to form, and such features have been confirmed, for example, through
CO observations of the Mira~AB system obtained with ALMA \citep{2014A&A...570L..14R}.
However, the real data
include additional surprises, such as  the presence of a large bubble not seen in the simulations. 
Mohamed suggested that this may be a result of the hot wind of Mira~A's white dwarf companion.

\subsection{Runaway Stars}
J. van den Eijnden (University of Warwick) spoke on the importance of current and future
radio wavelength observations for the identification and study of bow shocks associated
with massive (O and B) ``runaway'' stars. Runaway stars are traveling supersonically through the ISM and have
escaped from the clusters in which they originally formed \citep[e.g.,][]{2024Natur.634..809S}.
His talk highlighted how such studies complement high-energy
($\gamma$-ray) studies of bow shocks and provide insights into how massive stars interact
with their local environments.
The strong stellar winds of OB stars and the high space velocities at which runaway stars are moving
mean that the latter interactions are often strong.

The primary band for identification of bow shocks associated with runaway OB stars is currently
the infrared, where the bow shock regions emit owing to thermal radiation from heated dust grains.
However, as noted by van den Eijnden, the mechanical energy budget available in these regions ($\sim10^{50}-10^{51}$~erg,
  comparable to that of a supernova explosion, but with emission over a much longer timescale)
  is also sufficient to power nonthermal
  high-energy physical processes, including diffuse shock acceleration, whereby charged particles
  bounce back and forth in a magnetic field; the particles cross the shock front and continue to gain
  energy with each passage. At radio wavelengths, this leads to the production of synchrotron emission, while
  at the same time, inverse Compton scattering gives rise to $\gamma$- and X-ray emission.
  However, van den Eijnden stressed that the identification of these nonthermal signatures is challenging; among hundreds
  of known infrared-emitting bow shocks associated with massive stars, until recently,
  only a single radio counterpart to one of these bow shocks had been detected
  \citep{2010A&A...517L..10B}. 
  
This has finally begun to change thanks to the wide fields-of-view and exquisite
surface brightness sensitivity available from facilities such as ASKAP and MeerKAT, and the sample of
detected radio bow shocks now stands at $\sim$10 (\citealt{2022MNRAS.512.5374V}a,
\citeyear{2022MNRAS.510..515V-A}b; \citealt{2024MNRAS.532.2920V-B};
and in prep.). Van den Eijnden further noted that this is likely to
  represent only the ``tip of the iceberg''. Based on the observed properties of the currently
  sample, it is predicted that future deep observations with MeerKAT (and eventually SKA-Mid; see Section~\ref{SKA}),
  are expected to detect synchrotron emission from the bow shock regions of nearly all currently known
  runaway OB stars and enable determinations of the radio spectral indices (which are
  lacking for the current sample) and polarization
  signatures (if present).
  He also highlighted the value of future synergistic observations
  with $\gamma$-ray observations [e.g., with the planned Cherenkov Telescope Array (CTA)]
  hold the promise to provide new insights into the stellar wind physics and particle
  acceleration processes in these sources.

  One challenge in interpreting radio continuum observations of bow shocks and extracting meaningful physics
  is ``proving'' that the radio emission is at least partly
  nonthermal. Van den Eijnden reported that one way of doing this is to combine radio and H$\alpha$ observations
  and evaluate whether both the radio flux density and H$\alpha$ surface brightness can be
  accounted for by temperatures and electron densities that are consistent with thermal emission.
 Another is to estimate the electron injection efficiency (\citealt{2022MNRAS.510..515V-A}b).

 Additional work related to stellar bow shocks was presented by
 V. Yanza Lopez (National Autonomous University of Mexico), who discussed the radio-emitting
 bow shock at the center of the ultra-compact \HII\ region NGC~6334A. This region has long been
 known to exhibit both a shell and a compact source that are visible in radio wavelengths
 (\citealt{1982ApJ...255..103R}; \citealt{2002AJ....123.2574C}).  However, the central radio source has remained an
 enigma, showing an arc-like shape, a negative spectral index (indicative of a synchrotron mechanism),
 and variability on the timescale of years, leading to
 the suggestion it was associated with a colliding wind region \citep{2014RMxAA..50....3R}.
 
 Using the highest resolution  VLA (`A') configuration, 
 Yanza Lopez and her colleagues performed multi-frequency (10, 22, and 33~GHz),
 multi-epoch observations of the NGC~6334A  central
 source in an attempt to distinguish whether the radio emission is arising from
 a colliding wind region or a  bow shock \citep{2025MNRAS.538.1314Y}. 
 The emission that was observed is spatially resolved, with a mean spectral index
 $\alpha\approx-0.5$. To explain the observed level of ionization 
 as a colliding wind region would require a binary comprising two massive stars,
 with approximate spectral types O7 and B0; however, at centimeter wavelengths there
 are no counterparts to these putative stars. Yanza Lopez's team also examined data
 at other wavelengths (optical, infrared, X-rays), but still no counterpart was
 seen. They then considered an alternative scenario in which the radio emission is 
 produced by the bow shock of a runaway massive star. Using archival multi-epoch VLA radio data
 they measured
 a proper motion for the compact radio source and derived a
 space velocity of 123$\pm$40~\kms, consistent with
 other runaway stars. This favors the bow shock scenario.
 Nonetheless, a remaining puzzle is that there is no apparent counterpart to the massive star
 that would have led to the formation of a bow shock. This scenario also
 cannot as readily explain the observed ionization of the surrounding shell. Yanza Lopez noted that tentatively,
 both issues could be resolved by assuming the progenitor is a 
 B0 star (identified in a {\it Herschel} 70~$\mu$m image) that has since moved away from the
 center of the shell structure. 

\section{Stellar Masers\protect\label{masers}}
A. Bartkiewicz (Toru\'n Institute of Astronomy, Nicolaus Copernicus University) presented an
invited overview of the role of cosmic molecular masers for studying various phases of stellar evolution. 
She described how owing
to its inherently beamed nature, maser emission is able to escape from dense regions and deeply embedded
sources, providing a diagnostic of the kinematics and physical conditions near
star-forming regions and evolved stars. 
The non-thermal, high brightness temperature nature of the maser emission also means that it
can be observed at ultra-high angular resolution using VLBI techniques (see below).

\subsection{Masers in High-Mass Star-Forming Regions\protect\label{YSOmasers}}
Bartkiewicz pointed to the recent review by \cite{2024IAUS..380..135U} and the study of
\cite{2024IAUS..380..218K} 
as illustrations of how different molecular
lines observed at radio wavelengths (both thermal lines and maser lines) can uniquely probe distinct phases
in the process of massive star formation ($M_{\star}\gsim 8M_{\odot}$).
She also highlighted the recent work by \cite{2024IAUS..380..230L}  suggesting that
H$_{2}$O masers appear even earlier in the process
of massive star formation than methanol masers and tend to trace outflow processes.

One point emphasized by Bartkiewicz is the value of multi-epoch monitoring of masers over timescales of months to years
as a means of tracing gas kinematics and documenting ongoing evolution of sources (e.g.,
\citealt{2020A&A...637A..15B}, \citeyear{Bart2024b}, \citeyear{Bart2024a}). For example, 
 in the nearest massive star-forming regions, multi-epoch observations
using VLBI techniques can achieve spatial resolution of
$<0.1$~au, sufficient to record gas clumps spiraling along magnetic field lines (e.g.,
\citealt{2010ApJ...708...80M}; \citealt{2022NatAs...6.1068M}).  However,
spatially unresolved maser monitoring with single-dish radio telescopes can also provide unique and valuable information. For example,
Bartkiewicz cited recent studies demonstrating that methanol maser flares at 6.7~GHz can be used to
identify episodic accretion onto high-mass protostars (e.g., \citealt{2023A&A...671A.135K};
\citealt{2023NatAs...7..557B}). 
This phenomenon had been predicted by previous theoretical studies (\citealt{2017MNRAS.464L..90M},
\citeyear{2021MNRAS.500.4448M}). 
For the case of the high-mass protostar G358-MM1, VLBI follow-up by \citeauthor{2023NatAs...7..557B}
 further revealed
that the methanol masers trace a Keplerian disk with a four-arm spiral structure that appears
to be feeding material onto the protostar.

Also on the topic of methanol masers, 
a poster by S. Paulson (Tata Institute of Fundamental Research)
focused  on the Galactic mid-infrared-emitting bubble N59, known to
host eight 6.7~GHz methanol masers. Such masers are signposts of narrow time window during the process of
massive star formation ($t\sim0.4-2\times10^{4}$~yr; \citealt{2007MNRAS.377..571E};
\citealt{2020MNRAS.499.2744B}).
Using multiwavelength observations of N59, including
archival radio observations obtained with the James Clerk Maxwell Telescope (JCMT), the Atacama Pathfinder
EXperiment (APEX) telescope, and the VLA, \cite{2024MNRAS.530.1516P}
 performed a comprehensive analysis of this region that led to the identification of multiple
\HII\ regions and massive YSOs in the region, together with a cluster of OB stars.

\subsection{Masers in AGB and Post-AGB Stars\protect\label{AGBmasers}}
The study of masers in evolved stars serves a unique role in investigating their
mass-loss process, since the masers enable probing regions within a few $R_{\star}$, where dust formation occurs and
where the wind is accelerated. They also complement other multi-wavelength observations.
Bartkiewicz showed spectacular high-resolution images of the AGB star W~Hya from
the recent study of \cite{2024A&A...691L..14O}, 
obtained contemporaneously in visible light with the Very Large Telescope (VLT)/Spectro-Polarimetric High-contrast
Exoplanet REsearch (SPHERE)-Zurich Imaging Polarimeter (ZIMPOL) and in two vibrationally excited H$_{2}$O lines with
ALMA. The  overlap between the polarized intensity maps and the H$_{2}$O emission suggests that both
dust and the H$_{2}$O emission (which is most likely masing), both originate  at $\sim2-3R_{\star}$ from cool, dense pockets of
gas within the inhomogeneous atmosphere of the star.

Another recent, novel result obtained through the study of masers was made by \cite{2020A&A...644A..61H}
using ATOMIUM data (Section~\ref{redgiantsurveys}). In the AGB star $\pi^{1}$~Gru,
\citeauthor{2020A&A...644A..61H}  found that on scales of $\lsim$\as{0}{5}
the SiO masers appear to trace the effects of a second, previously unknown
companion on the inner wind, revealing an apparent flow of gas accelerating from the AGB star surface onto
the companion.

M. Lewis (Leiden Observatory) reported on the use of data from the Bulge
Asymmetries and Dynamical Evolution (BAaDE) survey of SiO masers (see Section~\ref{probes}) to study the statistical properties
of the SiO maser-emitting stars themselves. She reminded us that
among the molecular maser species commonly found in oxygen-rich AGB stars, the SiO masers arise
closest to the stellar core and that while SiO masers are detected in evolved stars
spanning a broad range of infrared colors, they are most prevalent in Mira-type variables with thin to moderately
thick circumstellar envelopes, and less common in the subclass of very red OH-IR stars with thick
envelopes. In her presentation Lewis focused on the analysis of the SiO maser properties of
subset of 28,000
AGB star candidates observed by the BAaDE project using the VLA with a bandpass tuned to encompass 7 lines from various
SiO transitions and isotopologues between 42.3--43.5~GHz. Detections of
the $^{28}$SiO $v$=1,2 $J=1-0$ lines were the most
common, being seen in over 50\% of the sample, while the $^{29}$SiO $v$=1, $J=1-0$ line was the rarest,
occurring in just 0.1\% of the targets. In a handful of cases, all seven maser lines were detected
in  a single star (see \citealt{2024IAUS..380..314L}). 

Lewis has been analyzing the line ratios of the various detected
maser lines, and highlighted in particular the importance of the $F_{^{28}{\rm SiO}~v=1}/F_{^{29}{\rm SiO}~v=0}$
flux ratio. She noted that in most cases this ratio is $\sim$10, but her team has identified
an interesting subset of stars for which
brightness of the $^{29}$SiO isotopologue is equal to or greater than that of the $^{28}$SiO line.
The cause of these unusual ratios is not yet entirely clear, but the effect appears to be time-variable
(\citealt{Lewis2020a}).

\begin{figure*} [ht!]
\plotone{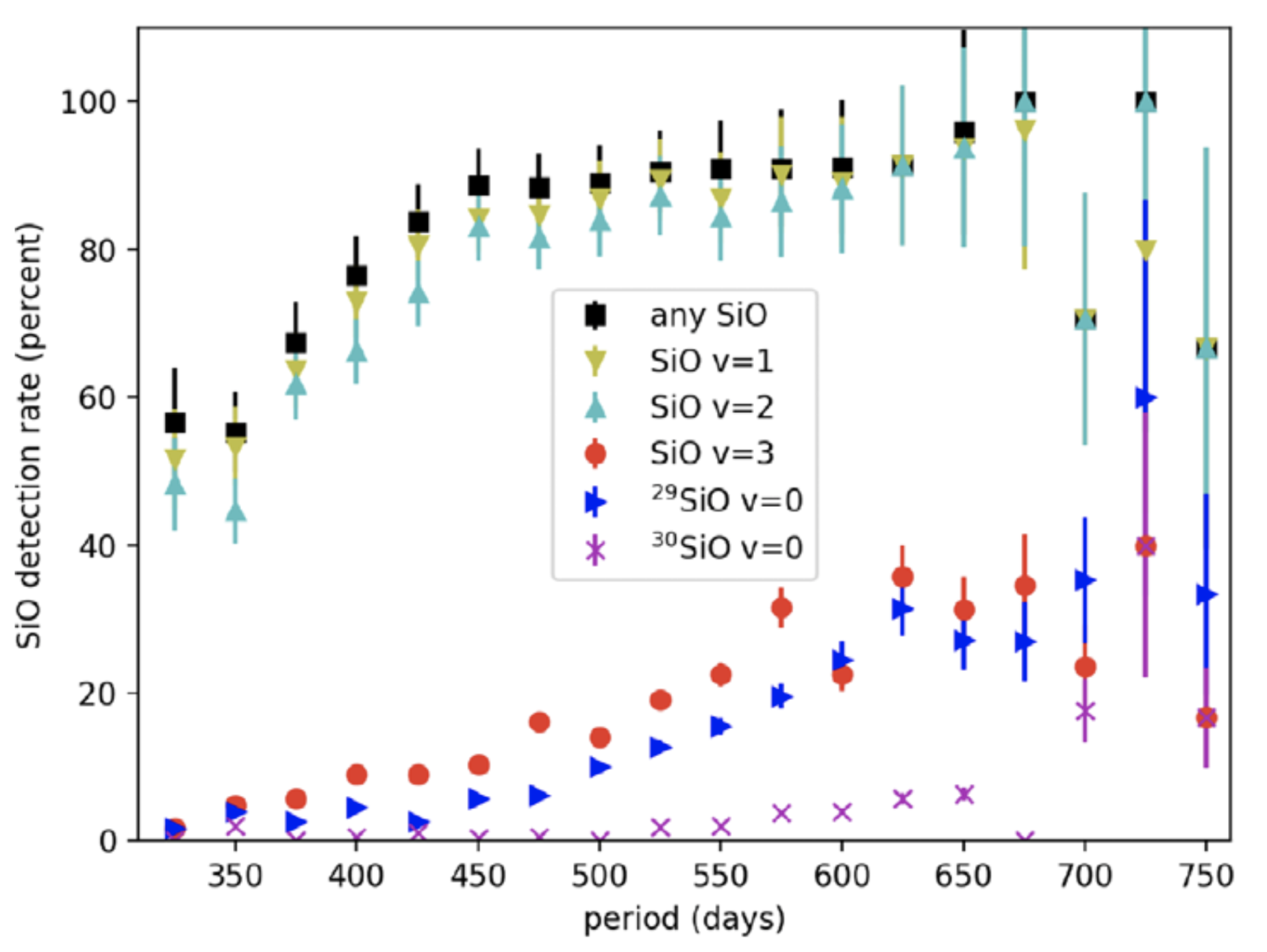} 
\caption{Detection rates  as a function of stellar pulsation period of the
  five  most commonly detected SiO transitions covered by the BAaDE
  survey sample of evolved stars
  (see Section~\ref{probes}). The SiO transition frequencies range from 42.3 to 43.4~GHz. Credit:
  \cite{2024IAUS..380..314L}. 
\label{fig:lewis-fig}}
\end{figure*}

Another aspect of the BAaDE SiO masers studied by Lewis and colleagues is variability. 
Incorporating multi-wavelength data, including optical light curve data from the Optical Gravitational Lensing Experiment
 (OGLE; \citealt{2022ApJS..260...46I}), 
Lewis and her team found a very strong correlation between SiO maser detectability and pulsation period,
with the longer-period stars ($P\gsim$450~days) having the highest detection fraction ($\sim$80\%; Figure~\ref{fig:lewis-fig}). 
Examining the SiO maser data as a function of light curve phase, they also found that all
maser transitions they observed show a statistically significant correlation with
stellar phase: maser detection rate is highest at maximum brightness of the infrared light
curve (when the stellar radius is the smallest), with the $^{28}$SiO $v$=3, $J=1-0$ line
being the most sensitive to phase. This is consistent with the predictions of theoretical
modeling that suggested that this transition arises from high-density environments
\citep{2014A&A...565A.127D}. 

Bartiewicz also stressed the importance of temporal information and
multi-epoch observations for studying masers in evolved stars,
noting that observing cadences of at least every few weeks are desirable.  She presented one recent example from the
East Asian VLBI Network (EAVN) Synthesis of Stellar Maser Animations project, which produced a
37-epoch study over 3 years of the H$_{2}$O masers in
the AGB star BX~Cam \citep{2022ApJ...941..105X}.  Unfortunately,  studies of this kind, particularly
when they involve high-resolution imaging, are extremely time-consuming and can be difficult in practice to
schedule at telescopes. However, the scientific value of such data sets is extraordinarily rich
(see also \citealt{2013MNRAS.433.3133G}) and Bartiewicz emphasized the value of undertaking additional studies of this kind
for larger samples of evolved stars.

Maser monitoring that spans  many years (or even decades) has a unique role to play in our
understanding of stellar sources, and as an example,
Bartkiewicz cited the recent study by \cite{2024IAUS..380..371E}
using the \nan\ Radio Telescope
to monitor OH masers (at 1612~MHz) in a sample of late AGB and post-AGB stars over timescales of more
than a decade. The short-lived post-AGB phase of stellar evolution is notoriously difficult to study, but
the results of the \nan\ survey in combination with near infrared photometry,
are establishing that with statistical samples it becomes possible to
study the AGB to post-AGB transition in real time.

\subsection{Masers in (Proto)-Planetary Nebulae (PNe)\protect\label{PNemasers}}
So-called ``water fountain'' sources are highly evolved sources (descended from low-to-intermediate
mass stars) that are found to be accompanied by high-velocity jets (velocities up to $\sim$500~\kms)
traced by
the 22.2~GHz H$_{2}$O maser line. First reported by \cite{2007IAUS..242..279I},
water fountain
sources are rare but important objects that are key to understanding the short-lived
post-AGB evolutionary phase during which an AGB star undergoes its transition to
a PN. As reported by Bartkiewicz, \cite{2024ApJ...964L..18O} recently discovered high-excitation
OH masers at 4.6 and
6.0~GHz in the water fountain source IRAS~18460$-$0151, providing a new window into
the kinematics and physical conditions of these sources. Other noteworthy recent developments in the study
of water fountain sources include the discovery of a second known case with
associated SiO masers by \cite{2024IAUS..380..359A},  which the authors argue to be a hallmark of
a unique evolutionary phase, and the discovery by \cite{2023ApJ...948...17U}
of a sudden, rapid growth in
the measured outflow velocity in the previously known water fountain source IRAS~18043$-$2116.
The outflow velocity in IRAS~18043$-$2116 was found to be $v_{\rm out}\sim$540~\kms, setting a record for
this type of object.

A small fraction of PNe are known to be associated with molecular maser
emission from H$_{2}$O and/or OH, and this presence of maser emission is thought to be a hallmark of
youth \citep[e.g.,][]{1989IAUS..131..210Z}.  Bartiewicz drew attention to the recent discovery of variability in
the 1612~MHz OH maser
line in the PN IRAS~07027$-$7934 \citep{2024IAUS..380..343C},  which has significantly weakened since the original
discovery by \cite{1991A&A...243L...9Z}.  \citeauthor{2024IAUS..380..343C} suggested that this may 
be related to the expansion of a  photoionized region during the early stages in the onset of
the PN phase. These authors  also found evidence that, unexpectedly,
low-mass, C-rich central stars may be common to PNe hosting OH and H$_{2}$O (i.e.,
oxygen-rich) masers and hypothesized that this may be reflective
of their status as post-common envelope binary systems.

\section{Circumstellar Chemistry\protect\label{chemistry}}
B. McGuire (MIT) presented an invited review on the topic of the
astrochemistry of circumstellar environments.
He defined the objective of astrochemistry as achieving an understanding on a molecule-by-molecule, chemical
reaction-by-reaction basis, the
steps required to form all of the known complex molecules throughout the Universe (including those fundamental
to life). Progress toward this goal
requires evaluating the chemical inventories in different environments in space and using that information
as part of understanding the physical processes and underlying chemistry that this resulted from.

McGuire reported that as of April 2024, the inventory of molecular species detected outside of our
Solar System\footnote{\url{https://bmcguir2.github.io/astromol}. See also \cite{2022ApJS..259...30M}.}
had surpassed
300. It is noteworthy that 95\%
of these have been detected through
radio observations, typically of rotationally excited transitions. Also, of the 300+ molecules,
roughly 30\% were discovered for the first
time in the circumstellar environments of red giants, and a significant fraction of those were in either the 
RSG star VY~CMa (see also Section~\ref{RSGs})
or the carbon-rich AGB star IRC+10216 (see below).  An interesting aspect of the CSE of
VY~CMa  is that it includes a number of lines from ``real'' metals, including
TiO and TiO$_{2}$
(\citealt{2013A&A...551A.113K}; \citealt{2015A&A...580A..36D}) and VO \citep{2019ApJ...874L..26H}. 
Furthermore,
McGuire noted that for line-rich sources, on average 30--40\% of molecules typically remain unidentified in current
studies. And while most of these are likely to be
vibrationally excited states or isotopologues of known molecules that have not been measured, some are
almost certainly new species awaiting discovery. 

\subsection{Carbon-Rich AGB Stars: IRC+10216}
As described by McGuire, the
carbon-rich AGB star IRC+10216 was discovered as an infrared source in the mid-1960s
\citep{1969ApJ...158L.133B}, 
and soon thereafter began yielding detections---through radio observations---of
a range of new molecular species seen for the first time outside the
solar system,
including CS \citep{1971ApJ...168L..53P},  SiS \citep{1975ApJ...199L..47M}, and C$_{2}$H$_{2}$
 \citep{1976Natur.264..345R}. 
This trend of discovery has continued almost linearly ever since, with the total number of molecules known in
the circumstellar environment of this star totaling 67 as of the time of the meeting
\citep{2022ApJS..259...30M}. 
McGuire stressed that access to large aperture, single-dish antennas has been crucial to these efforts,
with the Institut de
Radioastronomie Millim\'etrique (IRAM) 30~m antenna contributing the largest number  of
the molecular detections in this source to date, followed by the Yebes 40~m antenna.
To underscore the importance of IRC+10216 for advancing astrochemistry,
McGuire recounted that 100\% of molecular species outside the solar system
containing the elements Na (3 molecules), Mg (14 molecules),
K (2 molecules),
Ca (1 molecule), and Fe (2 molecules) were discovered for
the first time in IRC+10216, along with 12 of the 13 known Si-containing molecules.

Expounding on the significance of few of these,
McGuire pointed out that the overall scarcity of Fe-rich molecules in circumstellar environments (with the exception of the CSE of
IRC+10216, where FeCN and FeC have been detected; \citealt{2011ApJ...733L..36Z};
\citealt{2023ApJ...958L...6K}) 
is somewhat puzzling given that this element it is fairly common in space. Consequently, knowledge of
such molecules in the gas phase---where we can study their chemistry---is particularly valuable.
Ca is another molecule of interest owing to its importance for human biology, but historically
its origin has been difficult to study in space, as it is typically found in grains, not in the gas phase.
Thanks to the recent detection of CaNC in IRC+10216 \citep{2019A&A...627L...4C}, we are just now
beginning to be able to study it in the gas phase through rotational line spectroscopy.

McGuire also showcased some of the recent results on IRC+10216 by PhD student M. Siebert (University of Virginia).
Based on past work, IRC+10216 is well known to be surrounded by a complex array of
molecule-rich circumstellar shells. The numerous
Si-rich species surrounding this star are of particular interest, arising predominantly from the inner $\lsim10''$
of the CSE (e.g., \citealt{2014MNRAS.445.3289F}; \citealt{2019A&A...629A.146V}). Mg-containing molecules are also
important, as these seem to require UV radiation (and the subsequent formation of certain ions)
as a catalyst for their formation. These latter
molecules are found predominantly in a shell with radius $\sim15''$, which corresponds to the radius at which
dust no longer obscures the UV background radiation. However, McGuire described
work by  \cite{2022ApJ...941...90S} that revealed
that there also seems to be some UV-driven chemistry going on interior to this radius, as is needed
to explain the presence of higher energy transitions of species such as CN and H$_{2}$CN. 
Modeling by \citeauthor{2022ApJ...941...90S} led to the  suggestion that the observed abundances of these species requires the
presence of UV radiation from a solar-type companion, coupled with a clumpy outflow containing gaps that
allow exposure of the
circumstellar gas to  UV radiation.
In effect, astrochemistry has provided strong but indirect evidence for the presence of a companion to this star.
The existence of such a companion 
had been previously hinted at by {\it HST} observations \citep{2015ApJ...804L..10K}  and by the
characteristic spacing of molecular shells around this star
\citep{2018A&A...610A...4G}. 

\subsection{Oxygen-Rich AGB Stars}
\begin{figure*} [ht!] 
\plotone{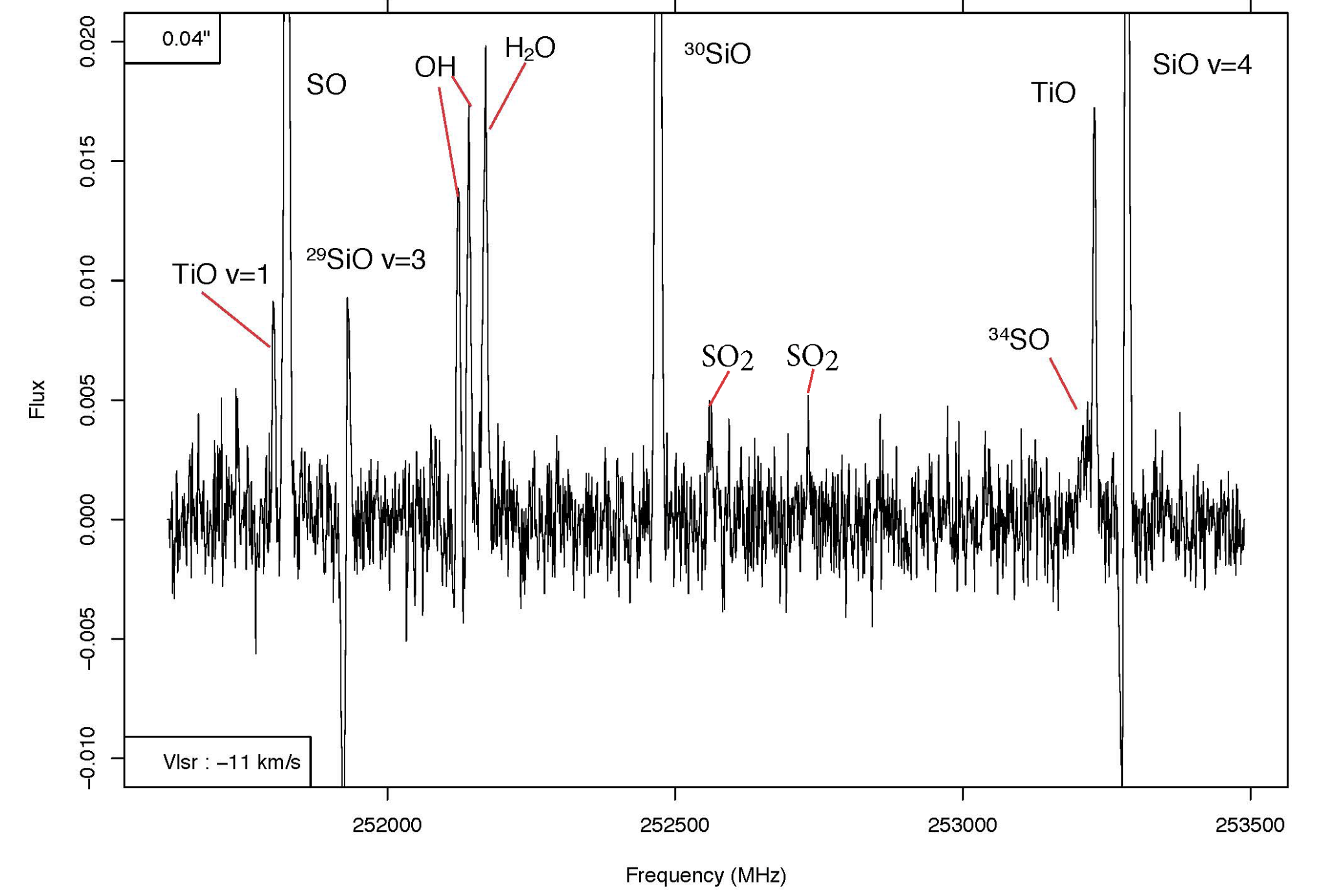}
\caption{A sample portion of the ALMA spectrum of the oxygen-rich AGB star R~Hya, obtained as part of the ATOMIUM project
  (see Section~\ref{redgiantsurveys}). The data were obtained using an extended array configuration, yielding
  angular resolution of $\sim$\as{0}{04}. Axes are flux density (in Jy) and frequency (in MHz).
  Identified molecular lines are labeled. Credit: L. Decin, C. A. Gottlieb, \& A. M. S. Richards on behalf of the
  ATOMIUM consortium (see also
  \citealt{2024A&A...681A..50W}).
\label{fig:gottlieb-fig}}
\end{figure*}

C. Gottlieb  summarized some of the wealth of
astrochemical data obtained for the circumstellar environments of evolved stars
through the ATOMIUM project (e.g., \citealt{2023A&A...674A.125B};  see also Section~\ref{redgiantsurveys}).
The
wide bandwidth covered by the ATOMIUM survey (214 to 269~GHz; Figure~\ref{fig:gottlieb-fig})
enabled efficient identification of a wide range of
molecular species, and in total, $\sim$300 rotational lines from 24
different molecules were observed, along with $\sim$30 unidentified lines
\citep{2024A&A...681A..50W}. 
Gottlieb noted that these data
are now being used to understand the physiochemical processes that convert diatomic and triatomic molecules into
grains within the inner winds of oxygen-rich evolved stars
(\citealt{2022A&A...660A..94G}; \citealt{2024A&A...681A..50W}).
This marks an important advance, as prior to ATOMIUM,  only a handful of oxygen-rich evolved stars
(e.g., VY~CMa; IK~Tau) had been 
the subject of dedicated line surveys (\citealt{2007Natur.447.1094Z};
 \citealt{2017A&A...597A..25V}). 

\subsection{Astrochemistry in the Environments of Massive Protostars\protect\label{protochem}}
In addition to discussing the astrochemistry of the environments of very old stars,
McGuire also touched on the roles of astrochemistry (and radio wavelength observations specifically) for 
studying massive protostars and the process of high-mass star formation (see also Sections~\ref{YSOCMEs},
\ref{YSOmasers}). Historically,
a persistent challenge has been penetrating the dust and
surrounding molecular cloud in which such sources are enveloped.
McGuire highlighted the work of \cite{2018ApJ...860..119G} 
who discovered a set of (initially unidentified)
molecular lines that trace a Keplerian disk around the high-mass young stellar object
Orion Source~I (see also \citealt{2010ApJ...708...80M}). With the help of McGuire, \cite{2019ApJ...872...54G} later identified these as various
vibrational levels of the salt molecules KCl and NaCl and their various isotopologues. Importantly, these
molecules trace the disk that can be observed without contamination from the surrounding molecular cloud. Furthermore,
these lines are bright and readily observable.
However, despite this, we do not fully understand how these molecules formed and how they give rise to the vibrationally
hot lines that are observed. In particular, it is not yet known whether they form in situ or are liberated from
grains. McGuire noted that answers to these questions would provide valuable new information about the immediate
environments of massive protostars.

Also on the topic of the astrochemistry of star-forming regions,
A. Burkhardt (Worcester State University) presented new survey results from a continuation of the ARKHAM (A Rigorous K-band
Hunt for Aromatic Molecules) survey conducted using the Green
Bank Telescope (GBT; \citealt{2021isms.confERC07B}).  This new work has expanded to $\sim$10 the number of
sources in star formation regions with detections of the aromatic molecule benzonitrile (C$_{6}$H$_{7}$CN) and established
that the star-forming region TMC-1 is not unique in containing aromatic molecules.
The results have implications for the long-debated question of whether aromatic molecules
form in the atmospheres of evolved carbon stars,
or in situ in star-forming regions (``top-down'' versus ``bottom-up'' formation, respectively).
At the time of the meeting the next major release of the ARKHAM survey was forthcoming.

\subsection{Future Astrochemistry Work}
McGuire concluded his presentation by emphasizing that the combined sensitivity, instantaneous bandwidth, and spatial
and spectral resolution of the ngVLA are expected to be a game-changer for astrochemistry (e.g.,
\citealt{2018arXiv181006586M}, \citeyear{2018arXiv181009550M}; Section~\ref{ngVLA}).
The wavebands covered by the ngVLA 
will enable the penetration of dust to probe the inner few arcseconds
around nearby AGB stars like IRC+10216. As described by A. Moullet, the ongoing Wideband Sensitivity Upgrade (WSU) at ALMA
is also expected to be extremely valuable for astrochemistry by enabling
ALMA to far more efficiently
observe wide swaths of spectrum in a given target with high spectral resolution (Section~\ref{ALMA2030}).

\section{VLBI Astrometry as a Tool for Stellar Astrophysics\protect\label{VLBI}}
An invited talk by G. Ortiz-Le\'on (Instituto National de Astrofisica, Mexico) summarized the
importance of VLBI astrometry as a tool for the study of stellar systems.
Ortiz-Le\'on began by highlighting the steady progress in achievable astrometric
accuracy over the years, from the level of $\sim1'$ reached by Hipparcus in 150~BC to the $\sim10\mu$m
levels that are now realized. For stellar astrophysics, such accuracy enables the derivation of fundamental
stellar parameters (including masses and ages) and the precise orbital characterization of binary
and planetary companions. 

While the space-based $Gaia$ mission has now provided astrometric measurements of $>$1 billion stars
\citep{2023A&A...674A...1G}, Ortiz-L\'eon pointed out that there are 
certain classes of stellar sources that are not readily accessible to $Gaia$,
including highly extincted young stars, close binaries (separations $<$\as{0}{2}; \citealt{2022MNRAS.517.2925C}),  
and red giants (AGB
and RSG stars) whose large angular diameters and dusty envelopes add large uncertainties to $Gaia$ measurements \citep{2022A&A...667A..74A}. 
As described by Ortiz-Le\'on, these are among the areas where VLBI astrometry---which
can currently achieve accuracies of $\sim10\mu$m for parallaxes and $\sim1\mu$m for proper motions
\citep{2014ARA&A..52..339R}---is particularly valuable.  A caveat, however, is that because of sensitivity
limitations, VLBI remains
limited to bright sources, including nonthermal continuum emitters such as YSOs, M dwarfs, and UCDs, and evolved stars with nonthermal (maser) lines (see Sections~\ref{UCDs}, \ref{masers}).

\begin{figure*} [ht!] 
\plotone{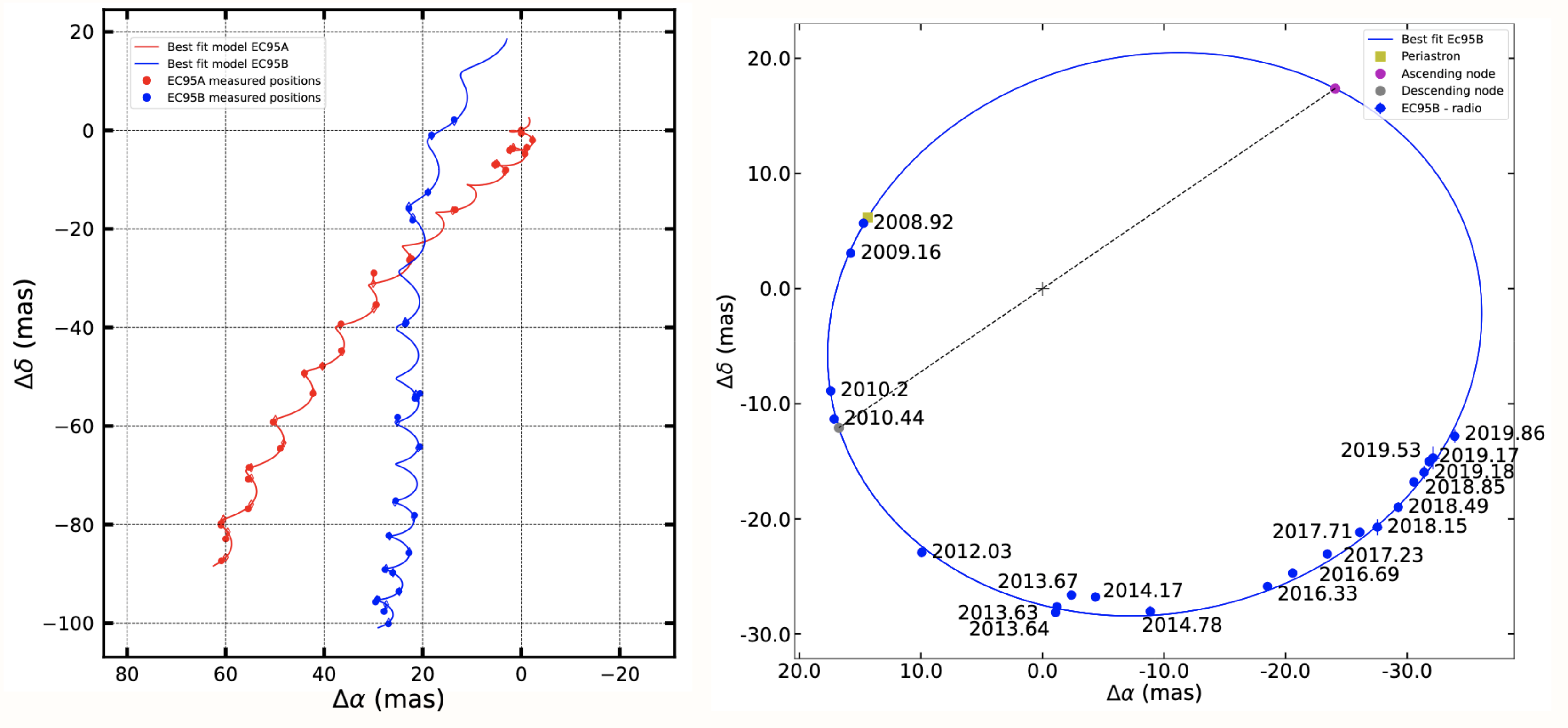}
\caption{{\it Left:} Positions of EC95A (red dots) and EC95B (blue dots), two components of a young multiple
  star system, as measured with the VLBA at 4.9~GHz as
  part of the DYNAMO-VLBA project. The positions are shown as offsets relative to the position of EC95A
  as previously measured on 2007 December 22. {\it Right:} Relative positions together with the resulting
  orbital fit model. These observations permit mass and orbit determinations for the individual binary components.
  Credit: \cite{2025MNRAS.540.2830O}. 
\label{fig:ordonez-toro-fig}}
\end{figure*}

One recent project discussed by Ortiz-Le\'on is known as
Dynamical Masses of Young Stellar Multiple Systems with the VLBA (DYNAMO–VLBA),
which has 
observed $\sim$20 protostellar systems using the VLBA. For the case of Oph-S1, a young binary in the Ophiuchus region,
  Ortiz-Le\'on and colleagues used 35 epochs of VLBA observations to model the astrometry and binary orbit and obtain dynamical
  mass estimates of the binary components \citep{2024AJ....167..108O}. Another DYNAMO–VLBA  result was featured  in the poster presentation of
J. Ord\'o\~nez-Toro (UNAM). She described work using the VLBA to study
the young triple stellar system EC~95 located in the Serpens star-forming region. Previous
radio observations of this system with the VLA and VLBA had played important roles in determining
its distance  and the nature of its
binary components  (e.g., \citealt{2005AJ....130..643E}; \citealt{2017ApJ...834..143O}). The multi-epoch
VLBA observations presented by Ord\'o\~nez-Toro (32 epochs spread over 12 years) expanded on this work
by enabling
precision measurements of the binary orbit, along with the the masses of the primary and secondary
components (\citealt{2025MNRAS.540.2830O};  Figure~\ref{fig:ordonez-toro-fig}).
Her group had also obtained radio measurements of the third component of the system during
4 epochs,
and efforts to determine its mass were ongoing.  

Another result showcased by Ortiz-Le\'on is the study by \cite{2024ApJ...967..112C}  of the close, ultracool
binary LP~349-25. Thanks to their multi-epoch VLBI astrometry, \citeauthor{2024ApJ...967..112C}  were able to obtain detailed orbital
parameters, along with dynamical masses
for both components, revealing that the secondary is a brown dwarf ($M\approx67~M_{\rm Jupiter}$)
rather than a hydrogen-burning star. In addition, through comparison with theoretical tracks, they were able
to derive ages, effective temperatures, and radii for both components and show that the secondary is
significantly less evolved (by $\sim$60~Myr) than the primary.

Ortiz-Le\'on and her
collaborators are also using their astrometric results to search for previously unseen
planetary mass companions in low-mass stars  within 10~pc. In the case of the system
TVLM~513-46546, the study by \cite{2020AJ....160...97C}  obtained an
astrometric detection of a planetary mass companion
($M\approx0.4~M_{\rm jup}$) to an M9 UCD---one of only a handful known and the first
to date detected using radio
wavelength astrometry. For a second example, the low-mass binary GJ~896, both M dwarf components were
detected in the radio and the VLBI astrometric data revealed the presence of an additional Jupiter-like
companion in the system \citep{2022AJ....164...93C}. The authors were also able to map the 3D orbital
architecture of all three components, revealing a surprisingly large mutual inclination angle (148$^{\circ}$)
between
the two orbital planes.

Ortiz-Le\'on emphasized that the future is expected to be bright for further
astrometric exoplanet searches using VLBI techniques (see also Section~\ref{exodirect}).
For example, with an expected astrometric accuracy of $\sim1\mu$m,
the ngVLA (see Section~\ref{ngVLA}) will have the potential to detect lower mass exoplanets and
also planets associated with more distant stars than in now possible with VLBI, including planetary mass
objects that fill a unique parameter space that is not accessible to $Gaia$ or other search methods  (see \citealt{NGVLA85}). 

\section{Stellar Explosions\protect\label{explosions}}
\subsection{Novae\protect\label{novae}}
\subsubsection{Background and Recent Work}
Classical novae are explosive events that
occur in binary systems containing a white dwarf component
that accretes material from its companion (either a main sequence star or a red giant),
leading to thermonuclear runaway on the white dwarf's surface. Characteristic
ejection velocities and ejection
masses of such events are $\sim$500-5000~\kms\ and $\sim10^{-7}$ to $10^{-3}~M_{\odot}$,
respectively, with the latter depending both on the accretion rate and the white dwarf mass.
Energies of these events are typically $\sim10^{45}$~erg (one millionth that of a supernova).
The importance
of radio wavelength observations for understanding the physics of classical novae was reviewed by L. Chomiuk
(Michigan State University).

Chomiuk noted that the longstanding picture to explain the electromagnetic
emission observed from novae is one
whereby residual nuclear burning is sustained on the white dwarf's surface for a period of days to weeks
at close to the Eddington luminosity ($10^{38}$~erg
s$^{-1}$). In this scenario, the radio emission from novae is expected to be thermal free-free
($S_{\nu}\propto\nu^{2}$), with properties similar to an expanding \HII\ region.
Initially, the source will be optically thick, and hence brighter at
higher radio frequencies. Higher frequencies then become increasingly optically thin until
an $S_{\nu}\propto\nu^{-1}$ spectral index is seen \citep[e.g.,][]{1996ApJ...470L.105H}.  Because higher ejection masses take longer to become optically thin, by studying radio
light curves it becomes possible to infer the mass of the ejecta \citep[e.g.,][]{2008clno.book..141S}.

Chomiuk described how radio light curves of novae pose challenges to this model, and
have provided some of the first clues that the ``expanding
\HII\ region'' picture is too simplistic. One of the first hints was the appearance  of a second
peak in the radio light curve of some novae. Examples can be seen in the compilation of radio-detected
novae (36 in total) published by \cite{Chomiuk2021a}. Estimates for the
brightness temperature of the emission during the two peaks systematically
point to values of $T_{B}>5\times10^{4}$~K for the initial peak, implying that
the emission is non-thermal (synchrotron) in nature, while the second peak is thermal. In total, $\sim$25\% of
the novae in the sample compiled by \citeauthor{Chomiuk2021a} show evidence for synchrotron emission.
She noted that this development is not surprising in light of other clear indicators for the role of shocks in novae,
such as the now routine detection of novae as GeV $\gamma$-ray sources (e.g.,
\citealt{2010Sci...329..817A}; \citealt{2014Sci...345..554A}; \citealt{2014Natur.514..339C},
\citeyear{Chomiuk2021b}).

VLBI observations, which are sensitive only to  nonthermal emission, allowed Chomiuk
and collaborators to pinpoint the regions where shock-induced synchrotron emission originates
\citep{2014Natur.514..339C}, 
and helped lead to the emergence of a working model
whereby the envelope of the nova expands slowly, primarily along
the orbital plane of the binary. Later,
nuclear burning on the white dwarf powers the emergence of a fast, bipolar wind, which interacts with
the denser equatorial disk/envelope and leads to synchrotron-producing shocks along the interaction region
(see also \citealt{2021MNRAS.501.1394N}).  Chomiuk cautioned that
these complexities, along with the possible presence of clumping in the thermal
emission, lead to uncertainties of at least a factor of a few in ejection masses derived from radio
observations.

The value of VLBI observations for nova studies was also highlighted in the presentation by M. Williams
(New Mexico Institute of Technology), who presented results of a recent study of nova V1674~Her (an intermediate
polar). V1674~Her  erupted in 2021 and was seen to
fade from its peak optical brightness by two magnitudes
in $\sim$1.1 days,  indicative of extremely rapid evolution; indeed, it is
the most rapidly evolving nova seen to date \citep{2023MNRAS.521.5453S}. 
Its detection in $\gamma$-rays suggested it was a good candidate for the
presence of nonthermal emission that could be followed up with VLBI. Motivated by this,
Williams's team observed V1674~Her with the VLBA during
4 epochs (21, 24, 36, and 71 days after eruption; \citealt{2023AAS...24210906W}).
Based on the morphology of the imaged radio emission,  V1674~Her appears different from other VLBI-imaged
novae (cf. \citealt{2014Natur.514..339C}; \citealt{2020A&A...638A.130G}; \citealt{2022A&A...666L...6M})
 in that it lacks the
prototypical double-lobe structure,
suggesting that the shock that formed in this system may be one-sided (or else that one side is hidden from view).
Using the revised equipartition theory of \cite{2005AN....326..414B}, Williams and colleagues were  able to estimate
from the VLBA measurements the brightness temperature and magnetic field strength during each epoch, enabling
a tracing of the evolution of the shock. They also derived a shock
expansion velocity of $\sim$5300~\kms.

Other areas of active, ongoing research on novae as highlighted by Chomiuk include
exploring how the properties of the radio light curves of novae
correlate with the properties of the $\gamma$-ray light curves and understanding how and where dust
forms in nova ejecta. It has been proposed that the same shocks responsible for the production
of $\gamma$-rays and synchrotron emission may also lead to dust formation
\citep{2017MNRAS.469.1314D}.  Supporting
this suggestion, Chomiuk pointed to the cases of novae (including V357~Mus and V1324~Sco)
where the synchrotron peak in the
radio light curve corresponds in time with a dip in the visual light curve that is believed to result from dust
formation.

While Chomiuk's review primarily focused on classical
novae (i.e., those with a main sequence companion star), I. Molina (Michigan State
University) provided details of a recent study of the recurrent
symbiotic nova V745~Sco, in which the companion is a red giant. Molina noted that for novae with
main sequence companions, the mass transfer is primarily through Roche lobe overflow, while in the case of
a red giant companion, wind accretion dominates, leading to the production 
of a denser and more copious circumstellar medium. V745~Sco is one of 10 recurrent novae
known in our Galaxy and one of only four with a red giant companion.

V745~Sco underwent its last outburst in 2014, during which it was observed with the VLA in several bands, including
1.2, 4.6, and 28.2~GHz. All frequencies were seen to peak at the same time and at comparable flux densities,
in contrast to classical novae where lower frequencies peak later and exhibit lower flux densities
(e.g., \citealt{Chomiuk2021a}).
Molina estimated the brightness temperature of the V745~Sco radio emission and found
$T_{B}>5\times10^{4}$~K, indicative of synchrotron emission. However, she and her colleagues found simple
models for the synchrotron emission unable to reproduce the behavior of the radio light curves
\citep{2024MNRAS.534.1227M}. 
Instead, the observed evolution of the spectral index points to a picture where the radio emission starts off optically
thick ($\alpha$=1.3), but then rather abruptly becomes optically thin ($\alpha\approx$0). This could occur if the
absorbing screen disappears suddenly, allowing the shock to break out abruptly, unveiling emission at all radio
frequencies at once. Molina noted that this idea is supported by X-ray observations which suggest a
rapid decrease in the column density of absorbing material \citep{2019MNRAS.490.3691D}.
One possible instigator,
explored by \citeauthor{2024MNRAS.534.1227M},
is a modulation of the optical depth by the presence of an equatorial density enhancement, as seen
in the hydrodynamic simulations of V745~Sco by \cite{2017MNRAS.464.5003O}. 

\subsubsection{Future Work}
In terms of future work on novae,
Chomiuk noted that on the theoretical side, more detailed modeling is needed to fully account for the
observed radio light curves, including complexities such as variations in mass-loss geometry.  She also
pointed to the anticipated power of the ngVLA (Section~\ref{ngVLA}), which
would allow observing both the thermal and nonthermal
radio emission from novae concurrently, as well as significantly increasing the number of novae each year
whose thermal and non-thermal radio properties can be studied in detail
(see \citealt{2018ASPC..517..271L}).  In the
nearer term,
M. Williams  pointed out that ongoing VLBA upgrades (see Section~\ref{VLBA}) will
improve the ability to study nova through improved sensitivity and better
imaging capabilities, including at Ka band where imaging with the VLBA has heretofore not been
possible.

\subsection{Supernovae\protect\label{SNe}}
Invited speaker D. Dong (NRAO)
reminded the audience that there is a considerable diversity in the types and properties
of progenitor systems that can give
rise to supernovae (SNe; e.g., \citealt{2005ASPC..342...87F}). That includes a wide range in mass-loss rates, leading to 
significant variants in the types of
explosions they can create and the ejecta they leave behind. Understanding the origin of this diversity
is relevant from the standpoint of
studying stellar evolution, but also for understanding the important role that SNe play in shaping their galactic
environments and driving new generations of star formation
(see e.g., \citealt{2022ApJ...925..165H}; \citealt{2022Natur.601..334Z}).  However, systematically understanding these relationships
and connecting
the properties of SNe to their progenitors requires
access to large sample sizes. As noted by Dong, historically this has meant studies based on samples
identified in optical surveys, which to date comprise  $\sim10^{4}$ SNe.
Unfortunately, those SNe samples have certain limitations. Optically identified SNe
span a relatively narrow range in peak luminosity ($\sim$ orders of magnitude; e.g.,
\citealt{2007ApJ...666.1116S}),  and optical
data are also limited by the timescales (days to months) and spatial scales (up to $<10^{15}$ to
$<10^{17}$~cm) to which
they are sensitive. On the other hand, as stressed by Dong, the inclusion of radio-emitting SNe, whose emission arises
predominantly from synchrotron emitted by the forward shock \citep[e.g.,][]{1998ApJ...499..810C}, can
significantly expand the luminosity range of SNe samples and enable
the study of larger-scale phenomena, including stellar mass loss on the
timescales of centuries to millenia before the supernova explosion.

As of 2020, $\sim$100 radio-detected SNe were known, the bulk of which had been targeted for radio wavelength
follow-up subsequent to optical discovery \citep{2021ApJ...908...75B}. Characterization of the radio light curves of
this sample was already seen
to increase by three orders
of magnitude the peak luminosity range known to occur among SNe.
However, this type of follow-up study is time-consuming and
resource-intensive. Fortunately, as reported by Dong, one of the outcomes of the latest generations of widefield
radio wavelength surveys such as VLASS (see also Section~\ref{surveys}) has
been the 
discovery of several dozen or more radio-detected SNe, both through cross-matching against optical
detections \citep{2021ApJ...923L..24S} and through the direct detection of new SNe in the radio (Dong et al., in prep). These
 SNe are preferentially luminous ones owing to the depth of the VLASS survey, but importantly
extend by 1--2 orders of magnitude to longer mass-loss timescales compared with pure optically selected samples. 
Dong noted that present-day widefield radio surveys, including VLASS, VLITE, VAST, and
others (see Section~\ref{surveys})
are sensitive to SNe luminosities of $\gsim$10$^{27}$ erg s$^{-1}$ Hz$^{-1}$. However, next-generation surveys with
facilities like the Deep Synoptic Array (DSA)
2000 are expected to reach $\sim$10$^{25}$ erg s$^{-1}$ Hz$^{-1}$, while pointed observations
with ngVLA or SKA (Section~\ref{futureisbright}) could reach as low as $\sim$10$^{23}$ erg s$^{-1}$ Hz$^{-1}$.

Another topic addressed by Dong is the question of what determines the radio luminosity of a supernova.
He noted that decreasing the energy of the explosion itself results in a sharp decrease in the radio luminosity---an
assumption that has been little discussed in the literature to date. Historically, low luminosities have instead
been attributed primarily to low densities alone. In addition, geometric effects can become important in the
optically thick regime. In such a case, assuming a spherical geometry,
free-free absorption can in principle absorb all of the radio emission.
However, an aspherical geometry (e.g., resulting from a binary interaction) can lead to the presence
of shock-heated, low-density regions along the poles where free-free absorption is no longer relevant
\citep{2021Sci...373.1125D}.

\section{Radio Star Surveys\protect\label{surveys}}
Most of the work done on radio stars over the past half century has targeted individual stars for study based on
known
properties at other wavebands. And while some limited attempts have been made to identify stellar sources
in existing radio surveys, distinguishing them from extragalactic background sources can be challenging.
However, as was
apparent from the number of related presentations at the RS3 meeting,  we
are now entering an exciting new era for radio survey-enabled stellar science.

\subsection{Survey Limitations and Biases\protect\label{surveybiases}}
As emphasized by G. Umana, essentially all current radio star surveys are subject to particular biases depending
on the search method used. For example, looking for
radio variability results in a bias toward detection of flare stars; targeting circular
polarization favors detections to stars with coherent emission; proper motion detections require radio emission to be
persistent; selection based on source classification requires extensive databases of supporting ancillary
information and good a priori knowledge of the source physics;  sky surveys that omit
low Galactic latitudes are biased against detection of types of stellar sources that are concentrated toward the
Galactic Plane.  For these reasons, the combined results
 from multiple survey approaches are crucial to obtaining a more comprehensive understanding of stellar radio emission. Several
  meeting presentations highlighted important strides in reducing previous limitations and biases.

One example described by Umana is the
GLObal view of STAR formation in the Milky Way
(GLOSTAR) survey in the Galactic Plane with the VLA, which combined higher sensitivity and angular resolution
than past centimeter wave surveys of the Plane. This survey
successfully identified more than 100 new PN
candidates and nearly 600 new radio star candidates \citep{2023A&A...680A..92Y}.  GLOSTAR also illustrated the power of
using databases of multiwavelength measurements to distinguish between different classes of sources.

A longstanding major challenge in identifying stellar sources in large radio sky surveys is contamination by
active galactic nuclei (AGN). However, Umana pointed out that we can take advantage of lessons learned from current and
ongoing sky surveys with various SKA precursors such as LOFAR, ASKAP, and MeerKAT to
address this challenge (see, e.g., \citealt{2017A&A...598A.104S}; \citealt{2021PASA...38...54M};
\citealt{2024MNRAS.531..649G}). 

One tool at our disposal to help distinguish stellar sources from background objects is
circular polarization (see Section~\ref{cmsurveys}). Another effective discriminant
is short-term variability; for example,
searching with a 15 minute cadence in ASKAP pilot survey data, \cite{2023MNRAS.523.5661W}  were
able to identify 38 transients, including 7 previously unknown radio stars (plus one known radio star).
Proper motions can also be used to find stars in widefield radio surveys with sufficiently long time baselines,
as was recently  demonstrated by \cite{2023PASA...40...36D}
using data from the VLA Faint Images of the Radio Sky at Twenty-cm (FIRST) survey
and the Rapid ASKAP Continuum Survey (RACS), combined with proper motion information from
the {\it Gaia} Data Release 3. Their search identified 2 new radio stars, plus several previously known radio stars and
43 candidate variable
radio stars. Additional results are discussed below. 

\subsection{Centimeter Wavelength Surveys\protect\label{cmsurveys}}
T. Murphy (University of Sydney)
provided an overview of a range of stellar science results emerging from surveys being conducted by 
ASKAP,
a telescope array in Western Australia comprising 36 12~m dishes and operating at frequencies from 700--1800~MHz.
Attributes of ASKAP data products that make them particularly well-suited to mining
for stellar science include
ASKAP's wide field-of-view (30 square degrees) and the availability of both  circular polarization
 and temporal information.  At the time of the conference,
 a year's worth of ASKAP data amassed from several different survey projects
 had been made publicly available to the worldwide community (e.g.,
 \citealt{2020PASA...37...48M}; \citealt{2021PASA...38...54M}; \citealt{2023PASA...40...34D}).

 As noted above, circular polarization is a useful tool for distinguishing stellar sources from background
 objects. The reason is that
only a few class of celestial objects--namely pulsars and certain classes of
stars---exhibit significant
circular polarization ($>$ a few percent). In contrast, active galactic nuclei (AGN),
whose source counts dominate centimeter-wave continuum surveys, typically
have circular polarization fractions of $\ll$1\%. Murphy stressed that
such polarization information is a game-changer compared
with earlier generations of radio surveys that lacked such information and where distinguishing stars
from AGN was extremely challenging.

As one example, Murphy highlighted the recent work by \citeauthor{2021MNRAS.502.5438P} 
(2021, \citeyear{2024MNRAS.529.1258P})  who identified
$\sim$70 radio stars in  ASKAP survey data, of which two-thirds had not  been detected previously.
A comparison between model predictions and derived brightness temperature limits allowed
constraining the radio emission mechanisms and suggests that radio emission from
active binaries such as RS CVn and Algol systems (see also Section~\ref{activebinaries})
tends to be consistent with gyrosynchrotron, whereas
certain other types of stars (including K and M dwarfs) require a coherent mechanism.
Another important finding was that each epoch of data 
revealed new radio stars. Furthermore,
the majority of the radio stars detected in each epoch were stars that were not
radio-bright in the previous one. An analysis of these data showed that
for  K and M dwarfs within 25~pc, the median fraction of stars that undergo
radio-bright bursting or flaring events at a given time is
$\gsim$10\% \citep{2024MNRAS.529.1258P}.  This rate is notably lower than  inferred previously from
samples of stars that were pre-selected 
based on activity levels at shorter wavelengths ($\sim$25\%; cf.
\citealt{1989ApJS...71..895W}; \citealt{2019ApJ...871..214V}). 
Nonetheless, an extrapolation of these results suggests that future all-sky surveys with SKA-Mid are
expected to result in the detection of $\sim$58,000 stars from just a single hour of data. 

\begin{figure*} [ht!] 
\plotone{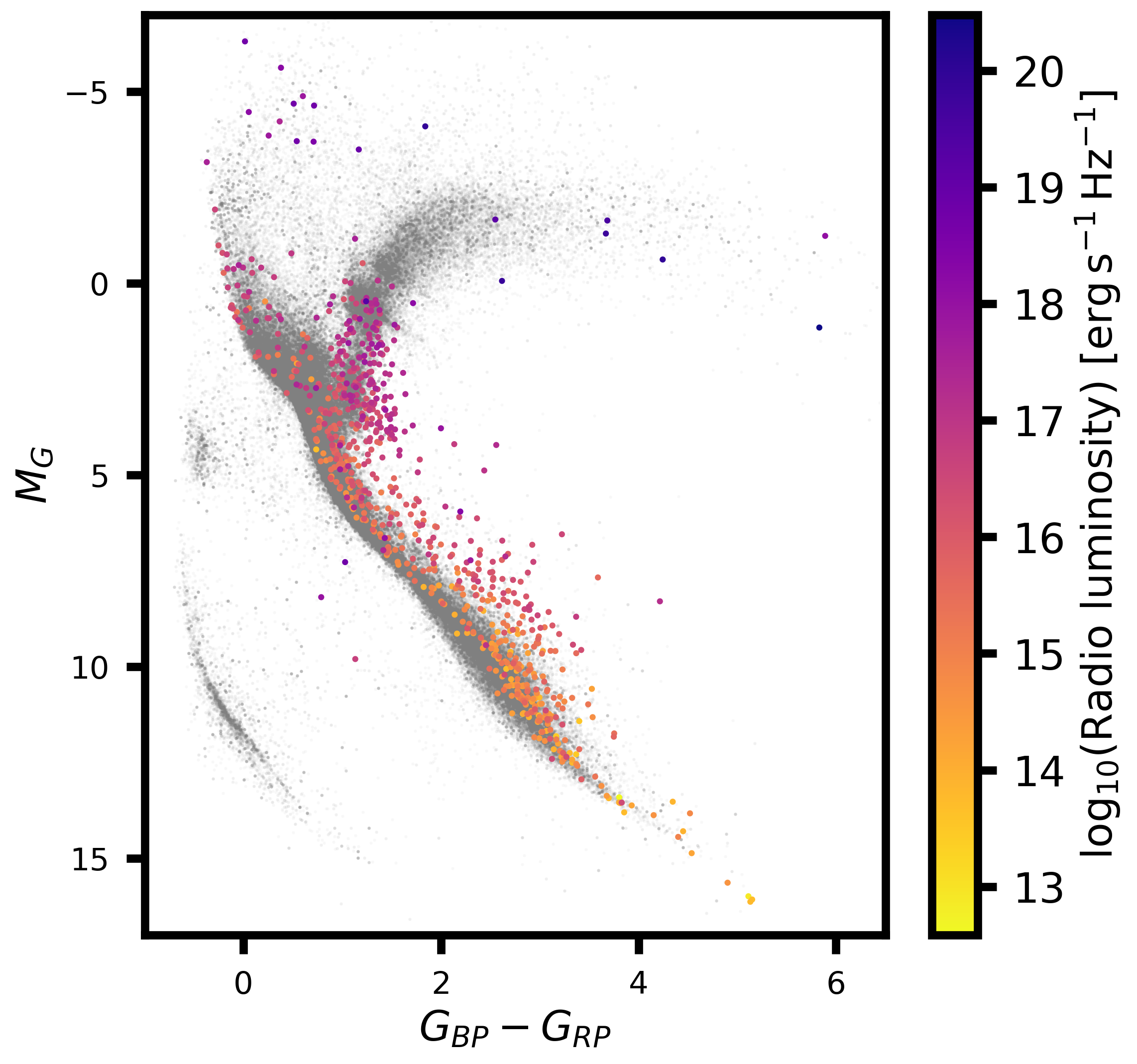}
\caption{Color magnitude diagram (CMD)  showing as colored symbols
  the radio stars in the SRSC compiled by Driessen et al. (2024). The color scale indicates
  the radio luminosity of each star assuming distances from $Gaia$.
  For reference, the  grey background points show the CMD from $Gaia$ Data Release 2, taken from Pedersen et al. (2019).
  Credit: \cite{2024PASA...41...84D}. 
\label{fig:driessenCMD}}
\end{figure*}

L. Driessen (University of Sydney) presented 
the Sydney Radio Star Catalogue (SRSC), a new radio star catalogue for $\nu<$3~GHz
based primarily, though not exclusively on ASKAP data, along with recently published data from
other SKA precursor instruments (\citealt{2024PASA...41...84D} and references therein).
The SRSC contains $\sim$850 radio stars representing types across 
the entire H-R diagram that were identified using a range of techniques,
including circular polarization searches, proper motions, variability searches, and cross-matching with
multi-wavelength data (Figure~\ref{fig:driessenCMD}). A significant fraction of  the SRSC stars ($\sim$600)
had not been identified as radio stars prior to ASKAP.

While a few small catalogues of radio stars have been
published over the years (e.g., \citealt{1990A&AS...86..357W}; \citealt{2009ApJ...701..535K}),  as noted by Driessen, there has not been a major radio star
catalogue published since the compilation of \citeauthor{1978AAHam..10....3W} (1978, \citeyear{1987A&AS...69...87W}, \citeyear{1995A&AS..109..177W}),  which was last updated in 2001
(see \citealt{2015yCat.8099....0W}). Driessen stressed that
new and more extensive catalogues have the potential to provide insights into a range of topics in stellar
astrophysics research, including the overall fraction of radio-bright stars of different types,
variability/flare rates of different stellar types, stellar rotation rates, and magnetic field
properties of cool stars. In turn, such statistics are relevant to a range of more general problems
including exoplanet habitability, the tying together of optical/radio reference frames,
and Galactic foreground removal for extragalactic and cosmological studies.
At the time of the meeting, work was underway to further expand the SRSC to include centimeter-detected
radio stars from other surveys, as well as new ASKAP detections. Contributions to the catalogue are also welcomed
from the community (see {\url{https://radiostars.org}}). 

Using soft X-ray data from from eROSITA, Driessen and her
collaborators have explored how well the SRCS sample conforms to the ``new'' G\"udel-Benz
relation of \cite{2014ApJ...785....9W}  and find that the bulk of the stars lie {\it below} this relation (i.e.,
for a given radio flux they are less bright in X-rays than predicted). One explanation could be that
a portion of the radio emission from many of the SRCS sample stars arises from a coherent
mechanism, but work is ongoing to better understand this result.

\begin{figure*} [ht!]
\plotone{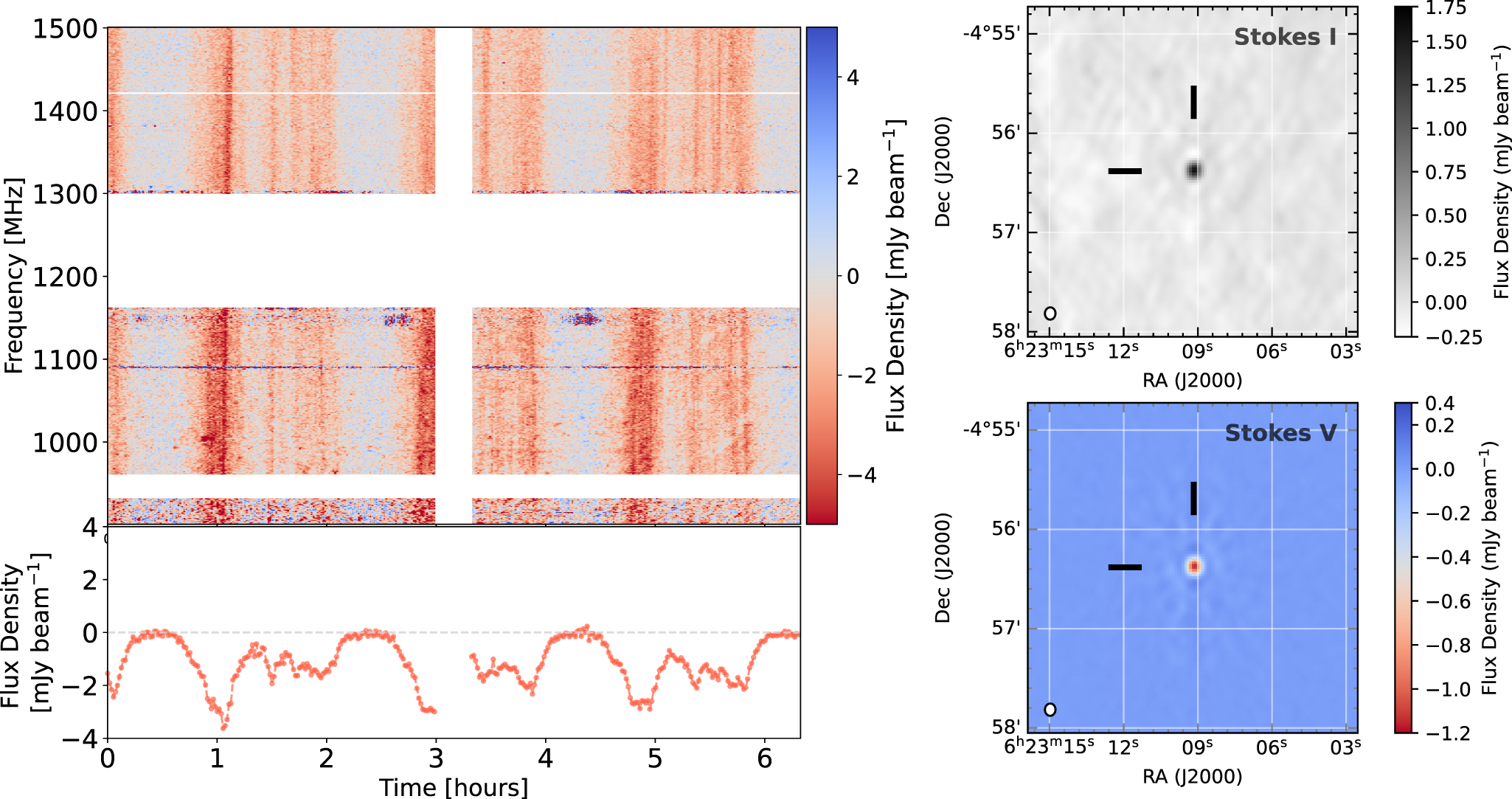} 
\caption{{\it Left:} Stokes V dynamic spectrum  (0.9–1.5~GHz) of the T8 dwarf WISE J062309.94$-$045624.6, obtained with
  MeerKAT. The Stokes V light curve is shown in the lower panel (based on 1~MHz frequency bins and 64~s time sampling).
  Both the light curve and the dynamic spectrum show periodic behavior with additional complex structure.
  The horizontal gaps correspond to frequencies flagged due to RFI, while the vertical gap corresponds in time
  to a calibration scan. {\it Right:} Stokes I (top) and V (bottom) continuum detection images from the same observation.
  Credit: \cite{2023ApJ...951L..43R}. 
\label{fig:roseT8}}
\end{figure*}

Another outcome of the survey data from ASKAP 
has been the identification of interesting individual sources
worthy of follow-up study. One example is the T8 brown dwarf WISE~J062309.94$-$045624.6 (Figure~\ref{fig:roseT8}), which was the subject
of a poster by K. Rose (University of Sydney;  see also Section~\ref{UCDs}).
This object was identified through its strongly
circularly polarized emission ($\sim$66\%) in the ASKAP RACS-mid survey at 0.9--2.0~GHz
and is the coolest dwarf yet detected in the radio. Consistent with other radio-bright UCDs (Section~\ref{UCDs}),
the high level of circular polarization suggests
that coherent emission generated via
the ECM instability dominates the radio signal. 
Follow-up observations with MeerKAT and Australian Telescope Compact Array (ATCA)
showed WISE~J062309.94$-$045624.6 to have rotationally modulated
coherent emission and enabled a determination of a lower limit on the magnetic field strength of $B>0.71$~kG \citep{2023ApJ...951L..43R}. A complementary poster by R. Kavanagh (ASTRON) showed how the complex
structure present in the periodic radio light curve of this object can be characterized using a purely
geometric model. In this model, the magnetic field is dipolar with a large obliquity (i.e., the angle between the
rotational and magnetic field axes), and the emission is beamed along two active field lines
\citep{2024A&A...692A..66K}. This work is expected to have future applications to other types of
objects as well, including  M dwarfs, magnetic massive stars (Section~\ref{mag-OB}),
or possible future radio-detected exoplanets.

Results of the largest non-targeted search for stellar radio transients at centimeter wavelengths undertaken to date
were summarized in  
a poster presented by C. Ayala (California Institute of Technology). This study was based on data
from  VLASS, which covered 33,885 deg$^{2}$ at 2--4~GHz and reached a sensitivity of $\sim140\mu$Jy (1$\sigma$).
Ayala and his colleagues identified a sample of point sources that were undetected in a first epoch of data
but detected at $>7\sigma$ in a second epoch obtained 2 years later and then cross-matched these
with known stellar sources from the $Gaia$ catalogue\footnote{\url{https://irsa.ipac.caltech.edu/Missions/gaia.html}}. After
correcting for proper motions and assigning a probability of false detection they found a final sample of 80
stellar radio transients. The classes of sources detected included tight binaries, pre-main sequence
stars, and M dwarfs, and in the majority of cases the radio emission appears linked with magnetic activity
(\citealt{2024AAS...24335933A};  see also \citealt{Davis2024a}).

\subsection{Future Stellar Science from Radio Surveys\protect\label{futuresurveys}}
Knowledge gained from current generations of radio star surveys is poised to help design
and optimize future radio star surveys with up-and-coming facilities, e.g., insuring that such efforts are, by
design, multi-epoch, full polarization surveys that include the Galactic Plane fields.
As suggested by M. Rupen
(National Research Council, Canada), building cross-referencing to other databases and catalogues into these
surveys could help to increase their value and usage given the enormous anticipated data volumes.
He also cautioned that compiling and processing such enormous future catalogues
will require investing in dedicated teams of people.

Rupen suggested that
source stacking, as is being recently employed by the Canadian Hydrogen Intensity Mapping Experiment (CHIME)
collaboration, could be another potential tool for identifying relatively weak stellar sources in survey data.
Also important will be well-developed source extraction algorithms, together with reliable, automated
classification tools. Finally, to maximize the scientific return of upcoming radio surveys, L.
 Driessen  emphasized the need for ``outreach''
 to astronomers working at other wavelengths, as well as making radio wavelength data products more accessible
 and their unique value more widely appreciated.
 
\section{Radio Stars as Probes of the Galaxy\protect\label{probes}}
As we were reminded by
L. Sjouwerman (NRAO), surveys of radio stars are useful not only for studying the properties of the stars themselves,
but for also gaining insights into our Galaxy.  A poster by Sjouwerman summarized recent results from the BAaDE
Survey, a large, long-term project that has been
using targeted VLA and ALMA observations of SiO masers in a large sample of evolved stars in the inner
Milky Way/bulge region to provide
information on the structure and evolution of the Milky Way \citep[e.g.,][]{2024IAUS..380..292S}. 
The radio measurements provide a powerful complement to other surveys
owing to their ability to penetrate regions obscured in the optical and infrared, as well as the provision
of important velocity information. BAaDE also complements the Bar and Spiral Structure Legacy
(BeSSEL) program, another Galactic structure survey
using masers \citep[e.g.,][]{2011AN....332..461B}; 
since BeSSEL focuses on methanol and H$_{2}$O masers from high-mass star-forming regions, it
primarily traces disk and spiral arm
regions---but it cannot trace the bulge. Some of the recent BAaDE highlights featured on
Sjouwerman's poster included  sample spectra showing commensal detections of
up to 7 SiO transitions in individual BAaDE target stars (see \citealt{2020ApJ...892...52L};
\citealt{2024IAUS..380..292S}; Section~\ref{AGBmasers}),
a determination of the distinct Galactic distributions of
oxygen-rich and carbon-rich AGB stars \citep{Lewis2020b}, a  map of the Milky Way's structure
as traced by BAaDE stars, and a new Galactic position-velocity diagram derived from SiO maser measurements
of $\sim$15,000 targets
over a full range of Galactic longitude \citep{2024ApJ...976..139X}. The latter sample has been used to measure
the gravitational potential of the Milky Way bar and enables
for the first time comparison with Galactic positions-velocity curves derived
using gas  (\HI\ or CO) measurements.

A complementary poster by
R. Weller (University of New Mexico) described a recently launched
effort to tie together the dynamics of the disk with
the older bulge and bar regions of the Milky Way using the SiO maser-emitting
RSG stars in the BAaDE sample. Up to 20\% of RSG-classified stars in their
sample were found to be SiO maser emitters.

Accurate distance determinations are a crucial piece of using the measurements accumulated by
the BAaDE project for Galactic science, and 
the poster presentation of R. Bhattacharya (University of New Mexico)
featured an exploration of supervised machine
learning as a tool to improve our ability to obtain distance estimates for large samples of Galactic AGB stars
from the BAaDE sample.
Using SEDs derived from infrared measurements of nearly 15,000 stars and assuming
that stars with similar intrinsic properties (e.g., initial mass, temperature, metallicity, etc.)
should have similar SEDs and similar luminosities, Bhattacharya and colleagues
created a set of distance-calibrated SED templates and
compared the inferred distances with direct distance measurements available 
for the subset of stars from $Gaia$ or VLBI parallaxes \citep{2024ApJ...969..109B}.  The results
appear promising and are expected to aid in mapping the structure and dynamics of the
inner Galaxy in addition to determining mass-loss rates and luminosities for large samples of AGB stars.

One additional topic addressed by M. Lewis was the use of the BAaDE SiO sample to derive a new period-luminosity
(P-L)
relation specific to maser-bearing stars. This is valuable for helping to overcome the extinction effects
that impact traditional P-L relations. This was done using a sample of sources believed
to lie in the Galactic Center region and with independently known distances. Applying this newly
derived relation to their sample they were able to derive new P-L distances to over 6000 stars (see
\citealt{2023A&A...677A.153L}).
This is complementary to the work presented in the poster by  Bhattacharya (see above). Access to new
and better distance estimates is expected to help with ongoing work on the SiO maser luminosity
function by Lewis and collaborators.
These
new distance estimates are also crucial to achieving BAaDE's science goals of measuring the structure and
dynamics of the Galaxy \citep[e.g.,][]{2024ApJ...976..139X}. 

Follow-up VLBI observations
provide an important means of tying down the calibration of indirect, statistical distance determination
methods for BAaDE stars.
As described in the presentation by Y. Pihlstr\"om (University of New Mexico),
VLBI methods are able to produce proper motion measurements
for SiO maser stars within the bulge even in cases where parallax measurements are not possible; these
in turn help to assess the Galactic orbits of these stars. However,
one impediment to VLBI follow-up studies of targets near the Galactic Plane is the lack of
calibration sources suitable for use at frequencies of $\gsim$43~GHz. To circumvent this issue,
Pihlstr\"om and her collaborators have been developing a technique termed ``shared astrometry'',
whereby the positions of multiple maser sources (each with its associated error ellipse) are used collectively
used to pin down the positions of targets of interest, effectively building a frame of reference sources
\citep[e.g.,][]{2022AAS...24021605Q}. 
Because of proper motions, sources move relative to one another over time, hence a second
step ties together the reference frames from different epochs. To date the shared astrometry approach has been
tested via both Monte Carlo simulations and pilot VLA observations, and the results appear promising.
While this method is too observationally intensive to apply to large samples, Pihlstr\"om stressed that it
could still serve as a means to calibrate other distance derivation methods used for BAaDE stars (see above).


\section{Degenerate Stars and Stellar Systems\protect\label{degnerates}}
\subsection{X-Ray Binaries}
A poster by E. Pattie (Texas Tech University) featured radio observations of the relativistic
jets associated with accreting black holes and neutron star binaries. Using archival ALMA and VLA data,
Pattie and his colleagues have derived power spectra and high-cadence light curves for a sample of such systems,
with the goal of
comparing the observed characteristics of
black hole binary and neutron star binary systems to assess their levels of similarity or
differences in terms of their jet launching mechanisms (see \citealt{2024ApJ...970..126P};
 \citealt{2025AAS...24534607P}). 
Unexpectedly, they find
that both types of binary systems
exhibit similar levels of variability despite large differences in radio luminosity, suggesting that
their underlying jet structures and launching mechanisms may be similar.

\subsection{Cataclysmic Variables}
\begin{figure*} [ht!]
\plotone{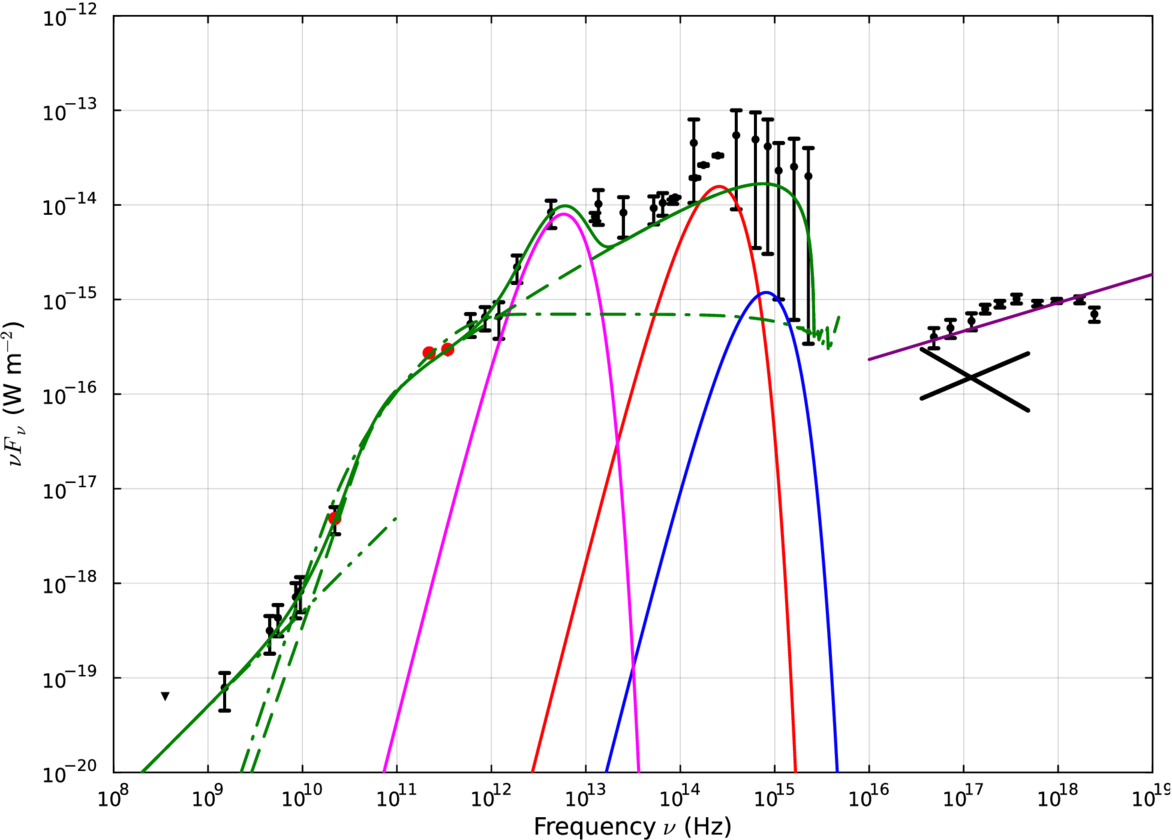} 
\caption{SED of the white dwarf pulsar AR Sco obtained from radio observations. The black points and lines are
  from previously published observations (see Barrett \& Gurwell 2025 for details).
  The three red points show 22~GHz VLA data
  along with 220 and 345~GHz SMA data. The magenta, red, and blue curves represent, respectively,
  blackbody emission from  cool (70~K) circumbinary dust, the red dwarf companion ($T_{\rm eff}$=3100~K),
  and the white dwarf primary ($T_{\rm eff}$=9750~K). The dashed and dot-dash green lines show, respectively,
  fast- and slow-cooling models for the synchrotron emission at $\nu>10$~GHz. The dash–double dot
  green line is the ECM emission at $\nu<10$~GHz, and the purple line indicates the
  X-ray emission (a power law). The solid green line shows the sum of all emission components.
  Credit: \cite{2025ApJ...986...78B}. 
\label{fig:barrett-fig}}
\end{figure*}

P. Barrett (George Washington University) presented new  observations at 220~GHz and 354~GHz of
the magnetic cataclysmic variable AR~Sco,
obtained with the Submillimeter Array (SMA; \citealt{2025ApJ...986...78B}).   AR~Sco is sometimes referred
to as a ``white dwarf pulsar''. Synchrotron emission dominates
the radio emission from this system at centimeter wavelengths, but the SMA observations have filled in a missing gap
in the previous SED and point to a break in the synchrotron emission
near $\nu\sim$200~GHz (Figure~\ref{fig:barrett-fig}). A periodigram analysis of the SMA measurements at 220~GHz shows modulation at approximately
twice the spin frequency of the white dwarf component of the binary and implies that the synchrotron emission
originates from matter accreted onto the white dwarf's magnetic dipole rather than the photosphere of the red dwarf
companion. A key conclusion of this study is that AR~Sco is a magnetic propeller.

\subsection{White Dwarf Binaries}
R. Ignace discussed the possibility of using radio wavelength observations
to search for the extensive disks
postulated to occur around double degenerate white dwarf mergers. Unlike most accretion disks, these putative
disks would not be thermally supported (which leads to a change in scale height with radius),
but would instead be very flat with a steep density gradient,
resulting in a signature kink or ``knee'' in the radio SED \citep{2024AAS...24440905I}. Ignace pointed out that 
analogous signatures have already been seen in the disks of Be stars and interpreted as a truncation of
their disks caused by a companion \citep{2017A&A...601A..74K}. 

\section{Future Prospects\protect\label{futureisbright}}
\subsection{New and Upgrade Facilities of Relevance for Stellar Science\protect\label{newtelescopes}}
\subsubsection{The Very Long Baseline Array (VLBA)\protect\label{VLBA}}
Planned upgrades to the 31-year-old VLBA that are of potential relevance for
stellar science were described in a
poster by J. Linford (NRAO). These included both near-term (next 1--2 years)
and longer-term ($\lsim$10 years) efforts, leading up the the ngVLA era (see Section~\ref{ngVLA}). Examples
include new synthesizers that will allow greater tuning flexibility, new digital back-ends (enabling
wider bandwidths, higher bit sampling rates, and improved timing stability),
and a possible new suite of
ultra-wide band (X--Ka band, i.e., 8-40~GHz)
receivers. Linford reported that a prototype of the latter receiver has already been tested at the VLBA Owens Valley
site. Other recent developments highlighted by Linford included improvements
in high-frequency (86~GHz) pointing and the possibility of performing real-time correlation
rather than the requirement to ship disk packs from individual stations to the central correlator in Socorro, New Mexico. 

\subsubsection{ALMA 2030 and the Wideband Sensitivity Upgrade (WSU)\protect\label{ALMA2030}}
Thanks to its combination of exquisite sensitivity (6600~m$^{2}$), unique dry, high-altitude site,
broad frequency coverage (35 to 950~GHz), and the ability to image emission on angular scales
from $\sim1'$
down to as fine as 
$\sim$5~mas,  ALMA, which began full science operations
in 2014, has opened vast new discovery
space for many areas of astrophysics. Stellar astrophysics is no exception, and numerous examples of
its contributions to this subject were highlighted at the meeting (see above).

As described in talks by G. Umana and A. Moullet, a series of major developments is presently
underway at ALMA under the umbrella of a program known as the Wideband Sensitivity Upgrade (WSU) to make
ALMA even more powerful.
The vision for the WSU was put forth in a document known as the ALMA 2030 Development Roadmap
(\citealt{2019arXiv190202856C}, \citeyear{ALMA621}),  and its 
top priority is expanding ALMA's IF bandwidth by at least a factor of two (to 16~GHz per polarization)---and
up to a factor of 4---across ALMA's entire frequency range. Additional WSU improvements will
include a new, more flexible ALMA correlator, improvements to the ALMA archive (to deal with larger data
volumes), receiver upgrades, and improvements in digital efficiency across
the entire signal chain. 
Essentially, with the exception of the antennas, the WSU will overhaul all components of ALMA by the end of the
current decade.

Some of the anticipated benefits of the WSU for stellar astrophysics were highlighted in Moullet's presentation. 
For instance, among the first bands to be fully upgraded will be Band~6, which is one of most important bands for
astrochemistry (Section~\ref{chemistry}). The improved spectral grasp will enable the simultaneous observations
of lines within
a spectral range of up to 32~GHz without sacrificing spectral resolution, and
it will become possible to simultaneously process up to 1.2 million spectral
channels, or 50$\times$ more  than currently. For AGB and RSG stars this will make it possible, for example,
to simultaneously
observe a full suite of CO isotopologues in a single observation  and enable simultaneously
measuring multiple time-varying maser lines during the course of the stellar pulsation cycle (see Section~\ref{probes}).
In the case of protostars, it will become possible to simultaneously observe at high spectral resolution lines
arising in {\it all} components of the protostellar environment (e.g., jet, outflows, warm inner envelope, disk,
disk wind, cold envelope; see \citealt{2021A&A...655A..65T}). 

As Moullet described, higher continuum sensitivities at ALMA resulting from the WSU
will also improve the ability to perform polarization measurements of protostars, evolved
stars, and PNe. Moreover,
for spatially resolved imaging of such sources, the higher bandwidths will result in
improved $uv$ coverage, and therefore improved imaging fidelity and dynamic range
(see \citealt{ALMA621}).  This will enhance
the ability to identify subtle features in the circumstellar environments of evolved, mass-losing stars, better
constrain gas/dust coupling in these environments, and benefit searches for signatures of
companion stars in radio data (see e.g., \citealt{2022MNRAS.513.4405A};
 \citealt{2024ApJ...965...21K}). 

The anticipated improvements in digital efficiency and receiver performance at ALMA
will result in a factors of $\sim$1.5--2.2$\times$ improvement in {\it spectral line} sensitivity as well. This will
provide
access to weaker molecular transitions and lines in more distant stars. In combination with the available
ultra-high spectral resolution ($\sim$10~m s$^{-1}$),
improved measurements of Zeeman-induced circular polarization patterns in stellar sources (cf.
\citealt{2019A&A...624L...7V}) should  also become possible. 

\subsubsection{Next Generation VLA (ngVLA)\protect\label{ngVLA}}
The game-changing potential of the planned ngVLA for stellar science
was spotlighted in a number of the presentations at the RS3 meeting (see above).
As G. Umana described, the ngVLA's goal is to
provide a factor of 10 improvement in both sensitivity and angular resolution  compared with the VLA or ALMA for the
study of thermal
emission across the frequency range 1.2--116~GHz. The ngVLA will comprise an array of 214 18~m dishes concentrated
in the Southwestern United States and northern Mexico. The array will include a dense core to provide
high surface brightness sensitivity at $\sim$1,000 mas resolution, spiral-shaped arms of antennas
for high-fidelity imaging at $\sim$10~mas scales, and longer arms to provide  baselines for imaging at $\sim$1~mas
scales. In addition, a continent-scale
long baseline array is proposed as part of the ngVLA and would replace the current VLBA (see Section~\ref{VLBA})
and achieve angular resolutions as high as $\sim$0.1~mas.
A more detailed description of the ngVLA design and examples of how it is expected to advance stellar science
can be found in \cite{2018ASPC..517.....M}. 
The
ngVLA was only new radio facility in the United States that was specifically recommended in the Astro2020 Decadal Survey
\citep{Astro2020}. However,
at the time of the RS3 meeting, the estimated cost of the
ngVLA was projected to be $\sim$2.3 billion and full funding for the project had not yet been secured.

\subsubsection{Square Kilometer Array (SKA)\protect\label{SKA}}
Another future interferometric array facility that received prominent mention at the RS3 conference
was the SKA. As described by G. Umana, at the time of the meeting the first phase of this project (known as SKA-1; \citealt{2019arXiv191212699B})
was about 80\% funded and construction was projected to be completed around the year 2030.
In the mean time, a taste of what is to come from the SKA for stellar science was already given at the meeting
through results already available from various SKA precursor telescopes,
including MeerKAT, ASKAP, and the MWA (see above).

The SKA will operate at lower frequencies compared to the ngVLA (Section~\ref{ngVLA}), although
there will be a significant region of overlap.
SKA-Low, to be sited in Western Australia, will operate at 50--350~MHz, while SKA-Mid, to be located in the Karoo
region of South Africa,
will cover 350~MHz--15.4~GHz. SKA-Low will be made up of more than 130,000 log-periodic antennas spread between
512 stations (equivalent collecting area of 419,000~m$^{2}$) and provide baselines of up to
74~km. SKA-Mid will comprise 197 parabolic antennas
of diameter 13.5~m on baselines of up to 150~km and have a total collecting area of 33,000~m$^{2}$.

The SKA's frequency coverage will make it less ideally suited compared with the ngVLA
for  spatially resolved
stellar imaging (see \citealt{2018ASPC..517..369C}; \citealt{2019arXiv191000013A})
and astrochemistry (see \citealt{2019BAAS...51c.233M}). 
On the other hand,  the SKA's enormous sensitivity, coupled with its primary operating mode as
a survey instrument, are expected to yield large numbers of new detections of radio stars, coupled
with a wealth of information on the radio light curves and SEDs of stellar radio sources
(see, e.g., \citealt{2004NewAR..48.1349M}; \citealt{2015aska.confE.118U}; Section~\ref{redgiants}). 

\subsection{Going to Space\protect\label{space}}
An invited talk by M. Knapp (MIT Haystack Observatory) summarized  some of the unique
advantages of pursuing future solar and stellar radio science from space. These include access to frequencies
not observable
through the Earth's atmosphere (i.e., $\lsim$15~MHz), access to longer baselines, and,
in the case of heliophysics, the ability to perform in situ measurements
of plasma.

Knapp noted that the
radio spectra of planets in our solar system obtained by the Voyager~2 satellite \citep{1992AdSpR..12h..99Z}  have long
informed our search for radio emission from planets {\it beyond} the solar system, including the push to go to
space to enable direct detection of these emissions. She then provided a brief overview of several recent and
planned space missions with relevance to solar and stellar science. For example,
the Sun Radio Interferometer Space Experiment (SunRISE) will form a low-frequency interferometer in space
(comprising 6 toaster-sized CubeSat spacecraft) to allow
studies of energetic particles produced by CMEs and other solar bursts \citep{2019AGUFMSH33A..02K}.  
Launch is planned for
2025 or later. The CubeSat Radio Interferometry Experiment (CURIE) is another space mission designed to study
low-frequency radio emission from CMEs (see Section~\ref{solarCMEs}), in this case utilizing a two-element interferometer \citep{2016AGUFMSH11C2271S}. \footnote{CURIE launched subsequent to the RS3 conference, in July 2024.}  Both of these missions are expected to be able to
follow CMEs
as they propagate out from the Sun, and where the plasma frequency falls to low-frequency bands that
can only be accessed from space.

Adding the list of proposed small spacecraft was the Auroral Emission Radio Observer (AERO)\footnote{At the time of
the meeting, two identical spacecraft, AERO and VISTA were envisioned. The mission concept was later scaled back to
a single spacecraft (AERO), but as of mid-2025 AERO had not been selected for development funding.},
led out of MIT Haystack Observatory, which
planned to operate at 100~kHz--5~MHz \citep{2019AGUFMSA44A..07E}.  A notable feature of AERO is that
it would make use of a vector sensor antenna, which has the ability to fully characterize incident
electromagnetic waves and measure the direction of arrival of auroral radio emission and image source regions,
and at the same time obtain polarimetric information.

Knapp also
described several space mission concepts that are in their early planning stages, including 
Farside Array for Radio Science Investigations of the Dark ages and Exoplanets
(FARSIDE; \citealt{2019arXiv191108649B}), a low radio frequency interferometric array proposed for the far side of the Moon
 which would comprise 128 dipole antennas spread across 10~km,
operating from 150~kHz--40~MHz. Building the array on the lunar
far side would shield it from terrestrial RFI, as well as other solar system
noise sources, enabling near-continuous searches for radio signatures of CMEs and energetic particles from nearby
stars, along with searches for exoplanet magnetospheres. An even more ambitious project, FarView, proposes
to build a dipole array on the lunar far side through in situ mining of materials
\citep{2024AdSpR..74..528P}. 
Lastly, Knapp introduced a project she is leading,
a concept for free-flying space-based radio observatory known as Great Observatory for Long Wavelengths (GO-LoW).
GO-LoW would comprise a large constellation of reconfigurable antenna nodes located at a stable Lagrange point (L4 or L5)
and operating between 300~kHz--15~MHz. The
number of spacecraft is driven by the goal of obtaining sufficient sensitivity to observe exoplanet magnetic
fields in the solar neighborhood, including those of terrestrial planets.
GO-LoW would be in the spirit of NASA's Great Observatory concept in that
it would open an entirely new area of the electromagnetic spectrum to exploration
\citep{2024arXiv240408432K}.  Knapp
speculated that a realistic timeline for building such an array may by $\sim$20--30~yr in the future.

\subsection{Closing Perspectives}
R. Mutel (University of Iowa) closed the RS3 conference by offering some perspectives and forward-looking remarks.
He reminded us that in spite of the capabilities of current radio instruments, 
the majority of  Milky Way stars are still undetectable as ``radio stars''. Thus
the stars that we do detect and study in the radio
are often ``unusual''. (As one example, with current telescopes, the radio Sun would be
undetectable out beyond a few pc). However, these limitations continue to fuel the desire  to build new and better
radio telescopes that enable observations over a broad range of radio frequencies. 

Mutel informally polled meeting attendees to compile
a list of some of the key outstanding questions
in solar/stellar astrophysics that are likely to be
prime targets for being answered in full or in part by future radio wavelength
measurements. The resulting list (sorted by topic) included:
\paragraph{$\bullet$ The Sun and Space Weather:}
(i)  {\it How is the solar corona heated?} The question of the source of the coronal heating dates
back more than 65 years \citep{1958ApJ...128..677P},  but  there is still no fully agreed upon answer. S. White noted that
nanoflares are now a prime candidate (see
Section~\ref{coronalheating}), but it is not yet clear that they alone are sufficient.
(ii) {\it Can we learn to reliably predict space weather (including CMEs)}? See e.g., Section~\ref{solarCMEs}
for some  discussion of this point.
\paragraph{$\bullet$ Exoplanets:} (i) {\it Do all exoplanets have significant magnetic fields?} (ii) {\it How close are we
  to directly detecting exoplanets
  using radio techniques (i.e., astrometric VLBI)?} While there are still no exoplanets directly
  detected using radio techniques (Section~\ref{exodirect}), Mutel flagged this is a burgeoning field and
  predicted that as new, more sensitive instruments come online, this is likely to be one of the big
  discoveries in the next decade.
\paragraph{$\bullet$ Star formation:}
(i) {\it What is the role of the magnetoplasma (especially in solving the angular momentum problem)?} Mutel
  stressed that it is clear that in many YSOs
 infall/outflow/jets are modulated by magnetic effects, yet many star formation
  studies do not account for the pressure of the magnetic field.
(ii) {\it Can we 
build a comprehensive picture of the complex astrochemistry in star-forming regions?}  This is
  a complex topic (see Section~\ref{protochem}), but Mutel noted that an upgraded ALMA   (see Section~\ref{ALMA2030})
  should help to build a more comprehensive understanding over the next decade.
  \paragraph{$\bullet$ Ultracool Stars:}
  {\it Why are the magnetic fields of many UCDs so strong?}
  Despite years of study
  (e.g., Section~\ref{UCDs}), it remains unclear what type of dynamo operates in UCDs \citep[e.g.,][]{2009A&A...496..229R}.  
  Mutel predicted that radio astronomy will play a key role in solving this puzzle.
  \paragraph{$\bullet$ Evolved Stars:} (i) {\it What is the Galactic distribution of evolved stars?} Some new
  insights were presented at the meeting thanks to a combination of new databases and analysis approaches
  (Section~\ref{probes}) and more are forthcoming.
  (ii)  {\it Why are PNe morphologies so complex and how do they evolve from main-sequence stars?} Some new clues have
    recently come from detailed ALMA observations of AGB stars (Section~\ref{redgiantsurveys}) and new observations
    of masers in PNe and water fountain sources (Section~\ref{PNemasers}).
(iii) {\it What accounts for the recent behavior of Betelgeuse?} See Section~\ref{RSGs}.
  \paragraph{$\bullet$ Active Stars:}
  (i) {\it Why are ``active'' (solar-type) stars so much more active than the Sun?}  (ii)
            {\it Why  is there not
a continuous range in radio luminosity for solar-type active stars, but rather an apparent ``jump''
in the activity levels?} See, e.g., \cite{1995A&A...302..775G} and
  Figure~6 of \cite{2002ARA&A..40..217G}. Possibilities mentioned by Mutel are that it is related to how the electron
            acceleration scales, or else 
            differences in the large-scale magnetic field strength. (iii) {\it Where is the energetic plasma in
              interacting binaries?} Typically it is unclear whether the radiating
            plasma is associated with the active star, the less active star, or
            somewhere in between (but see Section~\ref{activebinaries}). Generally
            interactive binaries like Algols and RS CVn and W Ursa Majoris systems  are $\sim$10,000$\times$
            more luminous than the Sun, requiring
intense magnetic fields with very energetic particles, but generally we do not  know exactly where they are arising from.
Mutel suggested that better astrometric VLBI (with phase referencing), and comparisons with $Gaia$ optical astrometry,
should  be able to answer
this question in the future.

\bigskip

There was general agreement that solving the above questions will require the aid of new and better instruments
(more sensitive; better angular resolution). However, as the volume and complexity of data continue
to increase over the next decade, Mutel stressed that
better analysis tools will play an equally important role [e.g.,
machine learning, high-performance computing, artificial intelligence (AI)].

Indeed, the dramatic growth in the use of
machine learning and AI in scientific research represents one of the major
changes since the previous Haystack Radio Stars conference in 2017 \citep{Matthews2019}. Although these topics were
not  prominently featured at RS3, their emergence as tools in stellar research was highlighted in a handful
of presentations (e.g., Sections~\ref{coronalheating}, \ref{probes}).
Applications of both machine learning and AI to stellar research are expected to grow significantly in the coming
years owing to the rapidly expanding data volumes anticipated from new generations or instruments,
for example, as aids to stellar source identification and extraction in large surveys (e.g., Section~\ref{futuresurveys}).
Another dramatic change compared with a decade ago is the emerging importance for stellar science of
a new set of low-frequency imaging arrays,
including LOFAR, MWA, ASKAP, MeerKAT, and uGMRT---alongside  the VLA, ALMA, and the VLBA, all of which remain
in heavy use for stellar science at radio wavelengths.

Mutel closed with some remarks on
the question of justification for the $>$\$1 billion+ price tags of some of the currently
proposed new radio telescope arrays (Section~\ref{newtelescopes}). He invoked the concept of the
observable ``hypervolume'', whose four axes are the spectrophotometric,
temporal, astrometric, and morphological domains (\citealt{1975QJRAS..16..378H}; \citealt{2014htu..conf..215D}), noting that
certain types of objects
will inhabit a part of this hypervolume that if not searched, will not be discovered.
Examples of recent discoveries in radio astronomy that were not made until a new part of parameter space was
sampled include fast radio bursts (no one had previously searched at such a fast time domain;
 \citealt{2007Sci...318..777L})  and UCDs (which were
not expected to be detectable radio sources until someone actually looked; \citealt{2001Natur.410..338B}). 
The meeting organizers look forward to seeing what similarly paradigm-shifting discoveries await at the next
Radio Stars Conference later this decade, as we approach the 50th anniversary of the first ever conference
dedicated to radio stars \citep{1979JRASC..73..271F}. 

\begin{acknowledgments}
The author gratefully
acknowledges the contributions of the other members of the RS3 Local Organizing Committee
(H. Johnson, D. Crowley,  N. W. Kotary, D. Tonnelli, J. Tsai) and the RS3 Scientific Organizing
Committee (R. Ignace, M. Rupen, L. Sjouwerman,
S. White). 
Financial support for the RS3 conference was provided by award AST-2332009
from the National Science Foundation (NSF). The author's scientific contributions to the meeting were supported
by grants from the NSF (AST-2107681) and from the Space Telescope Science Institute (HST-GO-16655.008-A).
The author also wishes to acknowledge helpful comments from an anonymous referee.
\end{acknowledgments}

\bibliography{bib}{}

\begin{thebibliography}{}
\expandafter\ifx\csname natexlab\endcsname\relax\def\natexlab#1{#1}\fi
\providecommand{\url}[1]{\href{#1}{#1}}
\providecommand{\dodoi}[1]{doi:~\href{http://doi.org/#1}{\nolinkurl{#1}}}
\providecommand{\doeprint}[1]{\href{http://ascl.net/#1}{\nolinkurl{http://ascl.net/#1}}}
\providecommand{\doarXiv}[1]{\href{https://arxiv.org/abs/#1}{\nolinkurl{https://arxiv.org/abs/#1}}}

\bibitem[{C. {Abate} {et~al.}(2013){Abate}, {Pols}, {Izzard}, {Mohamed}, \& {de
  Mink}}]{2013A&A...552A..26A}
{Abate}, C., {Pols}, O.~R., {Izzard}, R.~G., {Mohamed}, S.~S., \& {de Mink},
  S.~E. 2013, \bibinfo{title}{{Wind Roche-lobe overflow: Application to
  carbon-enhanced metal-poor stars},} \aap, 552, A26,
  \dodoi{10.1051/0004-6361/201220007}

\bibitem[{A.~A. {Abdo} {et~al.}(2010){Abdo}, {Ackermann}, {Ajello}, {Atwood},
  {Baldini}, {Ballet}, {Barbiellini}, {Bastieri}, {Bechtol}, {Bellazzini},
  {Berenji}, {Blandford}, {Bloom}, {Bonamente}, {Borgland}, {Bouvier},
  {Brandt}, {Bregeon}, {Brez}, {Brigida}, {Bruel}, {Buehler}, {Burnett},
  {Buson}, {Caliandro}, {Cameron}, {Caraveo}, {Carrigan}, {Casandjian},
  {Cecchi}, {Celik}, {Charles}, {Chaty}, {Chekhtman}, {Cheung}, {Chiang},
  {Ciprini}, {Claus}, {Cohen-Tanugi}, {Conrad}, {Corbel}, {Corbet}, {DeCesar},
  {den Hartog}, {Dermer}, {de Palma}, {Digel}, {Donato}, {do Couto e Silva},
  {Drell}, {Dubois}, {Dubus}, {Dumora}, {Favuzzi}, {Fegan}, {Ferrara},
  {Fortin}, {Frailis}, {Fuhrmann}, {Fukazawa}, {Funk}, {Fusco}, {Gargano},
  {Gasparrini}, {Gehrels}, {Germani}, {Giglietto}, {Giordano}, {Giroletti},
  {Glanzman}, {Godfrey}, {Grenier}, {Grondin}, {Grove}, {Guiriec}, {Hadasch},
  {Harding}, {Hayashida}, {Hays}, {Healey}, {Hill}, {Horan}, {Hughes}, {Itoh},
  {Jean}, {J{\'o}hannesson}, {Johnson}, {Johnson}, {Johnson}, {Johnson},
  {Kamae}, {Katagiri}, {Kataoka}, {Kerr}, {Kn{\"o}dlseder}, {Koerding}, {Kuss},
  {Lande}, {Latronico}, {Lee}, {Lemoine-Goumard}, {Garde}, {Longo}, {Loparco},
  {Lott}, {Lovellette}, {Lubrano}, {Makeev}, {Mazziotta}, {McConville},
  {McEnery}, {Mehault}, {Michelson}, {Mizuno}, {Moiseev}, {Monte}, {Monzani},
  {Morselli}, {Moskalenko}, {Murgia}, {Nakamori}, {Naumann-Godo}, {Nestoras},
  {Nolan}, {Norris}, {Nuss}, {Ohno}, {Ohsugi}, {Okumura}, {Omodei}, {Orlando},
  {Ormes}, {Ozaki}, {Paneque}, {Panetta}, {Parent}, {Pelassa}, {Pepe},
  {Pesce-Rollins}, {Piron}, {Porter}, {Rain{\'o}}, {Rando}, {Ray}, {Razzano},
  {Razzaque}, {Rea}, {Reimer}, {Reimer}, {Reposeur}, {Ripken}, {Ritz},
  {Romani}, {Roth}, {Sadrozinski}, {Sander}, {Parkinson}, {Scargle},
  {Schinzel}, {Sgr{\`o}}, {Shaw}, {Siskind}, {Smith}, {Smith}, {Sokolovsky},
  {Spandre}, {Spinelli}, {Stawarz}, {Strickman}, {Suson}, {Takahashi},
  {Takahashi}, {Tanaka}, {Tanaka}, {Thayer}, {Thayer}, {Thompson}, {Torres},
  {Tosti}, {Tramacere}, {Uchiyama}, {Usher}, {Vandenbroucke}, {Vasileiou},
  {Vilchez}, {Vitale}, {Waite}, {Wallace}, {Wang}, {Winer}, {Wolff}, {Wood},
  {Yang}, {Ylinen}, {Ziegler}, {Maehara}, {Nishiyama}, {Kabashima}, {Bach}, \&
  {Bower}}]{2010Sci...329..817A}
{Abdo}, A.~A., {Ackermann}, M., {Ajello}, M., {et~al.} 2010,
  \bibinfo{title}{{Gamma-Ray Emission Concurrent with the Nova in the Symbiotic
  Binary V407 Cygni},} Science, 329, 817, \dodoi{10.1126/science.1192537}

\bibitem[{M. {Ackermann} {et~al.}(2014){Ackermann}, {Ajello}, {Albert},
  {Baldini}, {Ballet}, {Barbiellini}, {Bastieri}, {Bellazzini}, {Bissaldi},
  {Blandford}, {Bloom}, {Bottacini}, {Brandt}, {Bregeon}, {Bruel}, {Buehler},
  {Buson}, {Caliandro}, {Cameron}, {Caragiulo}, {Caraveo}, {Cavazzuti},
  {Charles}, {Chekhtman}, {Cheung}, {Chiang}, {Chiaro}, {Ciprini}, {Claus},
  {Cohen-Tanugi}, {Conrad}, {Corbel}, {D'Ammando}, {de Angelis}, {den Hartog},
  {de Palma}, {Dermer}, {Desiante}, {Digel}, {Di Venere}, {do Couto e Silva},
  {Donato}, {Drell}, {Drlica-Wagner}, {Favuzzi}, {Ferrara}, {Focke},
  {Franckowiak}, {Fuhrmann}, {Fukazawa}, {Fusco}, {Gargano}, {Gasparrini},
  {Germani}, {Giglietto}, {Giordano}, {Giroletti}, {Glanzman}, {Godfrey},
  {Grenier}, {Grove}, {Guiriec}, {Hadasch}, {Harding}, {Hayashida}, {Hays},
  {Hewitt}, {Hill}, {Hou}, {Jean}, {Jogler}, {J{\'o}hannesson}, {Johnson},
  {Johnson}, {Kerr}, {Kn{\"o}dlseder}, {Kuss}, {Larsson}, {Latronico},
  {Lemoine-Goumard}, {Longo}, {Loparco}, {Lott}, {Lovellette}, {Lubrano},
  {Manfreda}, {Martin}, {Massaro}, {Mayer}, {Mazziotta}, {McEnery},
  {Michelson}, {Mitthumsiri}, {Mizuno}, {Monzani}, {Morselli}, {Moskalenko},
  {Murgia}, {Nemmen}, {Nuss}, {Ohsugi}, {Omodei}, {Orienti}, {Orlando},
  {Ormes}, {Paneque}, {Panetta}, {Perkins}, {Pesce-Rollins}, {Piron}, {Pivato},
  {Porter}, {Rain{\`o}}, {Rando}, {Razzano}, {Razzaque}, {Reimer}, {Reimer},
  {Reposeur}, {Saz Parkinson}, {Schaal}, {Schulz}, {Sgr{\`o}}, {Siskind},
  {Spandre}, {Spinelli}, {Stawarz}, {Suson}, {Takahashi}, {Tanaka}, {Thayer},
  {Thayer}, {Thompson}, {Tibaldo}, {Tinivella}, {Torres}, {Tosti}, {Troja},
  {Uchiyama}, {Vianello}, {Winer}, {Wolff}, {Wood}, {Wood}, {Wood},
  {Charbonnel}, {Corbet}, {De Gennaro Aquino}, {Edlin}, {Mason}, {Schwarz},
  {Shore}, {Starrfield}, {Teyssier}, \& {Fermi-LAT
  Collaboration}}]{2014Sci...345..554A}
{Ackermann}, M., {Ajello}, M., {Albert}, A., {et~al.} 2014,
  \bibinfo{title}{{Fermi establishes classical novae as a distinct class of
  gamma-ray sources},} Science, 345, 554, \dodoi{10.1126/science.1253947}

\bibitem[{K. {Akabane} \& M.~H. {Cohen}(1961){Akabane} \&
  {Cohen}}]{1961ApJ...133..258A}
{Akabane}, K., \& {Cohen}, M.~H. 1961, \bibinfo{title}{{Polarization
  Measurements of Type III Bursts and Faraday Rotation in the Corona.},} \apj,
  133, 258, \dodoi{10.1086/147021}

\bibitem[{K. {Akiyama} \& L.~D. {Matthews}(2019){Akiyama} \&
  {Matthews}}]{2019arXiv191000013A}
{Akiyama}, K., \& {Matthews}, L.~D. 2019, \bibinfo{title}{{Exploring
  Regularized Maximum Likelihood Reconstruction for Stellar Imaging with the
  ngVLA},} arXiv e-prints, arXiv:1910.00013, \dodoi{10.48550/arXiv.1910.00013}

\bibitem[{J.~D. {Alvarado-G{\'o}mez} {et~al.}(2018){Alvarado-G{\'o}mez},
  {Drake}, {Cohen}, {Moschou}, \& {Garraffo}}]{2018ApJ...862...93A}
{Alvarado-G{\'o}mez}, J.~D., {Drake}, J.~J., {Cohen}, O., {Moschou}, S.~P., \&
  {Garraffo}, C. 2018, \bibinfo{title}{{Suppression of Coronal Mass Ejections
  in Active Stars by an Overlying Large-scale Magnetic Field: A Numerical
  Study},} \apj, 862, 93, \dodoi{10.3847/1538-4357/aacb7f}

\bibitem[{K. {Amada} {et~al.}(2024){Amada}, {Imai}, {Hamae}, {Nakashima},
  {Shum}, {Tafoya}, {Uscanga}, {G{\'o}mez}, {Orosz}, \&
  {Burns}}]{2024IAUS..380..359A}
{Amada}, K., {Imai}, H., {Hamae}, Y., {et~al.} 2024, \bibinfo{title}{{Discovery
  of SiO masers in the ``Water Fountain'' source, IRAS 16552-3050},} in IAU
  Symposium, Vol. 380, Cosmic Masers: Proper Motion Toward the Next-Generation
  Large Projects, ed. T.~{Hirota}, H.~{Imai}, K.~{Menten}, \&
  Y.~{Pihlstr{\"o}m}, 359--361, \dodoi{10.1017/S1743921323002739}

\bibitem[{M. {Andriantsaralaza} {et~al.}(2022){Andriantsaralaza}, {Ramstedt},
  {Vlemmings}, \& {De Beck}}]{2022A&A...667A..74A}
{Andriantsaralaza}, M., {Ramstedt}, S., {Vlemmings}, W.~H.~T., \& {De Beck}, E.
  2022, \bibinfo{title}{{Distance estimates for AGB stars from parallax
  measurements},} \aap, 667, A74, \dodoi{10.1051/0004-6361/202243670}

\bibitem[{Y. {Asaki} {et~al.}(2020){Asaki}, {Maud}, {Fomalont}, {Phillips},
  {Hirota}, {Sawada}, {Barcos-Mu{\~n}oz}, {Richards}, {Dent}, {Takahashi},
  {Corder}, {Carpenter}, {Villard}, \& {Humphreys}}]{2020ApJS..247...23A}
{Asaki}, Y., {Maud}, L.~T., {Fomalont}, E.~B., {et~al.} 2020,
  \bibinfo{title}{{ALMA High-frequency Long Baseline Campaign in 2017:
  Band-to-band Phase Referencing in Submillimeter Waves},} \apjs, 247, 23,
  \dodoi{10.3847/1538-4365/ab6b20}

\bibitem[{Y. {Asaki} {et~al.}(2023){Asaki}, {Maud}, {Francke}, {Nagai},
  {Petry}, {Fomalont}, {Humphreys}, {Richards}, {Wong}, {Dent}, {Hirota},
  {Fernandez}, {Takahashi}, \& {Hales}}]{2023ApJ...958...86A}
{Asaki}, Y., {Maud}, L.~T., {Francke}, H., {et~al.} 2023, \bibinfo{title}{{ALMA
  High-frequency Long Baseline Campaign in 2021: Highest Angular Resolution
  Submillimeter Wave Images for the Carbon-rich Star R Lep},} \apj, 958, 86,
  \dodoi{10.3847/1538-4357/acf619}

\bibitem[{C. {Ayala} {et~al.}(2024){Ayala}, {Dong}, {Hallinan}, {Davis},
  {Huang}, \& {Law}}]{2024AAS...24335933A}
{Ayala}, C., {Dong}, D., {Hallinan}, G., {et~al.} 2024,
  \bibinfo{title}{{Stellar Radio Transients in the VLA Sky Survey},} in
  American Astronomical Society Meeting Abstracts, Vol. 243, American
  Astronomical Society Meeting Abstracts, 359.33

\bibitem[{E. {Aydi} \& S. {Mohamed}(2022){Aydi} \&
  {Mohamed}}]{2022MNRAS.513.4405A}
{Aydi}, E., \& {Mohamed}, S. 2022, \bibinfo{title}{{3D models of the
  circumstellar environments of evolved stars: Formation of multiple spiral
  structures},} \mnras, 513, 4405, \dodoi{10.1093/mnras/stac749}

\bibitem[{P.~E. {Barrett} \& M.~A. {Gurwell}(2025){Barrett} \&
  {Gurwell}}]{2025ApJ...986...78B}
{Barrett}, P.~E., \& {Gurwell}, M.~A. 2025, \bibinfo{title}{{Submillimeter
  Observations of the White Dwarf Pulsar AR Sco},} \apj, 986, 78,
  \dodoi{10.3847/1538-4357/add725}

\bibitem[{A. {Bartkiewicz} {et~al.}(2020){Bartkiewicz}, {Sanna}, {Szymczak},
  {Moscadelli}, {van Langevelde}, \& {Wolak}}]{2020A&A...637A..15B}
{Bartkiewicz}, A., {Sanna}, A., {Szymczak}, M., {et~al.} 2020,
  \bibinfo{title}{{The nature of the methanol maser ring G23.657-00.127. II.
  Expansion of the maser structure},} \aap, 637, A15,
  \dodoi{10.1051/0004-6361/202037562}

\bibitem[{A. {Bartkiewicz} {et~al.}(2024{\natexlab{a}}){Bartkiewicz}, {Sanna},
  {Szymczak}, {Moscadelli}, {van Langevelde}, {Wolak}, {Kobak}, \&
  {Durjasz}}]{Bart2024b}
{Bartkiewicz}, A., {Sanna}, A., {Szymczak}, M., {et~al.} 2024{\natexlab{a}},
  \bibinfo{title}{{Proper motion study of the 6.7 GHz methanol maser rings. I.
  A sample of sources with little variation},} \aap, 686, A275,
  \dodoi{10.1051/0004-6361/202449491}

\bibitem[{A. {Bartkiewicz} {et~al.}(2024{\natexlab{b}}){Bartkiewicz},
  {Szymczak}, {Kobak}, \& {Aramowicz}}]{Bart2024a}
{Bartkiewicz}, A., {Szymczak}, M., {Kobak}, A., \& {Aramowicz}, M.
  2024{\natexlab{b}}, \bibinfo{title}{{Methanol and excited OH masers in W49N
  as observed using EVN},} in IAU Symposium, Vol. 380, Cosmic Masers: Proper
  Motion Toward the Next-Generation Large Projects, ed. T.~{Hirota}, H.~{Imai},
  K.~{Menten}, \& Y.~{Pihlstr{\"o}m}, 207--209,
  \dodoi{10.1017/S1743921323002466}

\bibitem[{T.~S. {Bastian} {et~al.}(2001){Bastian}, {Pick}, {Kerdraon}, {Maia},
  \& {Vourlidas}}]{2001ApJ...558L..65B}
{Bastian}, T.~S., {Pick}, M., {Kerdraon}, A., {Maia}, D., \& {Vourlidas}, A.
  2001, \bibinfo{title}{{The Coronal Mass Ejection of 1998 April 20: Direct
  Imaging at Radio Wavelengths},} \apjl, 558, L65, \dodoi{10.1086/323421}

\bibitem[{A. {Baudry} {et~al.}(2023){Baudry}, {Wong}, {Etoka}, {Richards},
  {M{\"u}ller}, {Herpin}, {Danilovich}, {Gray}, {Wallstr{\"o}m}, {Gobrecht},
  {Khouri}, {Decin}, {Gottlieb}, {Menten}, {Homan}, {Millar}, {Montarg{\`e}s},
  {Pimpanuwat}, {Plane}, \& {Kervella}}]{2023A&A...674A.125B}
{Baudry}, A., {Wong}, K.~T., {Etoka}, S., {et~al.} 2023,
  \bibinfo{title}{{ATOMIUM: Probing the inner wind of evolved O-rich stars with
  new, highly excited H$_{2}$O and OH lines},} \aap, 674, A125,
  \dodoi{10.1051/0004-6361/202245193}

\bibitem[{S. {Bawaji} {et~al.}(2023){Bawaji}, {Alam}, {Mondal}, {Oberoi}, \&
  {Biswas}}]{2023ApJ...954...39B}
{Bawaji}, S., {Alam}, U., {Mondal}, S., {Oberoi}, D., \& {Biswas}, A. 2023,
  \bibinfo{title}{{An Unsupervised Machine Learning-based Algorithm for
  Detecting Weak Impulsive Narrowband Quiet Sun Emissions and Characterizing
  Their Morphology},} \apj, 954, 39, \dodoi{10.3847/1538-4357/ace042}

\bibitem[{R. {Beck} \& M. {Krause}(2005){Beck} \&
  {Krause}}]{2005AN....326..414B}
{Beck}, R., \& {Krause}, M. 2005, \bibinfo{title}{{Revised equipartition and
  minimum energy formula for magnetic field strength estimates from radio
  synchrotron observations},} Astronomische Nachrichten, 326, 414,
  \dodoi{10.1002/asna.200510366}

\bibitem[{E.~E. {Becklin} {et~al.}(1969){Becklin}, {Frogel}, {Hyland},
  {Kristian}, \& {Neugebauer}}]{1969ApJ...158L.133B}
{Becklin}, E.~E., {Frogel}, J.~A., {Hyland}, A.~R., {Kristian}, J., \&
  {Neugebauer}, G. 1969, \bibinfo{title}{{The Unusual Infrared Object
  IRC+10216},} \apjl, 158, L133, \dodoi{10.1086/180450}

\bibitem[{P. {Benaglia} {et~al.}(2010){Benaglia}, {Romero}, {Mart{\'\i}},
  {Peri}, \& {Araudo}}]{2010A&A...517L..10B}
{Benaglia}, P., {Romero}, G.~E., {Mart{\'\i}}, J., {Peri}, C.~S., \& {Araudo},
  A.~T. 2010, \bibinfo{title}{{Detection of nonthermal emission from the bow
  shock of a massive runaway star},} \aap, 517, L10,
  \dodoi{10.1051/0004-6361/201015232}

\bibitem[{A.~O. {Benz} \& M. {Guedel}(1994){Benz} \&
  {Guedel}}]{1994A&A...285..621B}
{Benz}, A.~O., \& {Guedel}, M. 1994, \bibinfo{title}{{X-ray/microwave ratio of
  flares and coronae},} \aap, 285, 621

\bibitem[{E. {Berger} {et~al.}(2001){Berger}, {Ball}, {Becker}, {Clarke},
  {Frail}, {Fukuda}, {Hoffman}, {Mellon}, {Momjian}, {Murphy}, {Teng},
  {Woodruff}, {Zauderer}, \& {Zavala}}]{2001Natur.410..338B}
{Berger}, E., {Ball}, S., {Becker}, K.~M., {et~al.} 2001,
  \bibinfo{title}{{Discovery of radio emission from the brown dwarf LP944-20},}
  \nat, 410, 338, \dodoi{10.1038/35066514}

\bibitem[{R. {Bhattacharya} {et~al.}(2024){Bhattacharya}, {Medina},
  {Pihlstr{\"o}m}, {Sjouwerman}, {Lewis}, {Sahai}, {Stroh},
  {Quiroga-Nu{\~n}ez}, {van Langevelde}, {Claussen}, \&
  {Weller}}]{2024ApJ...969..109B}
{Bhattacharya}, R., {Medina}, B.~M., {Pihlstr{\"o}m}, Y.~M., {et~al.} 2024,
  \bibinfo{title}{{Distance Estimate Method for Asymptotic Giant Branch Stars
  Using Infrared Spectral Energy Distributions},} \apj, 969, 109,
  \dodoi{10.3847/1538-4357/ad463e}

\bibitem[{R.~V. {Bhonsle} \& L.~R. {McNarry}(1964){Bhonsle} \&
  {McNarry}}]{1964ApJ...139.1312B}
{Bhonsle}, R.~V., \& {McNarry}, L.~R. 1964, \bibinfo{title}{{Polarization
  Characteristics of Type III Solar Radio Bursts at 74 Mc/s.},} \apj, 139,
  1312, \dodoi{10.1086/147866}

\bibitem[{S. {Bhunia} {et~al.}(2023){Bhunia}, {Carley}, {Oberoi}, \&
  {Gallagher}}]{2023A&A...670A.169B}
{Bhunia}, S., {Carley}, E.~P., {Oberoi}, D., \& {Gallagher}, P.~T. 2023,
  \bibinfo{title}{{Imaging-spectroscopy of a band-split type II solar radio
  burst with the Murchison Widefield Array},} \aap, 670, A169,
  \dodoi{10.1051/0004-6361/202244456}

\bibitem[{E. {Biermann} {et~al.}(2025){Biermann}, {Li}, {Naess}, {Choi},
  {Clark}, {Devlin}, {Dunkley}, {Gallardo}, {Guan}, {Foster}, {Hasselfield},
  {Herv{\'\i}as-Caimapo}, {Hilton}, {Hincks}, {Ho}, {Hood}, {Huffenberger},
  {Kosowsky}, {Niemack}, {Orlowski-Scherer}, {Page}, {Partridge}, {Salatino},
  {Sif{\'o}n}, {Staggs}, {Vargas}, \& {Wollack}}]{2025ApJ...986....7B}
{Biermann}, E., {Li}, Y., {Naess}, S., {et~al.} 2025, \bibinfo{title}{{The
  Atacama Cosmology Telescope: Systematic Transient Search of Single
  Observation Maps},} \apj, 986, 7, \dodoi{10.3847/1538-4357/adce70}

\bibitem[{M.~F. {Bietenholz} {et~al.}(2021){Bietenholz}, {Bartel}, {Argo},
  {Dua}, {Ryder}, \& {Soderberg}}]{2021ApJ...908...75B}
{Bietenholz}, M.~F., {Bartel}, N., {Argo}, M., {et~al.} 2021,
  \bibinfo{title}{{The Radio Luminosity-risetime Function of Core-collapse
  Supernovae},} \apj, 908, 75, \dodoi{10.3847/1538-4357/abccd9}

\bibitem[{S.~J. {Billington} {et~al.}(2020){Billington}, {Urquhart},
  {K{\"o}nig}, {Beuther}, {Breen}, {Menten}, {Campbell-White}, {Ellingsen},
  {Thompson}, {Moore}, {Eden}, {Kim}, \& {Leurini}}]{2020MNRAS.499.2744B}
{Billington}, S.~J., {Urquhart}, J.~S., {K{\"o}nig}, C., {et~al.} 2020,
  \bibinfo{title}{{ATLASGAL - relationship between dense star-forming clumps
  and interstellar masers},} \mnras, 499, 2744, \dodoi{10.1093/mnras/staa2936}

\bibitem[{A. {Biswas} {et~al.}(2023){Biswas}, {Das}, {Chandra}, {Wade},
  {Shultz}, {Cavallaro}, {Petit}, {Woudt}, \& {Alecian}}]{2023MNRAS.523.5155B}
{Biswas}, A., {Das}, B., {Chandra}, P., {et~al.} 2023,
  \bibinfo{title}{{Discovery of magnetospheric interactions in the doubly
  magnetic hot binary ϵ Lupi},} \mnras, 523, 5155,
  \dodoi{10.1093/mnras/stad1756}

\bibitem[{S. {Bloot} {et~al.}(2025){Bloot}, {Vedantham}, {Kavanagh},
  {Callingham}, \& {Pope}}]{2025A&A...695A.176B}
{Bloot}, S., {Vedantham}, H.~K., {Kavanagh}, R.~D., {Callingham}, J.~R., \&
  {Pope}, B.~J.~S. 2025, \bibinfo{title}{{Catching the wisps: Stellar mass-loss
  limits from low-frequency radio observations},} \aap, 695, A176,
  \dodoi{10.1051/0004-6361/202452722}

\bibitem[{A. {Boischot} \& A. {Lecacheux}(1975){Boischot} \&
  {Lecacheux}}]{1975A&A....40...55B}
{Boischot}, A., \& {Lecacheux}, A. 1975, \bibinfo{title}{{Linear polarization
  in meter and decameter solar radiobursts.},} \aap, 40, 55

\bibitem[{B. {Bojnordi Arbab} {et~al.}(2024){Bojnordi Arbab}, {Vlemmings},
  {Khouri}, \& {H{\"o}fner}}]{2024ApJ...976..138B}
{Bojnordi Arbab}, B., {Vlemmings}, W., {Khouri}, T., \& {H{\"o}fner}, S. 2024,
  \bibinfo{title}{{Probing the Extended Atmospheres of AGB Stars. I. Synthetic
  Imaging of 1D Hydrodynamical Models at Radio and (Sub-)millimeter
  Wavelengths},} \apj, 976, 138, \dodoi{10.3847/1538-4357/ad83b8}

\bibitem[{C. {Bordiu} {et~al.}(2022){Bordiu}, {Rizzo}, {Bufano},
  {Quintana-Lacaci}, {Buemi}, {Leto}, {Cavallaro}, {Cerrigone}, {Ingallinera},
  {Loru}, {Riggi}, {Trigilio}, {Umana}, \& {Sciacca}}]{2022ApJ...939L..30B}
{Bordiu}, C., {Rizzo}, J.~R., {Bufano}, F., {et~al.} 2022,
  \bibinfo{title}{{First Detection of Silicon-bearing Molecules in
  {\ensuremath{\eta}} Car},} \apjl, 939, L30, \dodoi{10.3847/2041-8213/ac9b10}

\bibitem[{C. {Bordiu} {et~al.}(2025){Bordiu}, {Riggi}, {Bufano}, {Cavallaro},
  {Cecconello}, {Camilo}, {Umana}, {Cotton}, {Thompson}, {Bietenholz},
  {Goedhart}, {Anderson}, {Buemi}, {Chibueze}, {Ingallinera}, {Leto}, {Loru},
  {Mutale}, {Rigby}, {Trigilio}, \& {Williams}}]{2025A&A...695A.144B}
{Bordiu}, C., {Riggi}, S., {Bufano}, F., {et~al.} 2025, \bibinfo{title}{{The
  SARAO MeerKAT Galactic Plane Survey extended source catalogue},} \aap, 695,
  A144, \dodoi{10.1051/0004-6361/202450356}

\bibitem[{P. {Boven} {et~al.}(2023){Boven}, {Vedantham}, {Callingham}, \& {van
  Langevelde}}]{2023elvb.confE..50B}
{Boven}, P., {Vedantham}, H.~K., {Callingham}, J., \& {van Langevelde}, H.~J.
  2023, \bibinfo{title}{{VLBI MultiView Astrometry of Radio Stars},} in 15th
  European VLBI Network Mini-Symposium and Users' Meeting, 50

\bibitem[{G.~C. {Bower} {et~al.}(2009){Bower}, {Bolatto}, {Ford}, \&
  {Kalas}}]{2009ApJ...701.1922B}
{Bower}, G.~C., {Bolatto}, A., {Ford}, E.~B., \& {Kalas}, P. 2009,
  \bibinfo{title}{{Radio Interferometric Planet Search. I. First Constraints On
  Planetary Companions For Nearby, Low-Mass Stars From Radio Astrometry},}
  \apj, 701, 1922, \dodoi{10.1088/0004-637X/701/2/1922}

\bibitem[{C.~E. {Brasseur} {et~al.}(2024){Brasseur}, {Jardine}, \&
  {Hussain}}]{2024MNRAS.530.2442B}
{Brasseur}, C.~E., {Jardine}, M.~M., \& {Hussain}, G.~A.~J. 2024,
  \bibinfo{title}{{Constraining the coronal properties of AB Dor in the radio
  regime},} \mnras, 530, 2442, \dodoi{10.1093/mnras/stae996}

\bibitem[{R. {Braun} {et~al.}(2019){Braun}, {Bonaldi}, {Bourke}, {Keane}, \&
  {Wagg}}]{2019arXiv191212699B}
{Braun}, R., {Bonaldi}, A., {Bourke}, T., {Keane}, E., \& {Wagg}, J. 2019,
  \bibinfo{title}{{Anticipated Performance of the Square Kilometre Array --
  Phase 1 (SKA1)},} arXiv e-prints, arXiv:1912.12699,
  \dodoi{10.48550/arXiv.1912.12699}

\bibitem[{A.~H. {Bridle} \& F.~R. {Schwab}(1989){Bridle} \&
  {Schwab}}]{1989ASPC....6..247B}
{Bridle}, A.~H., \& {Schwab}, F.~R. 1989, \bibinfo{title}{{Wide Field Imaging
  I: Bandwidth and Time-Average Smearing},} in Astronomical Society of the
  Pacific Conference Series, Vol.~6, Synthesis Imaging in Radio Astronomy, ed.
  R.~A. {Perley}, F.~R. {Schwab}, \& A.~H. {Bridle}, 247

\bibitem[{A. {Brunthaler} {et~al.}(2011){Brunthaler}, {Reid}, {Menten},
  {Zheng}, {Bartkiewicz}, {Choi}, {Dame}, {Hachisuka}, {Immer}, {Moellenbrock},
  {Moscadelli}, {Rygl}, {Sanna}, {Sato}, {Wu}, {Xu}, \&
  {Zhang}}]{2011AN....332..461B}
{Brunthaler}, A., {Reid}, M.~J., {Menten}, K.~M., {et~al.} 2011,
  \bibinfo{title}{{The Bar and Spiral Structure Legacy (BeSSeL) survey: Mapping
  the Milky Way with VLBI astrometry},} Astronomische Nachrichten, 332, 461,
  \dodoi{10.1002/asna.201111560}

\bibitem[{A.~M. {Burkhardt} {et~al.}(2021){Burkhardt}, {Loomis},
  {Shingledecker}, {Lee}, {Remijan}, {McCarthy}, \&
  {McGuire}}]{2021isms.confERC07B}
{Burkhardt}, A.~M., {Loomis}, R., {Shingledecker}, C.~N., {et~al.} 2021,
  \bibinfo{title}{{a Rigorous Ka-Band Hunt for Aromatic Molecules (arkham):
  Ubiquitous Aromatic Carbon Chemistry at the Earliest Stages of Star
  Formation},} in 2021 International Symposium on Molecular Spectroscopy,
  \dodoi{10.15278/isms.2021.RC07}

\bibitem[{J.~O. {Burns} {et~al.}(2019){Burns}, {Hallinan}, {Lux}, {Teitelbaum},
  {Kocz}, {MacDowall}, {Bradley}, {Rapetti}, {Wu}, {Furlanetto}, {Austin},
  {Romero-Wolf}, {Chang}, {Bowman}, {Kasper}, {Anderson}, {Zhen}, {Pober}, \&
  {Mirocha}}]{2019arXiv191108649B}
{Burns}, J.~O., {Hallinan}, G., {Lux}, J., {et~al.} 2019, \bibinfo{title}{{NASA
  Probe Study Report: Farside Array for Radio Science Investigations of the
  Dark ages and Exoplanets (FARSIDE)},} arXiv e-prints, arXiv:1911.08649,
  \dodoi{10.48550/arXiv.1911.08649}

\bibitem[{R.~A. {Burns} {et~al.}(2023){Burns}, {Uno}, {Sakai}, {Blanchard},
  {Rosli}, {Orosz}, {Yonekura}, {Tanabe}, {Sugiyama}, {Hirota}, {Kim},
  {Aberfelds}, {Volvach}, {Bartkiewicz}, {Caratti o Garatti}, {Sobolev},
  {Stecklum}, {Brogan}, {Phillips}, {Ladeyschikov}, {Johnstone}, {Surcis},
  {MacLeod}, {Linz}, {Chibueze}, {Brand}, {Eisl{\"o}ffel}, {Hyland}, {Uscanga},
  {Olech}, {Durjasz}, {Bayandina}, {Breen}, {Ellingsen}, {van den Heever},
  {Hunter}, \& {Chen}}]{2023NatAs...7..557B}
{Burns}, R.~A., {Uno}, Y., {Sakai}, N., {et~al.} 2023, \bibinfo{title}{{A
  Keplerian disk with a four-arm spiral birthing an episodically accreting
  high-mass protostar},} Nature Astronomy, 7, 557,
  \dodoi{10.1038/s41550-023-01899-w}

\bibitem[{R.~A. {Cala} {et~al.}(2024){Cala}, {G{\'o}mez}, \&
  {Miranda}}]{2024IAUS..380..343C}
{Cala}, R.~A., {G{\'o}mez}, J.~F., \& {Miranda}, L.~F. 2024,
  \bibinfo{title}{{Nascent planetary nebulae: new identifications and
  extraordinary evolution},} in IAU Symposium, Vol. 380, Cosmic Masers: Proper
  Motion Toward the Next-Generation Large Projects, ed. T.~{Hirota}, H.~{Imai},
  K.~{Menten}, \& Y.~{Pihlstr{\"o}m}, 343--346,
  \dodoi{10.1017/S1743921323001965}

\bibitem[{J.~R. {Callingham} {et~al.}(2024){Callingham}, {Pope}, {Kavanagh},
  {Bellotti}, {Daley-Yates}, {Damasso}, {Grie{\ss}meier}, {G{\"u}del},
  {G{\"u}nther}, {Kao}, {Klein}, {Mahadevan}, {Morin}, {Nichols}, {Osten},
  {P{\'e}rez-Torres}, {Pineda}, {Rigney}, {Saur}, {Stef{\'a}nsson}, {Turner},
  {Vedantham}, {Vidotto}, {Villadsen}, \& {Zarka}}]{2024NatAs...8.1359C}
{Callingham}, J.~R., {Pope}, B.~J.~S., {Kavanagh}, R.~D., {et~al.} 2024,
  \bibinfo{title}{{Radio signatures of star-planet interactions, exoplanets and
  space weather},} Nature Astronomy, 8, 1359,
  \dodoi{10.1038/s41550-024-02405-6}

\bibitem[{C.~L. {Carilli} {et~al.}(2018){Carilli}, {Butler}, {Golap},
  {Carilli}, \& {White}}]{2018ASPC..517..369C}
{Carilli}, C.~L., {Butler}, B., {Golap}, K., {Carilli}, M.~T., \& {White},
  S.~M. 2018, \bibinfo{title}{{Imaging Stellar Radio Photospheres with the Next
  Generation Very Large Array},} in Astronomical Society of the Pacific
  Conference Series, Vol. 517, Science with a Next Generation Very Large Array,
  ed. E.~{Murphy}, 369, \dodoi{10.48550/arXiv.1810.05055}

\bibitem[{J. {Carpenter} {et~al.}(2022){Carpenter}, {Brogan}, {Iono}, \&
  {Mroczkowski}}]{ALMA621}
{Carpenter}, J., {Brogan}, C., {Iono}, D., \& {Mroczkowski}, T. 2022,
  \bibinfo{title}{{The ALMA2030 Wideband Sensitivity Upgrade},} ALMA Memo
  Series, 621

\bibitem[{J. {Carpenter} {et~al.}(2019){Carpenter}, {Iono}, {Testi}, {Whyborn},
  {Wootten}, \& {Evans}}]{2019arXiv190202856C}
{Carpenter}, J., {Iono}, D., {Testi}, L., {et~al.} 2019, \bibinfo{title}{{The
  ALMA Development Roadmap},} arXiv e-prints, arXiv:1902.02856,
  \dodoi{10.48550/arXiv.1902.02856}

\bibitem[{P. {Carral} {et~al.}(2002){Carral}, {Kurtz}, {Rodr{\'\i}guez},
  {Menten}, {Cant{\'o}}, \& {Arceo}}]{2002AJ....123.2574C}
{Carral}, P., {Kurtz}, S.~E., {Rodr{\'\i}guez}, L.~F., {et~al.} 2002,
  \bibinfo{title}{{Detection of the Winds from the Exciting Sources of Shell H
  II Regions in NGC 6334},} \aj, 123, 2574, \dodoi{10.1086/339701}

\bibitem[{J.~I. {Castor} {et~al.}(1975){Castor}, {Abbott}, \&
  {Klein}}]{1975ApJ...195..157C}
{Castor}, J.~I., {Abbott}, D.~C., \& {Klein}, R.~I. 1975,
  \bibinfo{title}{{Radiation-driven winds in Of stars.},} \apj, 195, 157,
  \dodoi{10.1086/153315}

\bibitem[{P.~W. {Cauley} {et~al.}(2019){Cauley}, {Shkolnik}, {Llama}, \&
  {Lanza}}]{2019NatAs...3.1128C}
{Cauley}, P.~W., {Shkolnik}, E.~L., {Llama}, J., \& {Lanza}, A.~F. 2019,
  \bibinfo{title}{{Magnetic field strengths of hot Jupiters from signals of
  star-planet interactions},} Nature Astronomy, 3, 1128,
  \dodoi{10.1038/s41550-019-0840-x}

\bibitem[{Y. {Cendes} {et~al.}(2022){Cendes}, {Williams}, \&
  {Berger}}]{2022AJ....163...15C}
{Cendes}, Y., {Williams}, P.~K.~G., \& {Berger}, E. 2022, \bibinfo{title}{{A
  Pilot Radio Search for Magnetic Activity in Directly Imaged Exoplanets},}
  \aj, 163, 15, \dodoi{10.3847/1538-3881/ac32c8}

\bibitem[{J. {Cernicharo} {et~al.}(2019){Cernicharo}, {Velilla-Prieto},
  {Ag{\'u}ndez}, {Pardo}, {Fonfr{\'\i}a}, {Quintana-Lacaci}, {Cabezas},
  {Berm{\'u}dez}, \& {Gu{\'e}lin}}]{2019A&A...627L...4C}
{Cernicharo}, J., {Velilla-Prieto}, L., {Ag{\'u}ndez}, M., {et~al.} 2019,
  \bibinfo{title}{{Discovery of the first Ca-bearing molecule in space: CaNC},}
  \aap, 627, L4, \dodoi{10.1051/0004-6361/201936040}

\bibitem[{R.~A. {Chevalier}(1998){Chevalier}}]{1998ApJ...499..810C}
{Chevalier}, R.~A. 1998, \bibinfo{title}{{Synchrotron Self-Absorption in Radio
  Supernovae},} \apj, 499, 810, \dodoi{10.1086/305676}

\bibitem[{A. {Chiavassa} {et~al.}(2011){Chiavassa}, {Pasquato}, {Jorissen},
  {Sacuto}, {Babusiaux}, {Freytag}, {Ludwig}, {Cruzal{\`e}bes}, {Rabbia},
  {Spang}, \& {Chesneau}}]{2011A&A...528A.120C}
{Chiavassa}, A., {Pasquato}, E., {Jorissen}, A., {et~al.} 2011,
  \bibinfo{title}{{Radiative hydrodynamic simulations of red supergiant stars.
  III. Spectro-photocentric variability, photometric variability, and
  consequences on Gaia measurements},} \aap, 528, A120,
  \dodoi{10.1051/0004-6361/201015768}

\bibitem[{L. {Chomiuk} {et~al.}(2021{\natexlab{a}}){Chomiuk}, {Metzger}, \&
  {Shen}}]{Chomiuk2021b}
{Chomiuk}, L., {Metzger}, B.~D., \& {Shen}, K.~J. 2021{\natexlab{a}},
  \bibinfo{title}{{New Insights into Classical Novae},} \araa, 59, 391,
  \dodoi{10.1146/annurev-astro-112420-114502}

\bibitem[{L. {Chomiuk} {et~al.}(2014){Chomiuk}, {Linford}, {Yang}, {O'Brien},
  {Paragi}, {Mioduszewski}, {Beswick}, {Cheung}, {Mukai}, {Nelson}, {Ribeiro},
  {Rupen}, {Sokoloski}, {Weston}, {Zheng}, {Bode}, {Eyres}, {Roy}, \&
  {Taylor}}]{2014Natur.514..339C}
{Chomiuk}, L., {Linford}, J.~D., {Yang}, J., {et~al.} 2014,
  \bibinfo{title}{{Binary orbits as the driver of {\ensuremath{\gamma}}-ray
  emission and mass ejection in classical novae},} \nat, 514, 339,
  \dodoi{10.1038/nature13773}

\bibitem[{L. {Chomiuk} {et~al.}(2021{\natexlab{b}}){Chomiuk}, {Linford},
  {Aydi}, {Bannister}, {Krauss}, {Mioduszewski}, {Mukai}, {Nelson}, {Rupen},
  {Ryder}, {Sokoloski}, {Sokolovsky}, {Strader}, {Filipovi{\'c}}, {Finzell},
  {Kawash}, {Kool}, {Metzger}, {Nyamai}, {Ribeiro}, {Roy}, {Urquhart}, \&
  {Weston}}]{Chomiuk2021a}
{Chomiuk}, L., {Linford}, J.~D., {Aydi}, E., {et~al.} 2021{\natexlab{b}},
  \bibinfo{title}{{Classical Novae at Radio Wavelengths},} \apjs, 257, 49,
  \dodoi{10.3847/1538-4365/ac24ab}

\bibitem[{U.~R. {Christensen} {et~al.}(2009){Christensen}, {Holzwarth}, \&
  {Reiners}}]{2009Natur.457..167C}
{Christensen}, U.~R., {Holzwarth}, V., \& {Reiners}, A. 2009,
  \bibinfo{title}{{Energy flux determines magnetic field strength of planets
  and stars},} \nat, 457, 167, \dodoi{10.1038/nature07626}

\bibitem[{N. {Chrysaphi} {et~al.}(2018){Chrysaphi}, {Kontar}, {Holman}, \&
  {Temmer}}]{2018ApJ...868...79C}
{Chrysaphi}, N., {Kontar}, E.~P., {Holman}, G.~D., \& {Temmer}, M. 2018,
  \bibinfo{title}{{CME-driven Shock and Type II Solar Radio Burst Band
  Splitting},} \apj, 868, 79, \dodoi{10.3847/1538-4357/aae9e5}

\bibitem[{D. {Chulkov} \& O. {Malkov}(2022){Chulkov} \&
  {Malkov}}]{2022MNRAS.517.2925C}
{Chulkov}, D., \& {Malkov}, O. 2022, \bibinfo{title}{{Visual binary stars with
  known orbits in Gaia EDR3},} \mnras, 517, 2925,
  \dodoi{10.1093/mnras/stac2827}

\bibitem[{T. {Clarke} {et~al.}(2015){Clarke}, {Kassim}, {Polisensky}, {Peters},
  {Giacintucci}, \& {Hyman}}]{2015fers.confE..19C}
{Clarke}, T., {Kassim}, N.~E., {Polisensky}, E., {et~al.} 2015,
  \bibinfo{title}{{The VLA Low Band Ionospheric and Transient Experiment
  (VLITE): A Commensal Sky Survey},} in The Many Facets of Extragalactic Radio
  Surveys: Towards New Scientific Challenges, 19, \dodoi{10.22323/1.267.0019}

\bibitem[{J.~B. {Climent} {et~al.}(2020){Climent}, {Guirado}, {Azulay},
  {Marcaide}, {Jauncey}, {Lestrade}, \& {Reynolds}}]{2020A&A...641A..90C}
{Climent}, J.~B., {Guirado}, J.~C., {Azulay}, R., {et~al.} 2020,
  \bibinfo{title}{{The milliarcsecond-scale radio structure of AB Doradus A},}
  \aap, 641, A90, \dodoi{10.1051/0004-6361/202037542}

\bibitem[{J.~B. {Climent} {et~al.}(2023){Climent}, {Guirado},
  {P{\'e}rez-Torres}, {Marcaide}, \&
  {Pe{\~n}a-Mo{\~n}ino}}]{2023Sci...381.1120C}
{Climent}, J.~B., {Guirado}, J.~C., {P{\'e}rez-Torres}, M., {Marcaide}, J.~M.,
  \& {Pe{\~n}a-Mo{\~n}ino}, L. 2023, \bibinfo{title}{{Evidence for a radiation
  belt around a brown dwarf},} Science, 381, 1120,
  \dodoi{10.1126/science.adg6635}

\bibitem[{ {Committee for a Decadal Survey on Astronomy and
  Astrophysics}(2023){Committee for a Decadal Survey on Astronomy and
  Astrophysics}}]{Astro2020}
{Committee for a Decadal Survey on Astronomy and Astrophysics}. 2023,
  \bibinfo{title}{{Pathways to Discovery in Astronomy and Astrophysics for the
  2020s},}

\bibitem[{C.~M. {Cordun} {et~al.}(2025){Cordun}, {Vedantham}, {Brentjens}, \&
  {van der Tak}}]{2025A&A...693A.162C}
{Cordun}, C.~M., {Vedantham}, H.~K., {Brentjens}, M.~A., \& {van der Tak},
  F.~F.~S. 2025, \bibinfo{title}{{Deep radio interferometric search for
  decametre radio emission from the exoplanet Tau Bo{\"o}tis b},} \aap, 693,
  A162, \dodoi{10.1051/0004-6361/202452868}

\bibitem[{S. {Costantino} {et~al.}(2023){Costantino}, {Costa}, \&
  {Noschese}}]{2023ATel16374....1C}
{Costantino}, S., {Costa}, C., \& {Noschese}, A. 2023, \bibinfo{title}{{The
  occultation of Betelgeuse by Leona: recovering the stellar surface brightness
  of a red supergiant, with a diffuse telescope, on Dec 12 1:12 UT},} The
  Astronomer's Telegram, 16374, 1

\bibitem[{S.~R. {Cranmer} \& S.~P. {Owocki}(1996){Cranmer} \&
  {Owocki}}]{1996ApJ...462..469C}
{Cranmer}, S.~R., \& {Owocki}, S.~P. 1996, \bibinfo{title}{{Hydrodynamical
  Simulations of Corotating Interaction Regions and Discrete Absorption
  Components in Rotating O-Star Winds},} \apj, 462, 469, \dodoi{10.1086/177166}

\bibitem[{S. {Curiel} {et~al.}(2024){Curiel}, {Ortiz-Le{\'o}n}, {Mioduszewski},
  \& {Arenas-Martinez}}]{2024ApJ...967..112C}
{Curiel}, S., {Ortiz-Le{\'o}n}, G.~N., {Mioduszewski}, A.~J., \&
  {Arenas-Martinez}, A.~B. 2024, \bibinfo{title}{{Precise Mass, Orbital Motion,
  and Stellar Properties of the M-dwarf Binary LP 349‑25AB},} \apj, 967, 112,
  \dodoi{10.3847/1538-4357/ad3df6}

\bibitem[{S. {Curiel} {et~al.}(2022){Curiel}, {Ortiz-Le{\'o}n}, {Mioduszewski},
  \& {Sanchez-Bermudez}}]{2022AJ....164...93C}
{Curiel}, S., {Ortiz-Le{\'o}n}, G.~N., {Mioduszewski}, A.~J., \&
  {Sanchez-Bermudez}, J. 2022, \bibinfo{title}{{3D Orbital Architecture of a
  Dwarf Binary System and Its Planetary Companion},} \aj, 164, 93,
  \dodoi{10.3847/1538-3881/ac7c66}

\bibitem[{S. {Curiel} {et~al.}(2020){Curiel}, {Ortiz-Le{\'o}n}, {Mioduszewski},
  \& {Torres}}]{2020AJ....160...97C}
{Curiel}, S., {Ortiz-Le{\'o}n}, G.~N., {Mioduszewski}, A.~J., \& {Torres},
  R.~M. 2020, \bibinfo{title}{{An Astrometric Planetary Companion Candidate to
  the M9 Dwarf TVLM 513-46546},} \aj, 160, 97, \dodoi{10.3847/1538-3881/ab9e6e}

\bibitem[{B. {Das} \& P. {Chandra}(2021){Das} \&
  {Chandra}}]{2021ApJ...921....9D}
{Das}, B., \& {Chandra}, P. 2021, \bibinfo{title}{{Ultra-wideband, Multiepoch
  Radio Study of the First Discovered ``Main-sequence Radio Pulse Emitter'' CU
  Vir},} \apj, 921, 9, \dodoi{10.3847/1538-4357/ac1075}

\bibitem[{B. {Das} {et~al.}(2024){Das}, {Chandra}, \&
  {Petit}}]{2024ApJ...974..267D}
{Das}, B., {Chandra}, P., \& {Petit}, V. 2024, \bibinfo{title}{{Coherent Radio
  Emission from ``Main-sequence Radio Pulse Emitters'': A New Stellar
  Diagnostic to Probe 3D Magnetospheric Structures},} \apj, 974, 267,
  \dodoi{10.3847/1538-4357/ad71c5}

\bibitem[{B. {Das} {et~al.}(2022{\natexlab{a}}){Das}, {Chandra}, {Shultz},
  {Leto}, {Mikul{\'a}{\v{s}}ek}, {Petit}, \& {Wade}}]{Das2022b}
{Das}, B., {Chandra}, P., {Shultz}, M.~E., {et~al.} 2022{\natexlab{a}},
  \bibinfo{title}{{Testing a scaling relation between coherent radio emission
  and physical parameters of hot magnetic stars},} \mnras, 517, 5756,
  \dodoi{10.1093/mnras/stac3123}

\bibitem[{B. {Das} {et~al.}(2020{\natexlab{a}}){Das}, {Chandra}, \&
  {Wade}}]{Das2020a}
{Das}, B., {Chandra}, P., \& {Wade}, G.~A. 2020{\natexlab{a}},
  \bibinfo{title}{{Unravelling the complex magnetosphere of the B star HD
  133880 via wideband observation of coherent radio emission},} \mnras, 499,
  702, \dodoi{10.1093/mnras/staa2499}

\bibitem[{B. {Das} {et~al.}(2020{\natexlab{b}}){Das}, {Mondal}, \&
  {Chandra}}]{Das2020b}
{Das}, B., {Mondal}, S., \& {Chandra}, P. 2020{\natexlab{b}},
  \bibinfo{title}{{A 3D Framework to Explore the Propagation Effects in Stars
  Exhibiting Electron Cyclotron Maser Emission},} \apj, 900, 156,
  \dodoi{10.3847/1538-4357/aba8fd}

\bibitem[{B. {Das} {et~al.}(2022{\natexlab{b}}){Das}, {Chandra}, {Shultz},
  {Wade}, {Sikora}, {Kochukhov}, {Neiner}, {Oksala}, \& {Alecian}}]{Das2022a}
{Das}, B., {Chandra}, P., {Shultz}, M.~E., {et~al.} 2022{\natexlab{b}},
  \bibinfo{title}{{Discovery of Eight ``Main-sequence Radio Pulse Emitters''
  Using the GMRT: Clues to the Onset of Coherent Radio Emission in Hot Magnetic
  Stars},} \apj, 925, 125, \dodoi{10.3847/1538-4357/ac2576}

\bibitem[{I. {Davis} {et~al.}(2024{\natexlab{a}}){Davis}, {Hallinan}, {Ayala},
  {Dong}, \& {Myers}}]{Davis2024a}
{Davis}, I., {Hallinan}, G., {Ayala}, C., {Dong}, D., \& {Myers}, S.
  2024{\natexlab{a}}, \bibinfo{title}{{Detection of Radio Emission from
  Super-flaring Solar-Type Stars in the VLA Sky Survey},} arXiv e-prints,
  arXiv:2408.14612, \dodoi{10.48550/arXiv.2408.14612}

\bibitem[{I. {Davis} {et~al.}(2023){Davis}, {Hallinan}, \&
  {Saini}}]{2023AAS...24134607D}
{Davis}, I., {Hallinan}, G., \& {Saini}, N. 2023, \bibinfo{title}{{Flarescope:
  Multi-wavelength Monitoring of Space Weather from Young, Sun-like Stars},} in
  American Astronomical Society Meeting Abstracts, Vol. 241, American
  Astronomical Society Meeting Abstracts, 346.07

\bibitem[{I. {Davis} {et~al.}(2024{\natexlab{b}}){Davis}, {Hallinan}, {Saini},
  \& {OVRO-LWA Collaboration}}]{Davis2024b}
{Davis}, I., {Hallinan}, G., {Saini}, N., \& {OVRO-LWA Collaboration}.
  2024{\natexlab{b}}, \bibinfo{title}{{A dedicated system for coordinated radio
  and optical monitoring of the space weather of young solar-type stars},} in
  American Astronomical Society Meeting Abstracts, Vol. 243, American
  Astronomical Society Meeting Abstracts, 355.08

\bibitem[{E. {De Beck} {et~al.}(2015){De Beck}, {Vlemmings}, {Muller}, {Black},
  {O'Gorman}, {Richards}, {Baudry}, {Maercker}, {Decin}, \&
  {Humphreys}}]{2015A&A...580A..36D}
{De Beck}, E., {Vlemmings}, W., {Muller}, S., {et~al.} 2015,
  \bibinfo{title}{{ALMA observations of TiO$_{2}$ around VY Canis Majoris},}
  \aap, 580, A36, \dodoi{10.1051/0004-6361/201525990}

\bibitem[{L. {Decin} {et~al.}(2020){Decin}, {Montarg{\`e}s}, {Richards},
  {Gottlieb}, {Homan}, {McDonald}, {El Mellah}, {Danilovich}, {Wallstr{\"o}m},
  {Zijlstra}, {Baudry}, {Bolte}, {Cannon}, {De Beck}, {De Ceuster}, {de Koter},
  {De Ridder}, {Etoka}, {Gobrecht}, {Gray}, {Herpin}, {Jeste}, {Lagadec},
  {Kervella}, {Khouri}, {Menten}, {Millar}, {M{\"u}ller}, {Plane}, {Sahai},
  {Sana}, {Van de Sande}, {Waters}, {Wong}, \& {Yates}}]{2020Sci...369.1497D}
{Decin}, L., {Montarg{\`e}s}, M., {Richards}, A.~M.~S., {et~al.} 2020,
  \bibinfo{title}{{(Sub)stellar companions shape the winds of evolved stars},}
  Science, 369, 1497, \dodoi{10.1126/science.abb1229}

\bibitem[{L. {Decin} {et~al.}(2022){Decin}, {Gottlieb}, {Richards}, {Baudry},
  {Danilovich}, {Cannon}, {Ceulemans}, {de Ceuster}, {de Koter}, {El Mellah},
  {Etoka}, {Gottlieb}, {Gray}, {Herpin}, {Homan}, {Jeste}, {Kervella}, {Maes},
  {Malfait}, {Marinho}, {Menten}, {Millar}, {McDonald}, {Montarg{\`e}s},
  {M{\"u}ller}, {Pimpanuwat}, {Plane}, {Sahai}, {van de Sande},
  {Wallstr{\"o}m}, {Wong}, \& {Atomium Consortium}}]{2022Msngr.189....3D}
{Decin}, L., {Gottlieb}, C., {Richards}, A., {et~al.} 2022,
  \bibinfo{title}{{ATOMIUM: ALMA Tracing the Origins of Molecules In dUst
  forming oxygen-rich M-type stars},} The Messenger, 189, 3,
  \dodoi{10.18727/0722-6691/5283}

\bibitem[{L. {Delgado} \& M. {Hernanz}(2019){Delgado} \&
  {Hernanz}}]{2019MNRAS.490.3691D}
{Delgado}, L., \& {Hernanz}, M. 2019, \bibinfo{title}{{Early multiwavelength
  analysis of the recurrent nova V745 Sco},} \mnras, 490, 3691,
  \dodoi{10.1093/mnras/stz2765}

\bibitem[{W.~R.~F. {Dent} {et~al.}(2024){Dent}, {Harper}, {Richards},
  {Kervella}, \& {Matthews}}]{2024ApJ...966L..13D}
{Dent}, W.~R.~F., {Harper}, G.~M., {Richards}, A.~M.~S., {Kervella}, P., \&
  {Matthews}, L.~D. 2024, \bibinfo{title}{{Detection of Rydberg Lines from the
  Atmosphere of Betelgeuse},} \apjl, 966, L13, \dodoi{10.3847/2041-8213/ad3afa}

\bibitem[{A.~M. {Derdzinski} {et~al.}(2017){Derdzinski}, {Metzger}, \&
  {Lazzati}}]{2017MNRAS.469.1314D}
{Derdzinski}, A.~M., {Metzger}, B.~D., \& {Lazzati}, D. 2017,
  \bibinfo{title}{{Radiative shocks create environments for dust formation in
  classical novae},} \mnras, 469, 1314, \dodoi{10.1093/mnras/stx829}

\bibitem[{J.~F. {Desmurs} {et~al.}(2014){Desmurs}, {Bujarrabal}, {Lindqvist},
  {Alcolea}, {Soria-Ruiz}, \& {Bergman}}]{2014A&A...565A.127D}
{Desmurs}, J.~F., {Bujarrabal}, V., {Lindqvist}, M., {et~al.} 2014,
  \bibinfo{title}{{SiO masers from AGB stars in the vibrationally excited v =
  1, v = 2, and v = 3 states},} \aap, 565, A127,
  \dodoi{10.1051/0004-6361/201423550}

\bibitem[{S. {Dey} {et~al.}(2022){Dey}, {Kansabanik}, \&
  {Oberoi}}]{2022AGUFMSH24A..01D}
{Dey}, S., {Kansabanik}, D., \& {Oberoi}, D. 2022, \bibinfo{title}{{First
  detailed polarimetric study of a group of type-III solar radio bursts with
  the Murchison Widefield Array},} in AGU Fall Meeting Abstracts, Vol. 2022,
  SH24A--01

\bibitem[{G. {Djorgovski}(2014){Djorgovski}}]{2014htu..conf..215D}
{Djorgovski}, G. 2014, \bibinfo{title}{{Astrophysics in the Era of Massive
  Time-Domain Surveys},} in The Third Hot-wiring the Transient Universe
  Workshop, ed. P.~R. {Wozniak}, M.~J. {Graham}, A.~A. {Mahabal}, \&
  R.~{Seaman}, 215--215

\bibitem[{L. {Doan} {et~al.}(2020){Doan}, {Ramstedt}, {Vlemmings}, {Mohamed},
  {H{\"o}fner}, {De Beck}, {Kerschbaum}, {Lindqvist}, {Maercker}, {Paladini},
  \& {Wittkowski}}]{2020A&A...633A..13D}
{Doan}, L., {Ramstedt}, S., {Vlemmings}, W.~H.~T., {et~al.} 2020,
  \bibinfo{title}{{The extended molecular envelope of the asymptotic giant
  branch star {\ensuremath{\pi}}$^{1}$ Gruis as seen by ALMA. II. The
  spiral-outflow observed at high-angular resolution},} \aap, 633, A13,
  \dodoi{10.1051/0004-6361/201935245}

\bibitem[{D.~Z. {Dong} {et~al.}(2021){Dong}, {Hallinan}, {Nakar}, {Ho},
  {Hughes}, {Hotokezaka}, {Myers}, {De}, {Mooley}, {Ravi}, {Horesh},
  {Kasliwal}, \& {Kulkarni}}]{2021Sci...373.1125D}
{Dong}, D.~Z., {Hallinan}, G., {Nakar}, E., {et~al.} 2021, \bibinfo{title}{{A
  transient radio source consistent with a merger-triggered core collapse
  supernova},} Science, 373, 1125, \dodoi{10.1126/science.abg6037}

\bibitem[{J.~J. {Drake} {et~al.}(2016){Drake}, {Cohen}, {Garraffo}, \&
  {Kashyap}}]{2016IAUS..320..196D}
{Drake}, J.~J., {Cohen}, O., {Garraffo}, C., \& {Kashyap}, V. 2016,
  \bibinfo{title}{{Stellar flares and the dark energy of CMEs},} in IAU
  Symposium, Vol. 320, Solar and Stellar Flares and their Effects on Planets,
  ed. A.~G. {Kosovichev}, S.~L. {Hawley}, \& P.~{Heinzel}, 196--201,
  \dodoi{10.1017/S1743921316000260}

\bibitem[{L.~N. {Driessen} {et~al.}(2023){Driessen}, {Heald}, {Duchesne},
  {Murphy}, {Lenc}, {Leung}, \& {Moss}}]{2023PASA...40...36D}
{Driessen}, L.~N., {Heald}, G., {Duchesne}, S.~W., {et~al.} 2023,
  \bibinfo{title}{{Detection of radio emission from stars via proper-motion
  searches},} \pasa, 40, e036, \dodoi{10.1017/pasa.2023.26}

\bibitem[{L.~N. {Driessen} {et~al.}(2024){Driessen}, {Pritchard}, {Murphy},
  {Heald}, {Robrade}, {Das}, {Duchesne}, {Kaplan}, {Lenc}, {Lynch},
  {Mitchell-Bolton}, {Pope}, {Rose}, {Stelzer}, {Wang}, \&
  {Zic}}]{2024PASA...41...84D}
{Driessen}, L.~N., {Pritchard}, J., {Murphy}, T., {et~al.} 2024,
  \bibinfo{title}{{The Sydney Radio Star Catalogue: Properties of radio stars
  at megahertz to gigahertz frequencies},} \pasa, 41, e084,
  \dodoi{10.1017/pasa.2024.72}

\bibitem[{S.~W. {Duchesne} {et~al.}(2023){Duchesne}, {Thomson}, {Pritchard},
  {Lenc}, {Moss}, {McConnell}, {Wieringa}, {Whiting}, {Wang}, {Wang}, {Rose},
  {Raja}, {Murphy}, {Leung}, {Huynh}, {Hotan}, {Hodgson}, \&
  {Heald}}]{2023PASA...40...34D}
{Duchesne}, S.~W., {Thomson}, A.~J.~M., {Pritchard}, J., {et~al.} 2023,
  \bibinfo{title}{{The Rapid ASKAP Continuum Survey IV: continuum imaging at
  1367.5 MHz and the first data release of RACS-mid},} \pasa, 40, e034,
  \dodoi{10.1017/pasa.2023.31}

\bibitem[{G.~A. {Dulk}(1985){Dulk}}]{1985ARA&A..23..169D}
{Dulk}, G.~A. 1985, \bibinfo{title}{{Radio emission from the sun and stars.},}
  \araa, 23, 169, \dodoi{10.1146/annurev.aa.23.090185.001125}

\bibitem[{A.~K. {Dupree} {et~al.}(2020){Dupree}, {Strassmeier}, {Matthews},
  {Uitenbroek}, {Calderwood}, {Granzer}, {Guinan}, {Leike}, {Montarg{\`e}s},
  {Richards}, {Wasatonic}, \& {Weber}}]{2020ApJ...899...68D}
{Dupree}, A.~K., {Strassmeier}, K.~G., {Matthews}, L.~D., {et~al.} 2020,
  \bibinfo{title}{{Spatially Resolved Ultraviolet Spectroscopy of the Great
  Dimming of Betelgeuse},} \apj, 899, 68, \dodoi{10.3847/1538-4357/aba516}

\bibitem[{C. {Eiroa} {et~al.}(2005){Eiroa}, {Torrelles}, {Curiel}, \&
  {Djupvik}}]{2005AJ....130..643E}
{Eiroa}, C., {Torrelles}, J.~M., {Curiel}, S., \& {Djupvik}, A.~A. 2005,
  \bibinfo{title}{{Very Large Array 3.5 cm Continuum Sources in the Serpens
  Cloud Core},} \aj, 130, 643, \dodoi{10.1086/431742}

\bibitem[{S.~P. {Ellingsen}(2007){Ellingsen}}]{2007MNRAS.377..571E}
{Ellingsen}, S.~P. 2007, \bibinfo{title}{{A GLIMPSE-based search for 6.7-GHz
  methanol masers and the lifetime of their spectral features},} \mnras, 377,
  571, \dodoi{10.1111/j.1365-2966.2007.11615.x}

\bibitem[{C. {Erba} \& R. {Ignace}(2022){Erba} \&
  {Ignace}}]{2022ApJ...932...12E}
{Erba}, C., \& {Ignace}, R. 2022, \bibinfo{title}{{Radio Spectral Energy
  Distributions for Single Massive Star Winds with Free-Free and Synchrotron
  Emission},} \apj, 932, 12, \dodoi{10.3847/1538-4357/ac6c90}

\bibitem[{P.~J. {Erickson} {et~al.}(2019){Erickson}, {Lind}, {Knapp}, {Hecht},
  {Crew}, {Volz}, {Labelle}, {Robey}, {Cahoy}, {Malphrus}, {Vierinen}, \&
  {Weatherwax}}]{2019AGUFMSA44A..07E}
{Erickson}, P.~J., {Lind}, F.~D., {Knapp}, M., {et~al.} 2019,
  \bibinfo{title}{{Exploring Earth's High Latitude Aurora in Radio: The AERO
  and VISTA Cubesat Missions},} in AGU Fall Meeting Abstracts, Vol. 2019,
  SA44A--07

\bibitem[{S. {Etoka} {et~al.}(2022){Etoka}, {Baudry}, {Richards}, {Gray},
  {Decin}, \& {Atomium Consortium}}]{2022IAUS..366..199E}
{Etoka}, S., {Baudry}, A., {Richards}, A. M.~S., {et~al.} 2022,
  \bibinfo{title}{{Tracing the inner regions of circumstellar envelopes via
  high-excitation water transitions},} in IAU Symposium, Vol. 366, The Origin
  of Outflows in Evolved Stars, ed. L.~{Decin}, A.~{Zijlstra}, \& C.~{Gielen},
  199--203, \dodoi{10.1017/S1743921322000680}

\bibitem[{S. {Etoka} {et~al.}(2024){Etoka}, {Engels}, {Ullrich},
  {Gonz{\'a}lez}, \& {L{\'o}pez-Mart{\'\i}}}]{2024IAUS..380..371E}
{Etoka}, S., {Engels}, D., {Ullrich}, T., {Gonz{\'a}lez}, J.~B., \&
  {L{\'o}pez-Mart{\'\i}}, B. 2024, \bibinfo{title}{{The loss of OH maser
  emission in the early stage of Post-AGB evolution},} in IAU Symposium, Vol.
  380, Cosmic Masers: Proper Motion Toward the Next-Generation Large Projects,
  ed. T.~{Hirota}, H.~{Imai}, K.~{Menten}, \& Y.~{Pihlstr{\"o}m}, 371--373,
  \dodoi{10.1017/S1743921323002004}

\bibitem[{P.~A. {Feldman} \& S. {Kwok}(1979){Feldman} \&
  {Kwok}}]{1979JRASC..73..271F}
{Feldman}, P.~A., \& {Kwok}, S. 1979, \bibinfo{title}{{Workshop on Radio
  Stars},} \jrasc, 73, 271

\bibitem[{A.~V. {Filippenko}(2005){Filippenko}}]{2005ASPC..342...87F}
{Filippenko}, A.~V. 2005, \bibinfo{title}{{An Observational Review of
  Core-Collapse Supernovae},} in Astronomical Society of the Pacific Conference
  Series, Vol. 342, 1604-2004: Supernovae as Cosmological Lighthouses, ed.
  M.~{Turatto}, S.~{Benetti}, L.~{Zampieri}, \& W.~{Shea}, 87

\bibitem[{B.~H. {Foing}(1990){Foing}}]{1990ASIC..319..363F}
{Foing}, B.~H. 1990, \bibinfo{title}{{Solar-Like Magnetic Structures in Active
  Close Binaries},} in NATO Advanced Study Institute (ASI) Series C, Vol. 319,
  Active Close Binaries Proceedings, NATO Advanced Study Institute, 363

\bibitem[{J.~P. {Fonfr{\'\i}a} {et~al.}(2014){Fonfr{\'\i}a},
  {Fern{\'a}ndez-L{\'o}pez}, {Ag{\'u}ndez}, {S{\'a}nchez-Contreras}, {Curiel},
  \& {Cernicharo}}]{2014MNRAS.445.3289F}
{Fonfr{\'\i}a}, J.~P., {Fern{\'a}ndez-L{\'o}pez}, M., {Ag{\'u}ndez}, M.,
  {et~al.} 2014, \bibinfo{title}{{The complex dust formation zone of the AGB
  star IRC+10216 probed with CARMA 0.25 arcsec angular resolution molecular
  observations},} \mnras, 445, 3289, \dodoi{10.1093/mnras/stu1968}

\bibitem[{J. {Forbrich} \& E. {Berger}(2009){Forbrich} \&
  {Berger}}]{2009ApJ...706L.205F}
{Forbrich}, J., \& {Berger}, E. 2009, \bibinfo{title}{{The First VLBI Detection
  of an Ultracool Dwarf: Implications for the Detectability of Sub-Stellar
  Companions},} \apjl, 706, L205, \dodoi{10.1088/0004-637X/706/2/L205}

\bibitem[{J. {Forbrich} {et~al.}(2021){Forbrich}, {Dzib}, {Reid}, \&
  {Menten}}]{2021ApJ...906...23F}
{Forbrich}, J., {Dzib}, S.~A., {Reid}, M.~J., \& {Menten}, K.~M. 2021,
  \bibinfo{title}{{A VLBA Survey of Radio Stars in the Orion Nebula Cluster. I.
  The Nonthermal Radio Population},} \apj, 906, 23,
  \dodoi{10.3847/1538-4357/abc68e}

\bibitem[{J. {Forbrich} {et~al.}(2017){Forbrich}, {Reid}, {Menten}, {Rivilla},
  {Wolk}, {Rau}, \& {Chandler}}]{2017ApJ...844..109F}
{Forbrich}, J., {Reid}, M.~J., {Menten}, K.~M., {et~al.} 2017,
  \bibinfo{title}{{Extreme Radio Flares and Associated X-Ray Variability from
  Young Stellar Objects in the Orion Nebula Cluster},} \apj, 844, 109,
  \dodoi{10.3847/1538-4357/aa7aa4}

\bibitem[{J. {Forbrich} {et~al.}(2016{\natexlab{a}}){Forbrich}, {Dupuy},
  {Reid}, {Berger}, {Rizzuto}, {Mann}, {Liu}, {Aller}, \&
  {Kraus}}]{Forbrich2016a}
{Forbrich}, J., {Dupuy}, T.~J., {Reid}, M.~J., {et~al.} 2016{\natexlab{a}},
  \bibinfo{title}{{High-precision Radio and Infrared Astrometry of LSPM
  J1314+1320AB. I. Parallax, Proper Motions, and Limits on Planets},} \apj,
  827, 22, \dodoi{10.3847/0004-637X/827/1/22}

\bibitem[{J. {Forbrich} {et~al.}(2016{\natexlab{b}}){Forbrich}, {Rivilla},
  {Menten}, {Reid}, {Chandler}, {Rau}, {Bhatnagar}, {Wolk}, \&
  {Meingast}}]{Forbrich2016b}
{Forbrich}, J., {Rivilla}, V.~M., {Menten}, K.~M., {et~al.} 2016{\natexlab{b}},
  \bibinfo{title}{{The Population of Compact Radio Sources in the Orion Nebula
  Cluster},} \apj, 822, 93, \dodoi{10.3847/0004-637X/822/2/93}

\bibitem[{B. {Freytag} {et~al.}(2012){Freytag}, {Steffen}, {Ludwig},
  {Wedemeyer-B{\"o}hm}, {Schaffenberger}, \& {Steiner}}]{2012JCoPh.231..919F}
{Freytag}, B., {Steffen}, M., {Ludwig}, H.~G., {et~al.} 2012,
  \bibinfo{title}{{Simulations of stellar convection with CO5BOLD},} Journal of
  Computational Physics, 231, 919, \dodoi{10.1016/j.jcp.2011.09.026}

\bibitem[{ {Gaia Collaboration} {et~al.}(2023){Gaia Collaboration},
  {Vallenari}, {Brown}, {Prusti}, {de Bruijne}, {Arenou}, {Babusiaux},
  {Biermann}, {Creevey}, {Ducourant}, {Evans}, {Eyer}, {Guerra}, {Hutton},
  {Jordi}, {Klioner}, {Lammers}, {Lindegren}, {Luri}, {Mignard}, {Panem},
  {Pourbaix}, {Randich}, {Sartoretti}, {Soubiran}, {Tanga}, {Walton},
  {Bailer-Jones}, {Bastian}, {Drimmel}, {Jansen}, {Katz}, {Lattanzi}, {van
  Leeuwen}, {Bakker}, {Cacciari}, {Casta{\~n}eda}, {De Angeli}, {Fabricius},
  {Fouesneau}, {Fr{\'e}mat}, {Galluccio}, {Guerrier}, {Heiter}, {Masana},
  {Messineo}, {Mowlavi}, {Nicolas}, {Nienartowicz}, {Pailler}, {Panuzzo},
  {Riclet}, {Roux}, {Seabroke}, {Sordo}, {Th{\'e}venin}, {Gracia-Abril},
  {Portell}, {Teyssier}, {Altmann}, {Andrae}, {Audard}, {Bellas-Velidis},
  {Benson}, {Berthier}, {Blomme}, {Burgess}, {Busonero}, {Busso},
  {C{\'a}novas}, {Carry}, {Cellino}, {Cheek}, {Clementini}, {Damerdji},
  {Davidson}, {de Teodoro}, {Nu{\~n}ez Campos}, {Delchambre}, {Dell'Oro},
  {Esquej}, {Fern{\'a}ndez-Hern{\'a}ndez}, {Fraile}, {Garabato},
  {Garc{\'\i}a-Lario}, {Gosset}, {Haigron}, {Halbwachs}, {Hambly}, {Harrison},
  {Hern{\'a}ndez}, {Hestroffer}, {Hodgkin}, {Holl}, {Jan{\ss}en}, {Jevardat de
  Fombelle}, {Jordan}, {Krone-Martins}, {Lanzafame}, {L{\"o}ffler}, {Marchal},
  {Marrese}, {Moitinho}, {Muinonen}, {Osborne}, {Pancino}, {Pauwels},
  {Recio-Blanco}, {Reyl{\'e}}, {Riello}, {Rimoldini}, {Roegiers}, {Rybizki},
  {Sarro}, {Siopis}, {Smith}, {Sozzetti}, {Utrilla}, {van Leeuwen}, {Abbas},
  {{\'A}brah{\'a}m}, {Abreu Aramburu}, {Aerts}, {Aguado}, {Ajaj},
  {Aldea-Montero}, {Altavilla}, {{\'A}lvarez}, {Alves}, {Anders}, {Anderson},
  {Anglada Varela}, {Antoja}, {Baines}, {Baker}, {Balaguer-N{\'u}{\~n}ez},
  {Balbinot}, {Balog}, {Barache}, {Barbato}, {Barros}, {Barstow},
  {Bartolom{\'e}}, {Bassilana}, {Bauchet}, {Becciani}, {Bellazzini},
  {Berihuete}, {Bernet}, {Bertone}, {Bianchi}, {Binnenfeld}, {Blanco-Cuaresma},
  {Blazere}, {Boch}, {Bombrun}, {Bossini}, {Bouquillon}, {Bragaglia},
  {Bramante}, {Breedt}, {Bressan}, {Brouillet}, {Brugaletta}, {Bucciarelli},
  {Burlacu}, {Butkevich}, {Buzzi}, {Caffau}, {Cancelliere}, {Cantat-Gaudin},
  {Carballo}, {Carlucci}, {Carnerero}, {Carrasco}, {Casamiquela}, {Castellani},
  {Castro-Ginard}, {Chaoul}, {Charlot}, {Chemin}, {Chiaramida}, {Chiavassa},
  {Chornay}, {Comoretto}, {Contursi}, {Cooper}, {Cornez}, {Cowell}, {Crifo},
  {Cropper}, {Crosta}, {Crowley}, {Dafonte}, {Dapergolas}, {David}, {David},
  {de Laverny}, {De Luise}, \& {De March}}]{2023A&A...674A...1G}
{Gaia Collaboration}, {Vallenari}, A., {Brown}, A.~G.~A., {et~al.} 2023,
  \bibinfo{title}{{Gaia Data Release 3. Summary of the content and survey
  properties},} \aap, 674, A1, \dodoi{10.1051/0004-6361/202243940}

\bibitem[{D.~E. {Gary}(2023){Gary}}]{2023ARA&A..61..427G}
{Gary}, D.~E. 2023, \bibinfo{title}{{New Insights from Imaging Spectroscopy of
  Solar Radio Emission},} \araa, 61, 427,
  \dodoi{10.1146/annurev-astro-071221-052744}

\bibitem[{K.~V. {Getman} {et~al.}(2024){Getman}, {Feigelson}, {Waggoner},
  {Cleeves}, {Forbrich}, {Ninan}, {Kochukhov}, {Airapetian}, {Dzib}, {Law}, \&
  {Rab}}]{2024ApJ...976..195G}
{Getman}, K.~V., {Feigelson}, E.~D., {Waggoner}, A.~R., {et~al.} 2024,
  \bibinfo{title}{{Multi-Observatory Research of Young Stellar Energetic Flares
  (MORYSEF): X-Ray-flare-related Phenomena and Multi-epoch Behavior},} \apj,
  976, 195, \dodoi{10.3847/1538-4357/ad8562}

\bibitem[{A. {Ginsburg} {et~al.}(2018){Ginsburg}, {Bally}, {Goddi}, {Plambeck},
  \& {Wright}}]{2018ApJ...860..119G}
{Ginsburg}, A., {Bally}, J., {Goddi}, C., {Plambeck}, R., \& {Wright}, M. 2018,
  \bibinfo{title}{{A Keplerian Disk around Orion SrCI, a {\ensuremath{\sim}} 15
  M $_{{\ensuremath{\odot}}}$ YSO},} \apj, 860, 119,
  \dodoi{10.3847/1538-4357/aac205}

\bibitem[{A. {Ginsburg} {et~al.}(2019){Ginsburg}, {McGuire}, {Plambeck},
  {Bally}, {Goddi}, \& {Wright}}]{2019ApJ...872...54G}
{Ginsburg}, A., {McGuire}, B., {Plambeck}, R., {et~al.} 2019,
  \bibinfo{title}{{Orion SrcI{\textquoteright}s Disk Is Salty},} \apj, 872, 54,
  \dodoi{10.3847/1538-4357/aafb71}

\bibitem[{M. {Giroletti} {et~al.}(2020){Giroletti}, {Munari}, {K{\"o}rding},
  {Mioduszewski}, {Sokoloski}, {Cheung}, {Corbel}, {Schinzel}, {Sokolovsky}, \&
  {O'Brien}}]{2020A&A...638A.130G}
{Giroletti}, M., {Munari}, U., {K{\"o}rding}, E., {et~al.} 2020,
  \bibinfo{title}{{Very long baseline interferometry imaging of the advancing
  ejecta in the first gamma-ray nova V407 Cygni},} \aap, 638, A130,
  \dodoi{10.1051/0004-6361/202038142}

\bibitem[{S. {Goedhart} {et~al.}(2024){Goedhart}, {Cotton}, {Camilo},
  {Thompson}, {Umana}, {Bietenholz}, {Woudt}, {Anderson}, {Bordiu}, {Buckley},
  {Buemi}, {Bufano}, {Cavallaro}, {Chen}, {Chibueze}, {Egbo}, {Frank}, {Hoare},
  {Ingallinera}, {Irabor}, {Kraan-Korteweg}, {Kurapati}, {Leto}, {Loru},
  {Mutale}, {Obonyo}, {Plavin}, {Rajohnson}, {Rigby}, {Riggi}, {Seidu},
  {Serra}, {Smart}, {Stappers}, {Steyn}, {Surnis}, {Trigilio}, {Williams},
  {Abbott}, {Adam}, {Asad}, {Baloyi}, {Bauermeister}, {Bennet}, {Bester},
  {Botha}, {Brederode}, {Buchner}, {Burger}, {Cheetham}, {Cloete}, {de
  Villiers}, {de Villiers}, {du Toit}, {Esterhuyse}, {Fanaroff}, {Fourie},
  {Gamatham}, {Gatsi}, {Geyer}, {Gouws}, {Gumede}, {Heywood}, {Hokwana},
  {Hoosen}, {Horn}, {Horrell}, {Hugo}, {Isaacson}, {J{\'o}zsa}, {Jonas},
  {Jordaan}, {Joubert}, {Julie}, {Kapp}, {Kriek}, {Kriel}, {Krishnan}, {Kusel},
  {Legodi}, {Lehmensiek}, {Lord}, {Macfarlane}, {Magnus}, {Magozore}, {Main},
  {Malan}, {Manley}, {Marais}, {Maree}, {Martens}, {Maruping}, {McAlpine},
  {Merry}, {Mgodeli}, {Millenaar}, {Mokone}, {Monama}, {New}, {Ngcebetsha},
  {Ngoasheng}, {Nicolson}, {Ockards}, {Oozeer}, {Passmoor}, {Patel},
  {Peens-Hough}, {Perkins}, {Ramaila}, {Ratcliffe}, {Renil}, {Richter},
  {Salie}, {Sambu}, {Schollar}, {Schwardt}, {Schwartz}, {Serylak}, {Siebrits},
  {Sirothia}, {Slabber}, {Smirnov}, {Tiplady}, {van Balla}, {van der Byl}, {Van
  Tonder}, {Venter}, {Venter}, {Welz}, \& {Williams}}]{2024MNRAS.531..649G}
{Goedhart}, S., {Cotton}, W.~D., {Camilo}, F., {et~al.} 2024,
  \bibinfo{title}{{The SARAO MeerKAT 1.3 GHz Galactic Plane Survey},} \mnras,
  531, 649, \dodoi{10.1093/mnras/stae1166}

\bibitem[{W.~W. {Golay} {et~al.}(2024){Golay}, {Mutel}, \&
  {Abbuhl}}]{2024ApJ...965...86G}
{Golay}, W.~W., {Mutel}, R.~L., \& {Abbuhl}, E.~E. 2024,
  \bibinfo{title}{{Time-lapse Very Long Baseline Interferometry Imaging of the
  Close Active Binary HR 1099},} \apj, 965, 86,
  \dodoi{10.3847/1538-4357/ad29fb}

\bibitem[{W.~W. {Golay} {et~al.}(2023){Golay}, {Mutel}, {Lipman}, \&
  {G{\"u}del}}]{2023MNRAS.522.1394G}
{Golay}, W.~W., {Mutel}, R.~L., {Lipman}, D., \& {G{\"u}del}, M. 2023,
  \bibinfo{title}{{A search for thermal gyro-synchrotron emission from hot
  stellar coronae},} \mnras, 522, 1394, \dodoi{10.1093/mnras/stad980}

\bibitem[{I. {Gonidakis} {et~al.}(2013){Gonidakis}, {Diamond}, \&
  {Kemball}}]{2013MNRAS.433.3133G}
{Gonidakis}, I., {Diamond}, P.~J., \& {Kemball}, A.~J. 2013, \bibinfo{title}{{A
  long-term VLBA monitoring campaign of the v = 1, J = 1 {\textrightarrow}0 SiO
  masers towards TX Cam - I. Morphology and shock waves},} \mnras, 433, 3133,
  \dodoi{10.1093/mnras/stt954}

\bibitem[{C.~A. {Gottlieb} {et~al.}(2022){Gottlieb}, {Decin}, {Richards}, {De
  Ceuster}, {Homan}, {Wallstr{\"o}m}, {Danilovich}, {Millar}, {Montarg{\`e}s},
  {Wong}, {McDonald}, {Baudry}, {Bolte}, {Cannon}, {De Beck}, {de Koter}, {El
  Mellah}, {Etoka}, {Gobrecht}, {Gray}, {Herpin}, {Jeste}, {Kervella},
  {Khouri}, {Lagadec}, {Maes}, {Malfait}, {Menten}, {M{\"u}ller}, {Pimpanuwat},
  {Plane}, {Sahai}, {Van de Sande}, {Waters}, {Yates}, \&
  {Zijlstra}}]{2022A&A...660A..94G}
{Gottlieb}, C.~A., {Decin}, L., {Richards}, A.~M.~S., {et~al.} 2022,
  \bibinfo{title}{{ATOMIUM: ALMA tracing the origins of molecules in dust
  forming oxygen rich M-type stars. Motivation, sample, calibration, and
  initial results},} \aap, 660, A94, \dodoi{10.1051/0004-6361/202140431}

\bibitem[{R.~J.~M. {Grognard} \& D.~J. {McLean}(1973){Grognard} \&
  {McLean}}]{1973SoPh...29..149G}
{Grognard}, R.~J.~M., \& {McLean}, D.~J. 1973, \bibinfo{title}{{Non-Existence
  of Linear Polarization in Type III Solar Bursts at 80 MHz},} \solphys, 29,
  149, \dodoi{10.1007/BF00153446}

\bibitem[{J.~H. {Grunhut} {et~al.}(2012){Grunhut}, {Wade}, \& {MiMeS
  Collaboration}}]{2012AIPC.1429...67G}
{Grunhut}, J.~H., {Wade}, G.~A., \& {MiMeS Collaboration}. 2012,
  \bibinfo{title}{{The incidence of magnetic fields in massive stars: An
  overview of the MiMeS survey component},} in American Institute of Physics
  Conference Series, Vol. 1429, Stellar Polarimetry: from Birth to Death, ed.
  J.~L. {Hoffman}, J.~{Bjorkman}, \& B.~{Whitney} (AIP), 67--74,
  \dodoi{10.1063/1.3701903}

\bibitem[{M. {G{\"u}del}(2002){G{\"u}del}}]{2002ARA&A..40..217G}
{G{\"u}del}, M. 2002, \bibinfo{title}{{Stellar Radio Astronomy: Probing Stellar
  Atmospheres from Protostars to Giants},} \araa, 40, 217,
  \dodoi{10.1146/annurev.astro.40.060401.093806}

\bibitem[{M. {Guedel} \& A.~O. {Benz}(1993){Guedel} \&
  {Benz}}]{1993ApJ...405L..63G}
{Guedel}, M., \& {Benz}, A.~O. 1993, \bibinfo{title}{{X-Ray/Microwave Relation
  of Different Types of Active Stars},} \apjl, 405, L63, \dodoi{10.1086/186766}

\bibitem[{M. {Guedel} {et~al.}(1995){Guedel}, {Schmitt}, \&
  {Benz}}]{1995A&A...302..775G}
{Guedel}, M., {Schmitt}, J.~H.~M.~M., \& {Benz}, A.~O. 1995,
  \bibinfo{title}{{Microwave emission from X-ray bright solar-like stars: the
  F-G main sequence and beyond.},} \aap, 302, 775

\bibitem[{M. {Gu{\'e}lin} {et~al.}(2018){Gu{\'e}lin}, {Patel}, {Bremer},
  {Cernicharo}, {Castro-Carrizo}, {Pety}, {Fonfr{\'\i}a}, {Ag{\'u}ndez},
  {Santander-Garc{\'\i}a}, {Quintana-Lacaci}, {Velilla Prieto}, {Blundell}, \&
  {Thaddeus}}]{2018A&A...610A...4G}
{Gu{\'e}lin}, M., {Patel}, N.~A., {Bremer}, M., {et~al.} 2018,
  \bibinfo{title}{{IRC +10 216 in 3D: morphology of a TP-AGB star envelope},}
  \aap, 610, A4, \dodoi{10.1051/0004-6361/201731619}

\bibitem[{E.~F. {Guinan} {et~al.}(2019){Guinan}, {Wasatonic}, \&
  {Calderwood}}]{2019ATel13341....1G}
{Guinan}, E.~F., {Wasatonic}, R.~J., \& {Calderwood}, T.~J. 2019,
  \bibinfo{title}{{The Fainting of the Nearby Red Supergiant Betelgeuse},} The
  Astronomer's Telegram, 13341, 1

\bibitem[{S. {Guns} {et~al.}(2021){Guns}, {Foster}, {Daley}, {Rahlin},
  {Whitehorn}, {Ade}, {Ahmed}, {Anderes}, {Anderson}, {Archipley}, {Avva},
  {Aylor}, {Balkenhol}, {Barry}, {Basu Thakur}, {Benabed}, {Bender}, {Benson},
  {Bianchini}, {Bleem}, {Bouchet}, {Bryant}, {Byrum}, {Carlstrom}, {Carter},
  {Cecil}, {Chang}, {Chaubal}, {Chen}, {Cho}, {Chou}, {Cliche}, {Crawford},
  {Cukierman}, {de Haan}, {Denison}, {Dibert}, {Ding}, {Dobbs}, {Dutcher},
  {Everett}, {Feng}, {Ferguson}, {Fu}, {Galli}, {Gambrel}, {Gardner},
  {Goeckner-Wald}, {Gualtieri}, {Gupta}, {Guyser}, {Halverson},
  {Harke-Hosemann}, {Harrington}, {Henning}, {Hilton}, {Hivon}, {Holder},
  {Holzapfel}, {Hood}, {Howe}, {Huang}, {Irwin}, {Jeong}, {Jonas}, {Jones},
  {Khaire}, {Knox}, {Kofman}, {Korman}, {Kubik}, {Kuhlmann}, {Kuo}, {Lee},
  {Leitch}, {Lowitz}, {Lu}, {Marrone}, {Meyer}, {Michalik}, {Millea},
  {Montgomery}, {Nadolski}, {Natoli}, {Nguyen}, {Noble}, {Novosad}, {Omori},
  {Padin}, {Pan}, {Paschos}, {Pearson}, {Phadke}, {Posada}, {Prabhu}, {Quan},
  {Reichardt}, {Riebel}, {Riedel}, {Rouble}, {Ruhl}, {Sayre}, {Schiappucci},
  {Shirokoff}, {Smecher}, {Sobrin}, {Stark}, {Stephen}, {Story}, {Suzuki},
  {Thompson}, {Thorne}, {Tucker}, {Umilta}, {Vale}, {Vieira}, {Wang}, {Wu},
  {Yefremenko}, {Yoon}, {Young}, \& {Zhang}}]{2021ApJ...916...98G}
{Guns}, S., {Foster}, A., {Daley}, C., {et~al.} 2021,
  \bibinfo{title}{{Detection of Galactic and Extragalactic
  Millimeter-wavelength Transient Sources with SPT-3G},} \apj, 916, 98,
  \dodoi{10.3847/1538-4357/ac06a3}

\bibitem[{G. {Hallinan} {et~al.}(2015){Hallinan}, {Littlefair}, {Cotter},
  {Bourke}, {Harding}, {Pineda}, {Butler}, {Golden}, {Basri}, {Doyle}, {Kao},
  {Berdyugina}, {Kuznetsov}, {Rupen}, \& {Antonova}}]{2015Natur.523..568H}
{Hallinan}, G., {Littlefair}, S.~P., {Cotter}, G., {et~al.} 2015,
  \bibinfo{title}{{Magnetospherically driven optical and radio aurorae at the
  end of the stellar main sequence},} \nat, 523, 568,
  \dodoi{10.1038/nature14619}

\bibitem[{M. {Harwit}(1975){Harwit}}]{1975QJRAS..16..378H}
{Harwit}, M. 1975, \bibinfo{title}{{The Number of Class A Phenomena
  Characterizing the Universe},} \qjras, 16, 378

\bibitem[{S. {Hess} {et~al.}(2008){Hess}, {Cecconi}, \&
  {Zarka}}]{2008GeoRL..3513107H}
{Hess}, S., {Cecconi}, B., \& {Zarka}, P. 2008, \bibinfo{title}{{Modeling of
  Io-Jupiter decameter arcs, emission beaming and energy source},} \grl, 35,
  L13107, \dodoi{10.1029/2008GL033656}

\bibitem[{I. {Heywood} {et~al.}(2022){Heywood}, {Rammala}, {Camilo}, {Cotton},
  {Yusef-Zadeh}, {Abbott}, {Adam}, {Adams}, {Aldera}, {Asad}, {Bauermeister},
  {Bennett}, {Bester}, {Bode}, {Botha}, {Botha}, {Brederode}, {Buchner},
  {Burger}, {Cheetham}, {de Villiers}, {Dikgale-Mahlakoana}, {du Toit},
  {Esterhuyse}, {Fanaroff}, {February}, {Fourie}, {Frank}, {Gamatham}, {Geyer},
  {Goedhart}, {Gouws}, {Gumede}, {Hlakola}, {Hokwana}, {Hoosen}, {Horrell},
  {Hugo}, {Isaacson}, {J{\'o}zsa}, {Jonas}, {Joubert}, {Julie}, {Kapp},
  {Kenyon}, {Kotz{\'e}}, {Kriek}, {Kriel}, {Krishnan}, {Lehmensiek},
  {Liebenberg}, {Lord}, {Lunsky}, {Madisa}, {Magnus}, {Mahgoub}, {Makhaba},
  {Makhathini}, {Malan}, {Manley}, {Marais}, {Martens}, {Mauch}, {Merry},
  {Millenaar}, {Mnyandu}, {Mokone}, {Monama}, {Mphego}, {New}, {Ngcebetsha},
  {Ngoasheng}, {Ockards}, {Oozeer}, {Otto}, {Passmoor}, {Patel}, {Peens-Hough},
  {Perkins}, {Ramaila}, {Ramanujam}, {Ramudzuli}, {Ratcliffe}, {Robyntjies},
  {Salie}, {Sambu}, {Schollar}, {Schwardt}, {Schwartz}, {Serylak}, {Siebrits},
  {Sirothia}, {Slabber}, {Smirnov}, {Sofeya}, {Taljaard}, {Tasse}, {Tiplady},
  {Toruvanda}, {Twum}, {van Balla}, {van der Byl}, {van der Merwe}, {Van
  Tonder}, {Van Wyk}, {Venter}, {Venter}, {Wallace}, {Welz}, {Williams}, \&
  {Xaia}}]{2022ApJ...925..165H}
{Heywood}, I., {Rammala}, I., {Camilo}, F., {et~al.} 2022, \bibinfo{title}{{The
  1.28 GHz MeerKAT Galactic Center Mosaic},} \apj, 925, 165,
  \dodoi{10.3847/1538-4357/ac449a}

\bibitem[{R.~M. {Hjellming}(1988){Hjellming}}]{Hjellming1988}
{Hjellming}, R.~M. 1988, \bibinfo{title}{{Radio stars.},} in Galactic and
  Extragalactic Radio Astronomy, ed. K.~I. {Kellermann} \& G.~L. {Verschuur},
  381--438

\bibitem[{R.~M. {Hjellming} {et~al.}(1996){Hjellming}, {Rupen}, {Shrader},
  {Campbell-Wilson}, {Hunstead}, \& {McKay}}]{1996ApJ...470L.105H}
{Hjellming}, R.~M., {Rupen}, M.~P., {Shrader}, C.~R., {et~al.} 1996,
  \bibinfo{title}{{Radio and X-Ray Flaring Events in X-Ray Nova Ophiuchi
  1993},} \apjl, 470, L105, \dodoi{10.1086/310312}

\bibitem[{S. {H{\"o}fner} {et~al.}(2016){H{\"o}fner}, {Bladh}, {Aringer}, \&
  {Ahuja}}]{2016A&A...594A.108H}
{H{\"o}fner}, S., {Bladh}, S., {Aringer}, B., \& {Ahuja}, R. 2016,
  \bibinfo{title}{{Dynamic atmospheres and winds of cool luminous giants. I.
  Al$_{2}$O$_{3}$ and silicate dust in the close vicinity of M-type AGB
  stars},} \aap, 594, A108, \dodoi{10.1051/0004-6361/201628424}

\bibitem[{S. {H{\"o}fner} {et~al.}(2022){H{\"o}fner}, {Bladh}, {Aringer}, \&
  {Eriksson}}]{2022A&A...657A.109H}
{H{\"o}fner}, S., {Bladh}, S., {Aringer}, B., \& {Eriksson}, K. 2022,
  \bibinfo{title}{{Dynamic atmospheres and winds of cool luminous giants. II.
  Gradual Fe enrichment of wind-driving silicate grains},} \aap, 657, A109,
  \dodoi{10.1051/0004-6361/202141224}

\bibitem[{S. {H{\"o}fner} \& H. {Olofsson}(2018){H{\"o}fner} \&
  {Olofsson}}]{2018A&ARv..26....1H}
{H{\"o}fner}, S., \& {Olofsson}, H. 2018, \bibinfo{title}{{Mass loss of stars
  on the asymptotic giant branch. Mechanisms, models and measurements},} \aapr,
  26, 1, \dodoi{10.1007/s00159-017-0106-5}

\bibitem[{W. {Homan} {et~al.}(2020){Homan}, {Montarg{\`e}s}, {Pimpanuwat},
  {Richards}, {Wallstr{\"o}m}, {Kervella}, {Decin}, {Zijlstra}, {Danilovich},
  {de Koter}, {Menten}, {Sahai}, {Plane}, {Lee}, {Waters}, {Baudry}, {Wong},
  {Millar}, {Van de Sande}, {Lagadec}, {Gobrecht}, {Yates}, {Price}, {Cannon},
  {Bolte}, {De Ceuster}, {Herpin}, {Nuth}, {Philip Sindel}, {Kee}, {Grey},
  {Etoka}, {Jeste}, {Gottlieb}, {Gottlieb}, {McDonald}, {El Mellah}, \&
  {M{\"u}ller}}]{2020A&A...644A..61H}
{Homan}, W., {Montarg{\`e}s}, M., {Pimpanuwat}, B., {et~al.} 2020,
  \bibinfo{title}{{ATOMIUM: A high-resolution view on the highly asymmetric
  wind of the AGB star {\ensuremath{\pi}}$^{1}$Gruis. I. First detection of a
  new companion and its effect on the inner wind},} \aap, 644, A61,
  \dodoi{10.1051/0004-6361/202039185}

\bibitem[{R.~M. {Humphreys} \& K. {Davidson}(1994){Humphreys} \&
  {Davidson}}]{1994PASP..106.1025H}
{Humphreys}, R.~M., \& {Davidson}, K. 1994, \bibinfo{title}{{The Luminous Blue
  Variables: Astrophysical Geysers},} \pasp, 106, 1025, \dodoi{10.1086/133478}

\bibitem[{R.~M. {Humphreys} {et~al.}(2024){Humphreys}, {Richards}, {Davidson},
  {Singh}, {Decin}, \& {Ziurys}}]{2024AJ....167...94H}
{Humphreys}, R.~M., {Richards}, A.~M.~S., {Davidson}, K., {et~al.} 2024,
  \bibinfo{title}{{The Hidden Clumps in VY CMa Uncovered by the Atacama Large
  Millimeter/submillimeter Array},} \aj, 167, 94,
  \dodoi{10.3847/1538-3881/ad1dd7}

\bibitem[{R.~M. {Humphreys} {et~al.}(2019){Humphreys}, {Ziurys}, {Bernal},
  {Gordon}, {Helton}, {Ishibashi}, {Jones}, {Richards}, \&
  {Vlemmings}}]{2019ApJ...874L..26H}
{Humphreys}, R.~M., {Ziurys}, L.~M., {Bernal}, J.~J., {et~al.} 2019,
  \bibinfo{title}{{The Unexpected Spectrum of the Innermost Ejecta of the Red
  Hypergiant VY CMa},} \apjl, 874, L26, \dodoi{10.3847/2041-8213/ab11e5}

\bibitem[{R. {Ignace}(2024){Ignace}}]{2024AAS...24440905I}
{Ignace}, R. 2024, \bibinfo{title}{{An Application of a Rigidly Rotating
  Magnetosphere (RRM) Model for the Broad Emission Lines of J005311},} in
  American Astronomical Society Meeting Abstracts, Vol. 244, American
  Astronomical Society Meeting Abstracts, 409.05

\bibitem[{R. {Ignace} {et~al.}(2020){Ignace}, {St-Louis}, \&
  {Prinja}}]{2020MNRAS.497.1127I}
{Ignace}, R., {St-Louis}, N., \& {Prinja}, R.~K. 2020, \bibinfo{title}{{Radio
  variability from corotating interaction regions threading Wolf-Rayet winds},}
  \mnras, 497, 1127, \dodoi{10.1093/mnras/staa2014}

\bibitem[{H. {Imai}(2007){Imai}}]{2007IAUS..242..279I}
{Imai}, H. 2007, \bibinfo{title}{{Stellar molecular jets traced by maser
  emission},} in IAU Symposium, Vol. 242, Astrophysical Masers and their
  Environments, ed. J.~M. {Chapman} \& W.~A. {Baan}, 279--286,
  \dodoi{10.1017/S1743921307013130}

\bibitem[{P. {Iwanek} {et~al.}(2022){Iwanek}, {Soszy{\'n}ski}, {Koz{\l}owski},
  {Poleski}, {Pietrukowicz}, {Skowron}, {Wrona}, {Mr{\'o}z}, {Udalski},
  {Szyma{\'n}ski}, {Skowron}, {Ulaczyk}, {Gromadzki}, {Rybicki}, \&
  {Ratajczak}}]{2022ApJS..260...46I}
{Iwanek}, P., {Soszy{\'n}ski}, I., {Koz{\l}owski}, S., {et~al.} 2022,
  \bibinfo{title}{{The OGLE Collection of Variable Stars: Nearly 66,000 Mira
  Stars in the Milky Way},} \apjs, 260, 46, \dodoi{10.3847/1538-4365/ac6676}

\bibitem[{T. {Kami{\'n}ski} {et~al.}(2013){Kami{\'n}ski}, {Gottlieb}, {Menten},
  {Patel}, {Young}, {Br{\"u}nken}, {M{\"u}ller}, {McCarthy}, {Winters}, \&
  {Decin}}]{2013A&A...551A.113K}
{Kami{\'n}ski}, T., {Gottlieb}, C.~A., {Menten}, K.~M., {et~al.} 2013,
  \bibinfo{title}{{Pure rotational spectra of TiO and TiO$_{2}$ in VY Canis
  Majoris},} \aap, 551, A113, \dodoi{10.1051/0004-6361/201220290}

\bibitem[{J. {Kang} {et~al.}(2024){Kang}, {Kim}, {Kim}, {Hirota}, \& {KaVA SF
  team}}]{2024IAUS..380..218K}
{Kang}, J., {Kim}, M., {Kim}, K.-T., {Hirota}, T., \& {KaVA SF team}. 2024,
  \bibinfo{title}{{Jet and Outflows in Massive Star Forming Region:
  G10.34-0.14},} in IAU Symposium, Vol. 380, Cosmic Masers: Proper Motion
  Toward the Next-Generation Large Projects, ed. T.~{Hirota}, H.~{Imai},
  K.~{Menten}, \& Y.~{Pihlstr{\"o}m}, 218--220,
  \dodoi{10.1017/S1743921323003332}

\bibitem[{D. {Kansabanik}(2022){Kansabanik}}]{Kansabanik2022a}
{Kansabanik}, D. 2022, \bibinfo{title}{{Working Principle of the Calibration
  Algorithm for High Dynamic Range Solar Imaging with the Square Kilometre
  Array Precursor},} \solphys, 297, 122, \dodoi{10.1007/s11207-022-02053-x}

\bibitem[{D. {Kansabanik} {et~al.}(2023{\natexlab{a}}){Kansabanik}, {Bera},
  {Oberoi}, \& {Mondal}}]{Kansabanik2023a}
{Kansabanik}, D., {Bera}, A., {Oberoi}, D., \& {Mondal}, S. 2023{\natexlab{a}},
  \bibinfo{title}{{Tackling the Unique Challenges of Low-frequency Solar
  Polarimetry with the Square Kilometre Array Low Precursor: Pipeline
  Implementation},} \apjs, 264, 47, \dodoi{10.3847/1538-4365/acac79}

\bibitem[{D. {Kansabanik} {et~al.}(2023{\natexlab{b}}){Kansabanik}, {Mondal},
  \& {Oberoi}}]{Kansabanik2023b}
{Kansabanik}, D., {Mondal}, S., \& {Oberoi}, D. 2023{\natexlab{b}},
  \bibinfo{title}{{Deciphering Faint Gyrosynchrotron Emission from a Coronal
  Mass Ejection Using Spectropolarimetric Radio Imaging},} \apj, 950, 164,
  \dodoi{10.3847/1538-4357/acc385}

\bibitem[{D. {Kansabanik} {et~al.}(2024{\natexlab{a}}){Kansabanik}, {Mondal},
  \& {Oberoi}}]{Kansabanik2024a}
{Kansabanik}, D., {Mondal}, S., \& {Oberoi}, D. 2024{\natexlab{a}},
  \bibinfo{title}{{Spectropolarimetric Radio Imaging of Faint Gyrosynchrotron
  Emission from a CME: A Possible Indication of the Insufficiency of
  Homogeneous Models},} \apj, 968, 55, \dodoi{10.3847/1538-4357/ad43e9}

\bibitem[{D. {Kansabanik} {et~al.}(2022){Kansabanik}, {Mondal}, {Oberoi},
  {Biswas}, \& {Bhunia}}]{Kansabanik2022b}
{Kansabanik}, D., {Mondal}, S., {Oberoi}, D., {Biswas}, A., \& {Bhunia}, S.
  2022, \bibinfo{title}{{Robust Absolute Solar Flux Density Calibration for the
  Murchison Widefield Array},} \apj, 927, 17, \dodoi{10.3847/1538-4357/ac4bba}

\bibitem[{D. {Kansabanik} {et~al.}(2024{\natexlab{b}}){Kansabanik}, {Mondal},
  {Oberoi}, {Chibueze}, {Engelbrecht}, {Strauss}, {Kontar}, {Botha}, {Steyn},
  \& {Nel}}]{Kansabanik2024b}
{Kansabanik}, D., {Mondal}, S., {Oberoi}, D., {et~al.} 2024{\natexlab{b}},
  \bibinfo{title}{{Spectroscopic Imaging of the Sun with MeerKAT: Opening a New
  Frontier in Solar Physics},} \apj, 961, 96, \dodoi{10.3847/1538-4357/ad0b7f}

\bibitem[{M.~M. {Kao} {et~al.}(2016){Kao}, {Hallinan}, {Pineda}, {Escala},
  {Burgasser}, {Bourke}, \& {Stevenson}}]{2016ApJ...818...24K}
{Kao}, M.~M., {Hallinan}, G., {Pineda}, J.~S., {et~al.} 2016,
  \bibinfo{title}{{Auroral Radio Emission from Late L and T Dwarfs: A New
  Constraint on Dynamo Theory in the Substellar Regime},} \apj, 818, 24,
  \dodoi{10.3847/0004-637X/818/1/24}

\bibitem[{M.~M. {Kao} {et~al.}(2023){Kao}, {Mioduszewski}, {Villadsen}, \&
  {Shkolnik}}]{2023Natur.619..272K}
{Kao}, M.~M., {Mioduszewski}, A.~J., {Villadsen}, J., \& {Shkolnik}, E.~L.
  2023, \bibinfo{title}{{Resolved imaging confirms a radiation belt around an
  ultracool dwarf},} \nat, 619, 272, \dodoi{10.1038/s41586-023-06138-w}

\bibitem[{M.~M. {Kao} \& J.~S. {Pineda}(2022){Kao} \&
  {Pineda}}]{2022ApJ...932...21K}
{Kao}, M.~M., \& {Pineda}, J.~S. 2022, \bibinfo{title}{{Radio Emission from
  Binary Ultracool Dwarf Systems},} \apj, 932, 21,
  \dodoi{10.3847/1538-4357/ac660b}

\bibitem[{M.~M. {Kao} \& J.~S. {Pineda}(2025){Kao} \&
  {Pineda}}]{2025MNRAS.539.2292K}
{Kao}, M.~M., \& {Pineda}, J.~S. 2025, \bibinfo{title}{{Binarity enhances the
  occurrence rate of radiation belt emissions in ultracool dwarfs},} \mnras,
  539, 2292, \dodoi{10.1093/mnras/stae905}

\bibitem[{M.~M. {Kao} \& E.~L. {Shkolnik}(2024){Kao} \&
  {Shkolnik}}]{2024MNRAS.527.6835K}
{Kao}, M.~M., \& {Shkolnik}, E.~L. 2024, \bibinfo{title}{{The occurrence rate
  of quiescent radio emission for ultracool dwarfs using a generalized
  semi-analytical Bayesian framework},} \mnras, 527, 6835,
  \dodoi{10.1093/mnras/stad2272}

\bibitem[{M. {Karovska} {et~al.}(1997){Karovska}, {Hack}, {Raymond}, \&
  {Guinan}}]{1997ApJ...482L.175K}
{Karovska}, M., {Hack}, W., {Raymond}, J., \& {Guinan}, E. 1997,
  \bibinfo{title}{{First Hubble Space Telescope Observations of Mira AB
  Wind-accreting Binary Systems},} \apjl, 482, L175, \dodoi{10.1086/310704}

\bibitem[{M. {Karovska} {et~al.}(2005){Karovska}, {Schlegel}, {Hack},
  {Raymond}, \& {Wood}}]{2005ApJ...623L.137K}
{Karovska}, M., {Schlegel}, E., {Hack}, W., {Raymond}, J.~C., \& {Wood}, B.~E.
  2005, \bibinfo{title}{{A Large X-Ray Outburst in Mira A},} \apjl, 623, L137,
  \dodoi{10.1086/430111}

\bibitem[{J.~C. {Kasper} {et~al.}(2019){Kasper}, {Lazio}, {Romero-Wolf},
  {Bain}, {Bastian}, {Cohen}, {Landi}, {Manchester}, {Hegedus}, {Schwadron},
  {Sokolov}, {Cecconi}, {Hallinan}, {Krupar}, {Maksimovic}, {Moschou},
  {Zaslavsky}, {Lux}, \& {Neilsen}}]{2019AGUFMSH33A..02K}
{Kasper}, J.~C., {Lazio}, J., {Romero-Wolf}, A., {et~al.} 2019,
  \bibinfo{title}{{The Sun Radio Interferometer Space Experiment (SunRISE)
  Mission Concept},} in AGU Fall Meeting Abstracts, Vol. 2019, SH33A--02

\bibitem[{J.~H. {Kastner} {et~al.}(2024){Kastner}, {Wilner}, {Moraga Baez},
  {Bublitz}, {De Marco}, {Sahai}, \& {Wootten}}]{2024ApJ...965...21K}
{Kastner}, J.~H., {Wilner}, D.~J., {Moraga Baez}, P., {et~al.} 2024,
  \bibinfo{title}{{The Molecular Exoskeleton of the Ring-like Planetary Nebula
  NGC 3132},} \apj, 965, 21, \dodoi{10.3847/1538-4357/ad2848}

\bibitem[{R.~D. {Kavanagh} {et~al.}(2024){Kavanagh}, {Vedantham}, {Rose}, \&
  {Bloot}}]{2024A&A...692A..66K}
{Kavanagh}, R.~D., {Vedantham}, H.~K., {Rose}, K., \& {Bloot}, S. 2024,
  \bibinfo{title}{{Unravelling sub-stellar magnetospheres},} \aap, 692, A66,
  \dodoi{10.1051/0004-6361/202452094}

\bibitem[{R.~D. {Kavanagh} {et~al.}(2021){Kavanagh}, {Vidotto}, {Klein},
  {Jardine}, {Donati}, \& {{\'O} Fionnag{\'a}in}}]{2021MNRAS.504.1511K}
{Kavanagh}, R.~D., {Vidotto}, A.~A., {Klein}, B., {et~al.} 2021,
  \bibinfo{title}{{Planet-induced radio emission from the coronae of M dwarfs:
  the case of Prox Cen and AU Mic},} \mnras, 504, 1511,
  \dodoi{10.1093/mnras/stab929}

\bibitem[{P. {Kervella} {et~al.}(2018){Kervella}, {Decin}, {Richards},
  {Harper}, {McDonald}, {O'Gorman}, {Montarg{\`e}s}, {Homan}, \&
  {Ohnaka}}]{2018A&A...609A..67K}
{Kervella}, P., {Decin}, L., {Richards}, A. M.~S., {et~al.} 2018,
  \bibinfo{title}{{The close circumstellar environment of Betelgeuse. V.
  Rotation velocity and molecular envelope properties from ALMA},} \aap, 609,
  A67, \dodoi{10.1051/0004-6361/201731761}

\bibitem[{Z. {Keszthelyi} {et~al.}(2021){Keszthelyi}, {Meynet}, {Martins}, {de
  Koter}, \& {David-Uraz}}]{2021MNRAS.504.2474K}
{Keszthelyi}, Z., {Meynet}, G., {Martins}, F., {de Koter}, A., \& {David-Uraz},
  A. 2021, \bibinfo{title}{{The effects of surface fossil magnetic fields on
  massive star evolution - III. The case of {\ensuremath{\tau}} Sco},} \mnras,
  504, 2474, \dodoi{10.1093/mnras/stab893}

\bibitem[{Z. {Keszthelyi} {et~al.}(2024){Keszthelyi}, {Puls}, {Chiaki},
  {Nagakura}, {ud-Doula}, {Takiwaki}, \& {Tominaga}}]{2024MNRAS.533.3457K}
{Keszthelyi}, Z., {Puls}, J., {Chiaki}, G., {et~al.} 2024, \bibinfo{title}{{The
  effects of surface fossil magnetic fields on massive star evolution: V.
  Models at low metallicity},} \mnras, 533, 3457,
  \dodoi{10.1093/mnras/stae1855}

\bibitem[{Z. {Keszthelyi} {et~al.}(2020){Keszthelyi}, {Meynet}, {Shultz},
  {David-Uraz}, {ud-Doula}, {Townsend}, {Wade}, {Georgy}, {Petit}, \&
  {Owocki}}]{2020MNRAS.493..518K}
{Keszthelyi}, Z., {Meynet}, G., {Shultz}, M.~E., {et~al.} 2020,
  \bibinfo{title}{{The effects of surface fossil magnetic fields on massive
  star evolution - II. Implementation of magnetic braking in MESA and
  implications for the evolution of surface rotation in OB stars},} \mnras,
  493, 518, \dodoi{10.1093/mnras/staa237}

\bibitem[{H. {Kim} {et~al.}(2015){Kim}, {Lee}, {Mauron}, \&
  {Chu}}]{2015ApJ...804L..10K}
{Kim}, H., {Lee}, H.-G., {Mauron}, N., \& {Chu}, Y.-H. 2015,
  \bibinfo{title}{{HST Images Reveal Dramatic Changes in the Core of
  IRC+10216},} \apjl, 804, L10, \dodoi{10.1088/2041-8205/804/1/L10}

\bibitem[{A.~E. {Kimball} {et~al.}(2009){Kimball}, {Knapp}, {Ivezi{\'c}},
  {West}, {Bochanski}, {Plotkin}, \& {Gordon}}]{2009ApJ...701..535K}
{Kimball}, A.~E., {Knapp}, G.~R., {Ivezi{\'c}}, {\v{Z}}., {et~al.} 2009,
  \bibinfo{title}{{A Sample of Candidate Radio Stars in First and SDSS},} \apj,
  701, 535, \dodoi{10.1088/0004-637X/701/1/535}

\bibitem[{R. {Klement} {et~al.}(2017){Klement}, {Carciofi}, {Rivinius},
  {Matthews}, {Vieira}, {Ignace}, {Bjorkman}, {Mota}, {Faes}, {Bratcher},
  {Cur{\'e}}, \& {{\v{S}}tefl}}]{2017A&A...601A..74K}
{Klement}, R., {Carciofi}, A.~C., {Rivinius}, T., {et~al.} 2017,
  \bibinfo{title}{{Revealing the structure of the outer disks of Be stars},}
  \aap, 601, A74, \dodoi{10.1051/0004-6361/201629932}

\bibitem[{M. {Knapp} {et~al.}(2024){Knapp}, {Paritsky}, {Kononov}, \&
  {Kao}}]{2024arXiv240408432K}
{Knapp}, M., {Paritsky}, L., {Kononov}, E., \& {Kao}, M.~M. 2024,
  \bibinfo{title}{{Great Observatory for Long Wavelengths (GO-LoW) NIAC Phase I
  Final Report},} arXiv e-prints, arXiv:2404.08432,
  \dodoi{10.48550/arXiv.2404.08432}

\bibitem[{A. {Kobak} {et~al.}(2023){Kobak}, {Bartkiewicz}, {Szymczak}, {Olech},
  {Durjasz}, {Wolak}, {Chibueze}, {Hirota}, {Eisl{\"o}ffel}, {Stecklum},
  {Sobolev}, {Bayandina}, {Orosz}, {Burns}, {Kim}, \& {van den
  Heever}}]{2023A&A...671A.135K}
{Kobak}, A., {Bartkiewicz}, A., {Szymczak}, M., {et~al.} 2023,
  \bibinfo{title}{{Multi-frequency VLBI observations of maser lines during the
  6.7 GHz maser flare in the high-mass young stellar object G24.33+0.14},}
  \aap, 671, A135, \dodoi{10.1051/0004-6361/202244772}

\bibitem[{L.~A. {Koelemay} \& L.~M. {Ziurys}(2023){Koelemay} \&
  {Ziurys}}]{2023ApJ...958L...6K}
{Koelemay}, L.~A., \& {Ziurys}, L.~M. 2023, \bibinfo{title}{{Elusive Iron:
  Detection of the FeC Radical (X $^{3}${\ensuremath{\Delta}}$_{ i }$) in the
  Envelope of IRC+10216},} \apjl, 958, L6, \dodoi{10.3847/2041-8213/ad0899}

\bibitem[{J.~E. {Kooi} {et~al.}(2021){Kooi}, {Ascione}, {Reyes-Rosa}, {Rier},
  \& {Ashas}}]{2021SoPh..296...11K}
{Kooi}, J.~E., {Ascione}, M.~L., {Reyes-Rosa}, L.~V., {Rier}, S.~K., \&
  {Ashas}, M. 2021, \bibinfo{title}{{VLA Measurements of Faraday Rotation
  Through a Coronal Mass Ejection Using Multiple Lines of Sight},} \solphys,
  296, 11, \dodoi{10.1007/s11207-020-01755-4}

\bibitem[{J.~E. {Kooi} {et~al.}(2017){Kooi}, {Fischer}, {Buffo}, \&
  {Spangler}}]{2017SoPh..292...56K}
{Kooi}, J.~E., {Fischer}, P.~D., {Buffo}, J.~J., \& {Spangler}, S.~R. 2017,
  \bibinfo{title}{{VLA Measurements of Faraday Rotation through Coronal Mass
  Ejections},} \solphys, 292, 56, \dodoi{10.1007/s11207-017-1074-7}

\bibitem[{S. {Krucker} \& A.~O. {Benz}(2000){Krucker} \&
  {Benz}}]{2000SoPh..191..341K}
{Krucker}, S., \& {Benz}, A.~O. 2000, \bibinfo{title}{{Are Heating Events in
  the Quiet Solar Corona Small Flares? Multiwavelength Observations of
  Individual Events},} \solphys, 191, 341, \dodoi{10.1023/A:1005255608792}

\bibitem[{M. {Lacy} {et~al.}(2020){Lacy}, {Baum}, {Chandler}, {Chatterjee},
  {Clarke}, {Deustua}, {English}, {Farnes}, {Gaensler}, {Gugliucci},
  {Hallinan}, {Kent}, {Kimball}, {Law}, {Lazio}, {Marvil}, {Mao}, {Medlin},
  {Mooley}, {Murphy}, {Myers}, {Osten}, {Richards}, {Rosolowsky}, {Rudnick},
  {Schinzel}, {Sivakoff}, {Sjouwerman}, {Taylor}, {White}, {Wrobel},
  {Andernach}, {Beasley}, {Berger}, {Bhatnager}, {Birkinshaw}, {Bower},
  {Brandt}, {Brown}, {Burke-Spolaor}, {Butler}, {Comerford}, {Demorest}, {Fu},
  {Giacintucci}, {Golap}, {G{\"u}th}, {Hales}, {Hiriart}, {Hodge}, {Horesh},
  {Ivezi{\'c}}, {Jarvis}, {Kamble}, {Kassim}, {Liu}, {Loinard}, {Lyons},
  {Masters}, {Mezcua}, {Moellenbrock}, {Mroczkowski}, {Nyland}, {O'Dea},
  {O'Sullivan}, {Peters}, {Radford}, {Rao}, {Robnett}, {Salcido}, {Shen},
  {Sobotka}, {Witz}, {Vaccari}, {van Weeren}, {Vargas}, {Williams}, \&
  {Yoon}}]{2020PASP..132c5001L}
{Lacy}, M., {Baum}, S.~A., {Chandler}, C.~J., {et~al.} 2020,
  \bibinfo{title}{{The Karl G. Jansky Very Large Array Sky Survey (VLASS).
  Science Case and Survey Design},} \pasp, 132, 035001,
  \dodoi{10.1088/1538-3873/ab63eb}

\bibitem[{D. {Ladeyschikov}(2024){Ladeyschikov}}]{2024IAUS..380..230L}
{Ladeyschikov}, D. 2024, \bibinfo{title}{{Early Star Formation Traced by Water
  Masers},} in IAU Symposium, Vol. 380, Cosmic Masers: Proper Motion Toward the
  Next-Generation Large Projects, ed. T.~{Hirota}, H.~{Imai}, K.~{Menten}, \&
  Y.~{Pihlstr{\"o}m}, 230--231, \dodoi{10.1017/S1743921323003290}

\bibitem[{P. {Leto} {et~al.}(2016){Leto}, {Trigilio}, {Buemi}, {Umana},
  {Ingallinera}, \& {Cerrigone}}]{2016MNRAS.459.1159L}
{Leto}, P., {Trigilio}, C., {Buemi}, C.~S., {et~al.} 2016, \bibinfo{title}{{3D
  modelling of stellar auroral radio emission},} \mnras, 459, 1159,
  \dodoi{10.1093/mnras/stw639}

\bibitem[{P. {Leto} {et~al.}(2017){Leto}, {Trigilio}, {Oskinova}, {Ignace},
  {Buemi}, {Umana}, {Ingallinera}, {Todt}, \& {Leone}}]{2017MNRAS.467.2820L}
{Leto}, P., {Trigilio}, C., {Oskinova}, L., {et~al.} 2017, \bibinfo{title}{{The
  detection of variable radio emission from the fast rotating magnetic hot
  B-star HR\textbackslashxA07355 and evidence for its X-ray aurorae},} \mnras,
  467, 2820, \dodoi{10.1093/mnras/stx267}

\bibitem[{P. {Leto} {et~al.}(2018){Leto}, {Trigilio}, {Oskinova}, {Ignace},
  {Buemi}, {Umana}, {Ingallinera}, {Leone}, {Phillips}, {Agliozzo}, {Todt}, \&
  {Cerrigone}}]{2018MNRAS.476..562L}
{Leto}, P., {Trigilio}, C., {Oskinova}, L.~M., {et~al.} 2018,
  \bibinfo{title}{{A combined multiwavelength VLA/ALMA/Chandra study unveils
  the complex magnetosphere of the B-type star HR5907},} \mnras, 476, 562,
  \dodoi{10.1093/mnras/sty244}

\bibitem[{P. {Leto} {et~al.}(2021){Leto}, {Trigilio}, {Krti{\v{c}}ka},
  {Fossati}, {Ignace}, {Shultz}, {Buemi}, {Cerrigone}, {Umana}, {Ingallinera},
  {Bordiu}, {Pillitteri}, {Bufano}, {Oskinova}, {Agliozzo}, {Cavallaro},
  {Riggi}, {Loru}, {Todt}, {Giarrusso}, {Phillips}, {Robrade}, \&
  {Leone}}]{2021MNRAS.507.1979L}
{Leto}, P., {Trigilio}, C., {Krti{\v{c}}ka}, J., {et~al.} 2021,
  \bibinfo{title}{{A scaling relationship for non-thermal radio emission from
  ordered magnetospheres: from the top of the main sequence to planets},}
  \mnras, 507, 1979, \dodoi{10.1093/mnras/stab2168}

\bibitem[{M.~O. {Lewis} {et~al.}(2023){Lewis}, {Bhattacharya}, {Sjouwerman},
  {Pihlstr{\"o}m}, {Pietrzy{\'n}ski}, {Sahai}, {Karczmarek}, \&
  {G{\'o}rski}}]{2023A&A...677A.153L}
{Lewis}, M.~O., {Bhattacharya}, R., {Sjouwerman}, L.~O., {et~al.} 2023,
  \bibinfo{title}{{Long-period maser-bearing Miras in the Galactic center.
  Period-luminosity relations and extinction estimates},} \aap, 677, A153,
  \dodoi{10.1051/0004-6361/202346568}

\bibitem[{M.~O. {Lewis} {et~al.}(2024){Lewis}, {Pihlstr{\"o}m}, \&
  {Sjouwerman}}]{2024IAUS..380..314L}
{Lewis}, M.~O., {Pihlstr{\"o}m}, Y.~M., \& {Sjouwerman}, L.~O. 2024,
  \bibinfo{title}{{SiO maser line ratios in the BAaDE Survey},} in IAU
  Symposium, Vol. 380, Cosmic Masers: Proper Motion Toward the Next-Generation
  Large Projects, ed. T.~{Hirota}, H.~{Imai}, K.~{Menten}, \&
  Y.~{Pihlstr{\"o}m}, 314--318, \dodoi{10.1017/S1743921323002375}

\bibitem[{M.~O. {Lewis} {et~al.}(2020{\natexlab{a}}){Lewis}, {Pihlstr{\"o}m},
  {Sjouwerman}, \& {Quiroga-Nu{\~n}ez}}]{Lewis2020b}
{Lewis}, M.~O., {Pihlstr{\"o}m}, Y.~M., {Sjouwerman}, L.~O., \&
  {Quiroga-Nu{\~n}ez}, L.~H. 2020{\natexlab{a}}, \bibinfo{title}{{Infrared
  Color Separation between Thin-shelled Oxygen-rich and Carbon-rich AGB
  Stars},} \apj, 901, 98, \dodoi{10.3847/1538-4357/abaf46}

\bibitem[{M.~O. {Lewis} {et~al.}(2020{\natexlab{b}}){Lewis}, {Pihlstr{\"o}m},
  {Sjouwerman}, {Stroh}, {Morris}, \& {BAaDE Collaboration}}]{Lewis2020a}
{Lewis}, M.~O., {Pihlstr{\"o}m}, Y.~M., {Sjouwerman}, L.~O., {et~al.}
  2020{\natexlab{b}}, \bibinfo{title}{{Carbon- and Oxygen-rich Asymptotic Giant
  Branch (AGB) Stars in the Bulge Asymmetries and Dynamical Evolution (BAaDE)
  Survey},} \apj, 892, 52, \dodoi{10.3847/1538-4357/ab7920}

\bibitem[{M.~O. {Lewis} {et~al.}(2020{\natexlab{c}}){Lewis}, {Pihlstr{\"o}m},
  {Sjouwerman}, {Stroh}, {Morris}, \& {BAaDE
  Collaboration}}]{2020ApJ...892...52L}
{Lewis}, M.~O., {Pihlstr{\"o}m}, Y.~M., {Sjouwerman}, L.~O., {et~al.}
  2020{\natexlab{c}}, \bibinfo{title}{{Carbon- and Oxygen-rich Asymptotic Giant
  Branch (AGB) Stars in the Bulge Asymmetries and Dynamical Evolution (BAaDE)
  Survey},} \apj, 892, 52, \dodoi{10.3847/1538-4357/ab7920}

\bibitem[{J. {Lim} {et~al.}(1998){Lim}, {Carilli}, {White}, {Beasley}, \&
  {Marson}}]{1998Natur.392..575L}
{Lim}, J., {Carilli}, C.~L., {White}, S.~M., {Beasley}, A.~J., \& {Marson},
  R.~G. 1998, \bibinfo{title}{{Large convection cells as the source of
  Betelgeuse's extended atmosphere},} \nat, 392, 575, \dodoi{10.1038/33352}

\bibitem[{J. {Lim} \& S.~M. {White}(1996){Lim} \&
  {White}}]{1996ApJ...462L..91L}
{Lim}, J., \& {White}, S.~M. 1996, \bibinfo{title}{{Limits to Mass Outflows
  from Late-Type Dwarf Stars},} \apjl, 462, L91, \dodoi{10.1086/310038}

\bibitem[{J.~D. {Linford} {et~al.}(2018){Linford}, {Chomiuk}, \&
  {Rupen}}]{2018ASPC..517..271L}
{Linford}, J.~D., {Chomiuk}, L., \& {Rupen}, M.~P. 2018, \bibinfo{title}{{ngVLA
  Studies of Classical Novae},} in Astronomical Society of the Pacific
  Conference Series, Vol. 517, Science with a Next Generation Very Large Array,
  ed. E.~{Murphy}, 271

\bibitem[{J.~L. {Linsky}(1996){Linsky}}]{Linsky1996}
{Linsky}, J.~L. 1996, \bibinfo{title}{{Steady Radio Emission from Stars:
  Observations and Emission Processes},} in Astronomical Society of the Pacific
  Conference Series, Vol.~93, Radio Emission from the Stars and the Sun, ed.
  A.~R. {Taylor} \& J.~M. {Paredes}, 439

\bibitem[{J.~L. {Linsky} {et~al.}(1992){Linsky}, {Drake}, \&
  {Bastian}}]{1992ApJ...393..341L}
{Linsky}, J.~L., {Drake}, S.~A., \& {Bastian}, T.~S. 1992,
  \bibinfo{title}{{Radio Emission from Chemically Peculiar Stars},} \apj, 393,
  341, \dodoi{10.1086/171509}

\bibitem[{D.~R. {Lorimer} {et~al.}(2007){Lorimer}, {Bailes}, {McLaughlin},
  {Narkevic}, \& {Crawford}}]{2007Sci...318..777L}
{Lorimer}, D.~R., {Bailes}, M., {McLaughlin}, M.~A., {Narkevic}, D.~J., \&
  {Crawford}, F. 2007, \bibinfo{title}{{A Bright Millisecond Radio Burst of
  Extragalactic Origin},} Science, 318, 777, \dodoi{10.1126/science.1147532}

\bibitem[{J.-Z. {Ma} {et~al.}(2024){Ma}, {Chiavassa}, {de Mink}, {Valli},
  {Justham}, \& {Freytag}}]{2024ApJ...962L..36M}
{Ma}, J.-Z., {Chiavassa}, A., {de Mink}, S.~E., {et~al.} 2024,
  \bibinfo{title}{{Is Betelgeuse Really Rotating? Synthetic ALMA Observations
  of Large-scale Convection in 3D Simulations of Red Supergiants},} \apjl, 962,
  L36, \dodoi{10.3847/2041-8213/ad24fd}

\bibitem[{A.~M. {MacGregor} {et~al.}(2020){MacGregor}, {Osten}, \&
  {Hughes}}]{2020ApJ...891...80M}
{MacGregor}, A.~M., {Osten}, R.~A., \& {Hughes}, A.~M. 2020,
  \bibinfo{title}{{Properties of M Dwarf Flares at Millimeter Wavelengths},}
  \apj, 891, 80, \dodoi{10.3847/1538-4357/ab711d}

\bibitem[{M.~A. {MacGregor} {et~al.}(2018){MacGregor}, {Weinberger}, {Wilner},
  {Kowalski}, \& {Cranmer}}]{2018ApJ...855L...2M}
{MacGregor}, M.~A., {Weinberger}, A.~J., {Wilner}, D.~J., {Kowalski}, A.~F., \&
  {Cranmer}, S.~R. 2018, \bibinfo{title}{{Detection of a Millimeter Flare from
  Proxima Centauri},} \apjl, 855, L2, \dodoi{10.3847/2041-8213/aaad6b}

\bibitem[{M.~A. {MacGregor} {et~al.}(2021){MacGregor}, {Weinberger}, {Loyd},
  {Shkolnik}, {Barclay}, {Howard}, {Zic}, {Osten}, {Cranmer}, {Kowalski},
  {Lenc}, {Youngblood}, {Estes}, {Wilner}, {Forbrich}, {Hughes}, {Law},
  {Murphy}, {Boley}, \& {Matthews}}]{2021ApJ...911L..25M}
{MacGregor}, M.~A., {Weinberger}, A.~J., {Loyd}, R.~O.~P., {et~al.} 2021,
  \bibinfo{title}{{Discovery of an Extremely Short Duration Flare from Proxima
  Centauri Using Millimeter through Far-ultraviolet Observations},} \apjl, 911,
  L25, \dodoi{10.3847/2041-8213/abf14c}

\bibitem[{M. {Maercker} {et~al.}(2012){Maercker}, {Mohamed}, {Vlemmings},
  {Ramstedt}, {Groenewegen}, {Humphreys}, {Kerschbaum}, {Lindqvist},
  {Olofsson}, {Paladini}, {Wittkowski}, {de Gregorio-Monsalvo}, \&
  {Nyman}}]{2012Natur.490..232M}
{Maercker}, M., {Mohamed}, S., {Vlemmings}, W.~H.~T., {et~al.} 2012,
  \bibinfo{title}{{Unexpectedly large mass loss during the thermal pulse cycle
  of the red giant star R Sculptoris},} \nat, 490, 232,
  \dodoi{10.1038/nature11511}

\bibitem[{J. {Magdaleni{\'c}} {et~al.}(2020){Magdaleni{\'c}}, {Marqu{\'e}},
  {Fallows}, {Mann}, {Vocks}, {Zucca}, {Dabrowski}, {Krankowski}, \&
  {Melnik}}]{2020ApJ...897L..15M}
{Magdaleni{\'c}}, J., {Marqu{\'e}}, C., {Fallows}, R.~A., {et~al.} 2020,
  \bibinfo{title}{{Fine Structure of a Solar Type II Radio Burst Observed by
  LOFAR},} \apjl, 897, L15, \dodoi{10.3847/2041-8213/ab9abc}

\bibitem[{K. {Marvel}(2004){Marvel}}]{2004NewAR..48.1349M}
{Marvel}, K. 2004, \bibinfo{title}{{Late stages of stellar evolution},} \nar,
  48, 1349, \dodoi{10.1016/j.newar.2004.09.042}

\bibitem[{L.~D. {Matthews}(2013){Matthews}}]{Matthews2013}
{Matthews}, L.~D. 2013, \bibinfo{title}{{Radio Stars and Their Lives in the
  Galaxy},} \pasp, 125, 313, \dodoi{10.1086/670019}

\bibitem[{L.~D. {Matthews}(2019){Matthews}}]{Matthews2019}
{Matthews}, L.~D. 2019, \bibinfo{title}{{Radio Stars: From kHz to THz},} \pasp,
  131, 016001, \dodoi{10.1088/1538-3873/aae856}

\bibitem[{L.~D. {Matthews} \& M.~J. {Claussen}(2018){Matthews} \&
  {Claussen}}]{2018ASPC..517..281M}
{Matthews}, L.~D., \& {Claussen}, M.~J. 2018, \bibinfo{title}{{Evolved Stars},}
  in Astronomical Society of the Pacific Conference Series, Vol. 517, Science
  with a Next Generation Very Large Array, ed. E.~{Murphy}, 281,
  \dodoi{10.48550/arXiv.1810.06666}

\bibitem[{L.~D. {Matthews} {et~al.}(2024){Matthews}, {Dupree}, \&
  {Akiyama}}]{2024AAS...24340901M}
{Matthews}, L.~D., {Dupree}, A., \& {Akiyama}, K. 2024, \bibinfo{title}{{New
  Spatially Resolved Observations of the Atmosphere of Betelgeuse with the VLA
  and ALMA},} in American Astronomical Society Meeting Abstracts, Vol. 243,
  American Astronomical Society Meeting Abstracts, 409.01

\bibitem[{L.~D. {Matthews} \& A.~K. {Dupree}(2022){Matthews} \&
  {Dupree}}]{2022ApJ...934..131M}
{Matthews}, L.~D., \& {Dupree}, A.~K. 2022, \bibinfo{title}{{Spatially Resolved
  Observations of Betelgeuse at {\ensuremath{\lambda}}7 mm and
  {\ensuremath{\lambda}}1.3 cm Just prior to the Great Dimming},} \apj, 934,
  131, \dodoi{10.3847/1538-4357/ac7726}

\bibitem[{L.~D. {Matthews} {et~al.}(2010){Matthews}, {Greenhill}, {Goddi},
  {Chandler}, {Humphreys}, \& {Kunz}}]{2010ApJ...708...80M}
{Matthews}, L.~D., {Greenhill}, L.~J., {Goddi}, C., {et~al.} 2010,
  \bibinfo{title}{{A Feature Movie of SiO Emission 20-100 AU from the Massive
  Young Stellar Object Orion Source I},} \apj, 708, 80,
  \dodoi{10.1088/0004-637X/708/1/80}

\bibitem[{L.~D. {Matthews} {et~al.}(2018){Matthews}, {Reid}, {Menten}, \&
  {Akiyama}}]{2018AJ....156...15M}
{Matthews}, L.~D., {Reid}, M.~J., {Menten}, K.~M., \& {Akiyama}, K. 2018,
  \bibinfo{title}{{The Evolving Radio Photospheres of Long-period Variable
  Stars},} \aj, 156, 15, \dodoi{10.3847/1538-3881/aac491}

\bibitem[{P.~I. {McCauley} {et~al.}(2019){McCauley}, {Cairns}, {White},
  {Mondal}, {Lenc}, {Morgan}, \& {Oberoi}}]{2019SoPh..294..106M}
{McCauley}, P.~I., {Cairns}, I.~H., {White}, S.~M., {et~al.} 2019,
  \bibinfo{title}{{The Low-Frequency Solar Corona in Circular Polarization},}
  \solphys, 294, 106, \dodoi{10.1007/s11207-019-1502-y}

\bibitem[{D. {McConnell} {et~al.}(2020){McConnell}, {Hale}, {Lenc}, {Banfield},
  {Heald}, {Hotan}, {Leung}, {Moss}, {Murphy}, {O'Brien}, {Pritchard}, {Raja},
  {Sadler}, {Stewart}, {Thomson}, {Whiting}, {Allison}, {Amy}, {Anderson},
  {Ball}, {Bannister}, {Bell}, {Bock}, {Bolton}, {Bunton}, {Chippendale},
  {Collier}, {Cooray}, {Cornwell}, {Diamond}, {Edwards}, {Gupta}, {Hayman},
  {Heywood}, {Jackson}, {Koribalski}, {Lee-Waddell}, {McClure-Griffiths}, {Ng},
  {Norris}, {Phillips}, {Reynolds}, {Roxby}, {Schinckel}, {Shields},
  {Tremblay}, {Tzioumis}, {Voronkov}, \& {Westmeier}}]{2020PASA...37...48M}
{McConnell}, D., {Hale}, C.~L., {Lenc}, E., {et~al.} 2020, \bibinfo{title}{{The
  Rapid ASKAP Continuum Survey I: Design and first results},} \pasa, 37, e048,
  \dodoi{10.1017/pasa.2020.41}

\bibitem[{B. {McGuire}(2019){McGuire}}]{2019BAAS...51c.233M}
{McGuire}, B. 2019, \bibinfo{title}{{Lifting the Veil on Aromatic Chemistry:
  Complex Carbon Across the Stellar Life Cycle from Birth to the Afterlife},}
  \baas, 51, 233

\bibitem[{B.~A. {McGuire}(2022){McGuire}}]{2022ApJS..259...30M}
{McGuire}, B.~A. 2022, \bibinfo{title}{{2021 Census of Interstellar,
  Circumstellar, Extragalactic, Protoplanetary Disk, and Exoplanetary
  Molecules},} \apjs, 259, 30, \dodoi{10.3847/1538-4365/ac2a48}

\bibitem[{B.~A. {McGuire} {et~al.}(2018{\natexlab{a}}){McGuire}, {Carroll}, \&
  {Garrod}}]{2018arXiv181006586M}
{McGuire}, B.~A., {Carroll}, P.~B., \& {Garrod}, R.~T. 2018{\natexlab{a}},
  \bibinfo{title}{{Science with an ngVLA: Prebiotic Molecules},} arXiv
  e-prints, arXiv:1810.06586, \dodoi{10.48550/arXiv.1810.06586}

\bibitem[{B.~A. {McGuire} {et~al.}(2018{\natexlab{b}}){McGuire}, {Bergin},
  {Blake}, {Burkhardt}, {Cleeves}, {Loomis}, {Remijan}, {Shingledecker}, \&
  {Willis}}]{2018arXiv181009550M}
{McGuire}, B.~A., {Bergin}, E., {Blake}, G.~A., {et~al.} 2018{\natexlab{b}},
  \bibinfo{title}{{Science with an ngVLA: Observing the Effects of Chemistry on
  Exoplanets and Planet Formation},} arXiv e-prints, arXiv:1810.09550,
  \dodoi{10.48550/arXiv.1810.09550}

\bibitem[{D.~J. {McLean}(1967){McLean}}]{1967PASA....1...47M}
{McLean}, D.~J. 1967, \bibinfo{title}{{Band Splitting in Type II Solar Radio
  Bursts},} \pasa, 1, 47, \dodoi{10.1017/S1323358000010468}

\bibitem[{C. {Melis}(2019){Melis}}]{NGVLA85}
{Melis}, C. 2019, \bibinfo{title}{{$\mu$as Astrometry with the ngVLA},} ngVLA
  Memo Series, 85

\bibitem[{C. {Mercier} \& G. {Trottet}(1997){Mercier} \&
  {Trottet}}]{1997ApJ...474L..65M}
{Mercier}, C., \& {Trottet}, G. 1997, \bibinfo{title}{{Coronal Radio Bursts: A
  Signature of Nanoflares?},} \apjl, 474, L65, \dodoi{10.1086/310422}

\bibitem[{D.~M.~A. {Meyer} {et~al.}(2021){Meyer}, {Vorobyov}, {Elbakyan},
  {Eisl{\"o}ffel}, {Sobolev}, \& {St{\"o}hr}}]{2021MNRAS.500.4448M}
{Meyer}, D.~M.~A., {Vorobyov}, E.~I., {Elbakyan}, V.~G., {et~al.} 2021,
  \bibinfo{title}{{Parameter study for the burst mode of accretion in massive
  star formation},} \mnras, 500, 4448, \dodoi{10.1093/mnras/staa3528}

\bibitem[{D.~M.~A. {Meyer} {et~al.}(2017){Meyer}, {Vorobyov}, {Kuiper}, \&
  {Kley}}]{2017MNRAS.464L..90M}
{Meyer}, D.~M.~A., {Vorobyov}, E.~I., {Kuiper}, R., \& {Kley}, W. 2017,
  \bibinfo{title}{{On the existence of accretion-driven bursts in massive star
  formation},} \mnras, 464, L90, \dodoi{10.1093/mnrasl/slw187}

\bibitem[{S. {Mohamed} \& P. {Podsiadlowski}(2012){Mohamed} \&
  {Podsiadlowski}}]{2012BaltA..21...88M}
{Mohamed}, S., \& {Podsiadlowski}, P. 2012, \bibinfo{title}{{Mass Transfer in
  Mira-type Binaries},} Baltic Astronomy, 21, 88,
  \dodoi{10.1515/astro-2017-0362}

\bibitem[{A. {Mohan} {et~al.}(2019){Mohan}, {Mondal}, {Oberoi}, \&
  {Lonsdale}}]{2019ApJ...875...98M}
{Mohan}, A., {Mondal}, S., {Oberoi}, D., \& {Lonsdale}, C.~J. 2019,
  \bibinfo{title}{{Evidence for Super-Alfv{\'e}nic Oscillations in Solar Type
  III Radio Burst Sources},} \apj, 875, 98, \dodoi{10.3847/1538-4357/ab0ae5}

\bibitem[{I. {Molina} {et~al.}(2024){Molina}, {Chomiuk}, {Linford}, {Aydi},
  {Mioduszewski}, {Mukai}, {Sokolovsky}, {Strader}, {Craig}, {Dong}, {Harris},
  {Nyamai}, {Rupen}, {Sokoloski}, {Walter}, {Weston}, \&
  {Williams}}]{2024MNRAS.534.1227M}
{Molina}, I., {Chomiuk}, L., {Linford}, J.~D., {et~al.} 2024,
  \bibinfo{title}{{The symbiotic recurrent nova V745 Sco at radio
  wavelengths},} \mnras, 534, 1227, \dodoi{10.1093/mnras/stae2093}

\bibitem[{S. {Mondal}(2021){Mondal}}]{2021SoPh..296..131M}
{Mondal}, S. 2021, \bibinfo{title}{{A Search for the Counterparts of Quiet-Sun
  Radio Transients in Extreme Ultraviolet Data},} \solphys, 296, 131,
  \dodoi{10.1007/s11207-021-01877-3}

\bibitem[{S. {Mondal} {et~al.}(2021){Mondal}, {Biswas}, {Oberoi}, \&
  {Kansabanik}}]{2021AGUFMSH15E2059M}
{Mondal}, S., {Biswas}, A., {Oberoi}, D., \& {Kansabanik}, D. 2021,
  \bibinfo{title}{{Characterising the Properties of Weak Impulsive Narrowband
  Quiet Sun Emissions and Their Relationship with EUV Bursts},} in AGU Fall
  Meeting Abstracts, Vol. 2021, SH15E--2059

\bibitem[{S. {Mondal} {et~al.}(2022){Mondal}, {Chen}, \&
  {Yu}}]{2022tess.conf40808M}
{Mondal}, S., {Chen}, B., \& {Yu}, S. 2022, \bibinfo{title}{{Studying the quiet
  sun radio transients using the Frequency Agile Solar Radiotelescope},} in
  Third Triennial Earth-Sun Summit (TESS), Vol.~54, 2022n7i408p08

\bibitem[{S. {Mondal} {et~al.}(2023{\natexlab{a}}){Mondal}, {Chen}, \&
  {Yu}}]{Mondal2023a}
{Mondal}, S., {Chen}, B., \& {Yu}, S. 2023{\natexlab{a}},
  \bibinfo{title}{{Multifrequency Microwave Imaging of Weak Transients from the
  Quiet Solar Corona},} \apj, 949, 56, \dodoi{10.3847/1538-4357/acc838}

\bibitem[{S. {Mondal} {et~al.}(2019){Mondal}, {Mohan}, {Oberoi}, {Morgan},
  {Benkevitch}, {Lonsdale}, {Crowley}, \& {Cairns}}]{2019ApJ...875...97M}
{Mondal}, S., {Mohan}, A., {Oberoi}, D., {et~al.} 2019,
  \bibinfo{title}{{Unsupervised Generation of High Dynamic Range Solar Images:
  A Novel Algorithm for Self-calibration of Interferometry Data},} \apj, 875,
  97, \dodoi{10.3847/1538-4357/ab0a01}

\bibitem[{S. {Mondal} {et~al.}(2020){Mondal}, {Oberoi}, \&
  {Vourlidas}}]{2020ApJ...893...28M}
{Mondal}, S., {Oberoi}, D., \& {Vourlidas}, A. 2020,
  \bibinfo{title}{{Estimation of the Physical Parameters of a CME at High
  Coronal Heights Using Low-frequency Radio Observations},} \apj, 893, 28,
  \dodoi{10.3847/1538-4357/ab7fab}

\bibitem[{S. {Mondal} {et~al.}(2023{\natexlab{b}}){Mondal}, {Chen}, {Gary},
  {Hallinan}, {Anderson}, {Davis}, {O'Donnell}, {Chhabra}, {Law}, {Huang}, \&
  {OVRO-LWA}}]{Mondal2023b}
{Mondal}, S., {Chen}, B., {Gary}, D., {et~al.} 2023{\natexlab{b}},
  \bibinfo{title}{{Measuring the density and temperature of the corona using
  high dynamic range low frequency images from the OVRO-LWA},} in AAS/Solar
  Physics Division Meeting, Vol.~55, 54th Meeting of the Solar Physics
  Division, 501.01

\bibitem[{D.~E. {Morosan} {et~al.}(2022){Morosan}, {R{\"a}s{\"a}nen}, {Kumari},
  {Kilpua}, {Bisi}, {Dabrowski}, {Krankowski}, {Magdaleni{\'c}}, {Mann},
  {Rothkaehl}, {Vocks}, \& {Zucca}}]{2022SoPh..297...47M}
{Morosan}, D.~E., {R{\"a}s{\"a}nen}, J.~E., {Kumari}, A., {et~al.} 2022,
  \bibinfo{title}{{Exploring the Circular Polarisation of Low-Frequency Solar
  Radio Bursts with LOFAR},} \solphys, 297, 47,
  \dodoi{10.1007/s11207-022-01976-9}

\bibitem[{M. {Morris} {et~al.}(1975){Morris}, {Gilmore}, {Palmer}, {Turner}, \&
  {Zuckerman}}]{1975ApJ...199L..47M}
{Morris}, M., {Gilmore}, W., {Palmer}, P., {Turner}, B.~E., \& {Zuckerman}, B.
  1975, \bibinfo{title}{{Detection of interstellar SiS and a study of the IRC
  +10216 molecular envelope.},} \apjl, 199, L47, \dodoi{10.1086/181846}

\bibitem[{L. {Moscadelli} {et~al.}(2022){Moscadelli}, {Sanna}, {Beuther},
  {Oliva}, \& {Kuiper}}]{2022NatAs...6.1068M}
{Moscadelli}, L., {Sanna}, A., {Beuther}, H., {Oliva}, A., \& {Kuiper}, R.
  2022, \bibinfo{title}{{Snapshot of a magnetohydrodynamic disk wind traced by
  water maser observations},} Nature Astronomy, 6, 1068,
  \dodoi{10.1038/s41550-022-01754-4}

\bibitem[{U. {Munari} {et~al.}(2022){Munari}, {Giroletti}, {Marcote},
  {O'Brien}, {Veres}, {Yang}, {Williams}, \& {Woudt}}]{2022A&A...666L...6M}
{Munari}, U., {Giroletti}, M., {Marcote}, B., {et~al.} 2022,
  \bibinfo{title}{{Radio interferometric imaging of RS Oph bipolar ejecta for
  the 2021 nova outburst},} \aap, 666, L6, \dodoi{10.1051/0004-6361/202244821}

\bibitem[{E. {Murphy}(2018){Murphy}}]{2018ASPC..517.....M}
{Murphy}, E., ed. 2018, Astronomical Society of the Pacific Conference Series,
  Vol. 517, {Science with a Next Generation Very Large Array}

\bibitem[{T. {Murphy} {et~al.}(2021){Murphy}, {Kaplan}, {Stewart}, {O'Brien},
  {Lenc}, {Pintaldi}, {Pritchard}, {Dobie}, {Fox}, {Leung}, {An}, {Bell},
  {Broderick}, {Chatterjee}, {Dai}, {d'Antonio}, {Doyle}, {Gaensler}, {Heald},
  {Horesh}, {Jones}, {McConnell}, {Moss}, {Raja}, {Ramsay}, {Ryder}, {Sadler},
  {Sivakoff}, {Wang}, {Wang}, {Wheatland}, {Whiting}, {Allison}, {Anderson},
  {Ball}, {Bannister}, {Bock}, {Bolton}, {Bunton}, {Chekkala}, {Chippendale},
  {Cooray}, {Gupta}, {Hayman}, {Jeganathan}, {Koribalski}, {Lee-Waddell},
  {Mahony}, {Marvil}, {McClure-Griffiths}, {Mirtschin}, {Ng}, {Pearce},
  {Phillips}, \& {Voronkov}}]{2021PASA...38...54M}
{Murphy}, T., {Kaplan}, D.~L., {Stewart}, A.~J., {et~al.} 2021,
  \bibinfo{title}{{The ASKAP Variables and Slow Transients (VAST) Pilot
  Survey},} \pasa, 38, e054, \dodoi{10.1017/pasa.2021.44}

\bibitem[{R.~L. {Mutel} {et~al.}(1985){Mutel}, {Lestrade}, {Preston}, \&
  {Phillips}}]{1985ApJ...289..262M}
{Mutel}, R.~L., {Lestrade}, J.~F., {Preston}, R.~A., \& {Phillips}, R.~B. 1985,
  \bibinfo{title}{{Dual polarization VLBI observations of stellar binary
  systems at 5 GHz.},} \apj, 289, 262, \dodoi{10.1086/162886}

\bibitem[{R.~L. {Mutel} {et~al.}(1987){Mutel}, {Morris}, {Doiron}, \&
  {Lestrade}}]{1987AJ.....93.1220M}
{Mutel}, R.~L., {Morris}, D.~H., {Doiron}, D.~J., \& {Lestrade}, J.~F. 1987,
  \bibinfo{title}{{Radio Emission from RS CVn Binaries. II. Polarization and
  Spectral Properties},} \aj, 93, 1220, \dodoi{10.1086/114402}

\bibitem[{M.~M. {Nyamai} {et~al.}(2021){Nyamai}, {Chomiuk}, {Ribeiro}, {Woudt},
  {Strader}, \& {Sokolovsky}}]{2021MNRAS.501.1394N}
{Nyamai}, M.~M., {Chomiuk}, L., {Ribeiro}, V.~A.~R.~M., {et~al.} 2021,
  \bibinfo{title}{{Radio light curves and imaging of the helium nova V445
  Puppis reveal seven years of synchrotron emission},} \mnras, 501, 1394,
  \dodoi{10.1093/mnras/staa3712}

\bibitem[{D. {Oberoi} {et~al.}(2023){Oberoi}, {Bisoi}, {Sasikumar Raja},
  {Kansabanik}, {Mohan}, {Mondal}, \& {Sharma}}]{2023JApA...44...40O}
{Oberoi}, D., {Bisoi}, S.~K., {Sasikumar Raja}, K., {et~al.} 2023,
  \bibinfo{title}{{Preparing for solar and heliospheric science with the SKAO:
  An Indian perspective},} Journal of Astrophysics and Astronomy, 44, 40,
  \dodoi{10.1007/s12036-023-09917-z}

\bibitem[{D. {Oberoi} {et~al.}(2011){Oberoi}, {Matthews}, {Cairns}, {Emrich},
  {Lobzin}, {Lonsdale}, {Morgan}, {Prabu}, {Vedantham}, {Wayth}, {Williams},
  {Williams}, {White}, {Allen}, {Arcus}, {Barnes}, {Benkevitch}, {Bernardi},
  {Bowman}, {Briggs}, {Bunton}, {Burns}, {Cappallo}, {Clark}, {Corey},
  {Dawson}, {DeBoer}, {De Gans}, {deSouza}, {Derome}, {Edgar}, {Elton},
  {Goeke}, {Gopalakrishna}, {Greenhill}, {Hazelton}, {Herne}, {Hewitt},
  {Kamini}, {Kaplan}, {Kasper}, {Kennedy}, {Kincaid}, {Kocz}, {Koeing},
  {Kowald}, {Lynch}, {Madhavi}, {McWhirter}, {Mitchell}, {Morales}, {Ng},
  {Ord}, {Pathikulangara}, {Rogers}, {Roshi}, {Salah}, {Sault}, {Schinckel},
  {Udaya Shankar}, {Srivani}, {Stevens}, {Subrahmanyan}, {Thakkar}, {Tingay},
  {Tuthill}, {Vaccarella}, {Waterson}, {Webster}, \&
  {Whitney}}]{2011ApJ...728L..27O}
{Oberoi}, D., {Matthews}, L.~D., {Cairns}, I.~H., {et~al.} 2011,
  \bibinfo{title}{{First Spectroscopic Imaging Observations of the Sun at Low
  Radio Frequencies with the Murchison Widefield Array Prototype},} \apjl, 728,
  L27, \dodoi{10.1088/2041-8205/728/2/L27}

\bibitem[{E. {O'Gorman} {et~al.}(2015){O'Gorman}, {Harper}, {Brown}, {Guinan},
  {Richards}, {Vlemmings}, \& {Wasatonic}}]{2015A&A...580A.101O}
{O'Gorman}, E., {Harper}, G.~M., {Brown}, A., {et~al.} 2015,
  \bibinfo{title}{{Temporal evolution of the size and temperature of
  Betelgeuse's extended atmosphere},} \aap, 580, A101,
  \dodoi{10.1051/0004-6361/201526136}

\bibitem[{E. {O'Gorman} {et~al.}(2017){O'Gorman}, {Kervella}, {Harper},
  {Richards}, {Decin}, {Montarg{\`e}s}, \& {McDonald}}]{2017A&A...602L..10O}
{O'Gorman}, E., {Kervella}, P., {Harper}, G.~M., {et~al.} 2017,
  \bibinfo{title}{{The inhomogeneous submillimeter atmosphere of Betelgeuse},}
  \aap, 602, L10, \dodoi{10.1051/0004-6361/201731171}

\bibitem[{E. {O'Gorman} {et~al.}(2020){O'Gorman}, {Harper}, {Ohnaka},
  {Feeney-Johansson}, {Wilkeneit-Braun}, {Brown}, {Guinan}, {Lim}, {Richards},
  {Ryde}, \& {Vlemmings}}]{2020A&A...638A..65O}
{O'Gorman}, E., {Harper}, G.~M., {Ohnaka}, K., {et~al.} 2020,
  \bibinfo{title}{{ALMA and VLA reveal the lukewarm chromospheres of the nearby
  red supergiants Antares and Betelgeuse},} \aap, 638, A65,
  \dodoi{10.1051/0004-6361/202037756}

\bibitem[{K. {Ohnaka} {et~al.}(2024){Ohnaka}, {Wong}, {Weigelt}, \&
  {Hofmann}}]{2024A&A...691L..14O}
{Ohnaka}, K., {Wong}, K.~T., {Weigelt}, G., \& {Hofmann}, K.~H. 2024,
  \bibinfo{title}{{Contemporaneous high-angular-resolution imaging of the AGB
  star W Hya in vibrationally excited H$_{2}$O lines and visible polarized
  light with ALMA and VLT/SPHERE-ZIMPOL},} \aap, 691, L14,
  \dodoi{10.1051/0004-6361/202451977}

\bibitem[{J. {Ord{\'o}{\~n}ez-Toro} {et~al.}(2024){Ord{\'o}{\~n}ez-Toro},
  {Dzib}, {Loinard}, {Ortiz-Le{\'o}n}, {Kounkel}, {Masqu{\'e}}, {Medina},
  {Galli}, {Dupuy}, {Rodr{\'\i}guez}, \&
  {Quiroga-Nu{\~n}ez}}]{2024AJ....167..108O}
{Ord{\'o}{\~n}ez-Toro}, J., {Dzib}, S.~A., {Loinard}, L., {et~al.} 2024,
  \bibinfo{title}{{Dynamical Mass of the Ophiuchus Intermediate-mass Stellar
  System S1 with DYNAMO-VLBA},} \aj, 167, 108, \dodoi{10.3847/1538-3881/ad1bd3}

\bibitem[{J. {Ord{\'o}{\~n}ez-Toro} {et~al.}(2025){Ord{\'o}{\~n}ez-Toro},
  {Dzib}, {Loinard}, {Ortiz-Le{\'o}n}, {Kounkel}, {Galli}, {Masqu{\'e}},
  {Quiroga-Nu{\~n}ez}, {Srinivasan}, {Medina}, \&
  {Rodr{\'\i}guez}}]{2025MNRAS.540.2830O}
{Ord{\'o}{\~n}ez-Toro}, J., {Dzib}, S.~A., {Loinard}, L., {et~al.} 2025,
  \bibinfo{title}{{Dynamical mass of the Serpens intermediate-mass young
  stellar system EC 95 with DYNAMO-VLBA},} \mnras, 540, 2830,
  \dodoi{10.1093/mnras/staf904}

\bibitem[{S. {Orlando} {et~al.}(2017){Orlando}, {Drake}, \&
  {Miceli}}]{2017MNRAS.464.5003O}
{Orlando}, S., {Drake}, J.~J., \& {Miceli}, M. 2017, \bibinfo{title}{{Origin of
  asymmetries in X-ray emission lines from the blast wave of the 2014 outburst
  of nova V745 Sco},} \mnras, 464, 5003, \dodoi{10.1093/mnras/stw2718}

\bibitem[{E. {Orr{\'u}} {et~al.}(2024){Orr{\'u}}, {Norden}, {Iacobelli}, {ter
  Veen}, \& {Ahmadi}}]{2024SPIE13098E..0RO}
{Orr{\'u}}, E., {Norden}, M.~J., {Iacobelli}, M., {ter Veen}, S., \& {Ahmadi},
  A. 2024, \bibinfo{title}{{LOFAR vs LOFAR2.0 operations: new challenges},} in
  Society of Photo-Optical Instrumentation Engineers (SPIE) Conference Series,
  Vol. 13098, Observatory Operations: Strategies, Processes, and Systems X, ed.
  C.~R. {Benn}, A.~{Chrysostomou}, \& L.~J. {Storrie-Lombardi}, 130980R,
  \dodoi{10.1117/12.3019032}

\bibitem[{K.~N. {Ortiz Ceballos} {et~al.}(2024){Ortiz Ceballos}, {Cendes},
  {Berger}, \& {Williams}}]{2024AJ....168..127O}
{Ortiz Ceballos}, K.~N., {Cendes}, Y., {Berger}, E., \& {Williams}, P. K.~G.
  2024, \bibinfo{title}{{A Volume-limited Radio Search for Magnetic Activity in
  140 Exoplanets with the Very Large Array},} \aj, 168, 127,
  \dodoi{10.3847/1538-3881/ad58be}

\bibitem[{G.~N. {Ortiz-Le{\'o}n} {et~al.}(2017){Ortiz-Le{\'o}n}, {Dzib},
  {Kounkel}, {Loinard}, {Mioduszewski}, {Rodr{\'\i}guez}, {Torres}, {Pech},
  {Rivera}, {Hartmann}, {Boden}, {Evans}, {Brice{\~n}o}, {Tobin}, \&
  {Galli}}]{2017ApJ...834..143O}
{Ortiz-Le{\'o}n}, G.~N., {Dzib}, S.~A., {Kounkel}, M.~A., {et~al.} 2017,
  \bibinfo{title}{{The Gould{\textquoteright}s Belt Distances Survey
  (GOBELINS). III. The Distance to the Serpens/Aquila Molecular Complex},}
  \apj, 834, 143, \dodoi{10.3847/1538-4357/834/2/143}

\bibitem[{X.-J. {Ouyang} {et~al.}(2024){Ouyang}, {Zhang}, {Li}, {Nakashima},
  {Chen}, \& {Qiao}}]{2024ApJ...964L..18O}
{Ouyang}, X.-J., {Zhang}, Y., {Li}, J., {et~al.} 2024,
  \bibinfo{title}{{Excited-state OH Masers in the Water Fountain Source IRAS
  18460-0151},} \apjl, 964, L18, \dodoi{10.3847/2041-8213/ad3338}

\bibitem[{S.~P. {Owocki} {et~al.}(2022){Owocki}, {Shultz}, {ud-Doula},
  {Chandra}, {Das}, \& {Leto}}]{2022MNRAS.513.1449O}
{Owocki}, S.~P., {Shultz}, M.~E., {ud-Doula}, A., {et~al.} 2022,
  \bibinfo{title}{{Centrifugal breakout reconnection as the electron
  acceleration mechanism powering the radio magnetospheres of early-type
  stars},} \mnras, 513, 1449, \dodoi{10.1093/mnras/stac341}

\bibitem[{N. {Panagia} \& M. {Felli}(1975){Panagia} \&
  {Felli}}]{1975A&A....39....1P}
{Panagia}, N., \& {Felli}, M. 1975, \bibinfo{title}{{The spectrum of the
  free-free radiation from extended envelopes.},} \aap, 39, 1

\bibitem[{J.~M. {Paredes}(2005){Paredes}}]{2005EAS....15..187P}
{Paredes}, J.~M. 2005, \bibinfo{title}{{Stellar radio astrophysics},} in EAS
  Publications Series, Vol.~15, EAS Publications Series, ed. L.~I. {Gurvits},
  S.~{Frey}, \& S.~{Rawlings}, 187--206, \dodoi{10.1051/eas:2005153}

\bibitem[{E.~N. {Parker}(1958){Parker}}]{1958ApJ...128..677P}
{Parker}, E.~N. 1958, \bibinfo{title}{{Suprathermal Particle Generation in the
  Solar Corona.},} \apj, 128, 677, \dodoi{10.1086/146580}

\bibitem[{E. {Pattie} \& T. {Maccarone}(2025){Pattie} \&
  {Maccarone}}]{2025AAS...24534607P}
{Pattie}, E., \& {Maccarone}, T. 2025, \bibinfo{title}{{A radio survey in the
  Galactic Bulge and radio jets of neutron star X-ray binaries},} in American
  Astronomical Society Meeting Abstracts, Vol. 245, American Astronomical
  Society Meeting Abstracts, 346.07D

\bibitem[{E.~C. {Pattie} {et~al.}(2024){Pattie}, {Maccarone}, {Tetarenko},
  {Miller-Jones}, {Pichardo Marcano}, \& {Rivera
  Sandoval}}]{2024ApJ...970..126P}
{Pattie}, E.~C., {Maccarone}, T.~J., {Tetarenko}, A.~J., {et~al.} 2024,
  \bibinfo{title}{{Variable Radio Emission of Neutron Star X-Ray Binary Ser
  X{\textendash}1 during Its Persistent Soft State},} \apj, 970, 126,
  \dodoi{10.3847/1538-4357/ad5842}

\bibitem[{S.~T. {Paulson} {et~al.}(2024){Paulson}, {Mallick}, \&
  {Ojha}}]{2024MNRAS.530.1516P}
{Paulson}, S.~T., {Mallick}, K.~K., \& {Ojha}, D.~K. 2024,
  \bibinfo{title}{{Unveiling the Cosmic Cradle: clustering and massive star
  formation in the enigmatic Galactic bubble N59},} \mnras, 530, 1516,
  \dodoi{10.1093/mnras/stae917}

\bibitem[{A.~A. {Penzias} {et~al.}(1971){Penzias}, {Solomon}, {Wilson}, \&
  {Jefferts}}]{1971ApJ...168L..53P}
{Penzias}, A.~A., {Solomon}, P.~M., {Wilson}, R.~W., \& {Jefferts}, K.~B. 1971,
  \bibinfo{title}{{Interstellar Carbon Monosulfide},} \apjl, 168, L53,
  \dodoi{10.1086/180784}

\bibitem[{M. {P{\'e}rez-Torres} {et~al.}(2021){P{\'e}rez-Torres}, {G{\'o}mez},
  {Ortiz}, {Leto}, {Anglada}, {G{\'o}mez}, {Rodr{\'\i}guez}, {Trigilio},
  {Amado}, {Alberdi}, {Anglada-Escud{\'e}}, {Osorio}, {Umana}, {Berdi{\~n}as},
  {L{\'o}pez-Gonz{\'a}lez}, {Morales}, {Rodr{\'\i}guez-L{\'o}pez}, \&
  {Chibueze}}]{2021A&A...645A..77P}
{P{\'e}rez-Torres}, M., {G{\'o}mez}, J.~F., {Ortiz}, J.~L., {et~al.} 2021,
  \bibinfo{title}{{Monitoring the radio emission of Proxima Centauri},} \aap,
  645, A77, \dodoi{10.1051/0004-6361/202039052}

\bibitem[{R.~B. {Phillips} \& J.~F. {Lestrade}(1988){Phillips} \&
  {Lestrade}}]{1988Natur.334..329P}
{Phillips}, R.~B., \& {Lestrade}, J.~F. 1988, \bibinfo{title}{{Compact
  non-thermal radio emission from B-peculiar stars},} \nat, 334, 329,
  \dodoi{10.1038/334329a0}

\bibitem[{B. {Pimpanuwat} {et~al.}(2024){Pimpanuwat}, {Richards}, {Gray},
  {Etoka}, \& {Decin}}]{2024IAUS..380..386P}
{Pimpanuwat}, B., {Richards}, A.~M.~S., {Gray}, M.~D., {Etoka}, S., \& {Decin},
  L. 2024, \bibinfo{title}{{Investigating the inner circumstellar envelopes of
  oxygen-rich evolved stars with ALMA observations of high-J SiO masers},} in
  IAU Symposium, Vol. 380, Cosmic Masers: Proper Motion Toward the
  Next-Generation Large Projects, ed. T.~{Hirota}, H.~{Imai}, K.~{Menten}, \&
  Y.~{Pihlstr{\"o}m}, 386--388, \dodoi{10.1017/S1743921323002235}

\bibitem[{J.~S. {Pineda} {et~al.}(2017){Pineda}, {Hallinan}, \&
  {Kao}}]{2017ApJ...846...75P}
{Pineda}, J.~S., {Hallinan}, G., \& {Kao}, M.~M. 2017, \bibinfo{title}{{A
  Panchromatic View of Brown Dwarf Aurorae},} \apj, 846, 75,
  \dodoi{10.3847/1538-4357/aa8596}

\bibitem[{J.~S. {Pineda} \& J. {Villadsen}(2023){Pineda} \&
  {Villadsen}}]{2023NatAs...7..569P}
{Pineda}, J.~S., \& {Villadsen}, J. 2023, \bibinfo{title}{{Coherent radio
  bursts from known M-dwarf planet-host YZ Ceti},} Nature Astronomy, 7, 569,
  \dodoi{10.1038/s41550-023-01914-0}

\bibitem[{J.~S. {Pineda} {et~al.}(2024){Pineda}, {Villadsen}, {Vidotto}, \&
  {Bellotti}}]{2024ESS.....561713P}
{Pineda}, J.~S., {Villadsen}, J., {Vidotto}, A., \& {Bellotti}, S. 2024,
  \bibinfo{title}{{Updates on YZ Ceti: Using Star-planet Interactions to
  Determine Planet Magnetic Fields},} in AAS/Division for Extreme Solar Systems
  Abstracts, Vol.~56, AASTCS10, Extreme Solar Systems V, 617.13

\bibitem[{R.~S. {Polidan} {et~al.}(2024){Polidan}, {Burns}, {Ignatiev},
  {Hegedus}, {Pober}, {Mahesh}, {Chang}, {Hallinan}, {Ning}, \&
  {Bowman}}]{2024AdSpR..74..528P}
{Polidan}, R.~S., {Burns}, J.~O., {Ignatiev}, A., {et~al.} 2024,
  \bibinfo{title}{{FarView: An in-situ manufactured lunar far side radio array
  concept for 21-cm Dark Ages cosmology},} Advances in Space Research, 74, 528,
  \dodoi{10.1016/j.asr.2024.04.008}

\bibitem[{E. {Polisensky} {et~al.}(2023){Polisensky}, {Das}, {Peters},
  {Shultz}, {Semenko}, \& {Clarke}}]{2023ApJ...958..152P}
{Polisensky}, E., {Das}, B., {Peters}, W., {et~al.} 2023,
  \bibinfo{title}{{Unstable Phenomena in Stable Magnetospheres: Searching for
  Radio Flares from Magnetic OBA Stars Using VCSS},} \apj, 958, 152,
  \dodoi{10.3847/1538-4357/ad0295}

\bibitem[{J. {Pritchard} {et~al.}(2024){Pritchard}, {Murphy}, {Heald},
  {Wheatland}, {Kaplan}, {Lenc}, {O'Brien}, \& {Wang}}]{2024MNRAS.529.1258P}
{Pritchard}, J., {Murphy}, T., {Heald}, G., {et~al.} 2024,
  \bibinfo{title}{{Multi-epoch sampling of the radio star population with the
  Australian SKA Pathfinder},} \mnras, 529, 1258, \dodoi{10.1093/mnras/stae127}

\bibitem[{J. {Pritchard} {et~al.}(2021){Pritchard}, {Murphy}, {Zic}, {Lynch},
  {Heald}, {Kaplan}, {Anderson}, {Banfield}, {Hale}, {Hotan}, {Lenc}, {Leung},
  {McConnell}, {Moss}, {Raja}, {Stewart}, \& {Whiting}}]{2021MNRAS.502.5438P}
{Pritchard}, J., {Murphy}, T., {Zic}, A., {et~al.} 2021, \bibinfo{title}{{A
  circular polarization survey for radio stars with the Australian SKA
  Pathfinder},} \mnras, 502, 5438, \dodoi{10.1093/mnras/stab299}

\bibitem[{L.~H. {Quiroga-Nunez} {et~al.}(2022){Quiroga-Nunez}, {Pihlstrom},
  {Sjouwerman}, {Van Langevelde}, \& {Brown}}]{2022AAS...24021605Q}
{Quiroga-Nunez}, L.~H., {Pihlstrom}, Y., {Sjouwerman}, L., {Van Langevelde},
  H., \& {Brown}, A. 2022, \bibinfo{title}{{Characterizing evolved Galactic
  stellar population: A challenge for astrometric measurements},} in American
  Astronomical Society Meeting Abstracts, Vol. 240, American Astronomical
  Society Meeting \#240, 216.05

\bibitem[{M.~M. {Rahman} {et~al.}(2019){Rahman}, {McCauley}, \&
  {Cairns}}]{2019SoPh..294....7R}
{Rahman}, M.~M., {McCauley}, P.~I., \& {Cairns}, I.~H. 2019,
  \bibinfo{title}{{On the Relative Brightness of Coronal Holes at Low
  Frequencies},} \solphys, 294, 7, \dodoi{10.1007/s11207-019-1396-8}

\bibitem[{R. {Ramesh} {et~al.}(2013){Ramesh}, {Sasikumar Raja}, {Kathiravan},
  \& {Narayanan}}]{2013ApJ...762...89R}
{Ramesh}, R., {Sasikumar Raja}, K., {Kathiravan}, C., \& {Narayanan}, A.~S.
  2013, \bibinfo{title}{{Low-frequency Radio Observations of Picoflare Category
  Energy Releases in the Solar Atmosphere},} \apj, 762, 89,
  \dodoi{10.1088/0004-637X/762/2/89}

\bibitem[{S. {Ramstedt} {et~al.}(2018){Ramstedt}, {Mohamed}, {Olander},
  {Vlemmings}, {Khouri}, \& {Liljegren}}]{2018A&A...616A..61R}
{Ramstedt}, S., {Mohamed}, S., {Olander}, T., {et~al.} 2018,
  \bibinfo{title}{{CO envelope of the symbiotic star R Aquarii seen by ALMA},}
  \aap, 616, A61, \dodoi{10.1051/0004-6361/201833394}

\bibitem[{S. {Ramstedt} {et~al.}(2014){Ramstedt}, {Mohamed}, {Vlemmings},
  {Maercker}, {Montez}, {Baudry}, {De Beck}, {Lindqvist}, {Olofsson},
  {Humphreys}, {Jorissen}, {Kerschbaum}, {Mayer}, {Wittkowski}, {Cox},
  {Lagadec}, {Leal-Ferreira}, {Paladini}, {P{\'e}rez-S{\'a}nchez}, \&
  {Sacuto}}]{2014A&A...570L..14R}
{Ramstedt}, S., {Mohamed}, S., {Vlemmings}, W.~H.~T., {et~al.} 2014,
  \bibinfo{title}{{The wonderful complexity of the Mira AB system},} \aap, 570,
  L14, \dodoi{10.1051/0004-6361/201425029}

\bibitem[{S. {Ramstedt} {et~al.}(2017){Ramstedt}, {Mohamed}, {Vlemmings},
  {Danilovich}, {Brunner}, {De Beck}, {Humphreys}, {Lindqvist}, {Maercker},
  {Olofsson}, {Kerschbaum}, \& {Quintana-Lacaci}}]{2017A&A...605A.126R}
{Ramstedt}, S., {Mohamed}, S., {Vlemmings}, W.~H.~T., {et~al.} 2017,
  \bibinfo{title}{{The circumstellar envelope around the S-type AGB star W Aql.
  Effects of an eccentric binary orbit},} \aap, 605, A126,
  \dodoi{10.1051/0004-6361/201730934}

\bibitem[{M.~J. {Reid} \& M. {Honma}(2014){Reid} \&
  {Honma}}]{2014ARA&A..52..339R}
{Reid}, M.~J., \& {Honma}, M. 2014, \bibinfo{title}{{Microarcsecond Radio
  Astrometry},} \araa, 52, 339, \dodoi{10.1146/annurev-astro-081913-040006}

\bibitem[{M.~J. {Reid} \& K.~M. {Menten}(1997){Reid} \&
  {Menten}}]{1997ApJ...476..327R}
{Reid}, M.~J., \& {Menten}, K.~M. 1997, \bibinfo{title}{{Radio Photospheres of
  Long-Period Variable Stars},} \apj, 476, 327, \dodoi{10.1086/303614}

\bibitem[{M.~J. {Reid} \& K.~M. {Menten}(2007){Reid} \&
  {Menten}}]{2007ApJ...671.2068R}
{Reid}, M.~J., \& {Menten}, K.~M. 2007, \bibinfo{title}{{Imaging the Radio
  Photospheres of Mira Variables},} \apj, 671, 2068, \dodoi{10.1086/523085}

\bibitem[{A.~M.~S. {Richards} {et~al.}(2024){Richards}, {Asaki}, {Baudry},
  {Brand}, {Decin}, {Etoka}, {Gray}, {Herpin}, {Humphreys}, {Pimpanuwat},
  {Singh}, {Yates}, \& {Ziurys}}]{2024IAUS..380..389R}
{Richards}, A.~M.~S., {Asaki}, Y., {Baudry}, A., {et~al.} 2024,
  \bibinfo{title}{{Water masers high resolution measurements of the diverse
  conditions in evolved star winds},} in IAU Symposium, Vol. 380, Cosmic
  Masers: Proper Motion Toward the Next-Generation Large Projects, ed.
  T.~{Hirota}, H.~{Imai}, K.~{Menten}, \& Y.~{Pihlstr{\"o}m}, 389--391,
  \dodoi{10.1017/S1743921323003095}

\bibitem[{S.~T. {Ridgway} {et~al.}(1976){Ridgway}, {Hall}, {Wojslaw},
  {Kleinmann}, \& {Weinberger}}]{1976Natur.264..345R}
{Ridgway}, S.~T., {Hall}, D.~N.~B., {Wojslaw}, R.~S., {Kleinmann}, S.~G., \&
  {Weinberger}, D.~A. 1976, \bibinfo{title}{{Circumstellar acetylene in the
  infrared spectrum of IRC +10216.},} \nat, 264, 345, \dodoi{10.1038/264345a0}

\bibitem[{M.~J. {Rioja} {et~al.}(2017){Rioja}, {Dodson}, {Orosz}, {Imai}, \&
  {Frey}}]{2017AJ....153..105R}
{Rioja}, M.~J., {Dodson}, R., {Orosz}, G., {Imai}, H., \& {Frey}, S. 2017,
  \bibinfo{title}{{MultiView High Precision VLBI Astrometry at Low
  Frequencies},} \aj, 153, 105, \dodoi{10.3847/1538-3881/153/3/105}

\bibitem[{J. {Robrade} \& J.~H.~M.~M. {Schmitt}(2009){Robrade} \&
  {Schmitt}}]{2009A&A...496..229R}
{Robrade}, J., \& {Schmitt}, J.~H.~M.~M. 2009, \bibinfo{title}{{X-ray emission
  from the M9 dwarf 1RXS J115928.5-524717. Quasi-quiescent coronal activity at
  the end of the main-sequence},} \aap, 496, 229,
  \dodoi{10.1051/0004-6361/200811224}

\bibitem[{L.~F. {Rodriguez} {et~al.}(1982){Rodriguez}, {Canto}, \&
  {Moran}}]{1982ApJ...255..103R}
{Rodriguez}, L.~F., {Canto}, J., \& {Moran}, J.~M. 1982, \bibinfo{title}{{Radio
  sources in NGC 6334.},} \apj, 255, 103, \dodoi{10.1086/159808}

\bibitem[{L.~F. {Rodr{\'\i}guez} {et~al.}(2014){Rodr{\'\i}guez}, {Masqu{\'e}},
  {Dzib}, {Loinard}, \& {Kurtz}}]{2014RMxAA..50....3R}
{Rodr{\'\i}guez}, L.~F., {Masqu{\'e}}, J.~M., {Dzib}, S.~A., {Loinard}, L., \&
  {Kurtz}, S.~E. 2014, \bibinfo{title}{{New Radio Continuum Observations of the
  Compact Source Projected Inside NGC 6334A},} \rmxaa, 50, 3,
  \dodoi{10.48550/arXiv.1309.4764}

\bibitem[{K. {Rose} {et~al.}(2023){Rose}, {Pritchard}, {Murphy}, {Caleb},
  {Dobie}, {Driessen}, {Duchesne}, {Kaplan}, {Lenc}, \&
  {Wang}}]{2023ApJ...951L..43R}
{Rose}, K., {Pritchard}, J., {Murphy}, T., {et~al.} 2023,
  \bibinfo{title}{{Periodic Radio Emission from the T8 Dwarf WISE
  J062309.94-045624.6},} \apjl, 951, L43, \dodoi{10.3847/2041-8213/ace188}

\bibitem[{J. {Saur} {et~al.}(2013){Saur}, {Grambusch}, {Duling}, {Neubauer}, \&
  {Simon}}]{2013A&A...552A.119S}
{Saur}, J., {Grambusch}, T., {Duling}, S., {Neubauer}, F.~M., \& {Simon}, S.
  2013, \bibinfo{title}{{Magnetic energy fluxes in sub-Alfv{\'e}nic planet star
  and moon planet interactions},} \aap, 552, A119,
  \dodoi{10.1051/0004-6361/201118179}

\bibitem[{J. {Schmid-Burgk}(1982){Schmid-Burgk}}]{1982A&A...108..169S}
{Schmid-Burgk}, J. 1982, \bibinfo{title}{{Nonspherical stellar envelopes and
  winds - Effects of structure on radiative fluxes and apparent mass loss
  rates},} \aap, 108, 169

\bibitem[{E.~R. {Seaquist} \& M.~F. {Bode}(2008){Seaquist} \&
  {Bode}}]{2008clno.book..141S}
{Seaquist}, E.~R., \& {Bode}, M.~F. 2008, \bibinfo{title}{{Radio emission from
  novae},} in Classical Novae, ed. M.~F. {Bode} \& A.~{Evans}, Vol.~43,
  141--166, \dodoi{10.1017/CBO9780511536168.009}

\bibitem[{R. {Sharma} \& D. {Oberoi}(2020){Sharma} \&
  {Oberoi}}]{2020ApJ...903..126S}
{Sharma}, R., \& {Oberoi}, D. 2020, \bibinfo{title}{{Propagation Effects in
  Quiet Sun Observations at Meter Wavelengths},} \apj, 903, 126,
  \dodoi{10.3847/1538-4357/abb949}

\bibitem[{R. {Sharma} {et~al.}(2022){Sharma}, {Oberoi}, {Battaglia}, \&
  {Krucker}}]{2022ApJ...937...99S}
{Sharma}, R., {Oberoi}, D., {Battaglia}, M., \& {Krucker}, S. 2022,
  \bibinfo{title}{{Detection of Ubiquitous Weak and Impulsive Nonthermal
  Emissions from the Solar Corona},} \apj, 937, 99,
  \dodoi{10.3847/1538-4357/ac87fc}

\bibitem[{S. {Shiber} {et~al.}(2024){Shiber}, {Chatzopoulos}, {Munson}, \&
  {Frank}}]{2024ApJ...962..168S}
{Shiber}, S., {Chatzopoulos}, E., {Munson}, B., \& {Frank}, J. 2024,
  \bibinfo{title}{{Betelgeuse as a Merger of a Massive Star with a Companion},}
  \apj, 962, 168, \dodoi{10.3847/1538-4357/ad0e0a}

\bibitem[{T.~W. {Shimwell} {et~al.}(2017){Shimwell}, {R{\"o}ttgering}, {Best},
  {Williams}, {Dijkema}, {de Gasperin}, {Hardcastle}, {Heald}, {Hoang},
  {Horneffer}, {Intema}, {Mahony}, {Mandal}, {Mechev}, {Morabito}, {Oonk},
  {Rafferty}, {Retana-Montenegro}, {Sabater}, {Tasse}, {van Weeren},
  {Br{\"u}ggen}, {Brunetti}, {Chy{\.z}y}, {Conway}, {Haverkorn}, {Jackson},
  {Jarvis}, {McKean}, {Miley}, {Morganti}, {White}, {Wise}, {van Bemmel},
  {Beck}, {Brienza}, {Bonafede}, {Calistro Rivera}, {Cassano}, {Clarke},
  {Cseh}, {Deller}, {Drabent}, {van Driel}, {Engels}, {Falcke}, {Ferrari},
  {Fr{\"o}hlich}, {Garrett}, {Harwood}, {Heesen}, {Hoeft}, {Horellou},
  {Israel}, {Kapi{\'n}ska}, {Kunert-Bajraszewska}, {McKay}, {Mohan},
  {Orr{\'u}}, {Pizzo}, {Prandoni}, {Schwarz}, {Shulevski}, {Sipior}, {Smith},
  {Sridhar}, {Steinmetz}, {Stroe}, {Varenius}, {van der Werf}, {Zensus}, \&
  {Zwart}}]{2017A&A...598A.104S}
{Shimwell}, T.~W., {R{\"o}ttgering}, H.~J.~A., {Best}, P.~N., {et~al.} 2017,
  \bibinfo{title}{{The LOFAR Two-metre Sky Survey. I. Survey description and
  preliminary data release},} \aap, 598, A104,
  \dodoi{10.1051/0004-6361/201629313}

\bibitem[{T.~W. {Shimwell} {et~al.}(2019){Shimwell}, {Tasse}, {Hardcastle},
  {Mechev}, {Williams}, {Best}, {R{\"o}ttgering}, {Callingham}, {Dijkema}, {de
  Gasperin}, {Hoang}, {Hugo}, {Mirmont}, {Oonk}, {Prandoni}, {Rafferty},
  {Sabater}, {Smirnov}, {van Weeren}, {White}, {Atemkeng}, {Bester},
  {Bonnassieux}, {Br{\"u}ggen}, {Brunetti}, {Chy{\.z}y}, {Cochrane}, {Conway},
  {Croston}, {Danezi}, {Duncan}, {Haverkorn}, {Heald}, {Iacobelli}, {Intema},
  {Jackson}, {Jamrozy}, {Jarvis}, {Lakhoo}, {Mevius}, {Miley}, {Morabito},
  {Morganti}, {Nisbet}, {Orr{\'u}}, {Perkins}, {Pizzo}, {Schrijvers}, {Smith},
  {Vermeulen}, {Wise}, {Alegre}, {Bacon}, {van Bemmel}, {Beswick}, {Bonafede},
  {Botteon}, {Bourke}, {Brienza}, {Calistro Rivera}, {Cassano}, {Clarke},
  {Conselice}, {Dettmar}, {Drabent}, {Dumba}, {Emig}, {En{\ss}lin}, {Ferrari},
  {Garrett}, {G{\'e}nova-Santos}, {Goyal}, {G{\"u}rkan}, {Hale}, {Harwood},
  {Heesen}, {Hoeft}, {Horellou}, {Jackson}, {Kokotanekov}, {Kondapally},
  {Kunert-Bajraszewska}, {Mahatma}, {Mahony}, {Mandal}, {McKean}, {Merloni},
  {Mingo}, {Miskolczi}, {Mooney}, {Nikiel-Wroczy{\'n}ski}, {O'Sullivan},
  {Quinn}, {Reich}, {Roskowi{\'n}ski}, {Rowlinson}, {Savini}, {Saxena},
  {Schwarz}, {Shulevski}, {Sridhar}, {Stacey}, {Urquhart}, {van der Wiel},
  {Varenius}, {Webster}, \& {Wilber}}]{2019A&A...622A...1S}
{Shimwell}, T.~W., {Tasse}, C., {Hardcastle}, M.~J., {et~al.} 2019,
  \bibinfo{title}{{The LOFAR Two-metre Sky Survey. II. First data release},}
  \aap, 622, A1, \dodoi{10.1051/0004-6361/201833559}

\bibitem[{M.~E. {Shultz} {et~al.}(2022){Shultz}, {Owocki}, {ud-Doula},
  {Biswas}, {Bohlender}, {Chandra}, {Das}, {David-Uraz}, {Khalack},
  {Kochukhov}, {Landstreet}, {Leto}, {Monin}, {Neiner}, {Rivinius}, \&
  {Wade}}]{2022MNRAS.513.1429S}
{Shultz}, M.~E., {Owocki}, S.~P., {ud-Doula}, A., {et~al.} 2022,
  \bibinfo{title}{{MOBSTER - VI. The crucial influence of rotation on the radio
  magnetospheres of hot stars},} \mnras, 513, 1429,
  \dodoi{10.1093/mnras/stac136}

\bibitem[{M.~A. {Siebert} {et~al.}(2022){Siebert}, {Van de Sande}, {Millar}, \&
  {Remijan}}]{2022ApJ...941...90S}
{Siebert}, M.~A., {Van de Sande}, M., {Millar}, T.~J., \& {Remijan}, A.~J.
  2022, \bibinfo{title}{{Investigating Anomalous Photochemistry in the Inner
  Wind of IRC+10216 through Interferometric Observations of HC$_{3}$N},} \apj,
  941, 90, \dodoi{10.3847/1538-4357/ac9e52}

\bibitem[{A.~P. {Singh} {et~al.}(2023){Singh}, {Richards}, {Humphreys},
  {Decin}, \& {Ziurys}}]{2023ApJ...954L...1S}
{Singh}, A.~P., {Richards}, A.~M.~S., {Humphreys}, R.~M., {Decin}, L., \&
  {Ziurys}, L.~M. 2023, \bibinfo{title}{{ALMA Reveals Hidden Morphologies in
  the Molecular Envelope of VY Canis Majoris},} \apjl, 954, L1,
  \dodoi{10.3847/2041-8213/ace7cb}

\bibitem[{L.~O. {Sjouwerman} {et~al.}(2024){Sjouwerman}, {Pihlstr{\"o}m},
  {Lewis}, {Bhattacharya}, {Claussen}, \& {BAaDE
  Collaboration}}]{2024IAUS..380..292S}
{Sjouwerman}, L.~O., {Pihlstr{\"o}m}, Y.~M., {Lewis}, M.~O., {et~al.} 2024,
  \bibinfo{title}{{Masers in evolved stars; the Bulge Asymmetries and Dynamical
  Evolution (BAaDE) Survey},} in IAU Symposium, Vol. 380, Cosmic Masers: Proper
  Motion Toward the Next-Generation Large Projects, ed. T.~{Hirota}, H.~{Imai},
  K.~{Menten}, \& Y.~{Pihlstr{\"o}m}, 292--299,
  \dodoi{10.1017/S1743921323002958}

\bibitem[{N. {Smith} {et~al.}(2007){Smith}, {Li}, {Foley}, {Wheeler}, {Pooley},
  {Chornock}, {Filippenko}, {Silverman}, {Quimby}, {Bloom}, \&
  {Hansen}}]{2007ApJ...666.1116S}
{Smith}, N., {Li}, W., {Foley}, R.~J., {et~al.} 2007, \bibinfo{title}{{SN
  2006gy: Discovery of the Most Luminous Supernova Ever Recorded, Powered by
  the Death of an Extremely Massive Star like {\ensuremath{\eta}} Carinae},}
  \apj, 666, 1116, \dodoi{10.1086/519949}

\bibitem[{K.~V. {Sokolovsky} {et~al.}(2023){Sokolovsky}, {Johnson}, {Buson},
  {Jean}, {Cheung}, {Mukai}, {Chomiuk}, {Aydi}, {Molina}, {Kawash}, {Linford},
  {Mioduszewski}, {Rupen}, {Sokoloski}, {Williams}, {Steinberg}, {Vurm},
  {Metzger}, {Page}, {Orio}, {Quimby}, {Shafter}, {Corbett}, {Bolzoni},
  {DeYoung}, {Menzies}, {Romanov}, {Richmond}, {Ulowetz}, {Vanmunster},
  {Williamson}, {Lane}, {Bartnik}, {Bellaver}, {Bruinsma}, {Dugan}, {Fedewa},
  {Gerhard}, {Painter}, {Peterson}, {Rodriguez}, {Smith}, {Sullivan}, \&
  {Watson}}]{2023MNRAS.521.5453S}
{Sokolovsky}, K.~V., {Johnson}, T.~J., {Buson}, S., {et~al.} 2023,
  \bibinfo{title}{{The multiwavelength view of shocks in the fastest nova V1674
  Her},} \mnras, 521, 5453, \dodoi{10.1093/mnras/stad887}

\bibitem[{J.~W. {Stock} {et~al.}(2022){Stock}, {Kitzmann}, \&
  {Patzer}}]{2022MNRAS.517.4070S}
{Stock}, J.~W., {Kitzmann}, D., \& {Patzer}, A. B.~C. 2022,
  \bibinfo{title}{{FASTCHEM 2 : an improved computer program to determine the
  gas-phase chemical equilibrium composition for arbitrary element
  distributions},} \mnras, 517, 4070, \dodoi{10.1093/mnras/stac2623}

\bibitem[{J.~W. {Stock} {et~al.}(2018){Stock}, {Kitzmann}, {Patzer}, \&
  {Sedlmayr}}]{2018MNRAS.479..865S}
{Stock}, J.~W., {Kitzmann}, D., {Patzer}, A. B.~C., \& {Sedlmayr}, E. 2018,
  \bibinfo{title}{{FastChem: A computer program for efficient complex chemical
  equilibrium calculations in the neutral/ionized gas phase with applications
  to stellar and planetary atmospheres},} \mnras, 479, 865,
  \dodoi{10.1093/mnras/sty1531}

\bibitem[{M. {Stoop} {et~al.}(2024){Stoop}, {de Koter}, {Kaper}, {Brands},
  {Portegies Zwart}, {Sana}, {Stoppa}, {Gieles}, {Mahy}, {Shenar}, {Guo},
  {Nelemans}, \& {Rieder}}]{2024Natur.634..809S}
{Stoop}, M., {de Koter}, A., {Kaper}, L., {et~al.} 2024, \bibinfo{title}{{Two
  waves of massive stars running away from the young cluster R136},} \nat, 634,
  809, \dodoi{10.1038/s41586-024-08013-8}

\bibitem[{M.~C. {Stroh} {et~al.}(2021){Stroh}, {Terreran}, {Coppejans},
  {Bright}, {Margutti}, {Bietenholz}, {De Colle}, {DeMarchi}, {Duran},
  {Milisavljevic}, {Murase}, {Paterson}, \& {Williams}}]{2021ApJ...923L..24S}
{Stroh}, M.~C., {Terreran}, G., {Coppejans}, D.~L., {et~al.} 2021,
  \bibinfo{title}{{Luminous Late-time Radio Emission from Supernovae Detected
  by the Karl G. Jansky Very Large Array Sky Survey (VLASS)},} \apjl, 923, L24,
  \dodoi{10.3847/2041-8213/ac375e}

\bibitem[{X. {Sun} {et~al.}(2022){Sun}, {T{\"o}r{\"o}k}, \&
  {DeRosa}}]{2022MNRAS.509.5075S}
{Sun}, X., {T{\"o}r{\"o}k}, T., \& {DeRosa}, M.~L. 2022,
  \bibinfo{title}{{Torus-stable zone above starspots},} \mnras, 509, 5075,
  \dodoi{10.1093/mnras/stab3249}

\bibitem[{D.~J. {Sundkvist} {et~al.}(2016){Sundkvist}, {Saint-Hilaire}, {Bain},
  {Bale}, {Bonnell}, {Hurford}, {Maruca}, {Martinez Oliveros}, \&
  {Pulupa}}]{2016AGUFMSH11C2271S}
{Sundkvist}, D.~J., {Saint-Hilaire}, P., {Bain}, H.~M., {et~al.} 2016,
  \bibinfo{title}{{CURIE: Cubesat Radio Interferometry Experiment},} in AGU
  Fall Meeting Abstracts, SH11C--2271

\bibitem[{C. {Tandoi} {et~al.}(2024){Tandoi}, {Guns}, {Foster}, {Ade},
  {Anderson}, {Ansarinejad}, {Archipley}, {Balkenhol}, {Benabed}, {Bender},
  {Benson}, {Bianchini}, {Bleem}, {Bouchet}, {Bryant}, {Camphuis}, {Carlstrom},
  {Cecil}, {Chang}, {Chaubal}, {Chichura}, {Chou}, {Coerver}, {Crawford},
  {Cukierman}, {Daley}, {de Haan}, {Dibert}, {Dobbs}, {Doussot}, {Dutcher},
  {Everett}, {Feng}, {Ferguson}, {Fichman}, {Galli}, {Gambrel}, {Gardner},
  {Ge}, {Goeckner-Wald}, {Gualtieri}, {Guidi}, {Halverson}, {Hivon}, {Holder},
  {Holzapfel}, {Hood}, {Huang}, {K{\'e}ruzor{\'e}}, {Knox}, {Korman},
  {Kornoelje}, {Kuo}, {Lee}, {Levy}, {Lowitz}, {Lu}, {Maniyar}, {Menanteau},
  {Millea}, {Montgomery}, {Moon}, {Nakato}, {Natoli}, {Noble}, {Novosad},
  {Omori}, {Padin}, {Pan}, {Paschos}, {Phadke}, {Prabhu}, {Qu}, {Quan},
  {Rahimi}, {Rahlin}, {Reichardt}, {Reuter}, {Rouble}, {Ruhl}, {Schiappucci},
  {Smecher}, {Sobrin}, {Stark}, {Stephen}, {Suzuki}, {Thompson}, {Thorne},
  {Trendafilova}, {Tucker}, {Umilta}, {Vieira}, {Wan}, {Wang}, {Whitehorn},
  {Wu}, {Yefremenko}, {Young}, \& {Zebrowski}}]{2024ApJ...972....6T}
{Tandoi}, C., {Guns}, S., {Foster}, A., {et~al.} 2024, \bibinfo{title}{{Flaring
  Stars in a Nontargeted Millimeter-wave Survey with SPT-3G},} \apj, 972, 6,
  \dodoi{10.3847/1538-4357/ad58db}

\bibitem[{J. {Tang} {et~al.}(2022){Tang}, {Tsai}, \&
  {Li}}]{2022RAA....22f5013T}
{Tang}, J., {Tsai}, C.-W., \& {Li}, D. 2022, \bibinfo{title}{{The Potential of
  Detecting Radio-flaring Ultracool Dwarfs at L band in the FAST Drift-scan
  Survey},} Research in Astronomy and Astrophysics, 22, 065013,
  \dodoi{10.1088/1674-4527/ac66bd}

\bibitem[{P. {Tiede}(2022){Tiede}}]{2022JOSS....7.4457T}
{Tiede}, P. 2022, \bibinfo{title}{{Comrade: Composable Modeling of Radio
  Emission},} The Journal of Open Source Software, 7, 4457,
  \dodoi{10.21105/joss.04457}

\bibitem[{S.~J. {Tingay} {et~al.}(2013){Tingay}, {Oberoi}, {Cairns}, {Donea},
  {Duffin}, {Arcus}, {Bernardi}, {Bowman}, {Briggs}, {Bunton}, {Cappallo},
  {Corey}, {Deshpande}, {deSouza}, {Emrich}, {Gaensler}, {R}, {Greenhill},
  {Hazelton}, {Herne}, {Hewitt}, {Johnston-Hollitt}, {Kaplan}, {Kasper},
  {Kennewell}, {Kincaid}, {Koenig}, {Kratzenberg}, {Lonsdale}, {Lynch},
  {McWhirter}, {Mitchell}, {Morales}, {Morgan}, {Ord}, {Pathikulangara},
  {Prabu}, {Remillard}, {Rogers}, {Roshi}, {Salah}, {Sault}, {Udaya-Shankar},
  {Srivani}, {Stevens}, {Subrahmanyan}, {Waterson}, {Wayth}, {Webster},
  {Whitney}, {Williams}, {Williams}, \& {Wyithe}}]{2013JPhCS.440a2033T}
{Tingay}, S.~J., {Oberoi}, D., {Cairns}, I., {et~al.} 2013,
  \bibinfo{title}{{The Murchison Widefield Array: solar science with the low
  frequency SKA Precursor},} in Journal of Physics Conference Series, Vol. 440,
  Journal of Physics Conference Series (IOP), 012033,
  \dodoi{10.1088/1742-6596/440/1/012033}

\bibitem[{R.~H.~D. {Townsend} \& S.~P. {Owocki}(2005){Townsend} \&
  {Owocki}}]{2005MNRAS.357..251T}
{Townsend}, R.~H.~D., \& {Owocki}, S.~P. 2005, \bibinfo{title}{{A rigidly
  rotating magnetosphere model for circumstellar emission from magnetic OB
  stars},} \mnras, 357, 251, \dodoi{10.1111/j.1365-2966.2005.08642.x}

\bibitem[{C. {Trigilio} {et~al.}(2000){Trigilio}, {Leto}, {Leone}, {Umana}, \&
  {Buemi}}]{2000A&A...362..281T}
{Trigilio}, C., {Leto}, P., {Leone}, F., {Umana}, G., \& {Buemi}, C. 2000,
  \bibinfo{title}{{Coherent radio emission from the magnetic chemically
  peculiar star CU Virginis},} \aap, 362, 281,
  \dodoi{10.48550/arXiv.astro-ph/0007097}

\bibitem[{C. {Trigilio} {et~al.}(2008){Trigilio}, {Leto}, {Umana}, {Buemi}, \&
  {Leone}}]{2008MNRAS.384.1437T}
{Trigilio}, C., {Leto}, P., {Umana}, G., {Buemi}, C.~S., \& {Leone}, F. 2008,
  \bibinfo{title}{{The radio lighthouse CU Virginis: the spin-down of a single
  main-sequence star},} \mnras, 384, 1437,
  \dodoi{10.1111/j.1365-2966.2007.12749.x}

\bibitem[{C. {Trigilio} {et~al.}(2011){Trigilio}, {Leto}, {Umana}, {Buemi}, \&
  {Leone}}]{2011ApJ...739L..10T}
{Trigilio}, C., {Leto}, P., {Umana}, G., {Buemi}, C.~S., \& {Leone}, F. 2011,
  \bibinfo{title}{{Auroral Radio Emission from Stars: The Case of CU
  Virginis},} \apjl, 739, L10, \dodoi{10.1088/2041-8205/739/1/L10}

\bibitem[{C. {Trigilio} {et~al.}(2023){Trigilio}, {Biswas}, {Leto}, {Umana},
  {Busa}, {Cavallaro}, {Das}, {Chandra}, {Perez-Torres}, {Wade}, {Bordiu},
  {Buemi}, {Bufano}, {Ingallinera}, {Loru}, \& {Riggi}}]{2023arXiv230500809T}
{Trigilio}, C., {Biswas}, A., {Leto}, P., {et~al.} 2023,
  \bibinfo{title}{{Star-Planet Interaction at radio wavelengths in YZ Ceti:
  Inferring planetary magnetic field},} arXiv e-prints, arXiv:2305.00809,
  \dodoi{10.48550/arXiv.2305.00809}

\bibitem[{B.~E. {Turner} \& L.~M. {Ziurys}(1988){Turner} \&
  {Ziurys}}]{1988gera.book..200T}
{Turner}, B.~E., \& {Ziurys}, L.~M. 1988, \bibinfo{title}{{Interstellar
  molecules and astrochemistry.},} in Galactic and Extragalactic Radio
  Astronomy, ed. K.~I. {Kellermann} \& G.~L. {Verschuur}, 200--254

\bibitem[{J.~D. {Turner} {et~al.}(2024){Turner}, {Grie{\ss}meier}, {Zarka},
  {Zhang}, \& {Mauduit}}]{2024A&A...688A..66T}
{Turner}, J.~D., {Grie{\ss}meier}, J.-M., {Zarka}, P., {Zhang}, X., \&
  {Mauduit}, E. 2024, \bibinfo{title}{{Follow-up LOFAR observations of the
  {\ensuremath{\tau}} Bo{\"o}tis exoplanetary system},} \aap, 688, A66,
  \dodoi{10.1051/0004-6361/202450095}

\bibitem[{J.~D. {Turner} {et~al.}(2021){Turner}, {Zarka}, {Grie{\ss}meier},
  {Lazio}, {Cecconi}, {Emilio Enriquez}, {Girard}, {Jayawardhana}, {Lamy},
  {Nichols}, \& {de Pater}}]{2021A&A...645A..59T}
{Turner}, J.~D., {Zarka}, P., {Grie{\ss}meier}, J.-M., {et~al.} 2021,
  \bibinfo{title}{{The search for radio emission from the exoplanetary systems
  55 Cancri, {\ensuremath{\upsilon}} Andromedae, and {\ensuremath{\tau}}
  Bo{\"o}tis using LOFAR beam-formed observations},} \aap, 645, A59,
  \dodoi{10.1051/0004-6361/201937201}

\bibitem[{{\L}. {Tychoniec} {et~al.}(2021){Tychoniec}, {van Dishoeck}, {van't
  Hoff}, {van Gelder}, {Tabone}, {Chen}, {Harsono}, {Hull}, {Hogerheijde},
  {Murillo}, \& {Tobin}}]{2021A&A...655A..65T}
{Tychoniec}, {\L}., {van Dishoeck}, E.~F., {van't Hoff}, M. L.~R., {et~al.}
  2021, \bibinfo{title}{{Which molecule traces what: Chemical diagnostics of
  protostellar sources},} \aap, 655, A65, \dodoi{10.1051/0004-6361/202140692}

\bibitem[{G. {Umana} {et~al.}(2011){Umana}, {Buemi}, {Trigilio}, {Leto},
  {Hora}, \& {Fazio}}]{2011BSRSL..80..335U}
{Umana}, G., {Buemi}, C.~S., {Trigilio}, C., {et~al.} 2011,
  \bibinfo{title}{{The nebulae around LBVs: a multiwavelength approach},}
  Bulletin de la Societe Royale des Sciences de Liege, 80, 335,
  \dodoi{10.48550/arXiv.1011.3730}

\bibitem[{G. {Umana} {et~al.}(2015){Umana}, {Trigilio}, {Cerrigone},
  {Cesaroni}, {Zijlstra}, {Hoare}, {Weis}, {Beasley}, {Bomans}, {Hallinan},
  {Molinari}, {Taylor}, {Testi}, \& {Thompson}}]{2015aska.confE.118U}
{Umana}, G., {Trigilio}, C., {Cerrigone}, L., {et~al.} 2015,
  \bibinfo{title}{{The impact of SKA on Galactic Radioastronomy: continuum
  observations},} in Advancing Astrophysics with the Square Kilometre Array
  (AASKA14), 118, \dodoi{10.22323/1.215.0118}

\bibitem[{J.~S. {Urquhart}(2024){Urquhart}}]{2024IAUS..380..135U}
{Urquhart}, J.~S. 2024, \bibinfo{title}{{Evolutionary Trends in Star
  Formation},} in IAU Symposium, Vol. 380, Cosmic Masers: Proper Motion Toward
  the Next-Generation Large Projects, ed. T.~{Hirota}, H.~{Imai}, K.~{Menten},
  \& Y.~{Pihlstr{\"o}m}, 135--151, \dodoi{10.1017/S1743921323002326}

\bibitem[{L. {Uscanga} {et~al.}(2023){Uscanga}, {Imai}, {G{\'o}mez}, {Tafoya},
  {Orosz}, {McCarthy}, {Hamae}, \& {Amada}}]{2023ApJ...948...17U}
{Uscanga}, L., {Imai}, H., {G{\'o}mez}, J.~F., {et~al.} 2023,
  \bibinfo{title}{{Evolution of the Outflow in the Water Fountain Source IRAS
  18043-2116},} \apj, 948, 17, \dodoi{10.3847/1538-4357/acc06f}

\bibitem[{J. {van den Eijnden} {et~al.}(2024){van den Eijnden}, {Mohamed},
  {Carotenuto}, {Motta}, {Saikia}, \&
  {Williams-Baldwin}}]{2024MNRAS.532.2920V-B}
{van den Eijnden}, J., {Mohamed}, S., {Carotenuto}, F., {et~al.} 2024,
  \bibinfo{title}{{Particle acceleration at the bow shock of runaway star LS
  2355: non-thermal radio emission but no {\ensuremath{\gamma}}-ray
  counterpart},} \mnras, 532, 2920, \dodoi{10.1093/mnras/stae1622}

\bibitem[{J. {Van den Eijnden} {et~al.}(2022){Van den Eijnden}, {Saikia}, \&
  {Mohamed}}]{2022MNRAS.512.5374V}
{Van den Eijnden}, J., {Saikia}, P., \& {Mohamed}, S. 2022,
  \bibinfo{title}{{Radio detections of IR-selected runaway stellar bow
  shocks},} \mnras, 512, 5374, \dodoi{10.1093/mnras/stac823}

\bibitem[{J. {van den Eijnden} {et~al.}(2022){van den Eijnden}, {Heywood},
  {Fender}, {Mohamed}, {Sivakoff}, {Saikia}, {Russell}, {Motta},
  {Miller-Jones}, \& {Woudt}}]{2022MNRAS.510..515V-A}
{van den Eijnden}, J., {Heywood}, I., {Fender}, R., {et~al.} 2022,
  \bibinfo{title}{{MeerKAT discovery of radio emission from the Vela X-1 bow
  shock},} \mnras, 510, 515, \dodoi{10.1093/mnras/stab3395}

\bibitem[{H.~K. {Vedantham}(2020){Vedantham}}]{2020A&A...639L...7V}
{Vedantham}, H.~K. 2020, \bibinfo{title}{{Prospects for radio detection of
  stellar plasma beams},} \aap, 639, L7, \dodoi{10.1051/0004-6361/202038576}

\bibitem[{H.~K. {Vedantham} {et~al.}(2022){Vedantham}, {Callingham},
  {Shimwell}, {Benz}, {Hajduk}, {Ray}, {Tasse}, \&
  {Drabent}}]{2022ApJ...926L..30V}
{Vedantham}, H.~K., {Callingham}, J.~R., {Shimwell}, T.~W., {et~al.} 2022,
  \bibinfo{title}{{Peculiar Radio-X-Ray Relationship in Active Stars},} \apjl,
  926, L30, \dodoi{10.3847/2041-8213/ac5115}

\bibitem[{H.~K. {Vedantham} {et~al.}(2023){Vedantham}, {Dupuy}, {Evans},
  {Sanghi}, {Callingham}, {Shimwell}, {Best}, {Liu}, \&
  {Zarka}}]{2023A&A...675L...6V}
{Vedantham}, H.~K., {Dupuy}, T.~J., {Evans}, E.~L., {et~al.} 2023,
  \bibinfo{title}{{Polarised radio pulsations from a new T-dwarf binary},}
  \aap, 675, L6, \dodoi{10.1051/0004-6361/202244965}

\bibitem[{L. {Velilla-Prieto} {et~al.}(2019){Velilla-Prieto}, {Cernicharo},
  {Ag{\'u}ndez}, {Fonfr{\'\i}a}, {Quintana-Lacaci}, {Marcelino}, \&
  {Castro-Carrizo}}]{2019A&A...629A.146V}
{Velilla-Prieto}, L., {Cernicharo}, J., {Ag{\'u}ndez}, M., {et~al.} 2019,
  \bibinfo{title}{{IRC + 10{\textdegree}216 mass loss properties through the
  study of {\ensuremath{\lambda}}3 mm emission. Large spatial scale
  distribution of SiO, SiS, and CS},} \aap, 629, A146,
  \dodoi{10.1051/0004-6361/201834717}

\bibitem[{L. {Velilla Prieto} {et~al.}(2017){Velilla Prieto}, {S{\'a}nchez
  Contreras}, {Cernicharo}, {Ag{\'u}ndez}, {Quintana-Lacaci}, {Bujarrabal},
  {Alcolea}, {Balan{\c{c}}a}, {Herpin}, {Menten}, \&
  {Wyrowski}}]{2017A&A...597A..25V}
{Velilla Prieto}, L., {S{\'a}nchez Contreras}, C., {Cernicharo}, J., {et~al.}
  2017, \bibinfo{title}{{The millimeter IRAM-30 m line survey toward
  <ASTROBJ>IK Tauri</ASTROBJ>},} \aap, 597, A25,
  \dodoi{10.1051/0004-6361/201628776}

\bibitem[{J. {Villadsen} \& G. {Hallinan}(2019){Villadsen} \&
  {Hallinan}}]{2019ApJ...871..214V}
{Villadsen}, J., \& {Hallinan}, G. 2019, \bibinfo{title}{{Ultra-wideband
  Detection of 22 Coherent Radio Bursts on M Dwarfs},} \apj, 871, 214,
  \dodoi{10.3847/1538-4357/aaf88e}

\bibitem[{J. {Villadsen} {et~al.}(2025){Villadsen}, {Pineda}, {Bellotti}, \&
  {Vidotto}}]{2025AAS...24541807V}
{Villadsen}, J., {Pineda}, J.~S., {Bellotti}, S., \& {Vidotto}, A. 2025,
  \bibinfo{title}{{Testing for Planetary Modulation of YZ Ceti's Radio
  Bursts},} in American Astronomical Society Meeting Abstracts, Vol. 245,
  American Astronomical Society Meeting Abstracts, 418.07

\bibitem[{W.~H.~T. {Vlemmings} {et~al.}(2019){Vlemmings}, {Lankhaar},
  {Cazzoletti}, {Ceccobello}, {Dall'Olio}, {van Dishoeck}, {Facchini},
  {Humphreys}, {Persson}, {Testi}, \& {Williams}}]{2019A&A...624L...7V}
{Vlemmings}, W.~H.~T., {Lankhaar}, B., {Cazzoletti}, P., {et~al.} 2019,
  \bibinfo{title}{{Stringent limits on the magnetic field strength in the disc
  of TW Hya. ALMA observations of CN polarisation},} \aap, 624, L7,
  \dodoi{10.1051/0004-6361/201935459}

\bibitem[{S.~S. {Vogt} {et~al.}(1999){Vogt}, {Hatzes}, {Misch}, \&
  {K{\"u}rster}}]{1999ApJS..121..547V}
{Vogt}, S.~S., {Hatzes}, A.~P., {Misch}, A.~A., \& {K{\"u}rster}, M. 1999,
  \bibinfo{title}{{Doppler Imagery of the Spotted RS Canum Venaticorum Star HR
  1099 (V711 Tauri) from 1981 to 1992},} \apjs, 121, 547,
  \dodoi{10.1086/313195}

\bibitem[{S.~H.~J. {Wallstr{\"o}m} {et~al.}(2024){Wallstr{\"o}m}, {Danilovich},
  {M{\"u}ller}, {Gottlieb}, {Maes}, {Van de Sande}, {Decin}, {Richards},
  {Baudry}, {Bolte}, {Ceulemans}, {De Ceuster}, {de Koter}, {El Mellah},
  {Esseldeurs}, {Etoka}, {Gobrecht}, {Gottlieb}, {Gray}, {Herpin}, {Jeste},
  {Kee}, {Kervella}, {Khouri}, {Lagadec}, {Malfait}, {Marinho}, {McDonald},
  {Menten}, {Millar}, {Montarg{\`e}s}, {Nuth}, {Plane}, {Sahai}, {Waters},
  {Wong}, {Yates}, \& {Zijlstra}}]{2024A&A...681A..50W}
{Wallstr{\"o}m}, S.~H.~J., {Danilovich}, T., {M{\"u}ller}, H.~S.~P., {et~al.}
  2024, \bibinfo{title}{{ATOMIUM: Molecular inventory of 17 oxygen-rich evolved
  stars observed with ALMA},} \aap, 681, A50,
  \dodoi{10.1051/0004-6361/202347632}

\bibitem[{H.~G. {Walter} {et~al.}(1990){Walter}, {Hering}, \& {de
  Vegt}}]{1990A&AS...86..357W}
{Walter}, H.~G., {Hering}, R., \& {de Vegt}, C. 1990, \bibinfo{title}{{An
  Astrometric Catalogue of Radio Stars},} \aaps, 86, 357

\bibitem[{Y. {Wang} {et~al.}(2023){Wang}, {Murphy}, {Lenc}, {Mercorelli},
  {Driessen}, {Pritchard}, {Lao}, {Kaplan}, {An}, {Bannister}, {Heald}, {Lu},
  {Tuntsov}, {Walker}, \& {Zic}}]{2023MNRAS.523.5661W}
{Wang}, Y., {Murphy}, T., {Lenc}, E., {et~al.} 2023, \bibinfo{title}{{Radio
  variable and transient sources on minute time-scales in the ASKAP pilot
  surveys},} \mnras, 523, 5661, \dodoi{10.1093/mnras/stad1727}

\bibitem[{H.~J. {Wendker}(1978){Wendker}}]{1978AAHam..10....3W}
{Wendker}, H.~J. 1978, \bibinfo{title}{{A catalogue of radio stars.},}
  Astronomische Abhandlungen der Hamburger Sternwarte, 10, 3

\bibitem[{H.~J. {Wendker}(1987){Wendker}}]{1987A&AS...69...87W}
{Wendker}, H.~J. 1987, \bibinfo{title}{{A catalogue of stars emitting radio
  continuum.},} \aaps, 69, 87

\bibitem[{H.~J. {Wendker}(1995){Wendker}}]{1995A&AS..109..177W}
{Wendker}, H.~J. 1995, \bibinfo{title}{{Radio continuum emission from stars: a
  catalogue update.},} \aaps, 109, 177

\bibitem[{H.~J. {Wendker}(2015){Wendker}}]{2015yCat.8099....0W}
{Wendker}, H.~J. 2015, {VizieR Online Data Catalog: Catalogue of Radio Stars
  (Wendker, 2001)},, VizieR On-line Data Catalog: VIII/99. Originally published
  in: 1995A\&AS..109..177W

\bibitem[{J.~C. {Wheeler} {et~al.}(2017){Wheeler}, {Nance}, {Diaz}, {Smith},
  {Hickey}, {Zhou}, {Koutoulaki}, {Sullivan}, \&
  {Fowler}}]{2017MNRAS.465.2654W}
{Wheeler}, J.~C., {Nance}, S., {Diaz}, M., {et~al.} 2017, \bibinfo{title}{{The
  Betelgeuse Project: constraints from rotation},} \mnras, 465, 2654,
  \dodoi{10.1093/mnras/stw2893}

\bibitem[{S.~M. {White}(2000){White}}]{2000riss.conf...86W}
{White}, S.~M. 2000, \bibinfo{title}{{The Contributions of the VLA to the Study
  of Radio Stars},} in Radio interferometry : the saga and the science, ed.
  D.~G. {Finley} \& W.~M. {Goss}, 86

\bibitem[{S.~M. {White} {et~al.}(1989){White}, {Jackson}, \&
  {Kundu}}]{1989ApJS...71..895W}
{White}, S.~M., {Jackson}, P.~D., \& {Kundu}, M.~R. 1989, \bibinfo{title}{{A
  VLA Survey of Nearby Flare Stars},} \apjs, 71, 895, \dodoi{10.1086/191401}

\bibitem[{S.~M. {White} {et~al.}(2024){White}, {Shimojo}, {Iwai}, {Bastian},
  {Fleishman}, {Gary}, {Magdalenic}, \& {Vourlidas}}]{2024ApJ...969....3W}
{White}, S.~M., {Shimojo}, M., {Iwai}, K., {et~al.} 2024,
  \bibinfo{title}{{Electron Cyclotron Maser Emission and the Brightest Solar
  Radio Bursts},} \apj, 969, 3, \dodoi{10.3847/1538-4357/ad4640}

\bibitem[{J.~P. {Wild}(1950){Wild}}]{1950AuSRA...3..399W}
{Wild}, J.~P. 1950, \bibinfo{title}{{Observations of the Spectrum of
  High-Intensity Solar Radiation at Metre Wavelengths. II. Outbursts},}
  Australian Journal of Scientific Research A Physical Sciences, 3, 399,
  \dodoi{10.1071/CH9500399}

\bibitem[{M. {Williams} {et~al.}(2023){Williams}, {Linford}, {Sokolovsky},
  {Chomiuk}, {Aydi}, {Sokoloski}, {Mukai}, {Kawash}, {Mioduszewski}, \&
  {Rupen}}]{2023AAS...24210906W}
{Williams}, M., {Linford}, J., {Sokolovsky}, K., {et~al.} 2023,
  \bibinfo{title}{{VLBA Images of the Fastest Nova V1674 Her},} in American
  Astronomical Society Meeting Abstracts, Vol. 242, American Astronomical
  Society Meeting Abstracts \#242, 109.06

\bibitem[{P.~K.~G. {Williams}(2018){Williams}}]{2018haex.bookE.171W}
{Williams}, P. K.~G. 2018, \bibinfo{title}{{Radio Emission from Ultracool
  Dwarfs},} in Handbook of Exoplanets, ed. H.~J. {Deeg} \& J.~A. {Belmonte},
  171, \dodoi{10.1007/978-3-319-55333-7_171}

\bibitem[{P.~K.~G. {Williams} {et~al.}(2013){Williams}, {Berger}, \&
  {Zauderer}}]{2013ApJ...767L..30W}
{Williams}, P. K.~G., {Berger}, E., \& {Zauderer}, B.~A. 2013,
  \bibinfo{title}{{Quasi-quiescent Radio Emission from the First Radio-emitting
  T Dwarf},} \apjl, 767, L30, \dodoi{10.1088/2041-8205/767/2/L30}

\bibitem[{P.~K.~G. {Williams} {et~al.}(2014){Williams}, {Cook}, \&
  {Berger}}]{2014ApJ...785....9W}
{Williams}, P.~K.~G., {Cook}, B.~A., \& {Berger}, E. 2014,
  \bibinfo{title}{{Trends in Ultracool Dwarf Magnetism. I. X-Ray Suppression
  and Radio Enhancement},} \apj, 785, 9, \dodoi{10.1088/0004-637X/785/1/9}

\bibitem[{P.~K.~G. {Williams} {et~al.}(2017){Williams}, {Gizis}, \&
  {Berger}}]{2017ApJ...834..117W}
{Williams}, P.~K.~G., {Gizis}, J.~E., \& {Berger}, E. 2017,
  \bibinfo{title}{{Variable and Polarized Radio Emission from the T6 Brown
  Dwarf WISEP J112254.73+255021.5},} \apj, 834, 117,
  \dodoi{10.3847/1538-4357/834/2/117}

\bibitem[{B.~E. {Wood} {et~al.}(2002){Wood}, {M{\"u}ller}, {Zank}, \&
  {Linsky}}]{2002ApJ...574..412W}
{Wood}, B.~E., {M{\"u}ller}, H.-R., {Zank}, G.~P., \& {Linsky}, J.~L. 2002,
  \bibinfo{title}{{Measured Mass-Loss Rates of Solar-like Stars as a Function
  of Age and Activity},} \apj, 574, 412, \dodoi{10.1086/340797}

\bibitem[{B.~E. {Wood} {et~al.}(2021){Wood}, {M{\"u}ller}, {Redfield}, {Konow},
  {Vannier}, {Linsky}, {Youngblood}, {Vidotto}, {Jardine},
  {Alvarado-G{\'o}mez}, \& {Drake}}]{2021ApJ...915...37W}
{Wood}, B.~E., {M{\"u}ller}, H.-R., {Redfield}, S., {et~al.} 2021,
  \bibinfo{title}{{New Observational Constraints on the Winds of M dwarf
  Stars},} \apj, 915, 37, \dodoi{10.3847/1538-4357/abfda5}

\bibitem[{A.~E. {Wright} \& M.~J. {Barlow}(1975){Wright} \&
  {Barlow}}]{1975MNRAS.170...41W}
{Wright}, A.~E., \& {Barlow}, M.~J. 1975, \bibinfo{title}{{The radio and
  infrared spectrum of early type stars undergoing mass loss.},} \mnras, 170,
  41, \dodoi{10.1093/mnras/170.1.41}

\bibitem[{T.-Y. {Xia} {et~al.}(2024){Xia}, {Shen}, {Li}, {Feng}, {Sjouwerman},
  {Pihlstr{\"o}m}, {Lewis}, \& {Stroh}}]{2024ApJ...976..139X}
{Xia}, T.-Y., {Shen}, J., {Li}, Z., {et~al.} 2024, \bibinfo{title}{{The Milky
  Way Bar Potential Constrained by the Kinematics of SiO Maser Stars in BAaDE
  Survey},} \apj, 976, 139, \dodoi{10.3847/1538-4357/ad834f}

\bibitem[{S. {Xu} {et~al.}(2022){Xu}, {Imai}, {Yun}, {Zhang}, {Rioja},
  {Dodson}, {Cho}, {Kim}, {Cui}, {Sobolev}, {Chibueze}, {Kim}, {Amada},
  {Nakashima}, {Orosz}, {Oyadomari}, {Oh}, {Yonekura}, {Sun}, {Mai}, {Zhang},
  {Wen}, \& {Jung}}]{2022ApJ...941..105X}
{Xu}, S., {Imai}, H., {Yun}, Y., {et~al.} 2022, \bibinfo{title}{{The
  Astrometric Animation of Water Masers toward the Mira Variable BX Cam},}
  \apj, 941, 105, \dodoi{10.3847/1538-4357/ac9599}

\bibitem[{R.~K. {Yadav} \& D.~P. {Thorngren}(2017){Yadav} \&
  {Thorngren}}]{2017ApJ...849L..12Y}
{Yadav}, R.~K., \& {Thorngren}, D.~P. 2017, \bibinfo{title}{{Estimating the
  Magnetic Field Strength in Hot Jupiters},} \apjl, 849, L12,
  \dodoi{10.3847/2041-8213/aa93fd}

\bibitem[{A.~Y. {Yang} {et~al.}(2023){Yang}, {Dzib}, {Urquhart}, {Brunthaler},
  {Medina}, {Menten}, {Wyrowski}, {Ortiz-Le{\'o}n}, {Cotton}, {Gong}, {Dokara},
  {Rugel}, {Beuther}, {Pandian}, {Csengeri}, {Veena}, {Roy}, {Nguyen},
  {Winkel}, {Ott}, {Carrasco-Gonzalez}, {Khan}, \&
  {Cheema}}]{2023A&A...680A..92Y}
{Yang}, A.~Y., {Dzib}, S.~A., {Urquhart}, J.~S., {et~al.} 2023,
  \bibinfo{title}{{A global view on star formation: The GLOSTAR Galactic plane
  survey. IX. Radio Source Catalog III: 2{\textdegree} < {\ensuremath{\ell}} <
  28{\textdegree}, 36{\textdegree} < {\ensuremath{\ell}} < 40{\textdegree},
  56{\textdegree} < {\ensuremath{\ell}} < 60{\textdegree} and |b| <
  1{\textdegree}, VLA B-configuration},} \aap, 680, A92,
  \dodoi{10.1051/0004-6361/202347563}

\bibitem[{V. {Yanza} {et~al.}(2025){Yanza}, {Dzib}, {Palau}, {Rodr{\'\i}guez},
  {Masqu{\'e}}, {Rivera-Ortiz}, \& {Medina}}]{2025MNRAS.538.1314Y}
{Yanza}, V., {Dzib}, S.~A., {Palau}, A., {et~al.} 2025, \bibinfo{title}{{The
  arc-shaped radio source at the centre of NGC 6334A: is it a colliding wind
  region of two young massive stars or the bow shock of a runaway star?},}
  \mnras, 538, 1314, \dodoi{10.1093/mnras/staf344}

\bibitem[{L.~N. {Zack} {et~al.}(2011){Zack}, {Halfen}, \&
  {Ziurys}}]{2011ApJ...733L..36Z}
{Zack}, L.~N., {Halfen}, D.~T., \& {Ziurys}, L.~M. 2011,
  \bibinfo{title}{{Detection of FeCN (x $^{4}${\ensuremath{\Delta}} $_{i}$ ) in
  IRC+10216: A New Interstellar Molecule},} \apjl, 733, L36,
  \dodoi{10.1088/2041-8205/733/2/L36}

\bibitem[{P. {Zarka}(1992){Zarka}}]{1992AdSpR..12h..99Z}
{Zarka}, P. 1992, \bibinfo{title}{{The auroral radio emissions from planetary
  magnetospheres: What do we know, what don't we know, what do we learn from
  them?},} Advances in Space Research, 12, 99,
  \dodoi{10.1016/0273-1177(92)90383-9}

\bibitem[{P. {Zarka}(1998){Zarka}}]{1998JGR...10320159Z}
{Zarka}, P. 1998, \bibinfo{title}{{Auroral radio emissions at the outer
  planets: Observations and theories},} \jgr, 103, 20159,
  \dodoi{10.1029/98JE01323}

\bibitem[{Q. {Zhang} {et~al.}(2020){Zhang}, {Hallinan}, {Brisken}, {Bourke}, \&
  {Golden}}]{2020ApJ...897...11Z}
{Zhang}, Q., {Hallinan}, G., {Brisken}, W., {Bourke}, S., \& {Golden}, A. 2020,
  \bibinfo{title}{{Multiepoch VLBI of L Dwarf Binary 2MASS J0746+2000AB:
  Precise Mass Measurements and Confirmation of Radio Emission from Both
  Components},} \apj, 897, 11, \dodoi{10.3847/1538-4357/ab9177}

\bibitem[{A. {Zijlstra} {et~al.}(1989){Zijlstra}, {Pottasch}, {te Lintel
  Hekkert}, \& {Bignell}}]{1989IAUS..131..210Z}
{Zijlstra}, A., {Pottasch}, S.~R., {te Lintel Hekkert}, P., \& {Bignell}, C.
  1989, \bibinfo{title}{{OH maser emission from young planetary nebulae.},} in
  IAU Symposium, Vol. 131, Planetary Nebulae, ed. S.~{Torres-Peimbert}, 210

\bibitem[{A.~A. {Zijlstra} {et~al.}(1991){Zijlstra}, {Gaylard}, {te Lintel
  Hekkert}, {Menzies}, {Nyman}, \& {Schwarz}}]{1991A&A...243L...9Z}
{Zijlstra}, A.~A., {Gaylard}, M.~J., {te Lintel Hekkert}, P., {et~al.} 1991,
  \bibinfo{title}{{IRAS 07027-7934 : the link between OH/IR stars and
  carbon-rich planetary nebulae.},} \aap, 243, L9

\bibitem[{H. {Zirin} {et~al.}(1991){Zirin}, {Baumert}, \&
  {Hurford}}]{1991ApJ...370..779Z}
{Zirin}, H., {Baumert}, B.~M., \& {Hurford}, G.~J. 1991, \bibinfo{title}{{The
  Microwave Brightness Temperature Spectrum of the Quiet Sun},} \apj, 370, 779,
  \dodoi{10.1086/169861}

\bibitem[{L.~M. {Ziurys} {et~al.}(2007){Ziurys}, {Milam}, {Apponi}, \&
  {Woolf}}]{2007Natur.447.1094Z}
{Ziurys}, L.~M., {Milam}, S.~N., {Apponi}, A.~J., \& {Woolf}, N.~J. 2007,
  \bibinfo{title}{{Chemical complexity in the winds of the oxygen-rich
  supergiant star VY Canis Majoris},} \nat, 447, 1094,
  \dodoi{10.1038/nature05905}

\bibitem[{C. {Zucker} {et~al.}(2022){Zucker}, {Goodman}, {Alves}, {Bialy},
  {Foley}, {Speagle}, {Gro{\^I}{\texttwosuperior}schedl}, {Finkbeiner},
  {Burkert}, {Khimey}, \& {Swiggum}}]{2022Natur.601..334Z}
{Zucker}, C., {Goodman}, A.~A., {Alves}, J., {et~al.} 2022,
  \bibinfo{title}{{Star formation near the Sun is driven by expansion of the
  Local Bubble},} \nat, 601, 334, \dodoi{10.1038/s41586-021-04286-5}

\end{thebibliography}
\bibliographystyle{aasjournalv7}


\end{document}